\documentclass[12pt]{report}

\usepackage{graphicx,amsmath,mathrsfs}
\usepackage{amssymb}
\usepackage{amsthm}
\usepackage{bm}
\usepackage{float}
\usepackage{epsfig}

\usepackage{color}

\usepackage[hardcopy]{Suthesis-2e}
\def\onlinecite{\cite}

\newcommand{\bs}[1]{\boldsymbol{#1}}

\newcommand{\ket}[1]{\left|#1\right\rangle}
\newcommand{\bra}[1]{\left\langle#1\right|}
\newcommand{\braket}[2]{\bigl\langle#1\bigl|\bigr.#2\bigr\rangle}

\def\vk{{\bf k}}
\def\vr{{\bf r}}

\renewcommand{\b}{\beta}

\def\bra#1{\left\langle#1\right|}
\def\ket#1{\left|#1\right\rangle}
\def\braket#1#2{\left\langle #1\right|\left.#2\right\rangle}
\def\vk{{\bf k}}
\def\vr{{\bf r}}

    \title{Exact mappings in condensed matter physics}
    \author{Ching Hua Lee}
    \dept{Physics}
    \principaladviser{Xiao-Liang Qi}
    \firstreader{Steven Kivelson}
    \secondreader{Srinivas Raghu}

\begin{document}

%% the \beforepreface command produces the title page
%% in the online version it skips the copyright (page 2) and signature (page 3) pages 
%% in the non-online version these would be included

%\titlep
\beforepreface

    \prefacesection{Abstract}
        Condensed matter systems are complex yet simple. Amidst their complexity, one often find order specified by not more than a few parameters. Key to such a reductionistic description is an appropriate choice of basis, two of which I shall describe in this thesis. 
        
        The first, an exact mapping known as the Wannier State Representation (WSR), provides an exact Hilbert space correspondence between two intensely-studied topological systems, the Fractional Quantum Hall (FQH) and Fractional Chern Insulator (FCI) systems. FQH states exist within the partially filled Landau levels of interacting 2D electron gases under strong magnetic fields, where quasiparticles exhibit topologically nontrivial braiding statistics. FCI systems, which are novel lattice realizations of FQH systems without orbital magnetic field, are still not completely understood and will benefit from a basis that explicitly connects them to the much better understood FQH systems. 
        
        The second basis mapping, the Exact Holographic Mapping (EHM), maps any lattice system to a holographic 'bulk' with an additional emergent dimension representing energy scale. Devised in the spirit of the highly popular AdS-CFT correspondence, it attempts to understand the relationship between a given system and its equivalent dual geometry. In particular, we found excellent theoretical agreement of certain dual geometries from EHM with those expected from the Ryu-Takanayagi formula relating bulk geometry with entanglement entropy. Additionally, the EHM also proves useful in providing a link between different topological quantities, such as relating the Chern number of the abovementioned Chern Insulators with the Axion angle of 3D topological insulators.

\prefacesection{Acknowledgements}
			
	First and foremost, I will like to thank my advisor Prof. Xiao-Liang Qi with whom I have had extensive stimulating physics discussions. It was a great privilege to be able to learn the ropes of research from one of the most outstanding scholars in the field. %Such were his amazing physical insights that 	many a time he would propose seemingly extemporaneous ideas that, with the hindsight of weeks of my hard work, turned out to be spot on. 
I am deeply appreciative of the all-rounded training that Xiao-Liang provided me through exposure to a variety of fields from Fractional Chern Insulators to Graphene to Holography. And it was through his introduction that I was able to work with many other brilliant physicists around the world, and partake in stimulating discussions in prestigious institutes like the Kavli Institute of Theoretical Physics (KITP).
	
I am also greatly indebted to Ronny Thomale who not only introduced me to the beautiful topic of Fractional Quantum Hall systems, but also guided me through the rigors of scientific writing. Clear and inspirational, he made even the most abstruse topics seem transparent and accessible. I am also very grateful to Ronny for inviting me to extended stays at the University of W\"{u}rzburg, where I had extremely fruitful and inspirational discussions on Quantum Hall physics and Conformal Field Theory with brilliant physicists like Dunghai Lee, Joe Bhaseen, Zlatko Papi\'{c} and Martin Greiter. Being immersed in an institute other than my own also broadened my physics horizons and opened me up to new bright ideas that I never knew existed.

It was indeed a great privilege to grow up as a physicist at Stanford University, where the thriving intellect environment never stops populating one's mind with interesting ideas. It was a place where the intense exchange of ideas occur amidst the carefree, pastoral landscape of the greater campus - affectionately known as ``the Farm''. Over the years, I had accrued much wisdom from discussions with physicists like Steve Kivelson, Sri Raghu, Leonard Susskind, Shamit Kachru, Renata Kallosh, Shoucheng Zhang and of course Douglas Osheroff who encouraged me to pursue physics in my formative undergraduate years. 

I am also grateful to the wonderful company of the many post-docs and fellow graduate students around me. They were the ones who taught me what most other people thought I already knew. Most enjoyable were the interesting discussions on topics both physical and metaphysical with my labmates Chao-Ming, Yingfei, Danny and Zhao Yang. I also thank Martin Claassen for his extremely bright insights and for being a wonderful friend and collaborator. Equally enjoyable were the intellectual banter with Gang Xu, Yong Xu, Yifan, Haijun, Xiao Zhang, Laimei, Akash, Pavan, Yesheng, Andy Lucas and many others, when innocent debates sometimes evolve into full-fledged research projects. 

I also owe the diversity of my experience as a graduate student to the many physicists I met and/or worked with beyond Stanford. I thank Hiroaki Matsueda for introducing me an alternative but insightful approach to holography, and Peng Ye for inviting me to bask in the vibrant intellectual environment of the Perimeter Institute of Theoretical Physics (PI). 

Last but not least, I will like to thank the Agency for Science, Technology and Research of Singapore for providing me with a scholarship that made my undergraduate and graduate studies at Stanford possible. I am also grateful to Ravi Vakil, Aharon Kapitulnik, Shamit Kachru, Douglas Osheroff and Steve Kivelson who benevolently oversaw my thesis defense. Finally, I thank my parents for their support over this unique epoch of my life away from home.

%% afterpreface produces a table of contents and any other tables
%% wanted. At the end pagenumbering changes from roman to arabic and
%% is restarted
\afterpreface

\chapter{Preface}

The beauty of physics lies in the simplicity of its powerful, far-reaching theories. Among systems of interest to physicists, condensed matter systems are, in particular, highly complicated systems involving very large number of interacting constituents. Yet their physical properties are often dictated by a few surprisingly simple rules. Vital in such reductionistic successes are appropriately chosen 'bases', or perspectives, from which to study the system. For instance, the theory of bandstructure owe its universality to the Wannier basis that put materials with very different electronic orbitals on equal footing. Motivated by that, this thesis will be about two different 'exact mappings' that illuminate the physics of lattice systems through suitable basis choices. 

The first mapping, the Wannier State Representation (WSR), draws an exact correspondence between the single-particle bases of two intensely studied classes of topological systems, the Fractional Quantum Hall (FQH) and Fractional Chern Insulator (FCI) systems. FQH states are realized in interacting 2D electron gases under a strong perpendicular magnetic field, where quasiparticles in the discrete Landau levels exhibit nontrivial fractional statistics. Since their experimental discovery in 1982, FQH states have been the paradigmatic example of topologically ordered states in condensed matter. More recently, however, there has been a surge of interest in realizing topological states in lattice systems without orbital magnetic field, i.e. the FCIs, which are novel lattice realizations of fractionalized topological states. The discovery of such states was built on the discovery of the famed Quantum Spin Hall and Quantum Anomalous Hall Topological Insulator states\cite{qi2008topological,qiRMP2011}. But due to the complicated interplay between single-body bandstructure features and interaction effects, a full theoretical understanding of the stability of FCI states is still lacking. The WSR attempts to bridge this theoretical gap by defining an exact mapping between the well-studied FQH states and the more elusive FCI states. This mapping is especially useful for the construction of FCI pseudopotential Hamiltonians, which are the lattice analogs of energy penalty terms that uniquely determine the nature of topological groundstates in FQH systems. 

The second mapping, the Exact Holographic Mapping (EHM), defines for any lattice system a holographic 'bulk' basis with one additional dimension representing energy scale. It fits squarely within the theme of holography (widely known as the AdS-CFT correspondence), which has attracted great interest among high energy and condensed matter physicists alike. As contrasted with the WSR which is defined only for gapped system, the EHM is also well-defined for gapless (critical) systems. Through the EHM, the original 'boundary' system is exactly mapped onto a new 'bulk' basis, whose correlation functions can be interpreted as that of a massive field in curved spacetime. The resultant bulk geometries for the simple cases studied are, at least in the long-wavelength limit, consistent with those expected from the Ryu-Takanayagi formula, i.e. the ground states of a critical system at zero and nonzero temperature are respectively mapped onto the AdS (Anti de-Sitter) spacetime and the BTZ black hole.

Culminating in the synthesis of the above two contexts is the application of the EHM to topological states like the 2+1-D quantum anomalous Hall (Chern Insulator or CI) state. Starting from a 2+1-D Chern Insulator, the EHM produces an equivalent 'bulk' 3+1-D TI with the original Chern number density identified as the gradient of the instanton (theta) parameter of the bulk TI. Very unlike conventional realizations of TIs, the TI obtained via the EHM is defined on a tree-like network (hyperbolic lattice) with varying vestigial translational symmetry within each 'level' of branches. Furthermore, it also lives in a curved space in the form of a capped AdS space. The new basis from the EHM allows one to define a entanglement partition that separates degrees of freedom at different scales. The resultant entanglement spectrum of a CI reveals the helical surface states of a TI, which is a direct manifestation of the parity anomaly of 2+1-D Dirac cones. At a deeper level, this construction forges a suggestive relationship between the two bulk-edge correspondences that has created much excitement in the physics community: Holographic Duality which relates a conformal field theory on the edge with a gravitational theory in the bulk, and Topological edge states which relates a conformal field theory on the edge with a nontrivial topological invariant of the bulk states.

\chapter{Mapping I: Wannier State Representation}
\begin{figure}[htb]
\centering
\includegraphics[width=.75\linewidth]{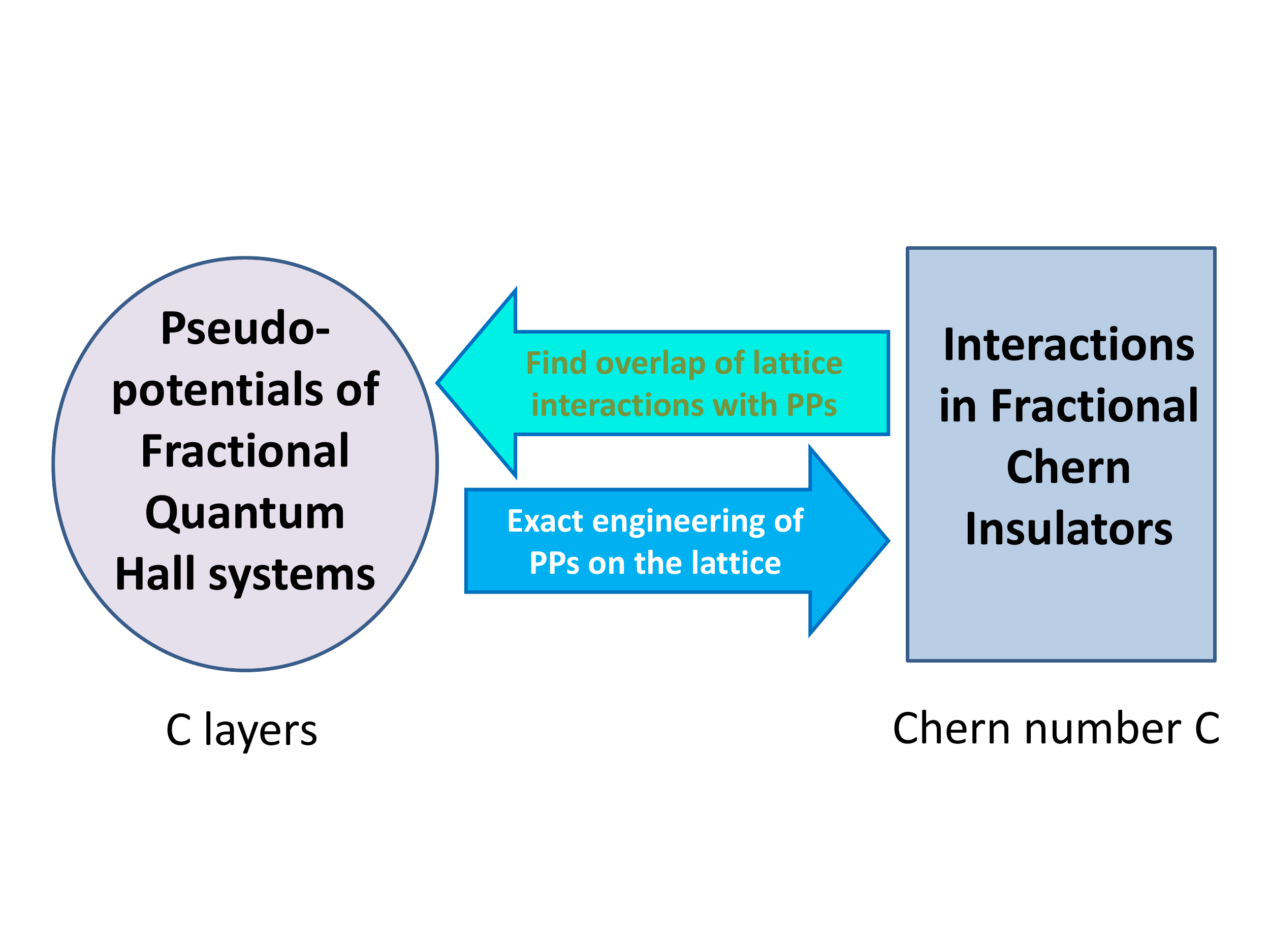}
%\captionsetup{justification=centerlast}
\caption{Schematic diagram of the two directions of application of the Wannier State Representation. An FQH system with $C$ layers corresponds to an FCI system with Chern number $C$ (to be defined in the subsequent sections).}
\label{fig:FQHFCI}
\end{figure}

The first mapping discussed in this thesis is the Wannier State Representation (WSR). It is an exact mapping between the Hilbert spaces of two different topological systems, the Fractional Quantum Hall (FQH) and Fractional Chern Insulator (FCI) systems. Through it, operators in one system can be directly compared with those of the other system. This is very useful in the context of understanding what constitutes a 'good' FCI model, i.e. one that robustly supports the lattice analog of FQH topological states. The WSR mapping will be studied from both directions, as described below:
\begin{itemize}
\item From FCI to FQH: Interaction terms in the lattice FCIs can be expanded in terms of the Pseudopotential operators of FQH systems, whose ground states are relatively well-understood.  
\item From FQH to FCI: Pseudopotential operators are exactly transcribed onto the lattice FCI systems, thereby explicitly constructing lattice interactions that are in principle guaranteed to stabilize desired topological ground states.
\end{itemize}

\section{Background: Fractional Quantum Hall (FQH) systems}

\subsection{Introduction to the FQH effect}

One of the most impressive emergent phenomena in condensed matter physics is the formation of incompressible quantum fluids in a 2-D electron gas under a strong perpendicular magnetic field~\cite{Tsui-PhysRevLett.48.1559}. Under the magnetic field, otherwise free electrons can no longer possess arbitrary kinetic energy, but fall into macroscopically degenerate quantized energy levels known as Landau levels (LLs). With kinetic energy quenched within each Landau level, interaction effects dominate, leading to an exciting playground for the emergence of interaction mediated topological order.

Experimental signatures of the quantized Landau levels date back to 1980, when 2-D electron gas systems of number density $n_0$ in a magnetic field $B$ were found to possess Hall conductivities plateaus at $\sigma_{xy}=\frac{\nu e^2}{h}$ where the number of filled LLs $\nu=\frac{hcn_0}{eB}$ takes on integer values. Known as the Integer Quantum Hall (IQH) effect, it was the earliest  experimental manifestation of topological order in condensed matter~\cite{vonklitzing}. Accompanying the quantized Hall conductivity plateaus is the vanishing of longitudinal resistivity, which together suggests a gapped phase robust against disorder. 

More excitement ensued with the discovery of the Fractional Quantum Hall (FQH) effect by Tsui, Stormer and Gossard~\cite{Tsui-PhysRevLett.48.1559}in 1982, when quantized Hall conductivity plateaus at $\sigma_{xy}=\frac{\nu e^2}{h}$ were also observed at certain \emph{fractional} filling fractions like $\nu=\frac1{3},\frac{2}{5}$, etc. Apparently, the stabilization of such fractionally-filled phases must be the result of the residual strong electron-electron interactions. 

A key theoretical insight to understanding the many-body nature of such phases was provided by Laughlin's wave function~\cite{Laughlin-PhysRevLett.50.1395} which gives, for $\nu=\frac1{3},\frac1{5}$, etc. the wavefunction
\begin{equation}
\psi_{\nu}(z_1,...,z_N) = \prod_{i<j} (z_i-z_j)^\frac1{\nu} \exp(-\sum_k |z_k|^2/4\ell_B^2).
\end{equation}
where $l_B=\sqrt{\frac{\hbar c}{eB}}$ is the magnetic length scale set by the system and $z_k=x_k-iy_k$ the position of the $k^{th}$ electron. The above wavefunction suggests a picture involving quasiparticles with fractional charges of $e\nu$, a picture that has since led to a wealth of interesting ideas on fractional and non-Abelian statistics that have more recently been appreciated for their relevance to topological quantum computing.

Shortly thereafter, another important theoretical development was made~\cite{haldane1983,trugman1985} when it was realized that such Laughlin wavefunctions are also exact zero-energy ground states of certain short-range Hamiltonians known as Haldane Psuedopotentials. Such Pseudopotential operators are positive semi-definite and assign an energy penalty for each pair of electrons in a state with certain chosen relative angular momentum (degree of polynomial part of $\psi_\nu$). For instance, the Pseudopotential corresponding to the $\nu=\frac1{3}$ Laughlin state penalizes states with relative angular momentum $l=1$ but not higher $l$, thereby setting the degree of $\psi_{\nu=\frac1{3}}$ to $3$. Such Psuedopotentials and their generalizations to more than two bodies will be used extensively in this thesis, where they are mapped onto the Hilbert spaces of lattice FCI systems for the search of analogous topologically nontrivial states.

\subsection{Landau gauge basis}

The Landau gauge basis is a convenient basis for the exact mapping of the Hilbert space of Quantum Hall systems to those of lattice systems. It can be easily derived from the single-particle Hamiltonian of an electron $e$ in a magnetic field $B$
\begin{equation}
H=\frac1{2m}\left(\vec{p} + \frac{e}{c}\vec{A}\right)^2 %+V(\vec{r}_1,\vec{r}_2,...,\vec{r}_N)
\end{equation}
where $\vec p$ is the momentum operator of the electron, and $m$ its effective mass. The Landau gauge is imposed by specializing to $\vec A=B  x \hat y $. The eigenstates of the above Hamiltonian can be found through trivial application of the Schrodinger's equation, with $y$ being a good quantum number. Periodic boundary conditions (PBCs) can be imposed, either along one direction (say $\hat y$ with system dimension $L_y$) or both directions, corresponding to cylinder and torus geometries. Under PBCs in one direction, the single-particle Hilbert space is spanned by the Landau gauge basis wave function labeled by an integer $n$ (or discrete momentum $k_y=2\pi n/L_y$):
\begin{equation}
\psi_n(x,y) \equiv  \langle \mathbf{r} | c_n^\dagger \ket{0} \propto e^{i\frac{\kappa}{\ell_B}ny}e^{-\frac{1}{2}\left(\frac{x}{\ell_B}-\kappa n \right)^2}=e^{ik_y y}e^{-(x-k_yl_B^2)^2/2l_B^2},
\label{LLL}
\end{equation}
where $c_n^\dagger$ is the second-quantized operator that creates a particle in the state $|n\rangle$. The important parameter $\kappa=2\pi \ell_B/L_y$ sets the effective separation between the one-body states in the $x$-direction, since each one-body state is a Gaussian packet approximately localized around $\kappa n$ in $x$-direction. The basis with toroidal boundary conditions can be obtained via a superposition of the cylindrical basis: $\psi_n^{\text{torus}}=\sum_{k\in \mathbb{Z}}\psi_{n+kN_\phi}$.
 
The Landau gauge basis affords an important simplification in the indexing of the states in the Hilbert space, being labeled by just one parameter $n$ despite spanning a 2-D system. We will soon see how it leads to an elegant representation of the Pseudopotential operators that is particularly well-suited for comparison with interactions on a lattice.

\subsection{Quantum Hall Pseudopotentials}

Within a given Landau level, the kinetic energy term in the quantum Hall (QH) Hamiltonian is ``quenched", i.e. is effectively a constant. Hence the remaining effective Hamiltonian only depends on the interaction between particles (e.g., the Coulomb potential) projected to the given Landau level~\cite{prangegirvin}. In the infinite plane, this lowest LL projection amounts to evaluating the matrix elements of the Coulomb interaction between the single-particle basis states. For this purpose, a more convenient basis is given by the symmetric gauge $\vec A=B(-y,x)/2$ which does not break the rotational invariance of the system; this will be elaborated in Appendix \ref{nbody}. The single-particle basis wavefunctions take the form %(Fig.~\ref{fig:disk}a) 
\begin{equation}\label{spwf}
z_j^m \exp(-|z_j|^2/4\ell_B^2),
\end{equation}
where $\ell_B=\sqrt{\hbar c/eB}$ is the magnetic length as before, and the system is assumed isotropic~\cite{HaldaneGeometry} so that one can simply write $z_j=x_j+i y_j$. The states in Eq.~\ref{spwf} are mutually orthonormal, and span the basis of the lowest Landau level. There $N_\phi$ of these states, which is also the number of magnetic flux quanta through the sample. %The above facts will be assumed throughout this paper, which is the case in strong magnetic fields, when the particle-hole excitations to other Landau levels are prohibited by the cyclotron energy gap $\hbar\omega_c$.

\subsubsection{Clustering properties of FQH wavefunctions}

Restricting to the lowest Landau level, we can classify different QH states by their \emph{clustering} properties. These are a set of rules which describe how the wave function vanishes as particles are brought together in real space. To define the clustering rules, it is essential first to understand the problem of two particles restricted to the lowest Landau level.% (Fig.~\ref{fig:disk}b). 
To proceed, we solve the two-body problem by transforming from coordinates $z_1,z_2$ into the center-of-mass (CM) frame containing the relative coordinate $z\equiv z_1-z_2$. In the new coordinates, the two-particle wave function decouples. As we are interested in translationally-invariant problems, only the relative wave function (which depends on $z$) will play a fundamental role in the following analysis. For any two particles, the relative wave function turns out to have an identical form to the single-particle wave function Eq. (\ref{spwf})
\begin{equation}\label{2pwf}
\phi_m(z \equiv z_1-z_2) \propto z^m \exp(-|z|^2/8\ell_B^2),
\end{equation}
up to the rescaling of the magnetic length $\ell_B$. An important difference between Eqs.~(\ref{2pwf}) and (\ref{spwf}) is the new meaning of $m$: since $z$ now represents the \emph{relative separation} between two particles, $m$ in Eq.~(\ref{2pwf}) is also related to particle statistics, and moreover encodes the clustering conditions. For spinless electrons, $m$ in Eq.~(\ref{2pwf}) is only allowed to take odd integer values since the wave function must be antisymmetric with respect to $z\to -z$, while for spinless bosons $m$ can be only be an even integer. Finally, $m$ is also the eigenvalue of the  angular momentum for two particles ($\hbar=1$), as we can directly confirm from
\begin{equation}
L^z = \sum_i z_i \frac{\partial}{\partial z_i}.
\end{equation}

After this discussion on two particles, we can introduce the notion of clustering properties for $N$-particle states. Let us pick a pair of coordinates $z_i$ and $z_j$ of indistiguishable particles in a many-particle wave function. We say that these particles are in a state $\psi^{ij}_m$ which obeys the clustering property with the power $m$
if $\psi^{ij}_m$ vanishes as a polynomial of total power $m$ as the coordinates of the two particles approach each other:
\begin{equation}\label{vanishing}
\psi_m^{ij} (z_i \to z_j) \sim  (z_i-z_j)^{m}.
\end{equation} 
Similarly as before, we can relate the exponent $m$ to the angular momentum $L^z$ if the latter is a conserved quantity.
Clustering conditions like this directly generalize to cases where more than $2$ particles approach each other, with the polynomial decay also specified by a power. For example, we say that an $N$-tuple of particles $z_1, z_2, \dots z_N$ is in a state $\psi^{N}_m$ with total relative angular momentum $m$ if $\psi_m^N$ vanishes as a polynomial of total degree $m$ as the coordinates of $N$ particles approach each other. Fixing an arbitrarily chosen reference particle of the $N$-tuple, i.e. $z_1=:z_r$, we have:
\begin{equation}\label{vanishing2}
\psi_m^{ij} (z_2,\dots, z_N \to z_r) \sim \prod_{j=2}^N (z_r-z_j)^{m_j},
\end{equation} 
 with $\sum_{j=2}^N m_j =m$ as all remaining $N-1$ particles approach the reference particle $z_r$.
If the system is rotationally invariant about at least a single axis (such as for a disk or a sphere) it directly follows that the state in Eq.~\ref{vanishing2} is also an eigenstate of the corresponding $L_z$ relative angular momentum operator with eigenvalue $m$. 

The simplest illustration of the clustering condition is the fully filled Landau level. The wave function for such a state is the single Slater determinant of states $\phi_m(z_i)$ in Eq.~\ref{spwf}. Due to the Vandermonde identity, this wave function can be expressed as
\begin{equation}
\psi_{\nu=1} = \det(\phi_{m_i}(z_j)) = \prod_{i<j} (z_i-z_j) \exp(-\sum_k |z_k|^2/4\ell_B^2).
\end{equation}
We see that when any pair of particles $z_i$ and $z_j$ is isolated, the relevant part of the wave function always contains $(z_i-z_j)$. Therefore, the wave function of the filled Landau level has the vanishing exponent $m=1$ as particles are brought together. This is the minimal clustering constraint that any spinless fermionic wave function must satisfy. %(As we will see below, the interesting many-body physics results from stronger clustering conditions on the wave function.)
\subsubsection{Pseudopotentials as projectors}

When properly orthogonalized, the $N$-body states $\{\psi_m\}$ form an orthonormal basis in the space of magnetic translation invariant QH states. They allow one to define the $N$-body \emph{Haldane pseudopotentials} (PPs)\cite{haldane1983} (PPs)
\begin{equation}
U^{m}= \ket{\psi_m}\bra{\psi_m},
\label{Um}
\end{equation}   
which obey the null space condition $U^{m'} \psi^m=0$ for $m' \neq m$. Since they are positive definite, the PPs $U^m$ give \emph{energy penalties} to $N$-body states with total relative angular momentum $m$. With a given many-body wave function, the Hamiltonian representation of $U^m$ will involve the sum over all $N$-tuple subsets of particles. 

For a given filling fraction, it can occur that a certain QH state is the unique \emph{densest} ground state lying in the null space (i.e. is annihilated by) of a certain linear combination of PPs. The requirement of being the densest state is necessary to render the finding non-trivial, as it is easy to construct additional zero modes of the given PP Hamiltonian by simply increasing the magnetic flux, i.e. by nucleating quasihole excitations.  
The most elementary examples are the Laughlin states at $1/m$ filling, which lie in the null space of 2-body $U^{m'}$ for all $m'<m$. As it also represents the densest configuration that is annihilated by the PP, the Laughlin state emerges as the unique ground state of a Hamiltonian at filling $1/m$, $H=\sum_{m'<m} c_{m'} U^{m'}$, where the coefficients are arbitrary as long as $c_{m'} > 0$. That is, the fermionic $1/3$ Laughlin state is the unique ground state of 2-body $U^1$, while the fermionic $1/5$ state is the unique groudstate of any linear combination of 2-body $U^1$ and $U^3$ with positive weights. Note that the PPs of even $m'$ are precluded by fermionic antisymmetry. 

More sophisticated combinations of Pseudopotentials involving three or more bodies and/or internal degrees of freedom admit much more interesting states like the Pfaffian and non-abelian NASS states. Indeed, given a desired state with known clustering properties, one can systematically derive a parent Hamiltonian, i.e. a combination of Pseudopotential operators that admit it as the densest ground state at the appropriate filling fraction. A brief recipe for doing so is outlined in Appendix \ref{sec:power}; the reader is invited to refer to Sect.~\ref{sec:spinful} for the construction of PPs with internal degrees of freedom and ~\cite{lee2015pp} for a more in-depth numerical study.

\subsubsection{Construction of QH Pseudopotentials}

As previously explained, kinetic energy is frozen out within a Landau level of a QH system, and the dynamics are effective controlled by an interaction Hamiltonian of the form (operators will be assumed to involve only 2 bodies here, except when otherwise indicated)
\begin{equation}
H_{int}=\sum_{i<j} V (\bs{r}_i-\bs{r}_j),
\end{equation}
where the sum extends over all pairs of particles. Here and below, it should be remembered that the Hamiltonian is projected to states in a Landau level, although for simplicity we will not write the projection explicitly.  Translation symmetry is assumed, so that interaction only depends on the relative position ${\bf r}_i-{\bf r}_j$. 

Since the PP operators $U^m$ in Eq. \ref{Um} form a complete orthonormal set, $H_{int}$ can be decomposed in terms of them:
\begin{eqnarray}
H_{int}= \sum_{m=0}^{\infty} V^m U^m= \sum_{m=0}^{\infty} V^m\ket{\psi_m}\bra{\psi_m}
%P^m_{ij}&=&L_m (-l_B^2 \nabla^2) \delta^2 (\bs{r}_i -\bs{r}_j)\nonumber
\end{eqnarray}
where $V^m$ are the expansion coefficients in this Pseudopotential basis. Knowledge of $V^m$ can yield crucial information about the propensity of $H_{int}$ in supporting certain ground states. To find $V^m$, the immediate task is to derive explicit expressions for $U^m$. 

Depending on the occasion, it will be helpful to obtain the Pseudopotentials in either a first-quantized (single-body) form, or a second-quantized (many-body) form. 

The former is most suitable for use on the infinite plane or, with minor modifications via stereoscopic projection, the sphere~\cite{lee2013}. A generic $N$-body first-quantized PP with total relative angular momentum $m$ is made up of a linear combinations of terms of Trugman-Kivelson type~\cite{trugman1985}
\begin{equation} \left(\prod^N_{j=1}\nabla_j^{2d_j}\right)\delta^2(\vec r_1-\vec r_2)...\delta^2(\vec r_{N-1}-\vec r_N)\end{equation}
such that $\sum_j d_j =m$. The exact determination of the linear combination coefficients can be computationally involved for large $N$ and $m$, and is the subject of Sect. \ref{sec:mbpp} as well as Appendix \ref{nbody}.

Second-quantized PP operators are understandably more complicated to write down, but are still dramatically simplified in the Landau gauge due the one-parameter labeling of the single-body states. They are most naturally written down in cylinder and torus geometries where $k_y$ is a good quantum number in the Landau gauge basis. An $N$-body interaction matrix element is labeled by $2N$ indices. Indeed, the matrix elements of any translationally-invariant interaction $H_{int}=V(\vec r_1,...,\vec r_N)$ projected to a Landau level are given by
\begin{eqnarray}
V_{\{ n_j \}, \{ n'_j \}} &=& \int d\vec r_1 \ldots d\vec r_N  \left( \prod_j \psi^\dagger_{n'_j}(\vec r_j)   \psi_{n_j}(\vec r_j) \right) V(\vec r_1,...,\vec r_N) , 
\label{Vint}
\end{eqnarray}
which is more clearly expressed as the second-quantized Hamiltonian
\begin{eqnarray}
\notag && H_{int} = \sum_\alpha \sum_{n} b^{\alpha\dagger}_n b^\alpha_n, \\
&& \notag b^\alpha_n \propto \sum_{ \sum_j \bar n_j = 0}p_\alpha(\kappa \bar n_1,...,\kappa \bar n_N)e^{-\frac{\kappa^2}{2}\sum_{j=1}^N \bar n_j^2}c_{n_1} \ldots c_{n_N}.
\label{bn}
\end{eqnarray}
Here $\bar n_j=n_j-n/N$ denotes the position of particle $j$ with respect to the center-of-mass (CM ), $n=n_1+...+n_N$, and $p_\alpha$ is a polynomial in the $N$ variables $\kappa \bar n_j$. From now on, $n$ will refer to the CM , and not the index of a single particle previously appearing in Eq. \ref{LLL}. $c_{n_j}^\dagger$ is the same operator as in Eq. \ref{LLL}, creating an electron in the state $|n_j\rangle$. The real polynomial $p_\alpha$ is derived in the next subsection and given explicitly in Table \ref{N2}. 

Eqs.~\ref{Vint} and~\ref{bn} represent a general translationally invariant Hamiltonian projected to a Landau level. This Hamiltonian has a rather special form: it decomposes into a linear combination of positive-definite operators $b^{\alpha\dagger}_n b^\alpha_n$, such that the form factor $p_\alpha e^{-\frac{\kappa^2}{2}\sum_{j=1}^N \bar n_j^2}$ in each $b^\alpha_n$ depends only on the relative coordinates. Any short-range Hamiltonian can be explictly expressed in this form~\cite{lee2004,seidel2005, lee2013}, which shows its Laplacian structure. 

\begin{figure}[htb]
\centering
\includegraphics[width=.65\linewidth]{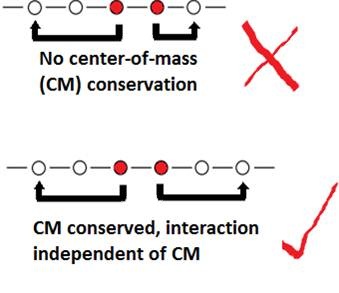}
\caption{ Disallowed and allowed hopping processes. Top: A hopping process that has different initial and final CM positions does not respect magnetic translation symmetry, and is disallowed. Bottom: An allowed pair hopping process preserves the CM position, and is also invariant under translations along the chain.}
\label{fig:hopping}
\end{figure}

Physically, a Landau-level-projected Hamiltonian can be described by a long-range interacting 1D chain, as illustrated in Fig.~\ref{fig:hopping}. The interaction terms can be interpreted as long-range (though Gaussian suppressed) hopping processes labeled by $\alpha$. For each $\alpha$, $N$ particles ``hop" from positions $n_j$ to positions $n'_j$, $j=1,...,N$ according to a CM independent amplitude given by $p_\alpha(\kappa \bar n_1,...,\kappa \bar n_N) e^{-\frac{\kappa^2}{2}\sum_{j=1}^N \bar n_j^2}$, such that the initial and final CM $n$ remains unchanged (lower diagram in Fig.~\ref{fig:hopping}). Note that although we use the term ``hopping'', there is no clear distinction between ``hopping'' and ``interaction'' in our case, in the same sense as in the Hubbard model. Rather, ``hopping'' is designated for any interaction term that is purely quantum (i.e., not of Hartree form). %The goal of the remainder of this paper is to show how to systematically construct the polynomial amplitudes $p_\alpha$ that need to be inserted into Eq.~(\ref{bn}) to obtain the parent Hamiltonian for the desired quantum Hall state, i.e. to accurately put the kernel of the Laplacian into a desired form.
Processes that do not respect magnetic translation symmetry either do not conserve the CM position along the chain, or are not translation invariant along the latter. One such example is illustrated in the upper diagram of Fig.~\ref{fig:hopping}. Note that such interaction Hamiltonians possess a special structure that gives rise to the symmetry under many-body translations which shifts every particle by $q$ orbitals at filling $\nu=p/q$. 

Further discussion on the relationship between, as well as the relative benefits/drawbacks of, the first and second-quantized representations of the Pseudopotential operators can be found in Appendix. \ref{sec:TK}. There exists a third representation of the PP operators in terms of a coherent state basis consisting of apposite linear combinations of the second-quantized Landau gauge basis states. It allows for more convenient tweaking of the locality of the PPs when mapped onto the lattice, and is explicitly described in Appendix \ref{ppcoherent}.

\subsection{Pseudopotentials beyond two bodies}
\label{sec:mbpp}

While we have only explored two-body interactions so far,
PP expansions are also well-suited for many-body interactions. From the established knowledge of FQHE systems, it follows that various
interesting FQAH liquids are located in the nullspace of certain many-body
PPs. In theory, we can construct many-body FCI
Hamiltonians that exhibit Pfaffian, Read-Rezayi etc. groundstates from
such
PPs~\cite{greiter-91prl3205,Read-99prb8084,simon2007generalized,mcs}.

The first task is to generalize Haldane's PPs for FCI
models to interactions beyond two bodies~\cite{simon2007}. For two particles in the Lowest Landau leve (LLL) and a translationally invariant potential $V(k)=\int
e^{-ik\cdot r} V(r)dr$, the expansion coefficient onto the sector with relative angular momentum $m$ is given by (see Eq.~\ref{Vm} and Appendix \ref{nbody})
\begin{equation*}
V^{m}=4\pi l_B^2\int\frac{d^2k}{(2\pi)^2}e^{-l_B^2k^2}L_m(l_B^2k^2)V(k)
%\label{simpler}
\end{equation*}
so that the $m$th pseudopotential (with $V^m\propto \delta_{mn}$) has the form
$U^m(k)\propto V_0L_m(l_B^2k^2)$. As before, $V_0$ is a constant with units
of energy. In the plane limit, the $U^m$s form an orthogonal basis
which one can use to expand a generic potential profile.

When an interaction involves more than two particles, additional
complications arise. To begin with, there are different ways of choosing to
assign relative distance variables, or,  angular momentum. (For
two-body interactions, there is a unique assignment, as one degree of
freedom drops out due the CM conservation.)
 When
there are $3$ particles, only one degree of freedom is eliminated due to CM
conservation. As such, an ambiguity remains in choosing the many-body
analog of relative angular momentum. This ambiguity is mathematically
manifest when one tries to generalize Eq.~\ref{Vm}. In the case
of $3$-body interactions, there will be integrals over both momenta
$k_1$ and $k_2$ in the above expression, and
one has  to chose the new expression to involve $L_m(k_1^2)$, $L_m(k_2^2)$, $L_m((k_1-k_2)^2)$, or a combination of these.

This formal ambiguity can be remedied in my formulatation of generalized Haldane pseudopotentials (GHPs) detailed in Appendix~\ref{nbody}. We shall
constrain ourselves to the application of GHPs to the total relative
angular momentum $N$-body PP Hamiltonians:
\begin{equation}
U^m(k)= L_m\left(\frac{k^2 l_B^2N}{2 (N-1)}\right).
\label{l2new}
\end{equation}
Here, $U^m(k)$ is the N-body interaction potential that has a total relative
angular momentum of $m$, with $k$ being the momentum conjugate to the total relative coordinate $w$ defined by
\begin{equation}
w= \frac{1}{N-1}\sum_{n=1}^{N-1}(z_n-z_N)=\frac{\sum_i^{N-1}z_i}{N-1}-z_N,
\label{nbodyw}
\end{equation}
where $z_i=x_i-iy_i$ are the complex coordinates of the $N$
particles.

In real space, the pseudopotential $U^m(w)\propto \int dk e^{ik\cdot
  w}U^m(k) $ depends explicitly on the positions of each of the $N$
particles. If we select one of the $N$ particles, the total relative
angular momentum is the sum of the relative angular momenta of the
other $N-1$ particles relative to the first one. Indeed, we see from
Eq.~\ref{nbodyw} that $w$ represents the relative seperation between
particle $N$ and the CM of the rest of the particles.
Note the appearance of the factor $\frac{N}{2(N-1)}$, which is essential
in obtaining the correct expressions for the PPs. (It will also be derived in detail in Appendix~\ref{nbody}.)

\subsubsection{Second-quantized many-body PPs through direct integration}

$N$-body PPs can be nicely expressed in terms of an second-quantized operator with $2N$ indices. One way to explicitly obtain the matrix elements is through direct brute-force integration. For instance, the 3-body PP can be obtained as 
\begin{eqnarray}
&&U^m_{n_1n_2n_3n_4n_5n_6} \nonumber \\
&\propto&\sum_\sigma\int d^2r_1 d^2r_2d^2r_3
\psi^\dagger_{n_1}(r_1)\psi^\dagger_{n_2}(r_2)\psi^\dagger_{n_3}(r_3)L_m\left(\frac{3}{4}l_B^2
  \nabla^2_{[(r_1+r_3)/2-r_2]}\right)\psi ^{\phantom{\dagger}}_{n_4}(r_3) \psi ^{\phantom{\dagger}}_{n_5}(r_2) \psi ^{\phantom{\dagger}}_{n_6}(r_1)
\nonumber \\
&\propto&\sum_\sigma\int \frac{d^2qd^2p}{(2\pi)^4}\int \prod_i^3d^2r_i
L_m\left(\frac{3}{4}(p-q)^2 l_B^2\right)e^{i q\cdot (r_1-r_2)}e^{i
  p\cdot
  (r_2-r_3)}\notag\\
&&\times\psi^\dagger_{n_1}(r_1)\psi^\dagger_{n_2}(r_2)\psi^\dagger_{n_3}(r_3)\psi ^{\phantom{\dagger}}_{n_4}(r_3)\psi ^{\phantom{\dagger}}_{n_5}(r_2)\psi ^{\phantom{\dagger}}_{n_6}(r_1)
.\nonumber \\
\label{pseudopotmaim}
\end{eqnarray}
Note that according to Eq.~\ref{l2new}, we have rescaled the magnetic length $l^2$ by $\frac{3}{4}l^2$. Hence, $U^m$ became $V_0L_m(\frac{3}{4}k^2)$, where $k$ is the momentum conjugate to the total relative coordinate $((r_1-r_2)+(r_3-r_2))/2$. The $\sum_\sigma$ sum refers to a symmetric (antisymmetric) sum over all permutations
$\sigma$ assuming the particles are bosons (fermions). We have $k=q-p$ because
\begin{eqnarray}
e^{i q\cdot (r_1-r_2)}e^{i p\cdot
  (r_2-r_3)}&=&e^{i(q-p)\cdot((r_1+r_3)/2-r_2)}e^{i(p+q)\cdot(r_1-r_3)/2}
\nonumber \\
&=&e^{i(q-p)\cdot w_2}e^{i(p+q)\cdot w_3},
\end{eqnarray}
where $w_2$ and $w_3$ are linear combinations of the original coordinates whose roles will be further expounded in Appendix~\ref{nbody}. Here, it is sufficient to understand that $k$ should be the momentum conjugate to $w_2$, the total relative angular momentum. As before,
$\psi_n(r)=\frac{1}{\sqrt{\sqrt{\pi}Ll_B}}e^{i\frac{\kappa}{l_B}ny}e^{-\frac{(x-\kappa
    n l_B)^2}{2l_B^2}}$ and $\kappa=\frac{2\pi l_B}{L_y}$ is a dimensionless ratio that is small in the limit of large magnetic fields.

The above integral can be systematically simplified into the form Eq. \ref{bn} after a reasonable amount of computation. This will be worked out in detail in Appendix \ref{nbody} for 3-boson PPs.

\subsubsection{Second-quantized many-body PPs through geometric orthogonalization}
\label{geometric}

The above approach in finding the explicit form of the PPs quickly becomes intractable as the number of bodies $N$ or the total degree $m$ are increased. Fortunately, a more elegant alternative exists. It is based on the observation that the PPs form a complete and orthonormal basis, and should be uniquely determined with the help of an appropriate inner product measure. 

Here we specialize Eq.~\ref{Vint} to $N$-body interactions that are Haldane PPs with no internal degrees of freedom, and present cases with internal degrees of freedom in Sect. \ref{sec:spinful}. The latter contains a much greater variety of possible PPs because the orbital and spin parts of the PPs can both possess arbitrary symmetry types, as long as they both conspire to result in an overall PP with bosonic or fermionic symmetry. 

%One appealing feature of the 2nd quantized form of Eq.~\ref{Vint} is that we can construct the  desired PPs from symmetry principles alone, without referring to the 1st quantized form of the interaction $V(\vec r_1,...,\vec r_N)$. 
The many-body PPs $U^m$ are, by definition, supposed to project onto orthogonal subspaces labeled by $m$, where $m$ is a composite index that includes the relative angular momentum and possibly some other quantum numbers: 
\begin{equation}
U^{m'}U^{m'}=\ket{\psi_{m'}}\bra{ \psi_{m'}}\ket{\psi_{m}}\bra{ \psi_{m}}=U^m\delta^{m'm}.
\label{Uortho}
\end{equation}
This requires that
\begin{eqnarray}
\bra{0}b^{m'\dagger}_nb^m_n\ket{0} &\propto& \sum_{\bar n_1+...+\bar n_N=0}p_{m'}(\kappa \bar n_1,...,\kappa \bar n_N) \notag \\
&\times& p_{m}(\kappa \bar n_1,...,\kappa \bar n_N) \exp{\left(-\kappa^2\sum_{j=1}^N \bar n_j^2\right)} \notag \\
&=& \delta_{m'm}.
\end{eqnarray}

If $\psi_m$ is to vanish with $m$th total power as $N$ particles approach each other, the polynomial $p_m$ must be of degree $m$. Hence the PPs will be \emph{completely} determined once we find a set of polynomials $\{p_m\}$ such that: (1)  $p_m$ is of total degree $m$;
(2) $p_m$ has the correct symmetry property under exchange of particles, i.e. is totally (anti)symmetric for bosonic (fermionic) particles; 
(3) the $p_m$'s are orthonormal under the inner product measure
\begin{eqnarray}
\langle p_{m'}, p_m\rangle &=& \sum_{\bar n_1+...+\bar n_N=0} p_{m'}(\kappa \bar n_1,...,\bar n_N) \notag \\ 
&\times& p_{m}(\kappa \bar n_1,...,\bar n_N)e^{-\kappa^2\sum_{j=1}^N \bar n_j^2}\notag\\
&\approx & \int_{\mathbb{R}^{N-1}} dW d\Omega p_{m'}(W ,\Omega) p_m(W ,\Omega) \notag \\
&\times& \exp{(-\frac{N-1}{N}W^2 )}J_{N-1}(W,\Omega)  \notag\\
&=& \delta_{m'm},
\label{innerproduct}
\end{eqnarray}
where (with a slightly abstract use of notation) $W$ and $\Omega$ are the radial and angular coordinates of a vector $\vec x\in \mathbb{R}^{N-1}$ representing the tuple $(\bar n_1,...,\bar n_N)$ in \emph{barycentric} coordinates (Appendix~\ref{sec:barycentric}), and $J_{N-1}$ is the Jacobian for spherical coordinates in $\mathbb{R}^{N-1}$. We have exploited the magnetic translation symmetry of the problem in quotienting out the CM coordinate $n$. It is desirable to quotient out $n=\sum_j^N n_j$, since $n$ takes values on an infinite set when the particles lie on the 2D infinite plane, and that complicates the definition of the inner product measure.
Each quotient space is most elegantly represented as an $N-1$-simplex in barycentric coordinates, where particle permutation symmetry (or subgroups of it) is manifest. 
Explicitly, the set of $\bar n_j$ can be encoded in the vector  
\begin{equation}
\vec x =\kappa \sum_{j=1}^N \bar n_j \vec\beta_j,
\end{equation}
where the $N$ basis vectors $\{\vec \beta_j\}$ form a linearly-dependent basis set that spans $\mathbb{R}^{N-1}$. A configuration $(\bar n_1,...,\bar n_N)$ is uniquely represented by a point $\vec x\in \mathbb{R}^{N-1}$ that is independent of $n=\sum_j^Nn_j$. Since $\vec x$ should not favor any particular $\bar n_j$, any pair of vectors in the basis $\{\vec \beta_j\}$ must make the same angle with each other. Specifically,
\begin{equation}
\vec \beta_j \cdot \vec \beta_k = \frac{N}{N-1}\delta_{jk} - \frac{1}{N-1},
\end{equation}
so that each vector points at angle of $\pi-\cos^{-1}\frac1{N-1}$ from another. 

With this parametrization, the Gaussian factor reduces to the simple form 
\begin{equation}
W^2=|\vec x|^2=\frac{N}{N-1}\kappa^2\sum_{j=1}^N\bar n_j^2.
\end{equation}
Further mathematical details can be found in the examples that follow, as well as in Appendix \ref{sec:barycentric}.
 
The integral approximation in the last line of Eq.~\ref{innerproduct} becomes exact in the infinite plane limit, and is still very accurate for values of $m$ where the characteristic inter-particle separation is smaller than the dimensions of the QH system. It does not affect the exact zero mode property of trial Hamiltonians constructed below. In the following, we will assume that the minimial size of the particle droplet that corresponds to the pseudopotential is smaller than either of the linear dimensions of the systems. If this is not true, there can be significant effects from the interaction of a particle with its periodic images. This was systematically studied in Appendix \ref{ortho} for $N=2$ bodies (see also Ref. \onlinecite{lee2013}).

In a nutshell, the second-quantized PP matrix elements can be found via the procedure outlined in Table~\ref{table:spinless}. This will be explicitly demonstrated below for $N=2,3$ and, to some extent, $N=4$-body interactions.  
\begin{table}[htb]
\centering
\renewcommand{\arraystretch}{2}
\begin{tabular}{|p{10.5cm}|}\hline
(1) Write down the allowed ``primitive" polynomials~\cite{simon2007,simon2007generalized, davenport2012} of degree $m'\leq m$ consistent with the symmetry of the particles; \\ \hline
(2) Orthogonalize this set of primitive polynomials according to the inner product measure in Eq.~\ref{innerproduct}. \\ \hline
\end{tabular}
%\captionsetup{justification=centerlast}
\caption{Summary of the PP construction for spinless particles.
}
\label{table:spinless}
\end{table}

\subsubsection{$2$-body PP matrix elements }
\label{n2}

For two-body interactions, we simply have 
\begin{equation}
\langle p_{m'},p_m\rangle = \int_{-\infty}^{\infty} p_{m'}( W)p_{m}( W)e^{-\frac1{2}W^2}dW, 
\label{inner2}
\end{equation}
where $W=-2\kappa\bar n_1=2\kappa\bar n_2$. For this $N=2$ case, we have allowed $W$ to take negative values as the angular direction spans the 1D circle, which consists of just two points. The primitive polynomials for bosons are $\{1,W^2,W^4,W^6,...\}$ while those for fermions are $\{ W,W^3,W^5,...\}$. After performing the Gram-Schmidt orthogonalization procedure, the $2$-body PPs are found to be $U^2_m\propto \kappa^3\sum_n b^{m\dagger}_n b^m_n$, where
\begin{equation} b^m_n =\sum_{\bar n_1+\bar n_2=0} p_m(\kappa \bar n_1,\kappa \bar n_2)e^{-\frac{1}{2}\kappa^2 (\bar n_1^2+\bar n_2^2)}c_{n/2+\bar n_1}c_{n/2+\bar n_2}, \label{b2}\end{equation}
$n/2 + \bar n_j$ are integers, and $p_m$ is a $m$th degree Hermite polynomial given in Table~\ref{N2}. In particular, we recover the Laughlin $\nu=1/2$ bosonic or $\nu=1/3$ fermionic state for $m=0$ or $m=1$ respectively.

\begin{table}[htb]
\begin{minipage}{0.99\linewidth}
\centering
\renewcommand{\arraystretch}{2}
\begin{tabular}{|l|l|l|}\hline
$m$ &\ Bosonic $p_m(W)$ &\ Fermionic $p_m(W)$ \\    \hline
0 &\ $1$ &\ 0 \\ \hline 
1 &\ $0$ &\ $W$ \\ \hline  
2 &\ $ \frac{1}{\sqrt{2!}}(-1+W^2)$ &\ 0 \\  \hline
3 &\ $0$ &\ $\frac{1}{\sqrt{3!}}(-3+W^2)W$ \\ \hline  
4 &\ $ \frac{1}{\sqrt{4!}}(3-6 W^2+W^4) $ &\ 0  \\ \hline 
5 &\ $0$ &\ $\frac{1}{\sqrt{5!}}(15-10 W^2+W^4)W$ \\ \hline  
6 &\ $ \frac{1}{\sqrt{6!}}(-15+45 W^2$ &\ $0$ \\ &\  $-15 W^4 +W^6)$ &\ \\ \hline 
7 &\ $0$  &\ $\frac{1}{\sqrt{7!}}(-105+105 W^2-21 W^4$ \\ &\ &\ $+ W^6)W$ \\ \hline  
\end{tabular}
\end{minipage}
%\captionsetup{justification=centerlast}
\caption{Representative polynomials $p_m$ for the first few $N=2$-body PPs for bosons and fermions. $p_m(W)e^{-W^2/4}$ represents the probability that two particles $W/\kappa$ sites apart are involved in a two-body hopping on the chain. }
\label{N2}
\end{table}
One can easily check that PPs become more delocalized in $W$-space as $m$ increases. Indeed,
\begin{eqnarray}
\notag \kappa^2\langle\sum_j \bar n_j^2\rangle&=&\frac1{2}\langle W^2\rangle =\frac1{2}\int_{-\infty}^{\infty}p_m(W)W^2e^{-\frac1{2}W^2} \notag\\
& =& m+\frac{1}{2}, 
\label{PPm}
\end{eqnarray}
which is reminiscent of the interpretation of $m$ as the angular momentum of a pair of particles in rotationally-invariant geometries.% (Fig.~\ref{fig:disk}).
Obtaining the pseudopotentials in this second quantized form is highly advantageous. In particular, note that this construction is free from ambiguities in the choice of $V(\vec r_1,...,\vec r_N)$, since several possible real-space interactions, e.g., those of the Trugman-Kivelson type~\cite{trugman1985}, can all belong to the same $m$ sector. This is discussed in more detail in Appendix~\ref{sec:TK}.

\subsubsection{$3$-body PP matrix elements}
\label{n3}

For $N=3$, the inner product measure takes the form
\begin{equation}
\langle p_{m'},p_m\rangle = \int_0^\infty \int_0^{2\pi}p_{m'}(W ,\theta) p_m(W ,\theta)e^{-\frac{2}{3}W^2}Wd\theta dW
\label{inner3}
\end{equation}
with
\begin{subequations}
\begin{align} \bar n_1 &=  \frac{2W}{3\kappa}\cos\theta \\
\bar  n_2 &=  \frac{W}{3\kappa}(\sqrt{3}\sin\theta-\cos\theta) \\
 \bar n_3 &=  \frac{W}{3\kappa}(-\sqrt{3}\sin\theta-\cos\theta) 
\end{align}\label{n3def}\end{subequations} 
Each of the $\bar n_j$'s are treated on equal footing, as one can easily check graphically. The above expressions are the simplest nontrivial cases of the general expressions for barycentric coordinates found in the appendix (Eqs.~\ref{barybasis}-\ref{eqg10}).

The bosonic primitive polynomials are made up of elementary symmetric polynomials $S_1,S_2,S_3$ in the variables $\kappa \bar n_j=\kappa (n_j-n/N)$. Since $S_1=\kappa \sum_j \bar n_j=0$, the only two symmetric primitive polynomials are
\begin{equation} 
-S_2=-\kappa^2\sum_{i<j}\bar n_i\bar n_j=\frac{S_1^2-2S_2}{2}=\frac{\kappa^2}{2}\sum_i \bar n_i^2 = \frac{1}{3}W^2, 
\end{equation} 
and
\begin{eqnarray} Y= S_3 &=&  \kappa^3\prod_i\bar n_i\notag\\
&=&\frac{\kappa^3}{27}(2n_1-n_2-n_3)(2n_2-n_3-n_1) \notag \\
&& (2n_3-n_1-n_2)\notag\\
&=& \frac{2}{27}W^3\cos 3 \theta. 
\label{Y}
\end{eqnarray}

The fermionic primitive polynomials are totally antisymmetric, and can always be written\cite{simon2007,davenport2012} as a symmetric polynomial multiplied by the Vandermonde determinant
\begin{eqnarray} 
\nonumber A &=& \kappa^3(n_1-n_2)(n_2-n_3)(n_3-n_1) \\
 &=& -\frac{2}{3\sqrt{3}}W^3 \sin 3\theta. 
\label{Anti}
\end{eqnarray}
Note that $W^2$ is of degree 2 while $A$ and $Y$ are of degree 3. All of them are independent of the CM coordinate $n$, as they should be. $N=3$ PPs were derived in Ref.~\onlinecite{lee2013} through tedious explicit integration, and the approach discussed here considerably simplifies those computations by exploiting symmetry.

To generate the fermionic (bosonic) PPs up to $U^3_m$, we need to orthogonalize the basis consisting of all possible (anti)symmetric primitive polynomials up to degree $m$. For instance, the first seven (up to $m=9$) 3-body fermionic PPs are generated from the primitive basis $\{ A, AW^2, AY,AW^4,AYW^2, AY^2, AW^3\}$. Note that the last two basis elements both contribute to the $m=9$ PP sector. 

The 3-body PPs are found to be $U^3_m\propto \kappa^3\sum_n b^{m\dagger}_nb^m_n$, where
\begin{equation} 
b^m_n = \sum_{n_1+n_2+n_3=n} p_m(W,Y,A)e^{-\frac1{3} W^2}c_{n_1}c_{n_2}c_{n_3}, \label{b3}
\end{equation}
with the polynomials $p_m$ are listed in Table~\ref{N3}. These results are fully compatible with those from Ref. \onlinecite{simon2007}. As mentioned, there can be more than one (anti)symmetric polynomial of the same degree for sufficiently large $m$. This leads to a degenerate PP subspace, which is discussed further in~\cite{lee2015pp}.

\begin{table*}%[H]
\centering
\renewcommand{\arraystretch}{2}
\begin{tabular}{|l|l|l|}\hline
$m$ &\ Bosonic $p_m(W ,Y)$ &\ Fermionic $p_m(W ,Y, A)$ \\    \hline
0 &\ $ 1 $ &\ 0 \\ \hline 
1 &\ $ 0$ &\ 0 \\  \hline
2 &\ $  1-\frac{2}{3}  W^2  $ &\ 0  \\ \hline 
3 &\ $ 3\sqrt{2}  Y$  &\ $A$ \\ \hline 
4 &\ $ 1-\frac{4}{3}  W^2+ \frac{2}{9}  W^4$ &\ 0 \\ \hline 
5 &\ $  \sqrt{2}  Y(6-  W^2)$  &\ $  A\left(2-\frac{1}{3}  W^2\right)  $ \\ \hline 
6 &\ \shortstack{(i)  $ 1-2  W^2 +\frac{2}{3} W^4 -\frac{4}{81}  W^6$  \\ (ii) $ \frac1{81}\sqrt{\frac{2}{5}}(2W^6-729Y^2)$ } &\ $ \frac{3}{\sqrt{5}}AY$ \\ \hline 
7 &\ $\frac{2}{3\sqrt{5}}(45-15W^2+W^4)Y$ &\ $ \sqrt{10}A\left(1-\frac{1}{3}  W^2 +\frac{1}{45}  W^4\right)$ \\ \hline
8 &\ (i)  $1+\frac{2}{243}W^2(W^2-6)(54-18W^2+W^4)$ &\ $  3 \sqrt{\frac{7}{5}}\left(1-\frac{2}{21} W^2\right)AY$ \\ &\ (ii) $\frac1{243}\sqrt{\frac{2}{35}}(2W^2-21)(2W^6-729Y^2)$ &\ \\ \hline
%9 &\ \begin{tabular}{@{}c@{}}(i) $\sqrt{\frac{6}{155}}Y(-30+15W^2-2W^4+18Y^2)$ \\ (ii) $\frac1{27}\sqrt{\frac{2}{1085}}Y(-5760W^2+756W^4-31W^6 +81(140+9Y^2))$ \end{tabular}&\  \begin{tabular}{@{}c@{}} (i) $\frac{  A}{\sqrt{5}} (-10+5  W^2-\frac{2}{3}  W^4+\frac{2}{81} W^6)$   \\  (ii) $\frac{ A}{81\sqrt{105}}\left(W^6-729 Y^2\right)$ \end{tabular}
9 &\ \shortstack{(i) $\sqrt{\frac{6}{155}}Y(-30+15W^2-2W^4+18Y^2)$ \\ (ii) $\frac1{27}\sqrt{\frac{2}{1085}}Y(-5760W^2+756W^4$\\$ -31W^6 +81(140+9Y^2))$ } &\ \shortstack{(i) $\frac{  A}{\sqrt{5}} (-10+5  W^2-\frac{2}{3}  W^4+\frac{2}{81} W^6)$ \\ (ii) $\frac{ A}{81\sqrt{105}}\left(W^6-729 Y^2\right)$}
 \\ \hline
%$\frac{  A}{\sqrt{5}} (-10+5  W^2-\frac{2}{3}  W^4
%+\frac{2}{81} W^6)$ \\ $ \frac{ A}{81\sqrt{105}}\left(W^6-729 Y^2\right)$ 
\end{tabular}
%\captionsetup{justification=centerlast}
\caption{The polynomials $p_m$ for $N=3$-body PPs for bosons and fermions up to $m=9$. 
Note that there is more than one possible PP for larger $m$, since there are multiple ways to build a homogeneous (anti)symmetric polynomial from the primitive and elementary symmetric polynomials. For instance, with $Y$ and $A$ defined in Eqs. \ref{Y} and \ref{Anti}, there are two possible bosonic PPs for $m=6$, since there are two ways ($W^6$ and $Y^2$) to build a homogeneous 6-degree polynomial from elementary symmetric polynomials. For the fermionic cases, we have explicitly kept only one factor of $A$, since even powers of $A$ can be expressed in terms of $W^2$ and $Y$. For all cases, the mean-square spread of the PPs also increases linearly with $m$, i.e. $\kappa^2\langle\sum_j^3 \bar n_j^2\rangle=m+1 $ (compare with the two-body case in Eq.~\ref{PPm}).  }
\label{N3}
\end{table*}

\subsubsection{$N\ge4$-body PP matrix elements}

For general PPs involving $N$ bodies, the inner product measure takes the form
\begin{eqnarray}
\nonumber \langle p_{m'},p_m\rangle &=& \int_0^\infty e^{-\frac{N-1}{N}W^2}W^{N-2} dW \\
\nonumber &&\times \int_0^{2\pi} d\phi_{N-2}\prod_{k=1}^{N-3}\int_0^\pi \sin^{N-2-k}(\phi_k) d\phi_k  \\
\nonumber && \times p_{m'}(W ,\phi_1,...,\phi_{N-2}) p_m(W ,\phi_1,...,\phi_{N-2}), 
\label{innerN}
\end{eqnarray}
where the Jacobian determinant from Eq. \ref{innerproduct} has already been explicitly included. One transforms the tuple $\{\bar n_1,...,\bar n_N\}$ into $(N-1)$-dim spherical coordinates via the barycentric coordinates detailed in Appendix~\ref{sec:barycentric}. 

\begin{figure}[ttt]
\includegraphics[width=\linewidth]{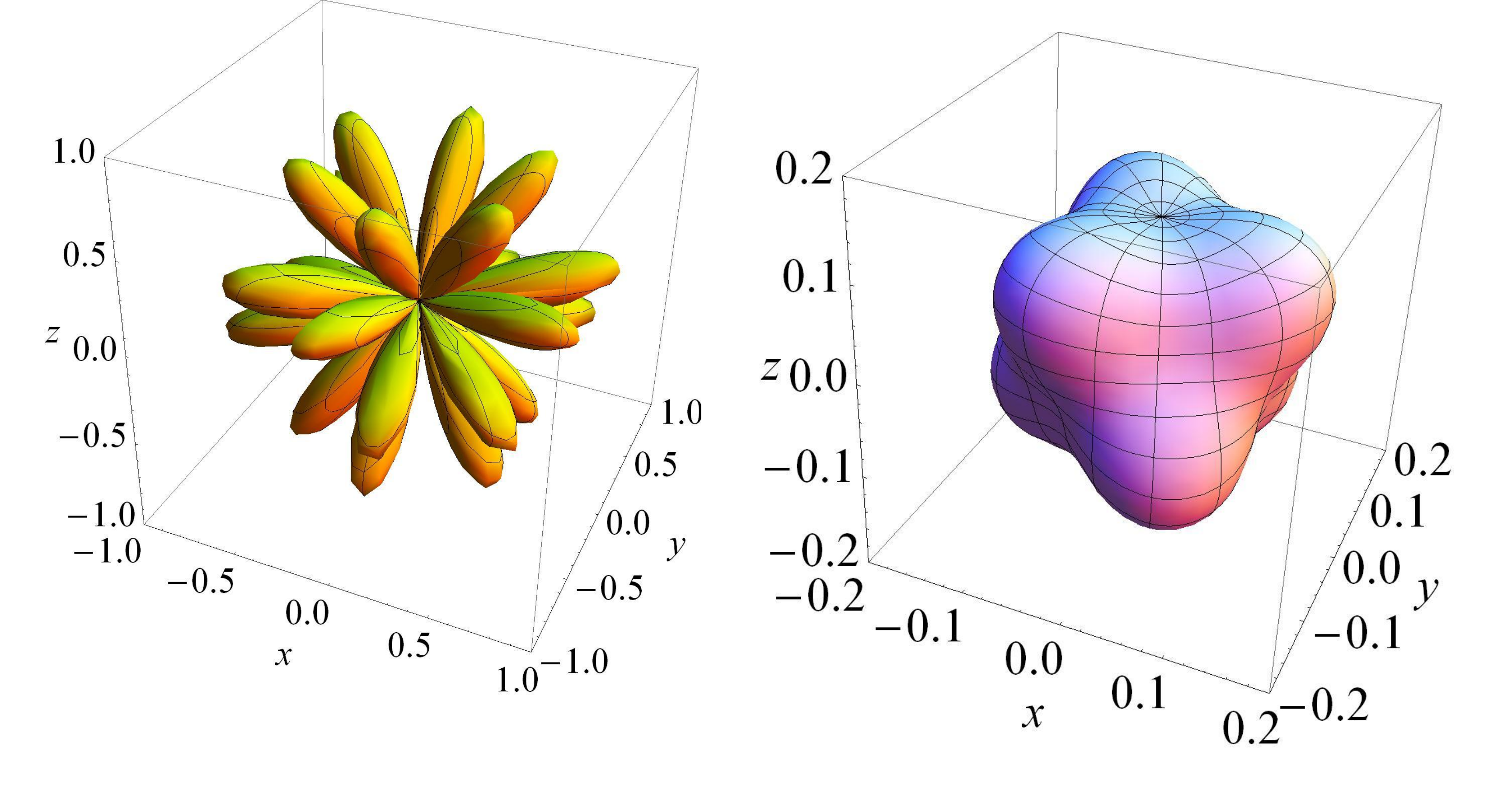}
%\captionsetup{justification=centerlast}
\caption{  The primitive polynomials have beautiful geometric shapes when plotted in $N-1$-dim spherical coordinates. Shown above are the constant $W$ plots of the antisymmetric polynomial $A=(\bar n_1-\bar n_2)(\bar n_1-\bar n_3)(\bar n_1-\bar n_4)(\bar n_2-\bar n_3)(\bar n_2-\bar n_4)(\bar n_3-\bar n_4)$ (left), and the degree $m=4$ symmetric expression in the orthonormalized space spanned by the elementary symmetric polynomials $1,S_2,S_3,S_4$ and $S^2_2$ (right). On the left, there are $24$ lobes that each maximally avoid the vertices of the tetrahedron (3-simplex). 
On the right, the $6$ lobes lie around the centers of the $6$ edges of the tetrahedron, where two of the $\bar n_i$'s are equal.   
}
\label{symantisymm4}
\end{figure}

The bosonic primitive basis is spanned by the elementary symmetric polynomials $\{S_2,S_3,...,S_N\}=\{\sum_{ij} \bar n_i\bar n_j,\sum_{ijk}\bar n_i\bar n_j\bar n_k,...,\prod_j\bar n_j\}$ and combinations thereof. For instance, with $N=5$ particles at degree $m=6$, there are $3$ possible primitive polynomials: $S_2^3,S_3^2$ and $S_2S_4$. The fermionic primitive basis is spanned by the all the symmetric polynomials as above, \emph{plus} the degree $\binom N{2}$ Vandermonde determinant $\prod_{i<j}(\bar n_i-\bar n_j)$ shown in Fig. \ref{symantisymm4}. 

From the examples above, one easily deduces the degeneracy of the PPs $U^N_m$ to be $P(m,N)-P(m-1,N)$ for bosons and $P\left(m-\binom N{2},N\right)-\left(m-1-\binom N{2},N\right)$ for fermions, where $P(m,N)$ is the number of partitions of the integer $N$ into at most $m$ parts\cite{simon2007,davenport2012}. In particular, the degeneracy is always nontrivial ($\geq 2$) whenever $N\geq 4$ and $m\geq 4$. 

\section{Background: Fractional Chern Insulators (FCIs)}
\subsection{Physical description}

Chern Insulators (CI) are two-dimensional electron systems which generalize Integer Quantum Hall (IQH) \cite{klitzing1980} states to band insulators. In CIs, a geometric gauge field is defined in momentum space by the adiabatic transport of Bloch states. The net flux of the gauge field in the Brillouin zone is always quantized in units of $2\pi$ times an integer $C_1$ which is known as the first Chern number\cite{thouless1982,haldane88prl2015}. The latter determines the quantized Hall conductance via $\sigma_H=\frac{e^2}{h}C_1$. %Soon after the discovery of IQH states, fractional quantum Hall (FQH) states with fractionally quantized Hall conductances were also realized experimentally, with Landau levels partially filled by interacting electrons. 
Recently, FQH states have also been generalized to lattice systems without orbital magnetic field, which are therefore known as fractional Chern insulators (FCI)\cite{Tang2011,Venderbos2011,sun2011,n2011}. Fractional Chern insulators are putatively realized in partially filled energy bands with narrow band width (almost flat bands) and commensurate filling, and evidence in support of them have been found in lattice analogs of various FQH states such as Laughlin $1/m$ states, hierarchy states and non-Abelian states\cite{Hu2011,Parameswaran2012,sheng2011,Tang2011,wang2011,wu2012,Goerbig2012,Roy2012,Murthy2011,Murthy2012,Neupert2011a,Neupert2012,Venderbos2012,sun2011,Kourtis2012,Wu2012a,Jain1989,Liu2013,Lauchli2012}.

More specifically, FCIs can be realized by taking a CI such as Haldane's
honeycomb model~\cite{haldane88prl2015} as a natural starting point, and then driving the system into the flat band limit where the chemical potential lies
within the band, e.g. at fractional one third filling, which is well
separated from the other bands and hence accomplishes a FQHE-type
lattice scenario\footnote{Unlike the quantum Hall case, the FCI filling is
not given by the ratio of electrons over magnetic flux quanta, but the
chemical potential of the lattice model.}. Different groups have
recently independently pursued this direction, proposing FCI models on
the honeycomb, kagome, square, and checkerboard
lattice~\cite{Tang2011,n2011,sun2011}.
 In different ways, the flattening of the Chern band can be
accomplished through geometric frustration (e.g. long-range
hopping)~\cite{Tang2011,n2011}, multi-band
effects~\cite{sun2011}, and multi-orbital
character~\cite{Venderbos2011}. While the $s$ and
$p$-type orbitals in previous candidates materials for topological insulators would assume only moderate interactions from small
hybridizations, $d$-orbital-type systems provide an arena for both strong correlations and topological band
structures~\cite{xiao-11ncomm596}.
First numerical investigations of
the FCI phases on a torus at one third band filling found indications
of a three-fold topologically degenerate ground state separated from the other energy levels by a
gap, where the flux insertion showed
level crossings with no level repulsion between them, and the Chern
numbers of these many-body ground states found to be $1/3$
each~\cite{sheng2011,n2011}.
While this
already gives a strong hint that a Laughlin-type fractional Chern
phase might be realized, this does not yet completely rule out a
competing charge density wave (CDW) state at this filling, which can
show similar fractional Chern numbers in the ground states, level
degeneracy, and a gap. Further evidence against a CDW
state, however, has been found by finite size scaling, entanglement measures, and the distribution of ground state momenta as a
function of cluster size\cite{regnault-11prx021014}. Compared to
the FQHE for which the joint perspective of energy and entanglement measures
generally gives a consistent and complementary picture, the current
stage of FCI models particularly calls for further investigation.

Generically, the Hamiltonian for a candidate FCI model takes the form
\begin{eqnarray}
H&=&H_{nonint}+ V_{int}\label{FCI}\notag\\
&=& \sum_{k\in BZ%,\alpha\beta=1,...,N
}H_{\alpha\beta}(k)c^\dagger_{\alpha k }c_{\beta k}+V_{int}
\end{eqnarray}
where $\alpha,\beta$ are spin/pseudospin indices with summation implied and $N$ is the number of bands. %We shall be primarily concerned in engineering $H(k)=H(k_x,k_y)$ in the single-particle contribution $H_{nonint}$. %., which contains information on the band topology and dispersion. 
Let $\epsilon_n$ and $|\phi_n\rangle$ label the eigenenergy and (periodic part of the) normalized Bloch eigenstate of the $n^{th}$ band of $H(k)$: $H(k)|\phi_n(k)\rangle = \epsilon_n(k)|\phi_n(k)\rangle$, $n=1,2,...,N$. The system is insulating when there are $N_f\leq N$ completely filled bands and no partially filled band. In this work, we shall only consider the case where $N_F=1$, so that the system is analogous to a FQH system involving only the lowest Landau level.

A first condition for a 'good' FCI is that the fractionally filled band has very little dispersion\footnote{Note that we can have lower-lying filled valence bands with significantly nonuniform dispersion, since they do not contribute to the dynamics.} in $\epsilon_1(k)$ relative to its gap to the lowest unoccupied band, so that the interaction term dominates just like in FQH. 
We can quantify the uniformity of the dispersion via the \emph{flatness ratio}
\begin{eqnarray}
f=\frac{\text{bandgap}}{\text{bandwidth}}=\frac{\text{min}(\epsilon_2(k))-\text{max}(\epsilon_1(k))}{\text{max}(\epsilon_1(k))-\text{min}(\epsilon_1(k))}
\end{eqnarray}  
where $\epsilon_{1,2}(k)$ are the dispersions of the filled(valence) and lowest conduction band respectively. Henceforth, the filled eigenstate $|\phi_1\rangle$ will be simply denoted as $|\phi\rangle$. As studied in~\cite{leearovasthomale2015} and proven in~\cite{seidel2014}, $f$ cannot be arbitrarily large for a topologically nontrivial system with finite-ranged real-space hoppings. Systematic approaches for improving $f$ have be devised \cite{leearovasthomale2015,leemartin2015}, but are not within the main focus of this thesis. 

Next to  be discussed is the band topology. The lowest (fractionally fileed) band of the Chern insulators is characterized by a nonzero integer called the Chern number
\begin{equation}
%C=\frac1{2\pi}\int tr F_{xy} d^2k=\frac1{2\pi}\int  \sum_n^{N_F}F^{nn}_{xy}d^2k
C_1=\frac1{2\pi}\int F_{xy} d^2k
\end{equation}
where
\begin{equation}
F_{xy}=\partial_x a_y - \partial_y a_x% + i[a_x,a_y]
\end{equation}
is the Berry curvature tensor with $a_j=-i\langle \phi|\partial_j\phi\rangle $ the gauge connection defined by the occupied state which we will from now simply denote as $|\phi\rangle$. The Chern number is the winding number of the map from the BZ, which is a 2-torus $T^2$, to the complex projective plane $\frac{U(N)}{U(1)\times U(N-1)}$, where the set of eigenstates resides\footnote{In the case of $N_F$ occupied bands, it will be a map from the torus $T^2$ to the complex Grassmannian $\frac{U(N)}{U(N_F)\times U(N-N_F)}$. }\cite{ryu2010}. This is most easily visualized in the case of a 2-band model $H(k)=\vec d(k)\cdot \vec \sigma$ with one occupied band, where $\vec \sigma=(\sigma_1,\sigma_2,\sigma_3)$ refers to the Pauli matrices. It maps $T^2$ to $\frac{U(2)}{U(1)\times U(1)}\sim S^2$ with Berry curvature $F_{xy}=\frac1{2}\vec d\cdot(\partial_x\vec d\times \partial_y \vec d)$. This is just the area on the Bloch sphere swept out by an unit area element on the BZ. 

As such, $ F_{xy}$ can be interpreted as the 'Jacobian' of this map, whose uniformity is quantified by the mean-square deviation
\begin{equation}
\langle(\Delta F)^2\rangle= \frac{1}{4\pi^2}\int \left( F_{xy} - \frac{C_1}{2\pi}\right)^2d^2k
\label{dFxy}
\end{equation}
It was shown in Ref. \onlinecite{Roy2012} that, to second order, the long wavelength limit of the FCI density algebra coincides with that of FQH if $\langle(\Delta F)^2\rangle=0$. Note that $\langle(\Delta F)^2\rangle$ has a finitely large lower bound for $N=2$ bands, because it is impossible to have a map $T^2\rightarrow S^2$ that has a constant Jacobian, as can be easily seen by drawing a grid on these manifolds. However, $\langle(\Delta F)^2\rangle\rightarrow 0$ is theoretically achievable for $N\geq 3$, as will be shown in~\cite{leemartin2015}.

Another important consideration is the \emph{quantum distance} $D(k,k+dk)$ between two points $k$ and $k+dk$ on the BZ of the FCI:
\begin{equation}
D(k,k+dk)=tr[\mathbb{I}-P_kP_{k+dk}]=g_{ij}dk_idk_j
\end{equation}
where $tr$ is the trace taken over all filled states, and $P_k=\sum_{n}^{N_F}|\phi_n(k)\rangle\langle \phi_n(k+dk)|$ is the projector onto the filled eigenstates at the point $k\in BZ$. In our case with $N_F=1$, $g_{ij}$ is the \emph{Fubini-Study metric} which characterizes the geometry. Intuitively, $D(k,k+dk)$ is a measure of how fast the projector $P_kP_{k+dk}$ differs from the identity as $dk$ is introduced. After taking the trace, it reduces to the usual definition of state overlap $D(k,k+dk)|_{N_F=1}=1-|\langle \phi(k)|\phi(k+dk)\rangle|^2$. Explicitly~\cite{marzari1997},
\begin{eqnarray}
g_{ij}&=&Re\sum_n\langle\partial_i\phi_n|\partial_j \phi_n\rangle-\sum_{mn}\langle \partial_i\phi_n| \phi_m\rangle\langle \phi_m|\partial_j\phi_n\rangle\notag\\
&=&Re\sum\langle\partial_i\phi|\partial_j \phi\rangle-a_ia_j
\label{gij}
\end{eqnarray} 
for one occupied band. 

As we shall see later, $Tr(g)=\sum_i g_{ii}$ enters in the lower bound of a certain locality measure of QH Pseudopotentials that are mapped onto FCIs. A more recent work also showed that FCI models which satisfy the so-called \emph{Ideal Droplet Condition}\cite{claassen2015} $2\sqrt{Det(g)}=Tr(g)$ possess an attractive dual first quantized description reminiscent of FQH systems.

\subsection{The Wannier basis}
\label{sec:wannierbasis}

A Chern Insulator possesses a Wannier basis that bears qualitative similarities to the Landau gauge basis introduced in the last section. This Wannier basis was first employed in the same context in %In this section, we review the Wannier state representation of FCIs proposed in Ref. \onlinecite{qi11prl126803}. The idea of this approach is to find a suitable single-particle basis, the one-dimensional (1D)Wannier state basis, and to use this basis to establish an exact mappingbetween FCI and FQH states. 
\onlinecite{qi2011} on a cylindrical geometry, but can also be formulated on the torus geometry\cite{yu2011} which shall be used in the following. (The torus formulation of the Wannier state representation has also been investigated independently in Ref. \cite{wu2012,gunnar}.)

Consider a band insulator with the Hamiltonian
\begin{eqnarray}
H=\sum_{i,j,\alpha,\beta}c_{i\alpha}^\dagger h_{ij}^{\alpha\beta}c_{j\beta},
\end{eqnarray}
with $i,j$ being the site indices of a two-dimensional lattice with periodic boundary conditions and $\alpha,\beta=1,2,..,N$ labeling internal states in each unit cell such as orbital and spin states. Due to translational symmetry $h_{ij}^{\alpha\beta}=h_{{\bf r}_j-{\bf r}_i}^{\alpha\beta}$, the Hamiltonian can be written in momentum space as
\begin{equation}
H=\sum_\vk c_{\vk \alpha}^\dagger h_\vk^{\alpha\beta}c_{\vk
  \beta},\label{Hk}
\end{equation}
with
\begin{eqnarray}
c_{\vk\alpha}&=&\frac1{\sqrt{L_xL_y}}\sum_{i}c_{i\alpha}e^{-i\vk\cdot{\bf r}_i},\nonumber\\
h_{{\bf r}_j-{\bf r}_i}^{\alpha\beta}&=&\frac1{L_xL_y}\sum_\vk h_{\vk}^{\alpha\beta}e^{-i{\vk\cdot(\vr_j-\vr_i)}}.\nonumber
\end{eqnarray}
We use $L_x,L_y$ to denote the number of lattice sites in $x$ and $y$ direction, respectively. The momentum $\vk$ takes values of $\left(\frac{2\pi n_x}{L_x},\frac{2\pi n_y}{L_y}\right)$, with $n_x=1,2,...,L_x,~n_y=1,2,...,L_y$ being  integers. The Hamiltonian matrix $h_\vk$ can be diagonalized to obtain the eigenstates
\begin{eqnarray}
h_\vk\ket{\phi(n,\vk)}=\epsilon_{n\vk}\ket{\phi(n,\vk)},~n=1,2,..,N.
\end{eqnarray}
We are interested in a system with a lowest energy band $\epsilon_{1\vk}$ occupied, and a gap separating this band from all other bands. Since only the lowest band will be involved, we will denote $\ket{\phi(1,\vk)}$ by $\ket{\phi(\vk)}$ for simplicity.

In the thermodynamic limit $L_x,L_y\rightarrow \infty$, ${\bf k}$ is a
good quantum number and the Berry's phase gauge field ${\bf
  a}=-i\bra{\phi(\vk)}\nabla_\vk\ket{\phi(\vk)}$ can be defined. It determines
the first Chern number as the flux of the gauge field in the Brillouin
zone: $C_1=\frac1{2\pi}\int_{BZ} d^2\vk \;\nabla\times{\bf a}$. For the
realization of FCIs, we are interested in a band with $C_1\neq 0$. More
specifically, we will primarily focus on $C_1=1$
systems. Moreover, for finite $L_x,L_y$, it is necessary to generalize
the definition of a Berry's phase gauge field and Chern number to
the case of $\ket{\phi(\vk)}$ with a discrete $\vk$ variable.

We start from the definition of 1D Wannier states
\begin{eqnarray}
\ket{W_{nk_y}}=\frac1{\sqrt{L_x}}\sum_{k_x}e^{-ik_xn}e^{i\varphi_\vk}\ket{\phi(\vk)},\label{eq:Wannier}
\end{eqnarray}
which is a Fourier transform of $\ket{\phi(\vk)}$ in the $x$-direction only, so that $k_y$ is still a good quantum number. Since the state $\ket{\phi(\vk)}$ is only determined by the Hamiltonian up to a phase, the phase factor $e^{i\varphi_\vk}$ is not pre-determined. As was discussed in Refs. \onlinecite{qi2011,kivelson1982}, the phase ambiguity can be fixed by defining the Wannier function to be an eigenstate of the projected position operator $\hat{x}=PxP$, with
$P=\sum_\vk\ket{\phi(\vk)}\bra{\phi(\vk)}$ the projection operator to the occupied
band, and $x=\sum_{i,\alpha}x_i\ket{i,\alpha}\bra{i,\alpha}$ the
position operator. However, in a system with periodic boundary
conditions in $x$-direction, it will be slightly more problematic to apply
this definition of $x$ due to the dependence on the choice of the boundary site. As pointed out in Ref.~\onlinecite{yu2011}, this problem can be resolved by defining a unitary operator
\begin{eqnarray}
X=\exp\left[ix\frac{2\pi}{L_x}\right]=\exp\left[i\sum_{i}\frac{2\pi x_i}{L_x}\sum_{\alpha}\ket{i\alpha}\bra{i\alpha}\right].
\end{eqnarray}
This definition preserves the periodicity $x\rightarrow x+L_x$. The eigenstates of this operator are the states localized on a given site $n$ in $x$-direction. We thus define the projected operator to be
\begin{eqnarray}
\hat{X}=PXP.
\end{eqnarray}
In momentum space, $\bra{\phi(\vk)}\hat{X}\ket{\phi(\vk')}=\bra{\phi(\vk)}X\ket{\phi(\vk')}$. It is easy to see that $X$ shifts the momentum $k_x$ by $2\pi/L_x$, since
\begin{eqnarray}
\ket{\phi(\vk)}&=&\frac1{\sqrt{L_xL_y}}\sum_{i,\alpha}u_{\vk \alpha}e^{ix_ik_i}\ket{i,\alpha}\nonumber,\\
X\ket{\phi(\vk)}&=&\sum_{i,\alpha}u_{\vk \alpha}e^{ix_i\left(k_x+\frac{2\pi}{L_x}\right)}\ket{i,\alpha},
\end{eqnarray}
with $u_{\vk \alpha}=\sqrt{L_xL_y}\braket{0,\alpha}{\phi(\vk)}$ the periodic part of the Bloch wave function.
Therefore, the only nonzero matrix element of $\bra{\phi(\vk)}X\ket{\phi(\vk')}$ is
\begin{eqnarray}
F_{k_xk_y}&\equiv&\bra{k_x+2\pi/L_x,k_y}X\ket{k_x,k_y}
\nonumber\\
&=&\sum_\alpha u_{k_x+\frac{2\pi}{L_x},k_y,\alpha}^*u_{k_x,k_y,\alpha}.
\end{eqnarray}
In the subspace of states with a fixed $k_y$, the matrix of $\hat{X}$ in the momentum basis is
\begin{eqnarray}
\hat{X}=\left(\begin{array}{ccccc}0&&..&&F_{2\pi}\\F_{2\pi/L_x}&0&&&\\&F_{4\pi/L_x}&&&..\\
..&&..&&\\&&&F_{(L_x-1)\pi/L_x}&0\end{array}\right),
\end{eqnarray}
where the omitted index $k_y$ is the same for all states.

\begin{figure}
\begin{minipage}{0.7\linewidth}
\includegraphics[width=\linewidth]{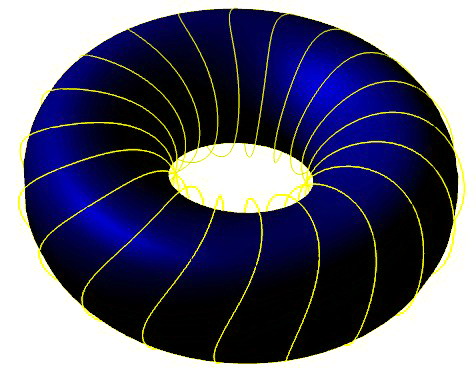}
\end{minipage}
\caption{  Shift of the Wannier polarization as $k_y$
  varies. The azimuthal direction represents the real space $x$ while
  the poloidal direction represents the periodic domain of $k_y$. For
  any complete cycle along $y$, i.e. $k_y\rightarrow k_y + 2\pi$, the
  Wannier state shifts by $x\rightarrow x+1$.}
\label{fig:wann_shift}
\end{figure}

Upon taking the thermodynamic limit $L_x\rightarrow \infty$, $F_{k_xk_y}\simeq 1-i\frac{2\pi}{L_x}a_{x}(\vk)$, with $a_x$ the $x$ component of the Berry's phase gauge field. For finite $L_x$, $\left|F_{k_xk_y}\right|$ should be close to but not exactly equal to $1$. Therefore, $\hat{X}$ is an approximately unitary matrix. To define the maximally localized Wannier states, we deform the $\hat{X}$ operator to a unitary operator by defining
\begin{eqnarray}
F_{k_xk_y}&=&\left|F_{k_xk_y}\right|e^{-iA_{k_xk_y}}\nonumber\\
\bar{X}&=&\left(\begin{array}{ccccc}0&&..&&e^{-iA_{2\pi}}\\e^{-iA_{2\pi/L_x}}&0&&&\\&e^{-iA_{4\pi/L_x}}&&&..\\
..&&..&&\\&&&e^{-iA_{(L_x-1)\pi/L_x}}&0\end{array}\right).\nonumber\\
\end{eqnarray}
Here, the index of the rows and columns are
$k_x=0,\frac{2\pi}{L_x},...,2\pi-\frac{2\pi}{L_x}$. The eigenstates of
the $\bar{X}$ operator form an orthogonal complete basis. Due to the simple form of $\bar{X}$ in momentum space, one can prove that the eigenstates of $\bar{X}$ are Wannier states defined in Eq. (\ref{eq:Wannier}), with the phase $\varphi_\vk$ defined by
\begin{eqnarray}
\ket{W_{nk_y}}&=&\frac1{\sqrt{L_x}}\sum_{k_x}e^{-ik_xn}e^{i\varphi_\vk}\ket{\phi(\vk)},\nonumber\\
\varphi_\vk&=&-\sum_{0\leq p_x< k_x}A_{p_xk_y}-k_xP_x(k_y),\label{eq:MLWF}
\end{eqnarray}
where
\begin{equation}
P_x(k_y)=-\frac1{2\pi}\sum_{0\leq p_x<2\pi}A_{p_xk_y}.\nonumber
\end{equation}
This definition is periodic in $k_x\rightarrow k_x+2\pi$. The corresponding eigenvalues are
\begin{eqnarray}
\bar{X}\ket{W_{nk_y}}=e^{i\frac{2\pi}{L_x}\left(n+P_x\right)}\ket{W_{nk_y}}.
\end{eqnarray}
Therefore, we see that the center-of-mass (CM) position of the state
$\ket{W_{nk_y}}$ is shifted by $P_x$ away from the lattice site
position $n$. This fact indicates that $P_x(k_y)$ has the physical
meaning of charge polarization\cite{kingsmith1993}. In the large $L_x$ limit, $A_{\vk}\rightarrow \frac{2\pi}{L_x}a_x$ and $P_x(k_y)=-\frac1{2\pi}\int_0^{2\pi}a_xdk_x$. Since $P_x(k_y)$ is a $U(1)$ phase for each $k_y$, one can define its winding number when $k_y$ goes from $0$ to $2\pi$:
\begin{equation}
C_1=\sum_{n=1}^{L_y}\frac{i}{2\pi}\log\left(e^{i2\pi\left[P_x(2\pi n/L_y)-P_x(2\pi(n-1)/L_y)\right]}\right).
\label{C1winding}
\end{equation}
The $\log$ function is defined to restrict the value of the exponential phase
difference to the region of $[-\pi,\pi)$. As long as $L_y$ is not so small that
$P_x(k_y)$ can jump by an integer between two neighboring $k_y$ values, the $C_1$ obtained above agrees with the Chern number in the large $L_y$ limit.

Some examples of the Wannier basis given by Eq. \ref{eq:MLWF}, as long as their corresponding polarizations, will be given in Appendix \ref{sec:wannpol}. There the connection between the Wannier polarization and the Berry curvature distribution in the BZ will also be explicitly discussed. The Wannier polarization is already related to the entanglement spectrum of the system to be discussed in Part II. This will be elaborated in Appendix \ref{wannES}.

There is a subtle point to be discussed before leaving this section. The definition of maximally localized Wannier states in
Eq.~\ref{eq:MLWF} leaves an ambiguity in the relative phase
between different $\ket{W_{nk_y}}$. If we redefine
$\ket{W_{nk_y}}\rightarrow e^{i\theta_{k_y}}\ket{W_{nk_y}}$ with any
phase $\theta_{k_y}$, all the results discussed above still holds. The WSR, however, depends on this choice and different choices of phase corresponds to physically different mappings between FCI and FQH systems. To preserve the locality in the mapping, a choice should be made which makes $\ket{W_{nk_y}}$ continuous in $k_y$ in the large $L_y$ limit. An example of the choice is the following~\cite{maissam}:
\begin{eqnarray}
\theta_{k_y}&=&-\sum_{0\leq p_y<k_y}A'_{p_y}-k_yP_{y0},\\
P_{y0}&=&-\frac1{2\pi}\sum_{0\leq p_y<2\pi}A'_{p_y},\nonumber\\
A'_{p_y}&=&-{\rm Im}\log\left(\sum_{\alpha}u_{0,p_y+\frac{2\pi}{L_y},\alpha}^*u_{0,p_y,\alpha}\right).\nonumber
\end{eqnarray}
In the $L_y\rightarrow \infty$ limit, $A'_{p_y}\simeq
a_y(0,p_y)\frac{2\pi}{L_y}$. This choice of $\theta_{k_y}$ corresponds
to a gauge transformation which makes $a_y({\bf k})$ uniform along the
$k_x=0$ line. Any other gauge choice
$\theta'_{k_y}=\theta_{k_y}+\delta\theta_{k_y}$ also works and
describes topologically equivalent states, as long as
$\delta\theta_{k_y}$ is a smooth periodic function of $k_y$ in the
large $L_y$ limit. While different gauge choices $\delta\theta_{k_y}$ of the Wannier states do not change the topological universality class of the associated state, it can be used as variational parameters in the many-body ground state and can be optimized numerically by the comparison of the Wannier ground state with the exact ground state.\cite{wu2012,gunnar}. This optimization will be performed later , in the maximization of the locality of reverse engineered PPs on the lattice.

\section{The Wannier State Representation (WSR) exact mapping between FQH and FCI systems}		
\label{sec:WSR}

\begin{figure}%[htb]
\includegraphics[width=\linewidth]{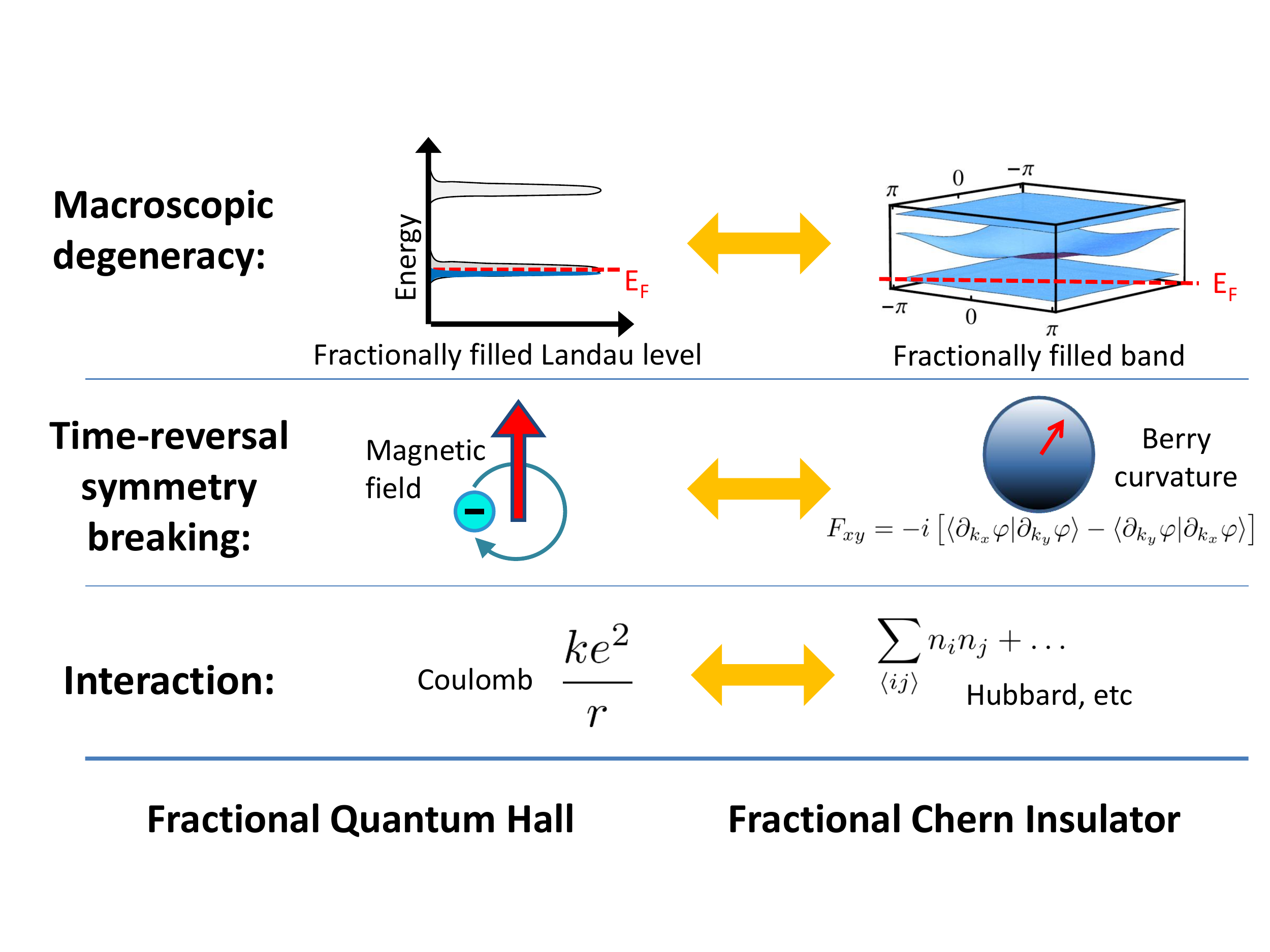}
\includegraphics[width=\linewidth]{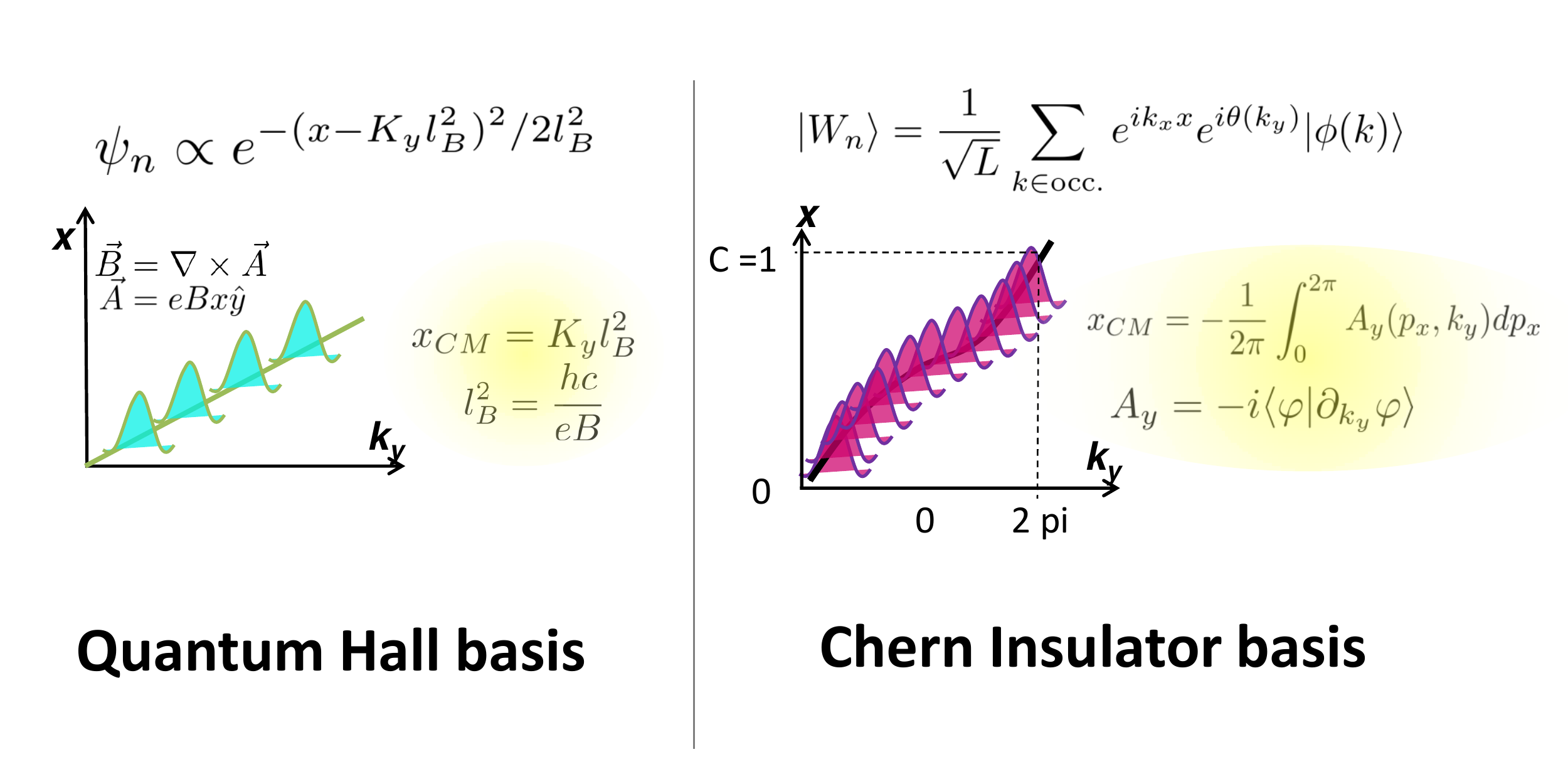}
%\captionsetup{justification=centerlast}
\caption{Top) Comparison between the physically analogous characteristics of the FQH and FCI systems. Bottom) Comparison between the salient similarities between their respective bases. Note that these two bases are qualitatively similar, but can never be exactly the same as the QH basis has a Gaussian profile, while the CI basis decays exponentially.}
\label{fig:FQHFCIcomparison}
\end{figure}

Having introduced both the FQH and FCI systems and their respective bases, it is now the time to describe the Wannier State Representation that provides an exact mapping between them. The key insight is that both bases exhibit a ``twisted boundary condition" in $k_y$, with the center-of-mass $x_{CM}=l_B^2 k_y$ (Eq. \ref{LLL}) for the quantum Hall system, and $x_{CM}=-P_x(k_y)$ (Eq. \ref{C1winding}) for the Chern insulator system. This is illustrated in Fig. \ref{fig:FQHFCIcomparison}.

But there is a caveat: The amounts of the twist per period of $k_y$ is not the same for both systems. This is because while the QH system has unbroken continuous translational symmetry, the CI system only has lattice translational symmetry. 

\begin{figure}
\begin{minipage}{0.99\linewidth}
\includegraphics[width=\linewidth]{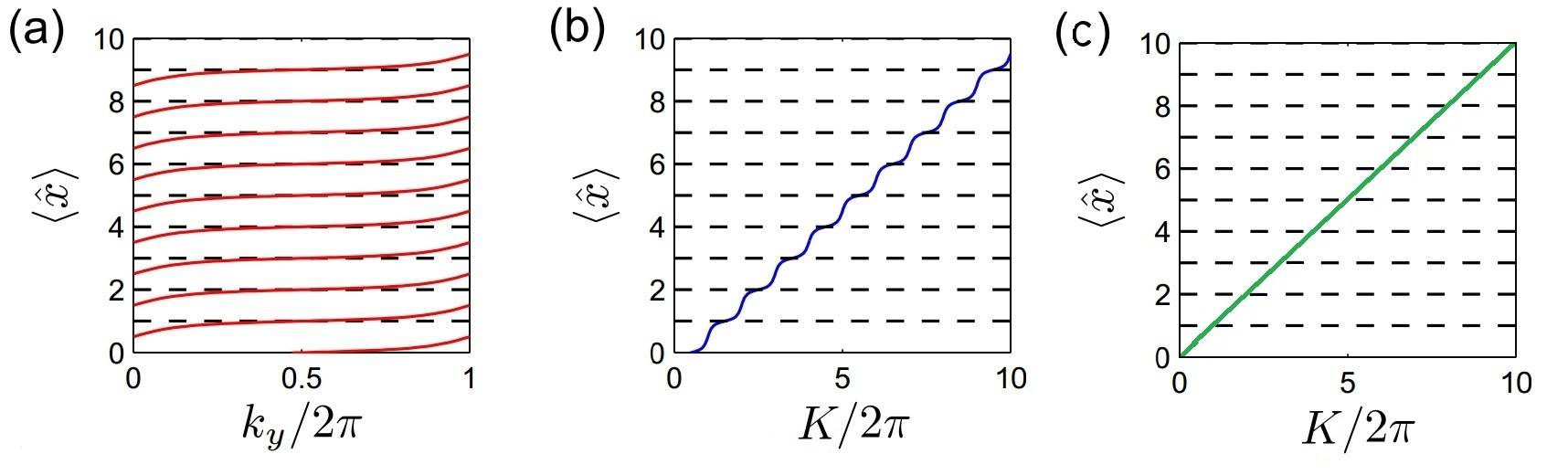}
\end{minipage}
\caption{  Wannier center evolution of the Dirac model used in Ref. \onlinecite{qi2011} from $x=0$ to
  $x=10$ as a function of (a) $k_y/2\pi$ and (b) of the extended wave vector $K/2\pi$. From (a) to (b), it becomes visible how the evolution
of $x$ changes
as a function of $K$, suggesting its similarity to Landau level wavefunctions (c).}
\label{fig:wann}
\end{figure}

The remedy is to concatenate the Wannier basis such that the end of the wavefunction on one site is joined with the beginning of the one on the next site. This is illustrated in Fig. \ref{fig:wann}. A more detailed explanation is as follows. The Wannier states $\ket{W_n(k_y)}$ have a ``twisted
boundary condition" in $k_y$, since $P_x(k_y+2\pi)=P_x(k_y)+C_1$, such
that $\ket{W_n(k_y+2\pi)}=\ket{W_{n+C_1}(k_y)}$. As is illustrated in
Fig.~\ref{fig:wann_shift}, for $C_1=1$ the Wannier state CM
position $x_n(k_y)=n+P_x(k_y)$ forms a helical curve on the parameter
space torus of  $x,k_y$. %The key observation which enables the Wannier state representation of FCI is the fact that the twisted boundary condition allows to label all Wannier states in such a 2D system by a 1D parameter. 
If we define
\begin{eqnarray}
\ket{W_{k_y+2\pi n}}=\ket{W_{n,k_y}},~\text{for~}k_y\in[0,2\pi),
\end{eqnarray}
we arrive at a $\ket{W_K}$ with $K=k_y+2\pi n$ a continuous function of $K$ (in the large $L_y$ limit). The CM position $x_{CM}=\bra{W_K}\bar{X}\ket{W_K}=e^{i\frac{2\pi}{L_x}
\left(n+P_x(k_y)\right)}$ for $K\in[2\pi n,2\pi(n+1))$ is continuous
in $K$ and satisfies $x_{K+2\pi}=x_{K}+1$. In this sense, $x_{CM}$ increases approximately linearly with $K$ (Fig.~\ref{fig:wann}).

Due to this behavior of $\ket{W_K}$, an exact mapping can be defined
between the Wannier states in the $C_1=1$ FCI and the LLL states in a
FQH problem. Consider a spinless fermion with the Hamiltonian
$H=\frac1{2m}({\bf p}-{\bf A})^2$ on a torus of the size $L_xl_B\times
L_yl_B$ with the uniform perpendicular magnetic field
$\nabla\times{\bf A}=B=2\pi/l_B^2$. The total number of flux is
$N_\Phi=L_xL_y$, so that the LLL contains the same dimension of
Hilbert space as the lattice model discussed above on a lattice of the
size $L_x\times L_y$. The lowest Landau level Landau gauge wave functions have the form
\begin{eqnarray}
\psi_K(x,y)&=&\frac1{\sqrt{2\pi l_B^2}}\sum_{n\in {\rm Z}}e^{iKy/l_B-\pi\left(\frac{x}{l_B}-\frac{K}{2\pi}-nL_x\right)^2}\nonumber\\
&\equiv &\frac1{\sqrt{2\pi l_B^2}}e^{iKy/l_B-\pi\left(\frac{x}{l_B}-\frac{K}{2\pi}\right)^2}\nonumber\\
& &\cdot\vartheta\left(-i{L_x}\left(\frac{x}{l_B}-\frac K{2\pi}\right)
  \vert {iL_x^2}\right)\label{torusLLWF}
\end{eqnarray}
with $\vartheta(z \vert \tau)$ the Jacobi theta function~\cite{Abra} which are
appropriate superpositions of the cylindric wave functions.

Notice that we have defined the momentum $K$ slightly differently from
 the usual definition used in Ref.~\onlinecite{lee2004,qi2011} and in the subsequent sections, so that here $K$ is dimensionless and given by $K=\frac{2\pi}{L_y}n,~n\in{\mathbb{Z}}$ on the $L_xl_B\times L_yl_B$ torus. This definition leads to identical results as the usual definition used later if we replace the $l_B$ here by $\sqrt{2\pi}l_B$.

For $L_x\gg 1$, this wave function has a Gaussian profilearound the CM position $x_{CM}=K\frac{l_B}{2\pi}$. Denoting $\ket{\psi_K}$ as the state corresponding to wave function $\psi_K(x,y)$, one can define a unitary mapping between the Hilbert spaces of the QH Landau level and the lattice $C_1$ Chern insulator:
\begin{eqnarray}
f:~{\rm H}_{\text{CI}}&\longrightarrow& {\rm H}_{\text{LLL}},\nonumber\\
f\left(\ket{W_K}\right)&=&\ket{\psi_K},\label{mapFQHFCI}
\end{eqnarray}
with ${\rm H}_{\text{CI}}$ and ${\rm H}_{\text{LLL}}$ denoting the Hilbert spaces of the CI and LLL, respectively.
Such a mapping preserves the continuity in $K$ and also the
topological properties of $\ket{W_K}$ and $\ket{\psi_K}$, i.e., their
winding while momentum $K$ is increased. Using the reverse map
$f^{-1}$, the many-body states of the LLL, such as
Laughlin states and other FQH states defined in the LLL, can all be mapped to corresponding states in the
FCI. Similarly, a Hamiltonian $H$ of a FQH system can also be mapped
to a corresponding Hamiltonian $H_{\text{FCI}}=f^{-1}Hf$. The main
purpose of the next two sections is to study the Hamiltonians
$H_{\text{FCI}}$ which are mapped from the PP Hamiltonians in the FQH system. One can also perform the reverse, mapping the FCI
Hamiltonian such as a Hubbard type interaction $H_U$ of the
lattice fermions to a FQH Hamiltonian
$H_{\text{FQH}}=fH_Uf^{-1}$. 

\section{Pseudopotential expansion of FCI interactions}

The WSR representation just described allows one to analyze interactions on a Fractional Chern Insulator in terms of the Pseudopotential operators on an FQH system. Specifically, one can consider a 2-body FCI interaction $H_{int}$, find its matrix elements in its Wannier basis, and compare them with the FQH Pseudopotential matrix elements in the Landau gauge basis. This will be described in the following few subsections.

\subsection{Expressing $H_{\text{int}}$ in the Wannier basis}

Consider for definiteness the most realistic class of 2-body interactions, which are Hubbard interactions. Define $H_{int}$ as below, such that the parameter $\lambda$ interpolates between the nearest-neighbor (NN) and next-nearest-neighbor (NNN) Hubbard interactions $n_i=c^\dagger_ic_i$. The first step is to perform a Fourier transform on $H_{\text{int}}$ since the Wannier basis has a $k_y$ Wannier index. Doing so, we find
\begin{eqnarray}
H_{\text{int}}&=&\lambda h_0\sum_{\langle ij \rangle}n_in_j + (1-\lambda)h_0\sum_{\langle \langle ij \rangle\rangle}n_in_j \nonumber\\
&=& \sum_qV^{\alpha\beta}(q)n_{-q\alpha}n_{q\beta} ,
\end{eqnarray}
where $q$ is an internal momentum variable, $\alpha,\beta$ are the sublattice
indices, and $n_{q\alpha} =
\sum_{k_xk_y}c^\dagger_{k+q,\alpha}c_{k\alpha}^{\phantom{\dagger}}$. $V^{\alpha \beta}(q)$ denotes the $q$th
Fourier component of the interaction between the sublattice index $\alpha$
and $\beta$. This is an expression quartic in the creation and annihilation operators $c^\dagger,c$ in the momentum/sublattice basis. Since we are only considering interactions within the flat band, we project out the upper band and keep only the matrix elements of $n_{q\alpha}$ in the flat band. After projecting out the upper band, the annihilation operator $c_{k\alpha}$ can be expanded in the Wannier state basis:
\begin{eqnarray}
c_{\vec{k}\alpha}&=&\sum_{n}\braket{\vec{k}\alpha}{W_{2\pi n/L_y}}a_n +\text{upper band contributions}\nonumber\\
&\rightarrow &\sum_{n}\braket{\vec{k}\alpha}{W_{2\pi n/L_y}}a_n \nonumber\\
&:=&\sum_{n} U_{xk_x\alpha k_y} a_{x,k_y},
\label{normbase}
\end{eqnarray}
since $\frac{2\pi n}{L_y}=K=k_y+  2\pi x C_1=k_y+  2\pi x $. In this representation, the density operator becomes
\begin{eqnarray}
n_{q\alpha} &=& \sum_{k_xk_y}L_x^{-1/2}c^\dagger_{k+q,\alpha}c_{k\alpha} \nonumber \\
&=&\sum_{k_xk_y}L_x^{-1/2}\sum_{x_1}U^*_{x_1,k_x+q_x,\alpha,k_y+q_y}a^\dagger_{x_1,k_y+q_y}\nonumber\\
&&\sum_{x_2}U_{x_2k_x\alpha k_y}a_{x_2k_y}\nonumber \\
&=&\sum_{k_y x_1x_2}N_{q\alpha x_1x_2k_y}a^\dagger_{x_1,k_y+q_y}a_{x_2k_y}^{\phantom{\dagger}},
\nonumber \\
\end{eqnarray}
where the normalization factors $N_{q\alpha x_1x_2k_y}$ follow from~\eqref{normbase}.
We obtain
\begin{eqnarray}
H_{\text{int}}&=&\sum_{q\alpha\beta}V^{\alpha \beta}_qn_{-q,\alpha}n_{q\beta} \nonumber \\
%&=&\sum_{q_x
%  q_y,k1_y,k2_y}\sum_{\{x_j\}}V^{ab}_qN_{-qx_1x_2k1_y}N_{+qx_3x_4k2_y}a^\dagger_{x_1k1_y-q_y}
%a_{x_2k1_y} a^\dagger_{x_3k2_y+q_y} a_{x_4k2_y} \nonumber \\
&=&\sum_{q_x}\sum_{\{x_j\},\{K_{jy}\},\alpha\beta}V^{\alpha\beta}_q\delta_{q_y-k_{2y}+k_{1y}}\delta_{k_{2y}-k_{1y}-k_{3y}+k_{4y}}N_{-q\alpha x_1 x_2k_{2y}}N_{q\beta x_3x_4k_{4y}}
a^\dagger_{K_1} a_{K_2}^{\phantom{\dagger}} a^\dagger_{K_3} a_{K_4}^{\phantom{\dagger}} \nonumber \\
&=&\sum_{K_1K_2K_3K_4}(h^{int})_{K_1K_2K_3K_4}a^\dagger_{K_1}
a_{K_2}^{\phantom{\dagger}} a^\dagger_{K_3} a_{K_4}^{\phantom{\dagger}} \nonumber \\
%&=&-\sum_{K_1K_2K_3K_4}(h^{int})_{K_1K_2K_3K_4}a^\dagger_{K_1}a^\dagger_{K_3}
%a_{K_2} a_{K_4} + quadratic \nonumber \\
&\simeq& -\sum_{n'n}\sum_{l_1l_2}f(n,n',l_1,l_2)(a_{(n-l_1)/2}a_{(n+l_1)/2})^\dagger(a_{(n'-l_2)/2}a_{(n'+l_2)/2}).
\label{basischange}
\end{eqnarray}
A quadratic term has been dropped in the final step because it
can be absorbed into the noninteracting part of the Hamiltonian. The latter is irrelevant for our current purpose of expressing the interaction operator in the Wannier basis. Note, however, that this quadratic term should not be omitted if we were to perform studies on energetics. Due to fermionic statistics of the $c_{n\pm l}$ operators, we can antisymmetrize $H_{\text{int}}$, leading to the matrix elements
\begin{eqnarray}
h(n,n',l_1,l_2)&=&f(n,n',l_1,l_2)-f(n,n',-l_1,l_2) \nonumber \\
&&-f(n,n',l_1,-l_2)+f(n,n',-l_1,-l_2),
\end{eqnarray}
which are manifestly antisymmetric in $l_1$ and $l_2$. We see from Eq.~\ref{basischange}
that $h(n,n',l_1,l_2)$ corresponds to a pair hopping interaction on a
line, analogous to the FQH Pseudopotentials. Two particles with the CM ``position" $n'$ separated by $l_2$ sites simultaneously hop onto new positions with CM position $n$ separated by $l_1$ sites (see also Fig.~\ref{fig:nonconservation}). More discussion on the physical interpretation of this interaction will be presented in Section~\ref{sec:CMMT}.

\subsection{Pseudopotentials and interactions in the Landau gauge basis}

The WSR mapping $f$ provides an exact mapping between the Landau gauge basis, and the Wannier basis for which Hubbard-type interaction matrix elements were previously derived. 

In the Landau gauge basis, a 2-body interaction $V(\bs{r}_i-\bs{r}_j)$ may be expanded in terms of the PPs $U^m$ which in first-quantized form involve Lauguerre polynomials $L_m$. A detailed derivation can be found in Appendix \ref{nbody}, where a more general treatment will be given. With that, the interaction is decomposed into the form 
\begin{eqnarray}
V(\bs{r}_i-\bs{r}_j)&=&\sum_m V^mU^m(\bs{r}_i-\bs{r}_j) \nonumber \\
&=& \sum_m l \int d^2\bs{p} V_m L_m(\bs{p}^2l^2)
e^{-\bs{p}^2 l^2/2} e^{i\bs{p}(\bs{r}_i-\bs{r}_j)} \nonumber \\
&=& \sum_m V^m L_m (-l_B^2 \nabla^2) \delta^2 (\bs{r}_i -\bs{r}_j).
%e^{-\vert \bs{r}_i -
%\bs{r}_j \vert^2/2l^2}.
\label{lag}
\end{eqnarray}
which is fully characterized by the coefficients $V^m$. This decomposition is valid for sufficiently short-ranged potentials $V(\bs{r_i}-\bs{r_j})$. Note that the functional form of Eq.~\ref{lag} is different
from that in some papers i.e. Ref.~\onlinecite{rezayi-94prb17199} because we have used ordinary coordinates instead of guiding
center coordinates. If one replaces all
coordinates including their derivatives with their guiding center
analogues, $l_B^2 \delta^2 (\bs{r}_i -\bs{r}_j)$ will be replaced by the exponential tail expression
$e^{-\vert \bs{r}_i - \bs{r}_j \vert^2/2l_B^2}$ in Ref.~\onlinecite{rezayi-94prb17199}. The details of this calculation are shown in Appendix~\ref{equi}.

To find the expansion coefficients $V^m$, one can Fourier transform Eq.~\ref{lag} and
invoke the orthogonality relation of the Laguerre polynomials (see also Appendix~\ref{nbody}) to obtain
\begin{equation}
V^m=4\pi l_B^2\int \frac{ d^2k}{(2\pi)^2} e^{-l_B^2 k^2}L_m(k^2l_B^2)V(k).
\label{Vm}
\end{equation}

The above expression, which will also be rederived in Appendix~\ref{nbody} as a special case of a much more general result obtained from first principles, enables us to compute the PP coefficients $V^m$ directly from a generic  potential $V(k)$. It is the starting point for the generalization to interactions involving more than two bodies, as is described in Section VI.

To apply the PP decomposition to an FCI system with periodic
boundary conditions in both directions, we have to compactify the open direction of
the cylinder. The single particle states on the torus are the
$\psi_K(x,y)$ defined in Eq.~\ref{torusLLWF}. In this basis, the $m$th PP Hamiltonian has  matrix elements
\begin{eqnarray}
U^m_{n_1,n_2,n_3,n_4}&=& \int d^2 \bs{r}_i d^2
\bs{r}_ j (\psi_{n_1}(\bs{r}_i)^* \psi_{n_2}(\bs{r}_j)^* -
\psi_{n_2}(\bs{r}_i)^* \psi_{n_1}(\bs{r}_j)^*)\notag\\
&&\times U^m(\bs{r}_i-\bs{r}_j) (\psi_{n_3}(\bs{r}_i) \psi_{n_4}(\bs{r}_j) -
\psi_{n_4}(\bs{r}_i) \psi_{n_3}(\bs{r}_j)).
\end{eqnarray}
Recall that $U^m$ refers to a normalized PP that has nonzero
projection only in the $m$th relative angular momentum sector, and form a basis in which a generic potential $V$ is expanded. To avoid confusion,
the $V^m$s which appear in Eq.~\ref{Vm} and other places below refer
to the overlap of $V$ with the PP $U^m$ in the relative angular momentum sector $m$. For simplicity, $\psi_{K=2\pi n/L_y}({\bf r})$ has been denoted interchangeably as
$\psi_n({\bf r})$, $n=1,2,...,L_xL_y$.

We can use the map
(\ref{mapFQHFCI}) to define the
corresponding PP Hamiltonian in FCI, which has the second-quantized form
\begin{eqnarray}
U^m&=&\sum_{n_1,n_2,n_3,n_4} a_{n_1}^{\dagger} a_{n_2}^{\dagger}
U^m_{n_1,n_2,n_3,n_4} a_{n_3}^{\phantom{\dagger}}
a_{n_4}^{\phantom{\dagger}}, \nonumber \\
&=&\sum_{n+l_1\in \mathbb{Z},n+l_2\in \mathbb{Z}} U_{l_1l_2}^{m} a_{n+l_1}^{\dagger} a_{n-l_1}^{\dagger}
 a_{n-l_2}^{\phantom{\dagger}}
a_{n+l_2}^{\phantom{\dagger}}. \label{1d}
\end{eqnarray}
Here,
\begin{eqnarray}
a_n=\sum_{i,\alpha}\braket{W_{2\pi n/L_y}}{i\alpha}c_{i\alpha}
\end{eqnarray}
 is the annihilation operator of the Wannier state $\ket{W_{2\pi n/L_y}}$. Note the change of notation: the matrix element $U^m_{n_1n_2n_3n_4}$ is rewritten in the form $U^m_{l_1l_2}=U^m_{n+l_1,n-l_1,n-l_2,n+l_2}$ consistent with magnetic translation symmetry. (More on issues regarding magnetic translation symmetry will be presented in Sect.~\ref{sec:CMMT}.) Depending on whether we consider the torus or cylinder geometry, the
sites along the main cylinder axis labelled by $n$ are assumed to obey
periodic or open boundary condition, respectively. For the cylindrical case, Eq.~\ref{1d} can be brought into a bosonic pair creation form given by
$U^m_{l_1 l_2}=g\kappa^3 b^m_{l_1}b^m_{l_2}$, where
\begin{equation}
b^{2j+1}_l = le^{-\kappa^{2}l^2}\sum_{p=0}^j \frac{(-2)^{3p-j}(\kappa l)^{2p}\sqrt{(2j+1)!}}{(j-p)!(2p+1)!}
\end{equation}
so that
\begin{equation}
U^m=g\kappa^3\sum_n \hat{b}^{m\dagger}_n \hat{b}^m_n,
\label{PP2}\end{equation}
where $\hat{b}_n^m=\sum_{n+l \in \mathbb{Z}} b ^m_{l}a_{n-l}a_{n+l}$. Here,
$g=\frac{4V_0}{(2\pi)^{3/2}}$ is a constant with units of energy and
$\kappa=\frac{2\pi l_B}{L_y}=\frac{1}{L_y}$. $l_B$ has been set to
$\frac{1}{2\pi}$ lattice constants in the latter equality in accordance to
Ref.~\onlinecite{qi2011}. In the following, $l_B$ will be expressed in
units of the lattice constant unless it appears in the combination
$l_B^2 \nabla^2$ or $l_B^2 q^2$, where $q$ is a momentum variable. The complete derivation of Eq.~\ref{PP2} can be found in
Appendix~\ref{2body}.

The Hamiltonian in~\eqref{1d} will be the
starting point of Sect.~\ref{sec:model} when we expand different FCI
models into this PP form. Its $m=1$ case has been previously used to define low-dimensional Mott-type models with bare onsite hardcore potentials at fractional filling~\cite{lee2004,seidel2005}.
The PP $U^m_{\text{tor}}$ on the torus can be found by summing
over all the periodic images of $U^m_{l_1l_2}$ (referred to as
$U^m_{\text{cyl}}$ in Appendix~\ref{ortho}) satisfying $l_1 +
l_2$ $\text{mod}$ $2L_xL_y=0$. This constraint can be generalized to the case with more
than two bodies, as shown in Appendix~\ref{3body}.

For
finite-size investigations on the cylinder or the torus, we have to
keep in mind that relative angular momentum is no longer a
well-defined quantum number, as opposed to the case of the sphere or the
plane.
The parameter $m$ in~\eqref{lag}, which corresponds
to the exact relative angular momentum as we take the planar limit,
determines the order of
the derivative acting on the hardcore potential via the degree of the Laguerre polynomial. This corresponds to a
Taylor expansion of the interaction in momentum
space~\cite{trugman1985}. While its interpretation as the
exact relative angular momentum is absent, it can still be employed as
an expansion parameter for short-range interactions on a sufficiently
large torus or cylinder.
To see this in terms of the Hilbert space basis, we describe
relative motion states on the torus by relative motion states on the
plane. The latter can be exactly classified via the relative angular
momentum $m$ which is
proportional to the interparticle distance in that relative state
$r_m$.
For $r_m/L_x, r_m/L_y << 1$, the overlap of the torus and planar
relative motion states goes to unity, effectively
reestablishing the notion of torus relative angular momentum for short distances.
Still, this approximation
becomes invalid for higher values of relative
angular momentum. At the Hamiltonian level, this is reflected by the overcompleteness of the PPs $U^m$.
This occurs because the interparticle distance $r_m\sim L_x,L_y$ that characterizes a relative angular momentum state
will no longer be well-defined when $r_m$ is comparable to the system lengths of the torus. A quantitative treatment of the
overcompleteness bounds can be found in Appendix~\ref{ortho}.

All in all, the properties of the pseudopotential expansion sets the stage for numerical investigations of short-ranged
interactions of FCIs as well as the defining of trial Hamiltonians for new quantum
Hall-type fractional Chern phases, both of which will be pursued in the following.

\subsection{Model FCIs for PP expansion}
\label{sec:model}

We apply the PP expansion of two model FCI Hamiltonians, the checkerboard (CB) model introduced in Ref.~\onlinecite{sun2011} and the honeycomb (HC) model introduced in Refs.~\onlinecite{wang2011,haldane88prl2015}.  Both models possess an almost flat (dispersionless) band which mimics the LLL in an FQH system. There, the Coulomb-type electron interactions lift the macroscopic degeneracy of the LLL, leading to a topologically degenerate groundstate. In the same spirit, we add Hubbard-type interaction terms to our model Hamiltonians such that
\begin{eqnarray}
H_{\text{int}}&=&\lambda H_{NN}+(1-\lambda)H_{NNN} \nonumber \\
&=&h_0 \left(\lambda \sum_{\langle ij \rangle}n_in_j +
  (1-\lambda)\sum_{\langle \langle ij
    \rangle\rangle}n_in_j\right),\nonumber \\
H&=&H_{0}+H_{\text{int}},
\label{hint_nonint}
\end{eqnarray}
where $\lambda$ characterizes the relative strengths of the nearest
neighbor (NN) and next-nearest-neighbor (NNN) interaction
terms. $h_0$, a parameter with units of energy, sets the overall
magnitude of $H_{\text{int}}$. The single-particle term $H_{0}$ gives
rise to the almost flat band and provides information for the
construction of the Wannier basis. This is the basis we will use for
expressing the two-body interacting term $H_{\text{int}}$ in the same
basis as the PP Hamiltonians $U^1, U^3,$ etc. as denoted
in~\eqref{1d}.

We  consider interactions that are much larger than the bandwidth
of the almost flat band, but much smaller than the interband gap. In
this limit, for a partially filled flat band, we can ignore the
coupling to the upper band and only study the effect of interactions
in the subspace of the flat band. With this picture in mind, we expand
$H_{\text{int}}$ in terms of the PPs in the Wannier basis of the
partially filled band. As long as the bandwidth of the almost flat band is much smaller than the interaction strength, we can ignore the bandwidth and consider only the interaction term.

\begin{figure}
\begin{minipage}{0.99\linewidth}
\includegraphics[width=0.35\linewidth]{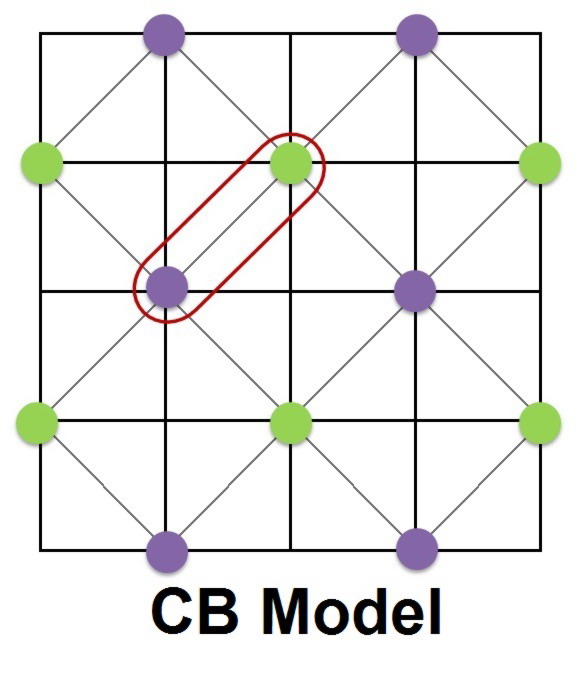}
\includegraphics[width=0.55\linewidth]{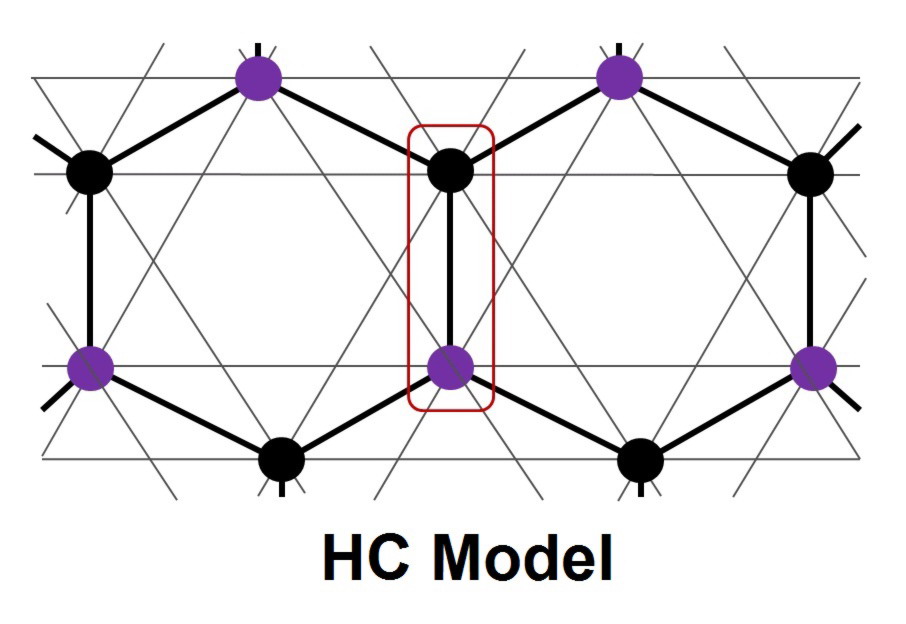}
\end{minipage}
\caption{  Lattice structure of the CB(left)  and HC (right)
  models. Sites colored differently belong to different
  sublattices. The unit cells are demarcated in red. For the CB model,
  NN interactions are between different sublattices while NNN and NNNN
  interactions occur within the same sublattice. For the HC model,
  both the NN and NNNN interactions occur between different
  sublattices, but the NNN interactions act within the same
  sublattice.}
\label{fig:CBHC}
\end{figure}

The checkerboard (CB) lattice model consists of two interlocking
square lattices displaced $(1/2,-1/2)$ sites relative to each
other (Fig.~\ref{fig:CBHC}). Its noninteracting Hamiltonian $H_0^{\text{CB}}$ consists of
NN, NNN and NNNN hopping terms parametrized by hopping
strengths $t$, $t'$, and $t''$ respectively\footnote{The original CB model\cite{sun2011} distinguishes between two types of
  NNN hoppings $t'_1$ and $t'_2$, but the constraint $t'=t'_1=-t'_2$ is
  sufficient to produce a band of maximum flatness.}. The NN hoppings exist between sites belonging to different sublattices and carry a phase $\phi$, giving rise to the time-reversal symmetry breaking necessary for a nonzero Chern number. Both the NN and NNNN hoppings exist between different sublattices, leading to off-diagonal terms in the single-particle (noninteracting) Hamiltonian. In sublattice space,
\begin{equation}
H^{\text{CB}}_{0}(k)=d_0 I + \sum_i d_i \sigma_i,
\label{modelhcb}
\end{equation}
where
\[d_1=-4t \cos \phi \cos \frac{k_x}{2} \cos \frac{k_y}{2},\]
\[d_2=-4t\sin \phi \sin\frac{k_x}{2}\sin \frac{k_y}{2},\]
\[d_3=-2t'(\cos k_x-\cos k_y).\]
 The expression for $d_0$ is irrelevant because it is not needed for
 the computation of the Wannier basis. We set $t=1$,
 $t'=-t''=1/(2+\sqrt{2})$ and $\phi=\pi/4$ as in
 Ref. \onlinecite{sun2011} to achieve the maximal the flatness ratio of $\sim 30$ for the
 bottom band. We can explicitly see why a nonzero $\phi$ is necessary
 for having a topologically nontrivial model: as the Chern
 number is given by $C_1=\frac{1}{4\pi}\int d^2k \vec \hat d \cdot
 (\partial_x\vec \hat d \times \partial_y\vec\hat d)$, it can only be nonzero if none of the $d_i$'s is identically zero.

Notice that $H^{\text{CB}}_{0}$ is not of Bloch form since the $d_i$'s
do not obey the periodicity of $2\pi$. This is because some sites are noninteger
lattice spacings away from each other (Fig.~\ref{fig:CBHC}). We can remedy this by shifting one sublattice site on top of the other within a unit cell. Mathematically, this corresponds to a gauge transformation of $c_{kB}^\dagger\rightarrow c_{kB}^\dagger e^{-i(k_x-k_y)/2}$ where $B$ refers to one of the sites within the sublattice. After the gauge transformation,
\[ d_1= -t[\cos\phi + \cos(k_x+k_y+\phi)+\cos(k_x-\phi)+\cos(k_y-\phi)], \]
\[ d_2= -t[\sin\phi + \sin(k_x+k_y+\phi)+\sin(k_x-\phi)+\sin(k_y-\phi)], \]
\[ d_3= -2t'(\cos k_x-\cos k_y).\]
The noninteracting part of the honeycomb model is defined
similarly. The unit cell consists of two adjacent sites. The phase
$\phi$ is carried between NNN sites, which lie in the same
sublattice. NNNN interactions which occur for diametral sites on the
same hexagon involve different sublattices (Fig.~\ref{fig:CBHC}). After an analogous gauge transformation,
\begin{equation}
H^{\text{HC}}_{0}(k)=d_0 I + \sum_i d_i \sigma_i,
\label{modelhhc}
\end{equation}
where
\[ d_1= -t(1+\cos k_x+\cos k_y)-t''(\cos(k_x+k_y)+2\cos(k_x-k_y)), \]
\[ d_2 = t(\sin k_x + \sin k_y) + t''\sin(k_x+k_y), \]
\[ d_3 = 2t'\sin\phi(\sin k_y-\sin k_x +\sin(k_x-k_y)). \]
The values for the
NN, NNN, and NNNN hoppings are given by $t=1$, $t'=0.6$, and
$t''=-0.58$, $\phi=0.4 \pi$ such that
the flatness ratio of the band is optimized to about $60$~\cite{wang2011}. We stress that while the optimization of these
flatband parameters is not necessary for performing the
PP expansion, it is physically relevant in
increasing the stability of an FQH state present in the system.

\subsection{Results of the Pseudopotential expansion}

While the PP matrix elements $U^m_{l_1l_2}$ depend only
on $l_1$ and $l_2$, the FCI interaction Hamiltonian matrix elements in
the Wannier basis $h(n_1,n_2,l_1,l_2)$ also depend on $n_1$ and
$n_2$. As a consequence, only a part of $h(n_1,n_2,l_1,l_2)$ can be expanded in terms of PPs. This important fact can be understood in terms of magnetic translation (MT) symmetry breaking, which will be analysed in depth in the next section. Here, we shall concern ourselves with the terms that can be expanded in PPs, defined by
\begin{equation}
H_p(n,n',l_1,l_2)=\frac{\delta_{n n'}}{2L_xL_y}\sum_{N=1}^{2L_y} h(N,N,l_1,l_2).
\label{hp}
\end{equation}
$H_p$ vanishes for $n \neq n'$ and does not depend on $n$ when $n_1=n_2=n$, as required. The sum runs from $1$ to $2L_y$ because $h(n,n,l_1,l_2)$ is periodic in $n$ with period $2L_xL_y$, as evident from the periodicity of $a_{(n\pm l)/2}$ in Eq.~\ref{basischange}.

We would like to expand $H_p$ in an orthonormal basis of
PPs $U^1$, $U^3$, etc. However, this expansion is only
unique and thus meaningful if we include PPs with $m$ bounded by a certain $m_{\rm max}$. This is because the inclusion of
higher PPs can yield an overcomplete operator basis, a
consequence of the finite size of the torus geometry. The truncated PP basis is no longer complete, but we can still perform a PP expansion of $H_p$ (now suitable normalized) by writing
\begin{equation}
H_p=\sum^{m_{max}}_{m} V^m U^m+H_{>} =:H_{\text{pseudo}} +H_{>},
\end{equation}
and finding the PP expansion coefficients $V^m$ that maximize the normalized
overlap $\langle H_p , H_{\text{pseudo}} \rangle$. The overlap is taken by
summing over all $|l_1|,|l_2|\leq L_xL_y$ since $H_p$ has a period of
$2L_xL_y$. Specifically, for any two Hamiltonians $H$ and $H'$ that respect MT symmetry,
\begin{equation}
\langle H,H'\rangle =
\frac{\sum_{l_1l_2}H_{l_1l_2}H'_{l_1l_2}}{\sqrt{\sum_{l_1l_2}H^2_{l_1l_2}\sum_{l_1l_2}H'^2_{l_1l_2}}}.
\label{over}
\end{equation}
The term $H_{>}$ consists of the part of $H_{\text{int}}$ that
does not break MT symmetry,  but still cannot be uniquely expressed in
terms of the PPs. It includes, for instance, hoppings
that occur over lengths comparable to the size of the torus.

When the $U^{m}$'s form an orthonormal basis, the $V^m$'s that maximize the normalized overlap $\langle H_p,H_{\text{pseudo}}\rangle $ can be determined as
\begin{equation}
V^m=\langle H_{p},U^{m}\rangle .
\end{equation}
\begin{figure}
\centering
\begin{minipage}{0.99\linewidth}
\includegraphics[width=.49\linewidth]{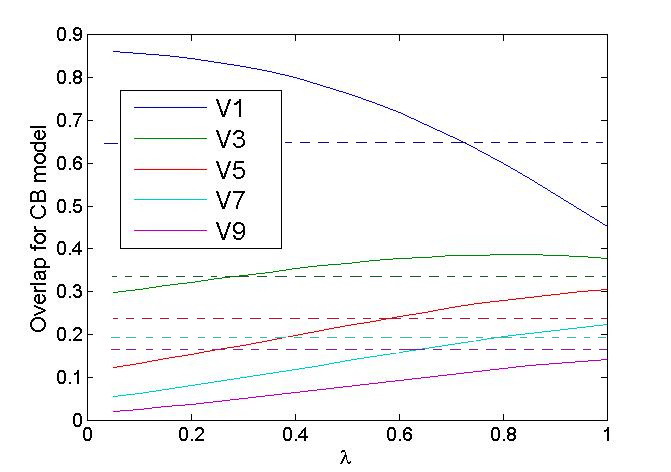}
\includegraphics[width=.49\linewidth]{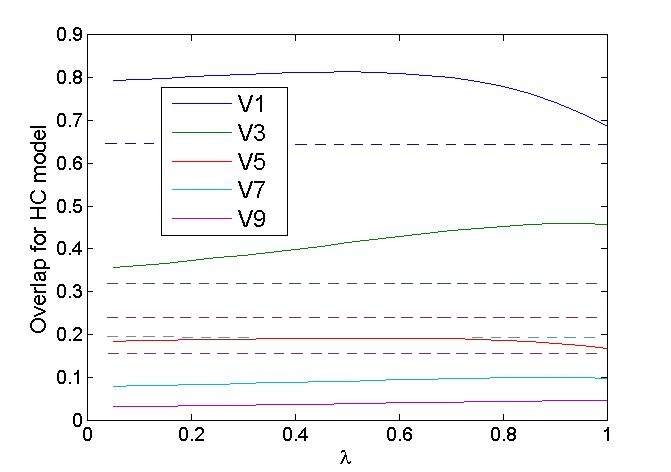}
\includegraphics[width=.9\linewidth]{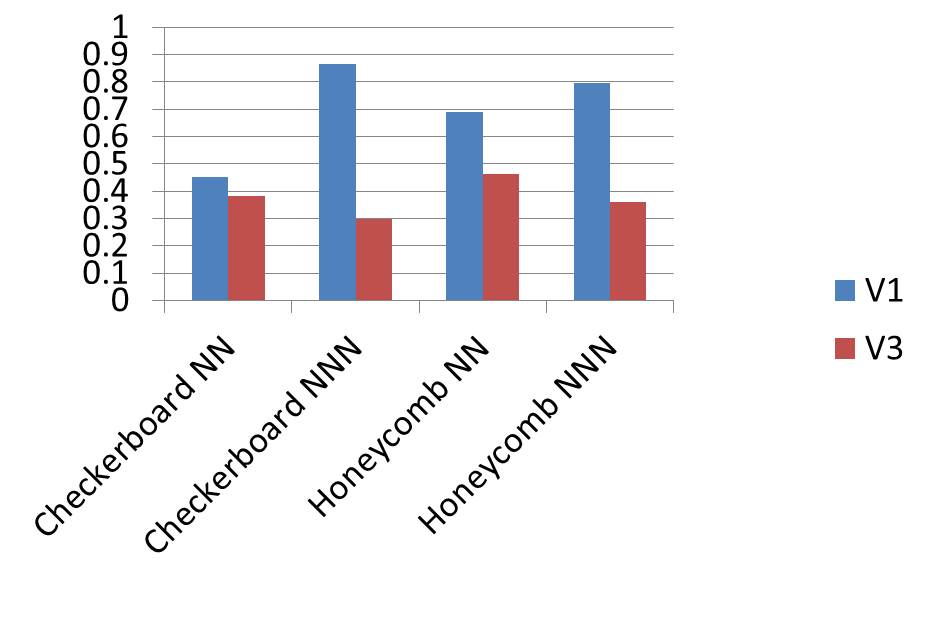}
\end{minipage}
\caption{  The pseudopotential expansion of the fermionic checkerboard (left) and the
  honeycomb (right) models for $L_x=L_y=6$. The normalized overlap $V^j$ is
  plotted against $\lambda$, where $H_{\text{int}}=\lambda H_{NN} +
  (1-\lambda) H_{NNN}$, so that we have the NNN limit on the left and the NN limit on the right.
  For the CB model, we see that the NN and NNN terms exhibit marked diferences in their pseudopotential expansions,
  with the MT symmetry conserving part of its NNN term consisting
  almost exclusively of the $V^1$ and $V^3$ terms. This can be
  understood by studying its distribution of matrix elements (Fig.~\ref{fig:l1l2plots}). For comparison, the PP coefficients $V^1$ to $V^9$ of the QH Coulomb interaction are plotted as dashed lines. The NNN interaction has a larger $V^1$ coefficient and smaller $V^3,V^5,...$ coefficients than the Coulomb interaction, and is hence even more likely to exhibit Laughlin groundstate. Bottom) Relative sizes of $V^1$ and $V^3$ at the $NN$ and $NNN$ limits.
 }
\label{fig:V13579}
\end{figure}
From Fig.~\ref{fig:V13579}, we see that the percentage of $H_p$ that
can be expanded as PPs
$\sum_{j=0}^{j_{max}}\left|V^j\right|^2$ has the maximum value for
$0.94$ for the CB NNN interaction (Fig.~\ref{fig:V13579}). As
expected, the first PP has the highest weight, which favors the possibility of simple FQH states such as Laughlin states. With the relative angular momentum $m$ being proportional to interparticle distance, $U^m$ is expected to
decay faster with $|l_1|,|l_2|$ as $m$ increases. This will be
shown in more detail in Appendix~\ref{2body}. It is notable, however, that the
second neighbor coupling leads to a better overlap with the first
PP than the nearest neighbor Hamiltonian. This
indicates that the PP Hamiltonians mapped to FCI systems
are not simple density-density interactions and their matrix elements
in real space lattice site basis can exhibit a nonmonotonic dependence
on distance. More specifically, this is because $U^1_{l_1l_2}\sim l_1l_2$ is not strongly peaked around  $l_1=\pm l_2$ in $l_1-l_2$ space, as shown in Fig.~\ref{fig:l1l2plots}, unlike the NN interaction. Since $l_1=\pm l_2$ corresponds to $q=0$ (as defined in Eq.~\ref{basischange}), we see that the NN terms are "too local" for a good overlap with $U^1$.
In general, the matrix elements $U^m_{l_1l_2}\sim (l_1l_2)^m$, so
$U^m$ becomes more localized at $l_1=\pm l_2$ for higher $m$.

For comparison, the pseudopotential coefficients for the Coulomb interaction in a QH system are also plotted in Fig.~\ref{fig:V13579}. They can be derived via Eq.~\ref{Vm}, where $V(k)=\frac{4\pi}{k}$. We see that the PP coefficients of the FCI interactions do not differ too much from those of the Coulomb interaction, and in fact have a larger $V^1$ coefficient in a large range of $\lambda$.

\begin{figure}
\begin{minipage}{0.99\linewidth}
\includegraphics[width=.28\linewidth]{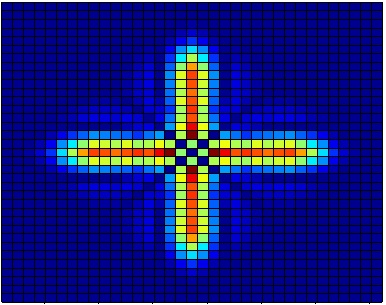}
\includegraphics[width=.28\linewidth]{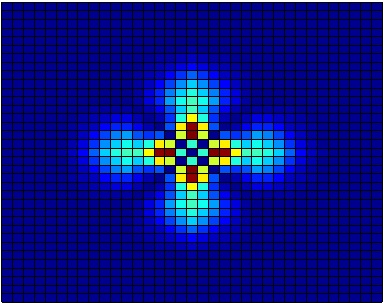}
\includegraphics[width=.28\linewidth]{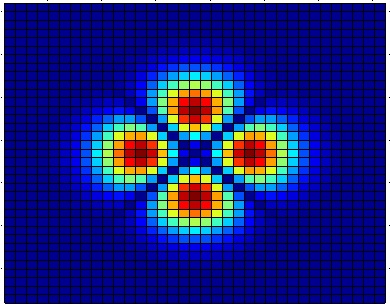}
\includegraphics[width=.1\linewidth]{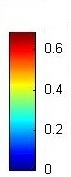}
\end{minipage}
\caption{  Plots of normalized $|H_p|$ for the NN (Left) and NNN (center)
  terms of  for the fermionic CB model. The horizontal and vertical axes
  represent $l_1-l_2$ and $l_1+l_2$ respectively. Regions with
  relatively large $|H_P|$ are colored red. $U^1$ (right) is plotted
  for comparison. The NNN term evidently bears more resemblance to
  $U^1$. }
\label{fig:l1l2plots}
\end{figure}

\subsubsection{Fermion-Boson Asymmetry}
In the quantum Hall effect, a Vandermode determinant allows to
equivalently switch from bosons to fermions which corresponds to
an additional attachment of one flux per particle. This symmetry is
broken in the fractional Chern insulator. We can see this explicitly by comparing the
PP coefficients of both the fermionic and bosonic HC model. The latter model is also studied in
other works like Ref. \onlinecite{wang2011}. The bosonic PPs are constructed analogously to the fermionic ones,
except that they are now symmetrized instead of antisymmetrized (refer to Appendix C for more details).

The comparison between the PPs of the bosonic and fermionic HC models are displayed in Fig. 5. The bosonic PP coefficients are in general closer to each other, with $V^0$ not larger than $V^2$. This is because of the large MT symmetry breaking (further described in the next section) that renders even the NN term rather nonlocal in the $l_1$, $l_2$ basis.

\begin{figure}
\begin{minipage}{0.99\linewidth}
\includegraphics[width=.97\linewidth]{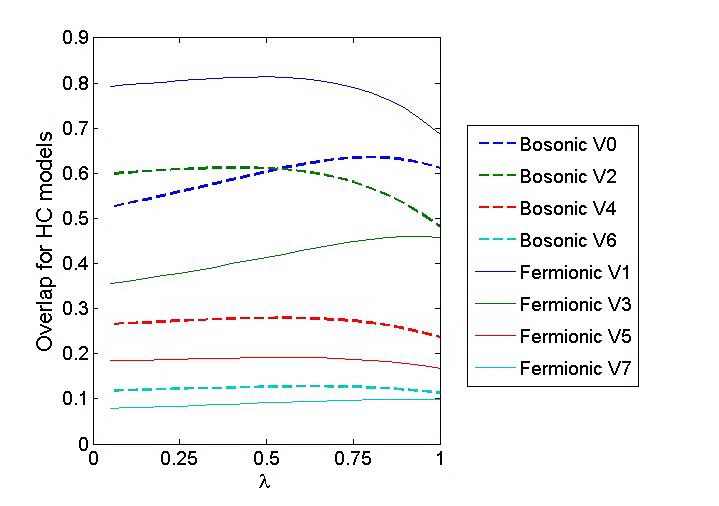}
\end{minipage}
\caption{  The pseudopotential expansion of the fermionic (solid line) and bosonic (dashed line) HC models for $L_x=L_y=6$.
The PP coefficient $V^j$ is
  plotted against $\lambda$, where $H_{\text{int}}=\lambda H_{NN} +
  (1-\lambda) H_{NNN}$, so that we have the NNN limit on the left and the NN limit on the right. The bosonic PP coefficients are in general closer to each other, with $V^0$ not larger than $V^2$ for some values of $\lambda$.}
  \end{figure}

\subsection{The effect of Magnetic Translation Symmetry breaking}
\label{sec:CMMT}

In this subsection, we analyze the origin of the terms in the FCI
Hamiltonian that cannot be expanded into PPs. We review how magnetic translation (MT) symmetry in FQH system
constrains the form of its two-body interaction terms, and investigate
how this picture is generalized to the FCI case.

\subsubsection{Origin of Magnetic Translation Symmetry breaking}

Consider an $L_xl_B\times L_yl_B$ torus geometry which has been discussed in Sect. II. In the Landau gauge $A_x=0,~A_y=Bx$, the covariant momentum operators are $P_x=-i\partial_x,~P_y=-i\partial_y-A_y$ which satisifes $\left[P_x,P_y\right]=iB$. The Hamiltonian $H=\frac1{2m}\left(P_x^2+P_y^2\right)$ has two translation symmetries $T_x^B,T_y^B$ defined by
\begin{eqnarray}
T_x^B&=&e^{i\frac{2\pi y}{L_yl_B}}e^{iP_x\frac{l_B}{L_y}},\nonumber \\
T_y^B&=&e^{iP_y\frac{l_B}{L_x}}.
\end{eqnarray}
$T_y^B$ is an ordinary translation while $T_x^B$ is a translation in $x$ direction by $l_B/L_y$ accompanied by a gauge transformation. The translation can only be defined in units of $l_B/L_y$ so that the change of gauge potential $A_y=Bx\rightarrow B(x+l_B/L_y)$ can be cancelled by a gauge transformation. The action of $T_x^B,T_y^B$ on the basis wavefunctions (\ref{torusLLWF}) is
\begin{eqnarray}
T_y^B\ket{\psi_K}&=&e^{iK/L_x}\ket{\psi_K},\nonumber \\
T_x^B\ket{\psi_K}&=&\ket{\psi_{K+\frac{2\pi}{L_y}}}.\label{magtrans}
\end{eqnarray}
For a general $2$-body interaction $H_{\rm int}=\sum_{n_1,n_2,n_3,n_4}a_{n_1}^\dagger a_{n_2}^\dagger
U^{n_1n_2n_3n_4}a_{n_3}a_{n_4}$,  the condition $\left[T_y^B,H_{\rm
    int}\right]=0$ requires $n_1+n_2=n_3+n_4$, since
\[{(T_y^B)}^{-1}a_{n_1}^\dagger a_{n_2}^\dagger
a_{n_3}a_{n_4}{T_y^B}=a_{n_1}^\dagger a_{n_2}^\dagger
a_{n_3}a_{n_4}e^{2\pi i(n_1+n_2-n_3-n_4)/L_xL_y}.\]
The condition
$\left[T_x^B,H_{\rm int}\right]=0$ requires
$U^{n_1n_2n_3n_4}=U^{n_1+1,n_2+1,n_3+1,n_4+1}$. Therefore, the magnetic
translation symmetry $T_x^B$ and $T_y^B$ determines the
CM conservation ($n_1+n_2=n_3+n_4$ or $n=n'$) and (one-dimensional)
translation symmetry of the interaction Hamiltonian in FQH states,
i.e., $n$-independence of the interaction matrix elements.

\begin{figure}
\begin{minipage}{0.99\linewidth}
\includegraphics[width=.75\linewidth]{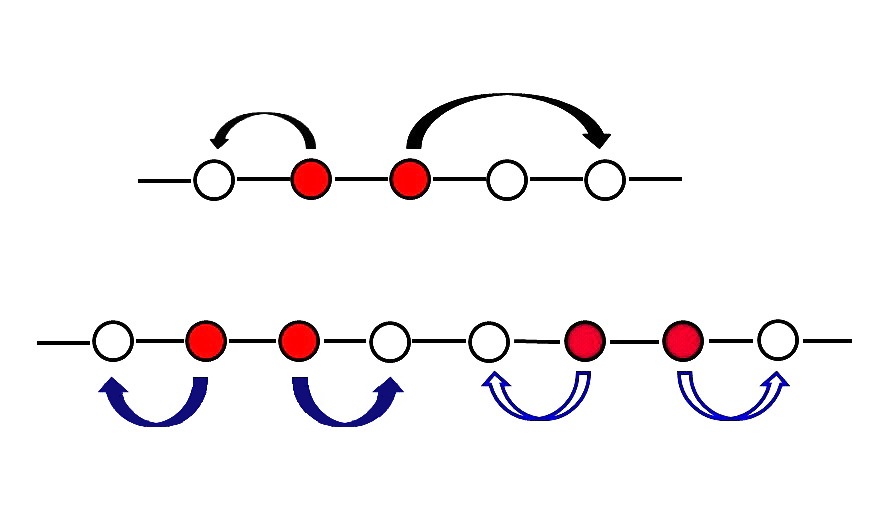}
\end{minipage}
\caption{  The two types of MT breaking hoppings. Top: A hopping process that changes the CM position. Bottom: Two hopping processes that preserve the CM position but break magnetic translation symmetry because the hopping at different CM position $n$ (solid and hollow arrows) have different amplitude.
}
\label{fig:nonconservation}
\end{figure}

By comparison, in the lattice model, we only have the lattice translation symmetries which commute with each other. The action of the lattice translation $T_x,T_y$ acts on the Wannier basis as
\begin{eqnarray}
T_y\ket{W_K}&=&e^{iK}\ket{W_K},\nonumber \\
T_x\ket{W_K}&=&\ket{W_{K+2\pi}}.\label{latticetrans}
\end{eqnarray}
Comparing Eq.~\ref{latticetrans} with Eq.~\ref{magtrans}, we see that in the mapping from FCI to FQH, $T_x, T_y$ is mapped to $\left(T_x^B\right)^{L_y}$ and $\left(T_y^B\right)^{L_x}$, respectively. Therefore, in the lattice model, the translation symmetries only require the matrix element of two-body interaction $U^{n_1n_2n_3n_4}$ to satisfy
\begin{eqnarray}
U^{n_1n_2n_3n_4}&=&U^{n_1+L_y,n_2+L_y,n_3+L_y,n_4+L_y},\nonumber \\
U^{n_1n_2n_3n_4}&=&0\text{~if~}n_1+n_2\neq n_3+n_4\text{~mod~}L_y.
\end{eqnarray}
The magnetic translation symmetry breaking in the lattice models (Fig.~\ref{fig:nonconservation}) is
also related to the non-uniform Berry curvature in momentum
space. As previously discussed, the CM
position of the Wannier state $\ket{W_K}$ is determined by the flux of
the Berry's phase gauge field $P_x(k_y)$. If the system has magnetic
translation symmetry, $\ket{W_K}$ and $\ket{W_{K+2\pi/L_y}}$ are
related by $T_x^B$, so that $P_x(k_y)$ must depend on $k_y$
linearly. As a result, we expect MT symmetry breaking whenever the Berry curvature is nonuniform in momentum space, which is the case in a generic CI.

In addition, MT symmetry breaking will still be present even in the hypothetical case of perfectly flat Berry curvature. This is because the Wannier basis is not perfectly local. Recall from~\eqref{basischange} that
\begin{equation} n_1+n_2=n{\phantom{'}}=k_{1y}+k_{2y}+(x_1+x_2)L_y, \label{eqn1}\end{equation}
\begin{equation} n_3+n_4=n'=k_{1y}+k_{2y} + (x_3 + x_4)L_y, \label{eqn2}\end{equation}
where the $x_i$s are the lattice sites of the original
$H_{\text{int}}$. CM nonconserving terms occur where $x_1+x_2 \neq x_3
+ x_4$, when $n$ and $n'$ differ by a multiple of $L_y$. These terms
do not appear in the original real-space basis where
$H_{\text{int}}\propto \sum_{ij}n_in_j = -\sum_{ij}c^\dagger_i
c^\dagger_j c_i c_j$ annihilates and creates two particles at the same
position. However, our Wannier basis functions generically have
exponentially decaying tails on both sides of their peak $\bar x$, which produce CM nonconserving and thus MT breaking contributions.

\subsubsection{Numerical results on MT symmetry breaking}

\begin{figure}
\centering
\begin{minipage}{.99\linewidth}
\includegraphics[width=.99\linewidth]{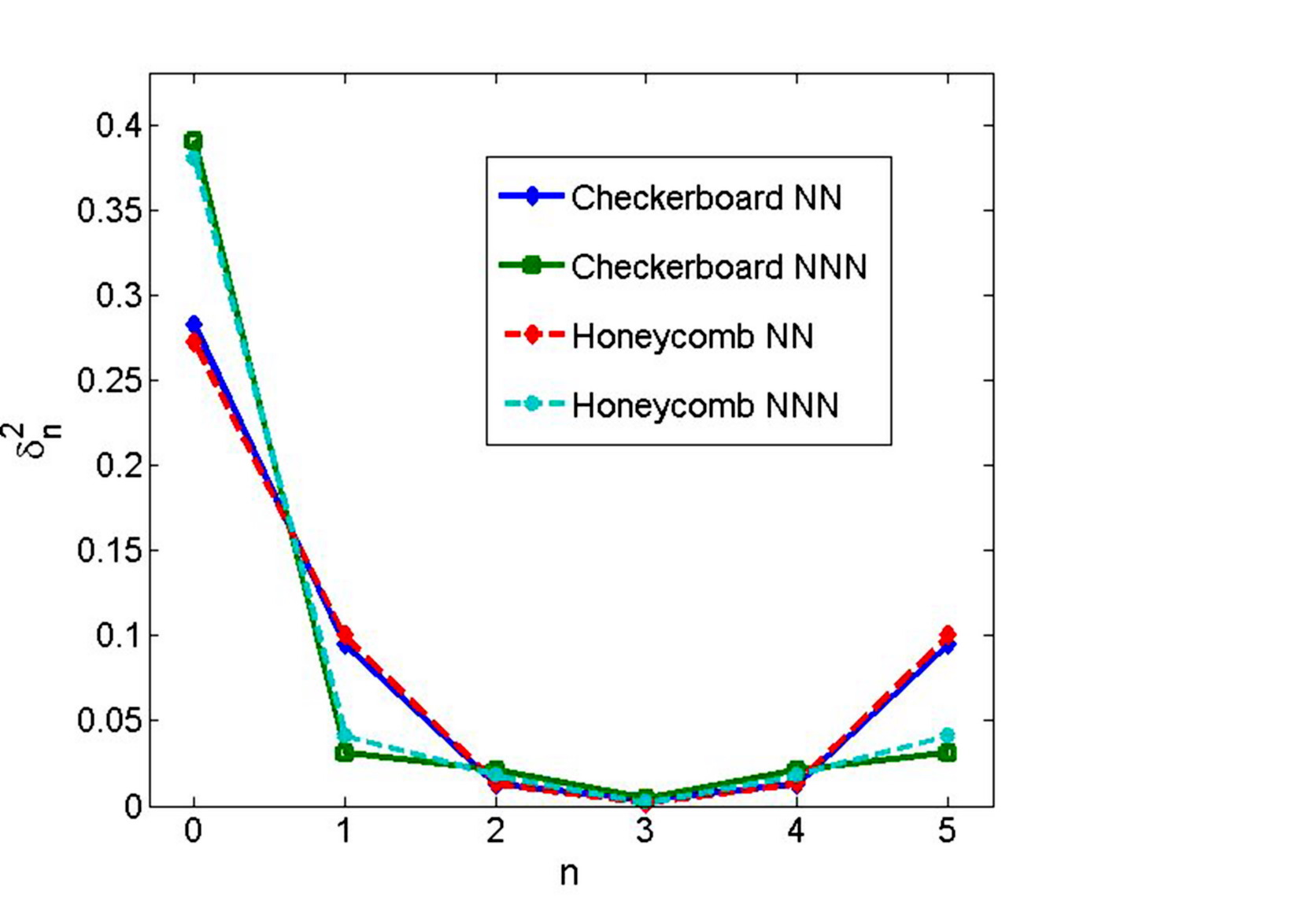}
\end{minipage}
\caption{  A plot of $\delta_n^2$ for $L_x=L_y=6$ for
  different model Hamiltonians. As evident from the dominance of the
  peak at $n=0$, the amount of MT symmetry breaking
  largely stems from CM conserving terms. There is little difference between the
  degree of MT symmetry breaking in the different models. }
\label{fig:error66}
\end{figure}

Here are the numerical results on MT symmetry breaking in our model Hamiltonians. Define the residual
\begin{equation}
H_{\text{res}}(n,n',l_1,l_2)=h(n,n',l_1,l_2)-H_p(l_1,l_2) \end{equation}
where, as before, $h(n,n',l_1,l_2)$ denotes the FCI interaction
Hamiltonian expressed in the Wannier basis. $H_{\text{res}}$ is the
part of $h(n,n',l_1,l_2)$ which does
not satisfy MT symmetry required by the PPs. Obviously, $H_{\text{res}}=0$ if $h$ is one of the PPs, since $h$ will then be equal to $H_p$. The quantity
\begin{equation}\delta^2_n = \frac{\sum_{l_1l_2,m}|H_{\text{res}}(m,m+nL_y,l_1,l_2)|^2 }{\sum_{l_1l_2,m,m'}|h(m,m',l_1,l_2)|^2 }\end{equation}
allows us to track the origin of MT nonconservation. $\delta^2_0$
comprises the elements of $H_{\text{res}}$ satisfying $n=n'$. As defined in
Eq.~\ref{hp}, these are the elements which are independent of
$n$. $\delta^2_0$ hence represents the fraction of matrix elements that
are CM conserving but MT symmetry breaking. For $n\neq 0$,
$\delta^2_n$ represents MT nonconserving contributions that likewise
do not respect CM conservation. $\delta^2_n$ is plotted in
Fig.~\ref{fig:error66} for various model Hamiltonians, for a system
size $L_x=L_y=6$. The results remain almost unchanged when $L_x$ and $L_y$
are varied as long as $L_x=L_y>3$.

From the enhanced peak at $n=0$, we conclude that most of the MT
symmetry breaking occurs when CM is conserved. This happens because
our maximally localized Wannier functions (WFs) are still mostly peaked at one
site. The subdominant contributions from $\delta_n^2$ for $n=\pm 1$
can be attributed to the finite tails of the WFs one site away from their center-of-mass. Indeed, $\delta_n^2$ becomes exponentially small for $|n|>1$. While the overall extent of MT symmetry breaking originates from the nonuniformity of the Berry curvature, its relative contribution to $\delta_n^2$ for different $n$ is dictated by the localization properties of the WFs.

\subsubsection{Discussion on MT symmetry breaking}
The decomposition of the FCI Hamiltonian into pseudopotentials is
only exact in the thermodynamic LLL limit of zero
bandwidth and homogeneous
Berry curvature. For the generic model,
the FCI Hamiltonian can only be partly decomposed into pseudopotentials, which we then discuss along general FQHE pseudopotentials on the cylinder or torus. From our calculations, the
deviations are significant, suggesting that at least for the spectrum
above the elementary low energy quasiparticle regime,
there is no clear similarity between FCI and
FQH systems. However, entanglement signatures of incompressible liquid phases, such as the entanglement spectrum~\cite{li-08prl010504} with
the emergence of an entanglement gap~\cite{thomale-10prl180502}, show
strong similarities of FCI ground states to their FQH analogues, even
in terms of the counting rule of low-lying
states~\cite{regnault-11prx021014,bernevig-12prb075128}. This is
astonishing from the viewpoint of PPs, as the FCI and FQH models at
the bare level
could only possibly agree to the extent of the PP decomposable
components of the FCI Hamiltonians.

Such an apparent discrepancy between analyses at the Hamiltonian and entanglement measure levels can be interpreted as a consequence of a renormalization group flow. As high energy modes are integrated out, the low energy physics of FCIs supposedly flows towards FQHE type scenarios, with reemergent symmetries
such as magnetic translation group which is conserved in the FQHE but
broken in the FCI at a bare level. Recast into pseudopotentials, it suggests
that the PP non-decomposable part of the FCI Hamiltonian at the
bare level should
decrease upon renormalization, while the ratios of pseudopotentials of
the PP-decomposable part of the bare interactions might deviate from the PP
ratios in the low energy theory. This implies that even if two FCI
Hamiltonians have similar PP ratios at the bare level, they can
still differ considerably in their low energy description, and hence
their propensity to host FQHE-type incompressible states
(Fig.~\ref{fig:error66}).
This interpretation
is consistent with the common theme from FQHE numerical studies that the data quality of entanglement spectra and its characterization of the bulk and edge mode properties is not significantly correlated to the spectral sharpness of the Hamiltonian spectrum, and
partly anticipates the energy spectral flow~\cite{thomale-10prl180502}. Ultimately, only the joint confirmation of both entanglement and
Hamiltonian measures will justify true evidence for a fractional
topologically ordered phase in the FCI models.
From a low energy perspective, the seemingly clean finding from entanglement measures
might not yet rule out that the
inhomogeneous berry curvature induces a flow to a liquid
different from FQHE, as may also be seen by hints such that the hierarchy
liquid construction
cannot be established for the FCIs as in the FQHE
case~\cite{andreihierarchy}. From the perspective of energetics, the effective pseudopotential weights of the FCI
models in a low-energy theory are likely to be strongly
modified due to "integrating out'' the PP
non-decomposable part of the bare model, which can also provide an
explanation for the parameter trends of the stability of FCI phases as
a function of system parameters~\cite{Grushin2012}.

\section{Reverse engineering of Pseudopotentials onto a lattice}		

We have previously seen how interactions from FCIs can be expanded in terms of the Pseudopotentials in FQH. In this section, the WSR mapping will be done in the reverse direction, namely from FQH to FCI. The first 2-body fermionic PP $U^1$ will be exactly mapped onto various FCI lattices, and the locality of the resultant ``reverse engineered'' PP will be optimized. 

\subsection{Basis setup}

In anticipation of the need to optimize the reverse engineering of the PPs, we will use a slightly generalized version of the Landau gauge basis (with the sum in the RHS enforcing toroidal boundary conditions)
\begin{equation} \psi_{K=\frac{2\pi n }{L}}(x,y)=\sum_{K'\in \mathbb{Z}}\frac{1}{\sqrt{\sqrt{\pi}Ll_B}}e^{i(K+K')y}e^{-A(\frac{x}{l_B}-l_B(K+K'))^2}
\label{LLLbasis}
\end{equation}
with $A$ the additional aspect ratio parameter that can be tuned through rescalings of the coordinates. When the rescaling is regarded as an active transformation, it can formally be represented by a redefinition of the intrinsic QH metric\cite{HaldaneGeometry,Yang2012a,Yang2012b}\footnote{We write the FQH Hamiltonian as $ H=\frac{1}{2m}g^{ab}\pi_a\pi_b $ where $\pi_a=p_a-\frac{e}{c}A_a(\vec r)$. $g^{ab}$ represents an intrinsic QH metric that reduces to the identity $\delta^{ab}$ in the usual isotropic case. Expressed a quadratic form, $g^{ab}$ specifies the eccentricity and orientation of the generically elliptical LL orbital in the symmetric gauge. For simplicity, we will consider the case of zero orientation shift, so that $g=diag(C_1/A,A/C_1)$ where $C_1$ is the Chern number and $A$ is an aspect ratio.}. When regarded as a passive transformation, it can be  interpreted as a LL basis redefinition which elegantly corresponds to Bogoliubov transformations of the QH second-quantized operators in the symmetric gauge\cite{Qiu2012}.

Since it will be desirable to control the locality of the $U^m$ PP operators in the FCI system, we shall first express them in a form where their real-space locality can be directly studied. This can be explicitly done in the QH case, after which the FCI case follows by simply replacing $a^\dagger_K$ with the WF creation operator $b^\dagger_K$, as mentioned below Eq. \ref{mapFQHFCI}. As shown in Appendix \ref{ppcoherent}, one can use Eqs. \ref{lag} and \ref{LLLbasis} to bring QH pseudopotential $U^m$ to the useful second-quantized form involving coherent states:\cite{qi2011,lee2013}
\begin{equation}
U^m\propto \int dz_1dz_2 \sum_r v^m_r \nabla_z^m (c^\dagger_z c_z ) \nabla_z^m (c^\dagger_z c_z )
\label{main}
\end{equation}
where $v^m_r$ are the coefficients of the Laguerre polynomials $ L_m(x)=\sum^{m}_{r=0} v^m_r x^r $, $z=z_1+iz_2$, and $\nabla_z$ is to be taken as a grad operator so that $\nabla_z^3 f=\nabla_z (\nabla_z\cdot \nabla_z f)$, etc.

The coherent state creation and annihilation operators are defined by\cite{qi2011}:
\begin{equation}
c^\dagger_z =  \sum_{K} e^{-iz_2K}e^{-\pi A(z_1-K/(2\pi))^2}a^\dagger_K
\label{coherentdef}
\end{equation}
where $a^\dagger_K$ is an annihilation operator for a Landau gauge basis state. As will be made evident in the next subsection, this way of expressing $U^m$ makes its locality explicit because the coherent state created by $c_z^\dagger$ has a local wavefunction in real space (in the sense of exponential decay which we will demonstrate shortly after). Hence our attempt to maximally localize the PPs is reduced to a search for a basis with the smallest coherent state spread.

Now let's perform the WSR mapping to the FCI system. For simplicity, we shall first consider the case with Chern number $C_1=1$ in the lattice system, which is a direct analog of a QH system with one LL. The coherent state $\ket{\Psi_z}=c^\dagger_z\ket{0}$ of the lattice system takes the form %via the replacement $a^\dagger_K\rightarrow b^\dagger_K$, in the following convenient form
\begin{eqnarray}
\Psi_z(x,y)&=&\sum_K e^{-ik_yz_2}e^{-\pi A(z_1-K/{2\pi})^2}W_K(x,y)\notag\\
&=&\sum_m\int d^2k e^{-ik_yz_2}e^{i\vec k \cdot \vec r}e^{-\pi A\left(z_1-(\frac{k_y}{2\pi}+m)\right)^2}e^{i\theta}\phi(k_x,k_y)\notag\\
\label{coherentformc1}
\end{eqnarray}
with $\vec{r}=(x,y), ~\vec{k}=(k_x,k_y)$ here and below. Note that the $a^\dagger_K$ operator in the FQH case is now replaced by one that creates $W_K$, the concatenated Wannier function
\begin{equation}
W_{K=k_y+2\pi x}(x,y)=\int dk_x e^{i(k_x x+ k_y y)}e^{i\theta(k_x,k_y)}\phi(k_x,k_y)
\label{wannierdef}
\end{equation}
with the gauge phase $e^{i\theta}$ arising from the intrinsic arbitrariness of the phase of the Bloch states. To fully identify $\ket{W_K}$ with the LLL Landau gauge wavefunction $\ket{\psi_K}$ of the QH system, one also has to set the effective magnetic length to be $l_B=\sqrt{\frac{1}{2\pi}}$ in units of the FCI lattice spacing. 
That $\Psi_z(x,y)$ is asymptotically exponentially decaying in $x,y$ follows from the fact that $\phi(k_x,k_y)$ in the integrand has a complex singularity at a finite distance from the real $k_x$ or $k_y$ axis \footnote{More generally, the fourier coefficients of a complex function decay exponentially at a rate $h$ whenever the function has a singularity at a distance $h$ from the real axis. This is proven in Ref. \onlinecite{he2001}, and also in \cite{leearovasthomale2015}}.

For lattice systems with general $C_1$, there will be $C_1$ coherent states "colors" in a multiplet\cite{Barkeshli2012,Wu2013}. This is because each $W_K$ will now shift by $C_1$ sites as $K$ (or $k_y$) is translated by $2\pi$, leading to $C_1$ inequivalent "layers". Indeed, we now have a map from a Chern number $C_1$ lattice system to a multilayer $QH$ system, with each layer having Chern number one and $l_B=\sqrt{\frac{C_1}{2\pi}}$.

For each layer (color), the expression for the coherent state is generalized to
\begin{eqnarray}
\Psi_z(x,y)&=&\sum_m\int d^2k e^{-ik_yz_2}e^{i\vec k \cdot \vec r}e^{-\pi A\left(\frac{z_1}{C_1}-(\frac{k_y}{2\pi}+m)\right)^2}e^{i\theta}\phi(k_x,k_y)
\label{coherentform}
\end{eqnarray}
which is manifestly expressed in terms of three tunable inputs: The aspect ratio $A$, the gauge phase $\theta$ and the Bloch states $\phi$. To minimize the proliferation of terms in the resultant PP $U^m$, we will need to maximize the locality of the coherent states, as evident from the relation Eq. \ref{main}. The optimal $A$ and $\theta$ will be derived in great detail in the next subsection.

Note that Eq. \ref{coherentform} reduces to a 2-dimensional fourier transform of the Bloch wavefunction in the absence of the two terms containing $z$. When $C_1\neq 0$, such a fourier transform cannot result in a localized state due to topological obstruction\cite{thonhauser2006a}. In this sense, the Gaussian term containing $z_2$ can also be regarded as a regularization factor that makes localization in both directions possible.

\subsubsection{Relabeling of the unit cell}
From Eq. \ref{coherentform}, it is evident that we can, by redefining the unit cell, change the bloch states and hence significantly change the form of the coherent states $\Psi_z(x,y)$ and the locality of the PPs. Let the original position coordinates and crystal momentum be labeled as $r=(x,y)$ and $k=(k_x,k_y)$. Under a redefinition of unit cell by a 2-by-2 matrix $M$,
\begin{eqnarray} r\rightarrow r'&=&Mr \\
 k\rightarrow k'&=&M^{-1}k \end{eqnarray}
with the scalar product $k\cdot r$ staying invariant. Under general transformations $M$, $\Psi_z(x,y)$ given by Eq. \ref{coherentform} will change unless $k_y'=k_y$ and the Bloch states $\phi(k)$ remain invariant under $M$. Manifest here is the asymmetry between the roles played by $k_x$ and $k_y$, an unavoidable feature of the nature of the Landau gauge wavefunctions and their analogs.

There is a special class of transformations $M$ where all the terms in the definition of $\Psi_z$ Eq. \ref{coherentform} remain invariant, except for the Bloch state $\phi(k)\rightarrow \phi(k')$. In other words, the PPs will be exactly that obtained for the same Hamiltonian with $k\rightarrow k'$, up to the rescaling of coordinates $r=(x,y)$. These transformations are given by
\begin{eqnarray}
x'&=&\alpha x \notag\\
y'&=&y-\beta x
\label{abredef1}
\end{eqnarray}
\begin{eqnarray}
k'_x&=&\frac{k_x+\beta k_y}{\alpha} \notag\\
k'_y&=&k_y
\label{abredef2}
\end{eqnarray}
We will revisit transformations of this type in Sect. \ref{sec:numerics}, where we compare the PPs of the same Dirac model after some coordinate redefinitions.

Before further investigations on the pseudopotential in Eq. \ref{main}, we will like to discuss its symmetry properties. Although the Chern band is topologically equivalent to a Landau level system, the latter possess a higher symmetry which includes all symmetries of the Chern band. More specifically, the pseudopotentials $U^m$ in FQH systems preserve the magnetic translation symmetry which which is a higher symmetry than the lattice translation symmetry of the lattice. In a FQH system spanned by the Landau gauge basis $\ket{\psi_K}$, the magnetic translation symmetry implies the translation symmetry of $K$ to $K+\frac{2\pi}L$. However, a generic Chern band does not have this symmetry due to its non-uniform Berry's curvature. This non-uniformity is clearly manifested in the nonlinearity of the center-of-mass of the Wannier basis, as detailed in Ref. \onlinecite{qi2011}. Therefore the psedopotentials $U^m$ that we defined in FCIs have a higher symmetry than a generic interaction term preserving the lattice symmetries. This is not a concern for our current work since our purpose is to explicitly write down pseudopotential interactions which can then be used in numerical studies for realizing certain topological states as ground states. 
%An alternative explicit expression for $U^m$ have also been derived in Ref. \onlinecite{lee2013}, where they take an elegant form in a one-dimensional basis proposed in Ref. \onlinecite{lee2004}. While that form allows for easy comparisons between different interactions at the operator level, the forms derived above (i.e. Eqs \ref{main},\ref{coherentform}) are more suitable for locality optimization, being explicitly defined in position space.

\subsection{Maximally localizing the coherent states basis}
\label{sec:MLCS}
\subsubsection{The optimal gauge phase}

Let's first show how to minimize the spread of the coherent states $\Psi_z$ so that the PPs will be most localized and be best approximated by a short-ranged interaction. For this purpose, we define the mean-squared range of the coherent state wavefunction $\Psi_z(x,y)$ by
\begin{eqnarray}
&&I[\theta]=\int_{-\frac{1}{2}}^{\frac{1}{2}}\int_{-\frac{1}{2}}^{\frac{1}{2}} d^2z\langle r^2\rangle_{\Psi_z}\nonumber\\
&=&\int_{-\frac{1}{2}}^{\frac{1}{2}}\int_{-\frac{1}{2}}^{\frac{1}{2}}d^2z \int dxdy (x^2+y^2) |\Psi_z(x,y)|^2 \nonumber \\
&=&\int_{-\frac{1}{2}}^{\frac{1}{2}}\int_{-\frac{1}{2}}^{\frac{1}{2}}d^2z \sum_m \int d^2k \left|\nabla_k \left (e^{-E_m-ik_yz_2}e^{i\theta }\phi(k_x,k_y)\right )\right |^2
\label{coherentI}
\end{eqnarray}
which is regarded as a functional over the $U(1)$ gauge phase $\theta(k_x,k_y)$. Here $E_m(k_x,k_y)=-\pi A \left(\frac{k_y}{2\pi}-\frac{z_1}{C_1}+m\right )^2$. For simplicity, we will only consider the case with one occupied band.
Since we fourier transformed over the concatenated momentum $K= k_y+ 2\pi m$ from the penultimate to the last line, the $m$ sum must be included. $A$ represents the QH aspect ratio that is yet to be optimized.

We had sought to minimize the spread $\langle r^2\rangle$ of $\Psi_z$ averaged for different $z$. This averaging is motivated by various reasons. Firstly, the optimal localization condition for each $\Psi_z$ will lead to a different condition on gauge choice $\theta$, and we desire for an unique optimal gauge choice. Secondly, it makes physical sense to optimize the average spread since we will require the locality of the coherent states over all $z$ when calculating the PP. Furthermore, performing the average over $z$ will greatly simplify the minimization problem, as we will soon see. As distinguished from the FQH case, in FCI the coherent states with different center-of-mass position $z$ generically have different shapes, since the translation symmetry is broken to discrete lattice translation. Since lattice translation symmetry guarantees that $\Psi_z$ has the same spread as $\Psi_{z+1}$ and $\Psi_{z+i}$, we will only need to average across the unit cell $(z_1,z_2)\in\left[-\frac12,\frac12\right]\times \left[-\frac12,\frac12\right]$.

Given a Bloch wavefunction $\phi$ with Berry connection $\vec{a}=-i\phi^\dagger \nabla_k \phi$, a gauge rotation $\phi\rightarrow e^{i\theta}\phi$ will produce a new connection $\vec{a}\rightarrow \vec{a}_{new}=\vec{a} + \nabla_k\theta$.
Performing the Euler-Lagrange minimization \begin{equation}\frac{\delta I}{\delta \theta}=0,\end{equation} whose detailed steps are shown in Appendix \ref{gaugederivation}, we arrive at the \textbf{Coulomb gauge condition}
\begin{equation}
\nabla_k \cdot\vec{a}_{new}=0
\label{eq:coulomb}
\end{equation}
We note that Eq.~\ref{eq:coulomb} coincide with the gauge choice made in Ref. \onlinecite{Wu2013}, although it was not obtained by maximal localization condition over there.

In this gauge choice, there exists a simple way to express $\vec{a}_{new}$ in terms of the Berry curvature $f$. This gauge is also consistent with the conditions for a $C_4$ symmetric basis in the case of $C_4$ symmetric systems, as shown in Ref. \onlinecite{Jian2013}. Since $\nabla_k\cdot\vec{a}_{new}=0$, we can write $\vec{a}_{new}=(-\partial_y\varphi,\partial_x \varphi)^T$, where
\begin{equation}\nabla_k^2 \varphi(k_x,k_y) = f(k_x,k_y)=\partial_xa_y-\partial_ya_x
\label{eq:poisson1}
\end{equation}
This is the Poisson's equation on a finite torus whose explicit solution is given by Eq. $9$ of Ref. \onlinecite{Wu2013}. When the dimensions of the torus $L_x,L_y>10$, as is the case for the calculations in this work, the above solution will be almost identical to that given by the electrostatics Green's function integral
\begin{equation} \varphi (k_x,k_y)= \int_{periodic} f(\vec{p})\log|\vec{p}-\vec{k}| d^2p
\label{eq:poisson2}
\end{equation}
where the contributions from the periodic images of $f$ are summed over. Here $\varphi$ and $f$ take the role of the electrical potential and the periodic electric charge distribution.

To calculate the PPs, however, we need to find the gauge phase $\theta$ and not just $\vec{a}_{new}=(-\partial_y\varphi,\partial_x \varphi)^T$. The optimal gauge phase $\theta=\theta_C$ can be found from the condition Eq. \ref{eq:coulomb}, which upon subtituting $\vec{a}_{new}=\vec{a} + \theta_C$ gives another Poisson's equation
\begin{equation}\nabla_k^2\theta_C =-\nabla_k\cdot \vec{a} \end{equation}
The solution to this equation is already given by Eq.~\ref{eq:poisson2} with $\varphi$ and $f$ replaced by $\theta_C$ and $\nabla_k \cdot \vec{a}$. Note that the solution $\theta$ obtained is unique, because the difference of two different solutions will satisfy the Laplace equation, and that must be identically zero due to toroidal BCs.

\begin{figure}[H]
\begin{minipage}{0.99\linewidth}
\includegraphics[width=.5\linewidth]{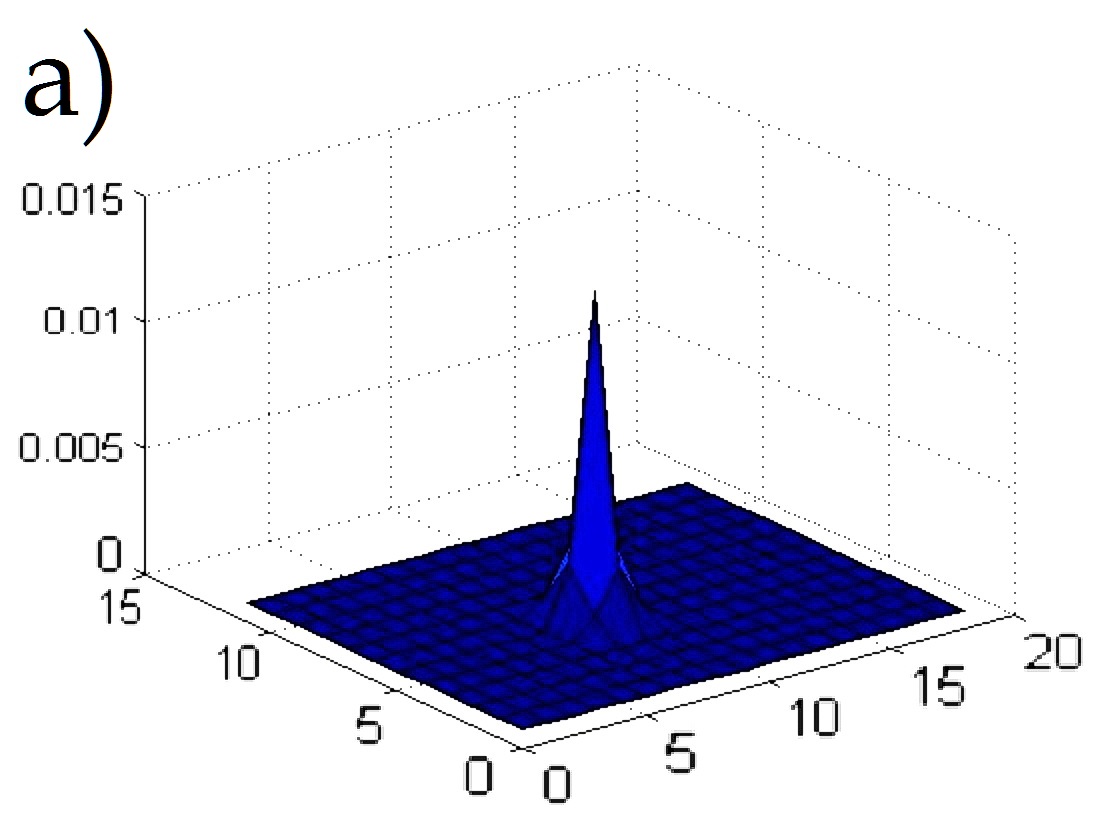}
\includegraphics[width=.5\linewidth]{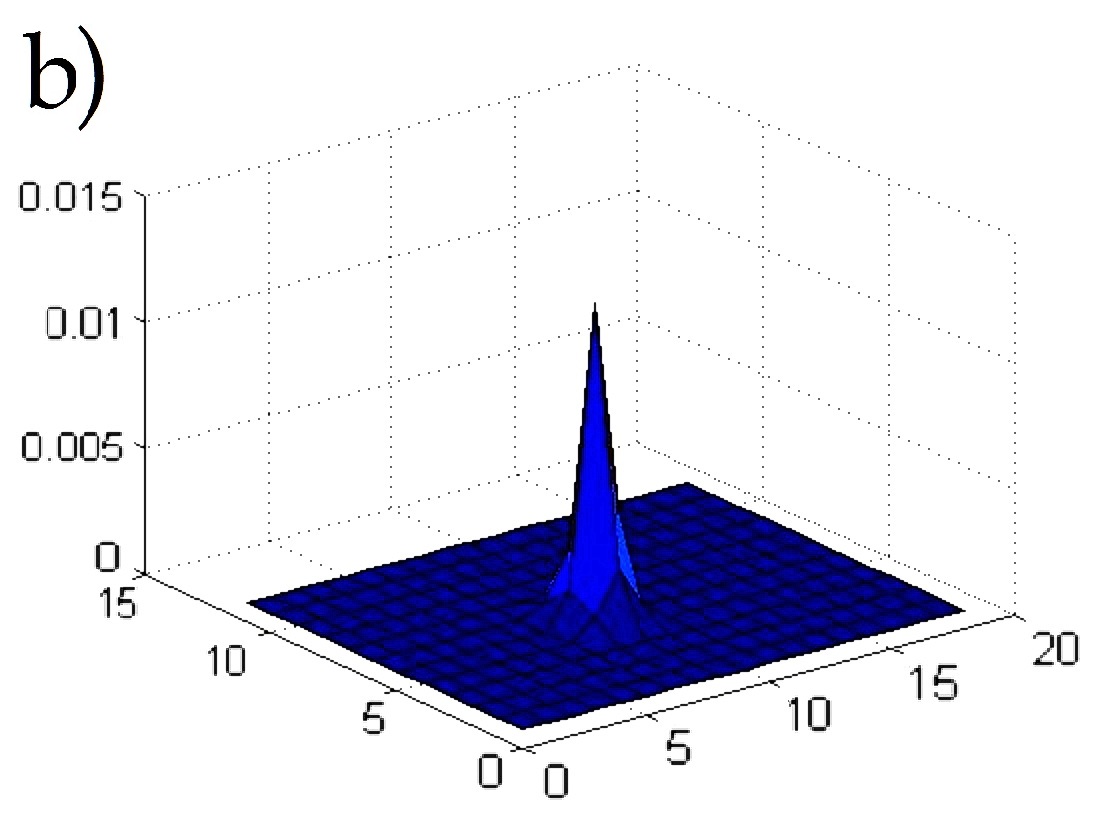}
\end{minipage}
\caption{  A typical coherent state $|\Psi_z|^2$ of the $C_1=2$ d-wave model (see Eq. \ref{interpol1}) computed with the gauge phase that maximizes the localization of (a) the Wannier states and (b) the coherent states.
We see that there is very little difference in their spread, as expected from the relatively flat gauge curvature. }
\label{fig:twogauges}
\end{figure}

There is a nice relation between Coulomb gauge $\theta_C$ (obtained by solving Eq.~\ref{eq:coulomb}) which maximally localizes the coherent states and $\theta_{W}$, the gauge that maximally localizes WFs. The latter, which was used in our previous work Ref. \onlinecite{qi2011}, is given by
\begin{equation}
\theta_{W}(k_x,k_y)=-\int_0^{k_x} a_x(p_x,k_y)dp_x +\frac{k_x}{2\pi}\int_0^{2\pi} a_x(p_x,k_y)dp_x
\label{eq:XL}
\end{equation}
We shall show that these two gauge phases differ only by a relatively small correction, at least for most sufficiently smooth Berry curvatures. Write
\begin{equation}
\theta_C= \theta_{W}+\theta_0
\end{equation}
where $\theta_C$ is the Coulomb gauge phase and $\theta_0$ is a relatively small correction. We have
\begin{eqnarray}
&&\nabla_k^2\theta_0(k_x,k_y)\nonumber\\
%&=&\nabla_k^2(\theta-\theta_W)\nonumber \\
&=& -\partial_x a_x -\left(-\partial_x a_x -\int_0^{k_x} \partial^2_y a_x dp_x +\frac{k_x}{2\pi}\int_0^{2\pi}\partial^2_y a_x dp_x \right)\nonumber \\
%&=& -\partial_x a_x -\left(-\partial_x a_x -\int_0^{k_x} \partial^2_y a_x(p_x,k_y)dp_x +\frac{k_x}{2\pi}\int_0^{2\pi}\partial^2_y a_x(p_x,k_y)dp_x \right)\nonumber \\
&=&\partial_y\left(\int_0^{k_x} f(p_x,k_y)dp_x -\frac{k_x}{2\pi}\int_0^{2\pi}f(p_x,k_y)dp_x \right),
\label{fy}
\end{eqnarray}
so that
\begin{equation} \nabla_k^2 (\partial_x\theta_0(k_x,k_y))= \partial_y (f(k_x,k_y)-\bar f (k_y))\end{equation}
where the Berry curvature $f=\partial_y a_x$ and $\bar f(k_y)=\frac1{2\pi}\int dk_x f(k_x,k_y)$ is $f$ averaged over $k_x$. Hence the correction $\theta_0$ is directly related to the nonuniformity of the Berry curvature, and disappears when the latter is constant w.r.t. to $k_x$ or $k_y$ (or both). We should thus expect it to be relatively small in magnitude. The comparison between the two gauge choices is shown in Fig. \ref{fig:twogauges}.

\subsubsection{Properties of the average coherent state spread}

Interestingly, the average spread of the coherent state $\int_{[-1/2,1/2]^2} d^2z\langle r^2\rangle_{\Psi_z}$ can be expressed in terms of the geometric quantity $\Omega_1$, to be defined shortly, and the Berry curvature $f$. As shown in Appendix \ref{spread}, the average spread of the coherent state can be rewritten as
\begin{eqnarray}
I[A]&=&\int_{[-1/2,1/2]^2} d^2z\langle r^2\rangle_{\Psi_z}\nonumber \\
&=&\frac{1}{\sqrt{2A}}\int d^2k \left(\left[|\nabla_k \phi|^2-|\vec{a}|^2\right]   + |\vec{a}_{new}|^2+\frac{1}{12} +\frac{A}{4\pi}\right) \nonumber \\
\label{eq:spread}
\end{eqnarray}
where $\vec{a}_{new}=\vec{a}+\nabla_k\theta_C$ is the Coulomb gauge connection that will minimize the spread. Note that $I[A]$ depends on $A$, the FQH aspect ratio that has yet to be chosen. Its optimal value will be the subject of the next subsection; here we will explore its $A$-independent properties.

Of central interest is the quantity $J$ given by
\begin{eqnarray}
J&=&\int d^2k\left(|\nabla_k \phi|^2-|\vec{a}|^2\right)   + |\vec{a}_{new}|^2
\label{J}
\end{eqnarray}
This quantity can be expressed in terms of the geometrical properties of the system. First, consider the bracketed terms
\begin{equation}
\Omega_1=\int d^2k \left(|\nabla_k \phi|^2-|\vec{a}|^2\right)
\label{omega1eq}
\end{equation}
$\Omega_1$ is actually a gauge invariant. This can be seen directly by replacing $\phi$ by $e^{i\theta}\phi$, or by rewriting $\Omega_1$ as a trace over certain position operators projected onto the occupied states. This will be shown in much more detail in Appendix \ref{omega1}, where its connection with the Wannier operator will also be made more explicit.

$\Omega_1$ can also be expressed in terms of the Fubini-Study metric $g$ associated with the quantum distance between two points on the BZ, i.e. $ds^2_{12}=1-|\langle \phi_1|\phi_2\rangle|^2$. Here $g_{ab}(k)=Re\langle \partial_a \phi|\partial_b \phi\rangle - \langle \partial_a \phi|\phi\rangle\langle\phi|\partial_b \phi\rangle $ (See Ref. \onlinecite{marzari1997} for a multiband expression). Referring back to Eq. \ref{omega1eq}, we see that
\begin{equation}
\Omega_1 = \frac{1}{4\pi^2}\int d^2k Tr g
\end{equation}
i.e. $\Omega_1$ is a trace over the metric integrated over the BZ. This is a geometrical quantity, and is obviously gauge-invariant. Physically, it is a measure of how orthogonal the Bloch states on neighboring points the BZ are: a large contribution to the integral arises at regions of the BZ where even nearby states quickly become orthogonal%\footnote{We can see this easily when there are two bands, and one is occupied. Write the Hamiltonian as $\sigma\cdot \vec d$, where $\vec d=(\sin\theta\cos\varphi,\sin\theta\sin\varphi,\cos\theta)$. Then the eigenstates $|\phi\rangle\propto (\sin\frac{\theta}{2}e^{i\varphi},\cos\frac{\theta}{2})^T$ and
% $Tr g= \sum_i(\partial_i\theta)^2+\sin^2\theta(\partial_i\varphi)^2$
% where $i=k_x,k_y$.
% We see that $Tr g$ indeed diverges more when the eigenstate changes faster.}
, i.e. $ds^2\rightarrow 1$.

The last term of $J$ as shown in Eq. \ref{J} is geometrical too, albeit not gauge-invariant. It can be expressed in terms of the Berry curvature of the system as we shall see below. In the optimal Coulomb gauge derived previously, $\nabla_k\cdot\vec{a}_{new}=0$. This permits us to write $\vec{a}_{new}=(-\partial_y\varphi,\partial_x \varphi)^T$, where
\begin{equation}
\nabla_k^2 \varphi(k_x,k_y) = f(k_x,k_y), 
\end{equation}
Hence $J$ can be written in terms of manifestly geometric quantities as follows:
\begin{eqnarray}
J&=&\int d^2k\left( [|\nabla_k \phi|^2-|\vec{a}|^2] + |\vec{a}_{new}|^2\right)\notag\\
&=& \Omega_1  + \int d^2k \left(|\partial_x \varphi|^2+|\partial_y \varphi|^2\right )\notag \\
&=& \Omega_1 - \int d^2k \varphi \nabla_k^2 \varphi =\Omega_1 - \int d^2k\varphi f   \notag \\
%&=& \Omega_1 - \int d^2k\varphi f   \notag \\
&=& \int d^2k \left [\frac{1}{4\pi^2}Tr g - \varphi f  \right]
\end{eqnarray}
On the last line, $\varphi$ is derived from the geometric Berry curvature $f$ via $\nabla_k^2\varphi = f$. Hence the optimal average spread in Eq. \ref{eq:spread} can be computed wholly from the Fubini-Study metric and the Berry Curvature of the system without first computing the Coulomb gauge connection.

Since we have specialized to the Coulomb gauge, $J$ can also be simplified via
\begin{eqnarray}
J&=& \int d^2k \left(|\nabla_k\phi_{new}|^2-|\vec{a}_{new}|^2\right)   + |\vec{a}_{new}|^2\notag \\
&=&\int d^2k |\nabla_k \phi_{new}|^2
\end{eqnarray}
where $\phi_{new}$ is the Bloch wavefunction in the Coulomb gauge. $J$ superficially looks like a real-space spread $\langle r^2\rangle$ of $\phi_{new}$. However, this cannot be strictly true, since for our regime of interest $C_1\neq 0$, $\phi_{new}$ is not periodic in the BZ and $\nabla_k^2$ cannot be fourier transformed into $r^2$. In fact, our coherent Wannier state construction circumvents this difficulty by involving Gaussian-weighted concatenated Wannier states that are periodic\footnote{If $C_1=0$, however, $\phi_{new}$ can be made periodic with a suitable choice of gauge and $J$ does indeed represent its real-space spread. In this case, $I[A]$ will be irrelevant as the coherent-state construction will be unnecessary.}. Note that $J$ is fundamentally different from $I[A]$, because the latter involves an averaging of coherent states over different centers-of-mass while the former involves only the Bloch states.

\subsubsection{The optimal aspect ratio $A$}

Setting $\frac{dI}{dA}=0$ in Eq.~\ref{eq:spread}, we obtain the optimal value of the aspect ratio $A$ as
\begin{eqnarray}
A_{opt}= \frac{1}{\pi}\int d^2k \left(\left[|\nabla_k \phi|^2-|\vec{a}|^2\right]   + |\vec{a}_{new}|^2+\frac{1}{12} \right) =  \frac{J}{\pi} + \frac{\pi}{3}\notag\\
\label{eq:aopt}
\end{eqnarray}
where $ J = \Omega_1 - \int d^2k\varphi f %$ or $ J
= \int d^2k |\nabla_k \phi_{new}|^2$.

For this optimal aspect ratio, the gaussian factor in the coherent state wavefunction (\ref{coherentformc1}) reads
\begin{eqnarray}
&&-\pi A\left(\frac{z_1}{C_1}-\frac{2\pi m +k_y}{2\pi}\right)^2\nonumber \\
&=&-\frac{1}{C_1^2}\left(J+\frac{\pi^2}{3}\right)\left(z_1-C_1\frac{2\pi m +k_y}{2\pi}\right)^2
\end{eqnarray}
Note that the $\frac{\pi^2}{3}$ term comes from the $dz_2$ integration, and can roughly speaking be regarded as a feature of the two-dimensionality of the coherent states. The above result agrees with the numerically obtained typical optimal $A$s of $\approx 2$ when $C=2$ and $\approx 1.4$ when $C=1$.

Substituting this optimal value of $A$ into the spread function $I$, we finally arrive at
\begin{equation}
I_{opt}=\int_{[-1/2,1/2]^2} d^2z\langle r^2\rangle_{\Psi_z}=\sqrt{2\pi \left(J+\frac{\pi^2}{3}\right)}
\end{equation}
As mentioned in the previous section, we can also optimize the locality of the PPs changing the Bloch basis themselves through coordinate redefinitions.
Now, we explicitly see how the locality depends on the smoothness of the Bloch states through
\begin{equation} J = \int d^2k |\nabla_k \phi_{new}|^2. \label{j}\end{equation}.
In the numerical results that follow, we see that systems with more bands generically have more complicated $\phi_{new}$ and hence poor locality of the PPs. This seems to be true for band insulators, although not for the QH LLs which always takes a similar functional form.

\subsection{Truncated Pseudopotential Hamiltonians on the lattice: Numerical results}
\label{sec:numerics}

Even after the abovementioned optimization procedure, the wavefunction of a coherent state still has a Gaussian decaying tail, which means the PP Hamiltonian (\ref{main}) still contains infinite number of terms when we expand it in the real space basis. To obtain a physical lattice Hamiltonian with finite interaction range, we can take a truncation and keep only a small number of dominant short-range terms. We have numerically studied the first PP Hamiltonian $U^1$ for four different models with Chern bands. The truncated PP Hamiltonians are summarized in Table \ref{table1}, and the definition and more details of these four models are presented in the rest of this section.

\begin{table}[htb]
\centering
\renewcommand{\arraystretch}{2}
\begin{tabular}{|l|l|l|}\hline
\textbf{Model} &\ $C_1$ &\ \textbf{The truncated first PP Hamiltonian $U^1$ } \\ \hline
Checkerboard  &\ 1 &\ $\displaystyle\sum_{<ij>}\rho_i\rho_j+0.93 \sum_{<<ij>>}\rho_i\rho_j+0.93\sum_{ijkl\in \square}c^\dagger_i c^\dagger_k c_jc_l+0.65\sum_{ijk\in \Delta}\rho_i c^\dagger_j c_k  $\\ \hline
Dirac  &\ 1 &\ $\displaystyle\sum_{<ij>}\rho_{i2}\rho_{j2}+0.4 \sum_{<<ij>>}\rho_{i2}\rho_{j2} +0.32 \sum_i \rho_{i1}\rho_{i2} -0.27 \sum_{<<<<ij>>>>}\rho_{i2}\rho_{j2}  $ \\   \hline
D-wave  &\ 2 &\ $\displaystyle\sum_{<<ij>>}\rho_{i1}\rho_{j2} +0.76 \sum_{<ij>} \rho_{i1}\rho_{j2} +[0.57e^{2.5i}\sum_{ijk\in \Delta}\rho_{i2}c^\dagger_{j1}c_{k2}+h.c.]$\\
 &\  &\ $ +0.44 \sum_{<<<ij>>>}\rho_{i2}\rho_{j2} -0.26\sum_{[ij]} \rho_{i2}\rho_{j2} $ \\ \hline
Triangular &\ 2 &\ $\displaystyle\sum_{<<AB>>}\rho_A\rho_B +0.8 \sum_{ABC\in \Gamma}\rho_C c^\dagger_A c_B  $ \\ \hline
\end{tabular}
\caption{The largest contributions to the first pseudopotential $U^1$ for various models in the lattice basis. Here $\rho_{i\sigma}=c^\dagger_{i\sigma} c_{i\sigma}$ where the $i$ is the position index and $\sigma=1,2$ is the spin index, if any. The brackets $<ij>$ refers to nearest-neighbor (NN) sites, $<<ij>>$ to next-nearest-neighbor (NNN) sites, etc. while $[ij]$ is the shorthand for eighth neighbor (NNNNNNNN) sites (see Fig. \ref{triangular}). The symbols $\square$,$\Delta$ and $\Gamma$ refers to special configurational patterns described in the main text. Note that $A,B,C$ are inequivalent sites in the triangular lattice model. }
\label{table1}
\end{table}

\subsubsection{Checkerboard model}

\begin{figure}
\centering
\begin{minipage}{0.99\linewidth}
\includegraphics[width=0.99\linewidth]{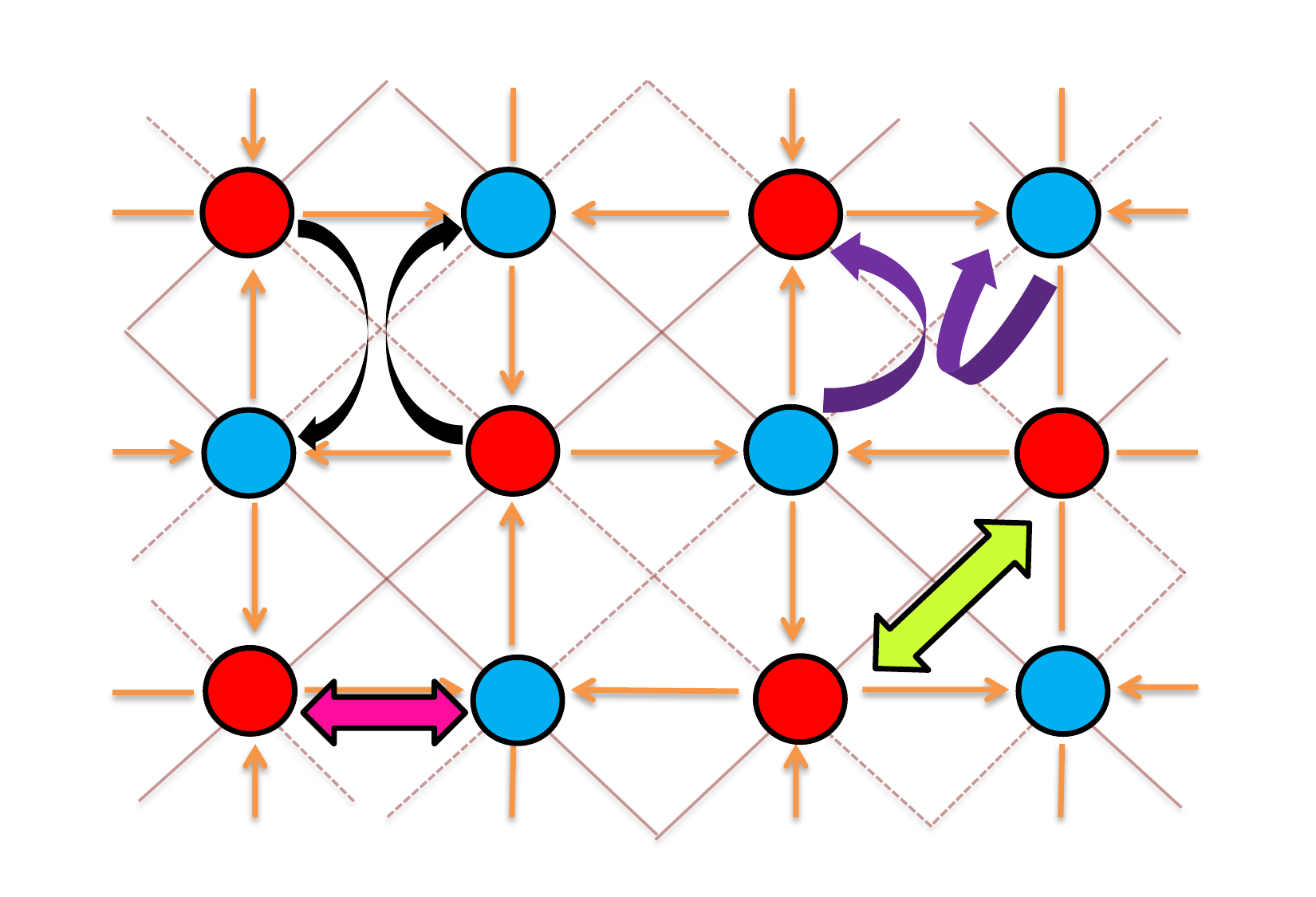}
\end{minipage}
\caption{  The truncated $U^1$ PP consists of two density-density interaction terms (pink and bright green of relative strengths $1$ and $0.93$) and two pair hopping terms (black and purple of strengths $0.93$ and $0.65$), as indicated in Eq. \ref{cbeq}. Density-density terms are represented by double-headed arrows. The black interaction involves two hoppings around a square plaquette. The purple interaction involves one self-hopping (i.e. density term) and one hopping among sites arranged in a triangle as shown. The red and blue sites sit on different sublattices. The thin arrows indicate the sign of the phase $\phi$ for the NN hoppings while the solid(dashed) diagonal lines indicate the sign of the NNN hoppings.}
\label{cb}
\end{figure}

The checkerboard (CB) model is a $C_1=1$ flat-band lattice model proposed in Ref. \onlinecite{sun2011} and also used in the previous subsection. As shown in Fig.~\ref{cb}, it consists of two square lattices interlocked in a checkerboard fashion. As defined below Eq. \ref{modelhcb} and in ~\cite{lee2013}, the NN interactions are parametrized by $t$ and exists between sites belonging to different sublattices. They carry the phase $\phi\neq \pi$ that produces the time-reversal symmetry breaking necessary for a nonzero Chern number. The NNN hoppings take values of $t'$ or $-t'$ depending on whether they are connected by a solid or dashed line as shown in Fig. \ref{cb}. 

We set the parameters to be $t=1$, $t'=1/(2+\sqrt{2})$ and $\phi=\pi/4$ as in
 Ref. \onlinecite{sun2011} to maximize the flatness of the
 occupied band. The dominant terms of $U^1$ are given by

\begin{eqnarray}
H&=&\displaystyle\sum_{<ij>}\rho_i\rho_j+0.93 \sum_{<<ij>>}\rho_i\rho_j+0.93\sum_{ijkl\in \square}c^\dagger_i c^\dagger_k c_jc_l\nonumber\\
& &+0.65\sum_{ijk\in \Delta}\rho_i c^\dagger_j c_k
\label{cbeq}
\end{eqnarray}
which consists of two density-density interaction ($\rho_i\rho_j$) terms and two pair hopping terms ($c^\dagger_i c^\dagger_k c_jc_l$ and $\rho_i c^\dagger_j c_k =c^\dagger_ic_i c^\dagger_j c_k $), as is illustrated in Fig. \ref{cb}.

\subsubsection{Dirac model}

\begin{figure}
\begin{minipage}{0.99\linewidth}
\includegraphics[width=0.99\linewidth]{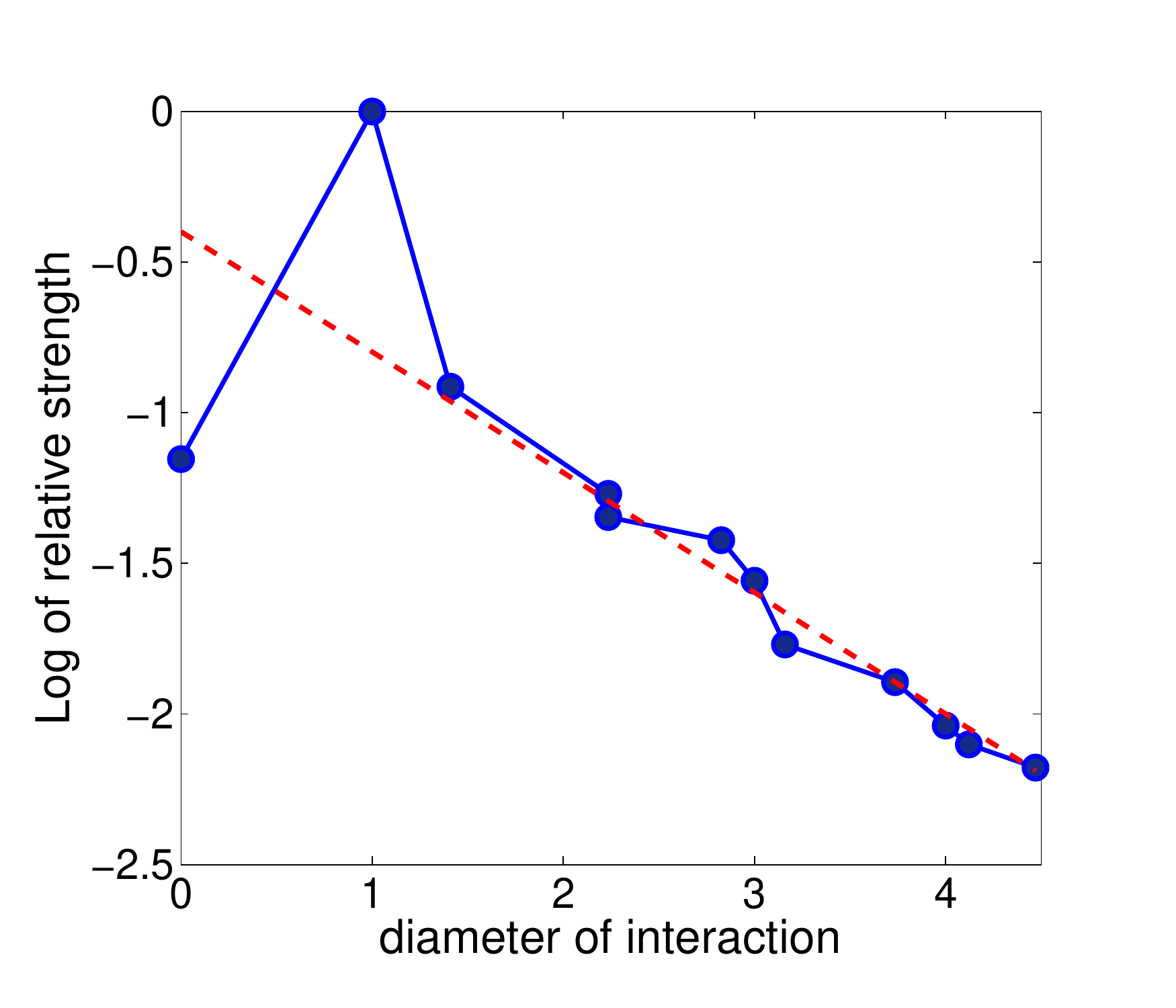}
\end{minipage}
\caption{  A Log plot of the relative strengths of the larger contributions to $U^1$ as a function of the diameter of the interaction contribution, i.e. spatial distance between the furthest pair in $c^\dagger_ic^\dagger_jc_kc_l$. The dominant term, which is normalized to have unit magnitude, is the NN density-density term with a diameter of $1$, while the term with smallest diameter is that of the same-site unequal spin density-density interaction. More than one type of interaction may have the same diameter, i.e. at diameter $\sqrt{5}\approx 2.23$. We observe an approximate spatial exponential decay of the interactions beyond the first few terms. The decay rate of $\approx 0.4$ depends on the model being studied.%The top 100 terms of $U^1$ for the Dirac model arranged by magnitude, inclusive of those related by symmetry. We see a dominant term that is $2.5$ times stronger than the subdominant contributions.
}
\label{diraclog}
\end{figure}

The lattice Dirac model\cite{Qi2006} provides one of the simplest realizations of a $C_1=1$ system on a lattice. Each site admits a spin-$1/2$ degree of freedom, and is connected to each other only through NN hoppings. In momentum space, the model is defined as
\begin{equation} H^{\text{Dirac}}(k)=d_0 I + \sum_i d_i \sigma_i \label{dirac01}\end{equation}
with $d_1=\sin k_x -\sin k_y$, $d_2=\sin k_x+ \sin k_y$ and $d_3=m+\cos k_x+\cos k_y$. For $0 < \pm m< 2 $, the Dirac cones at $(k_x,k_y)=(n_x\pi,n_y\pi)$, $n_x,n_y\in \mathbb{Z}$ lead to a Chern number of $\pm1$ for the lower band. The band can be made approximately flat by adjusting $d_0$. % and retaining a sufficient number of real-space terms.
It should be noted that the $d$-vector here has been chosen slightly differently from Ref. \onlinecite{Qi2006} for later convenience, but the two choices are equivalent by a simple basis rotation. The leading contributions of $U^1$ are given by
\begin{eqnarray}
H&=&\displaystyle\sum_{<ij>}\rho_{i2}\rho_{j2}+0.4 \sum_{<<ij>>}\rho_{i2}\rho_{j2}\notag \\
&& +0.32 \sum_i \rho_{i1}\rho_{i2} -0.27 \sum_{<<<<ij>>>>}\rho_{i2}\rho_{j2}
\label{diraceq}
\end{eqnarray}
From this equation we can see that the $U^1$ PP is dominated by the first term (see Fig. \ref{diraclog} (a)) which is a nearest neighbor density-density interaction between electrons with down spin or pseudospin. Although in Eq. (\ref{diraceq}) we have also kept other subleading terms, it is reasonable to conjecture that only the nearest neighbor term is already a good approximation to the PP Hamiltonian, which admits a Laughlin ground state at $1/3$ filling.

\begin{figure}
\begin{minipage}{0.99\linewidth}
\includegraphics[width=0.99\linewidth]{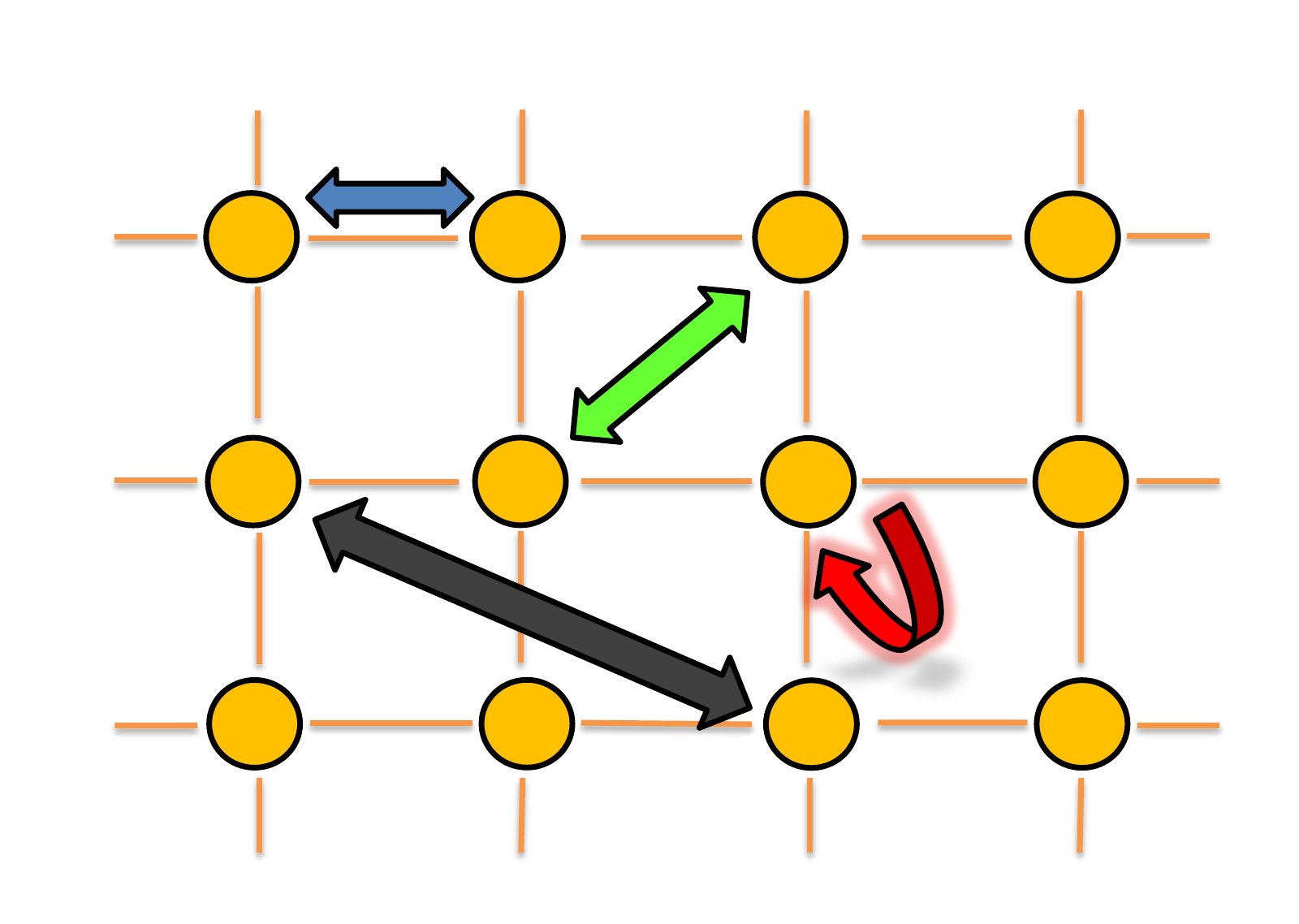}
\end{minipage}
\caption{  The truncated $U^1$ PP of the Dirac Model consists of four density-density interaction terms (blue, bright green,red and grey corresponding to relative strengths of $1$,$0.4$,$0.32$ and $0.27$), as indicated in Eq. \ref{diraceq}. The red arrow represents a density-density interaction between the up and down spins on the same site, while the other arrows all involve the down spins only.}
\label{diracmodel}
\end{figure}

\subsubsection{The Chern number $2$ D-wave model and its interpolation to two decoupled Dirac models}

Here we proceed to models with Chern number $2$, where the FCI system is mapped onto a bilayer QH system with decoupled layers. We consider a natural generalization of the Dirac model:
\begin{eqnarray}
H^d(k)&=&\sigma_x\left(\cos k_x-\cos k_y\right)+\sigma_z\left[\cos(k_x-k_y)\right.\nonumber\\
& &\left.-\cos(k_x+k_y)\right]+\sigma_y\left(\cos k_x+\cos k_y\right)\label{Hdwave}
\end{eqnarray}
This model can be viewed as a $d$-wave version of the lattice Dirac model, which is obtained by replacing the spin-dependent hopping terms $\sin k_x, \sin k_y$ in the Dirac model by the terms $\cos k_x-\cos k_y$ and $\cos(k_x-k_y)-\cos (k_x+k_y)$ with $d$-wave symmetry. The $d$-wave symmetry can be seen clearly by expanding this model near $(k_x,k_y)=(0,0)$, which leads to
\begin{eqnarray}
H^d(k)&\simeq &-\frac12\left(k_x^2-k_y^2\right)\sigma_x+2k_xk_y\sigma_z+2\sigma_y
\end{eqnarray}
The model has a Chern number $2$ for its lower band, which can be understood as coming from two quadratic touching points at $(0,0)$ and $(\pi,\pi)$ with the degeneracy lifted by the $\sigma_y$ term\cite{Chong2008,Sun2009}.

As is discussed in the previous subsection, the Chern number $2$ model (on an even-by-even sized lattice) is mapped to two decoupled ``layers" of Landau levels. Two sets of coherent states can be constructed for these two layers, and pseudopotential Hamiltonians can be defined for layer-decoupled FCI states, {\it i.e.}, direct product of Laughlin states in each layer. For more generic states such as the Halperin $(mnl)$ states\cite{Halperin1983} with interlayer coupling, the pseudopotential Hamiltonian is not known, except the special case of $n=m,~l=n-1$ which is a singlet state\cite{Halperin1983}. Here we will focus on the first PP Hamiltonian for each layer, which has $(330)$ state as its ground state. This state is of particular interest since it is a topological nematic state\cite{Barkeshli2012} in which lattice dislocations become non-Abelian defects, and has not been realized in any existing FCI models.

Before computing the pseudopotential Hamiltonian for this model, it is interesting to note a relation of this model to the Dirac model discussed in the last subsection. Since Chern number is the only topological invariant for a 2D energy band, a model with a $C_1=2$ band can always be adiabatically deformed into one with two decoupled $C_1=1$ bands, as is demonstrated by the Wannier state mapping. For the model in Eq.(\ref{Hdwave}), this equivalence can also be shown explicitly by an adiabatic deformation of the Hamiltonian to that of two decoupled copies of Dirac model in Eq.(\ref{diracmodel}). For this purpose, consider the following parameterized Hamiltonian

\begin{figure}[H]
\begin{minipage}{\linewidth}
\includegraphics[width=.32\linewidth]{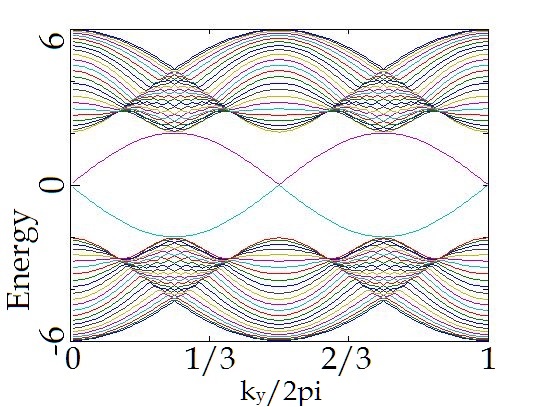}
\includegraphics[width=.32\linewidth]{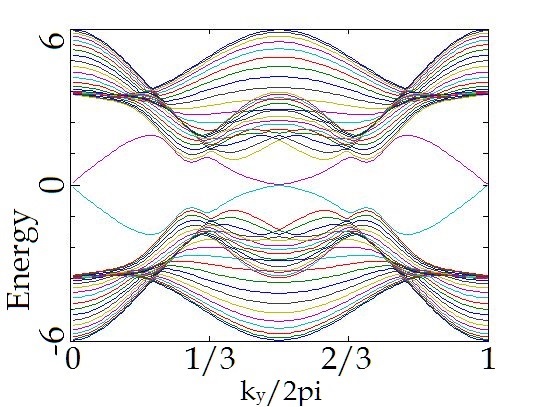}
\includegraphics[width=.32\linewidth]{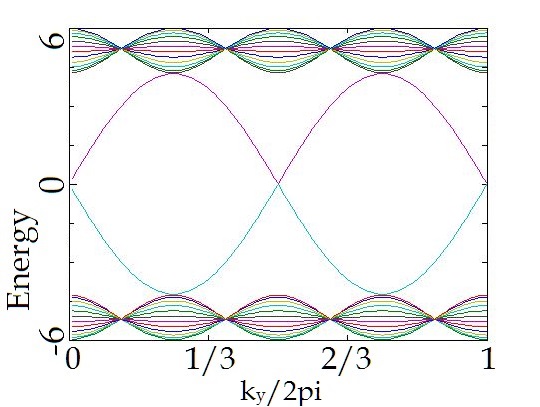}
\end{minipage}
\caption{Left to Right) The bandstructures of $H(0)$, $H(0.5)$ and $H(1)$ as a function of $k_y$. The edge states remains qualitatively the same during the entire interpolation, with no gap closure.}\label{interbands}\end{figure}
\begin{eqnarray}
 H_\lambda(k)&=&\sigma_x \left[(1-\lambda)[\sin (k_x+k_y)-\sin(k_x-k_y)]\right.\nonumber\\
 & &\left.+\lambda (\sin k_x -\sin k_y)\right]\nonumber\\
 && +\sigma_y\left[(1-\lambda)[\sin (k_x+k_y)+\sin(k_x-k_y)]\right.\nonumber\\
 & &\left.+\lambda (\sin k_x +\sin k_y)\right] \nonumber\\
 &&+ \sigma_z[(1-\lambda)+\cos(k_x+k_y)+\cos(k_x-k_y)]\nonumber \\
\label{interpol1}
\end{eqnarray}
$H_1(k)$ is equivalent to $H^d$ in Eq. (\ref{Hdwave}) by a translation $(k_x,k_y)\rightarrow \left(k_x+\frac{\pi}2,k_y+\frac{\pi}2\right)$.
$H_0(k)$ only contains second neighbor couplings, so that the two sublattices are decoupled, as is illustrated in Fig. \ref{interpol1} (Left panel). Restricted to each sublattice, this model is actually the Dirac model Eq. (\ref{dirac01}) with $m=1$. For all $\lambda\in[0,1]$, the model is gapped with $C_1=2$ in the occupied band, as is shown in Fig. \ref{interbands} by the energy spectrum on a cylindrical geometry. %, while $H(1)$ describes a $C=2$ system with a d-wave type singularity in the d-vector. Indeed,
%\begin{equation}
%H_1(k)\sim \sigma_x \left(\frac{p_y^2-p_x^2}{2}\right)+2\sigma_y + 2p_xp_y\sigma_z
%\end{equation}
%
%for small $\vec p = \vec k -(\pi/2,\pi/2)$. Indeed, the d-vector will precess by $2(2\pi)=4\pi$ as it is brought around $(\pi/2,\pi/2)$.

%As shown in the left illustration of Fig. \ref{interpol}, the decoupled Dirac model consists of two interlocking inequivalent Dirac systems. It has only NN hoppings within one subsystem, or equivalently NNN hoppings on the whole lattice.
If we construct the Wannier states of this model in the ordinary way by Fourier transforming the Bloch state along $k_x$ direction with fixed $k_y$ at $\lambda=0$, the Wannier states and their correspondingly coherent states do not have a direct relation with those of the Dirac model since the translation along $x$ direction exchanges the two sublattices. From the last subsection, we have observed that the PP Hamiltonian of the lattice Dirac model has a nice short-ranged form. We will thus like to define an alternative Wannier basis for the $C=2$ model which is localized along the diagonal $(1,1)$ direction.

\begin{figure}
\begin{minipage}{0.99\linewidth}
\includegraphics[width=0.49\linewidth]{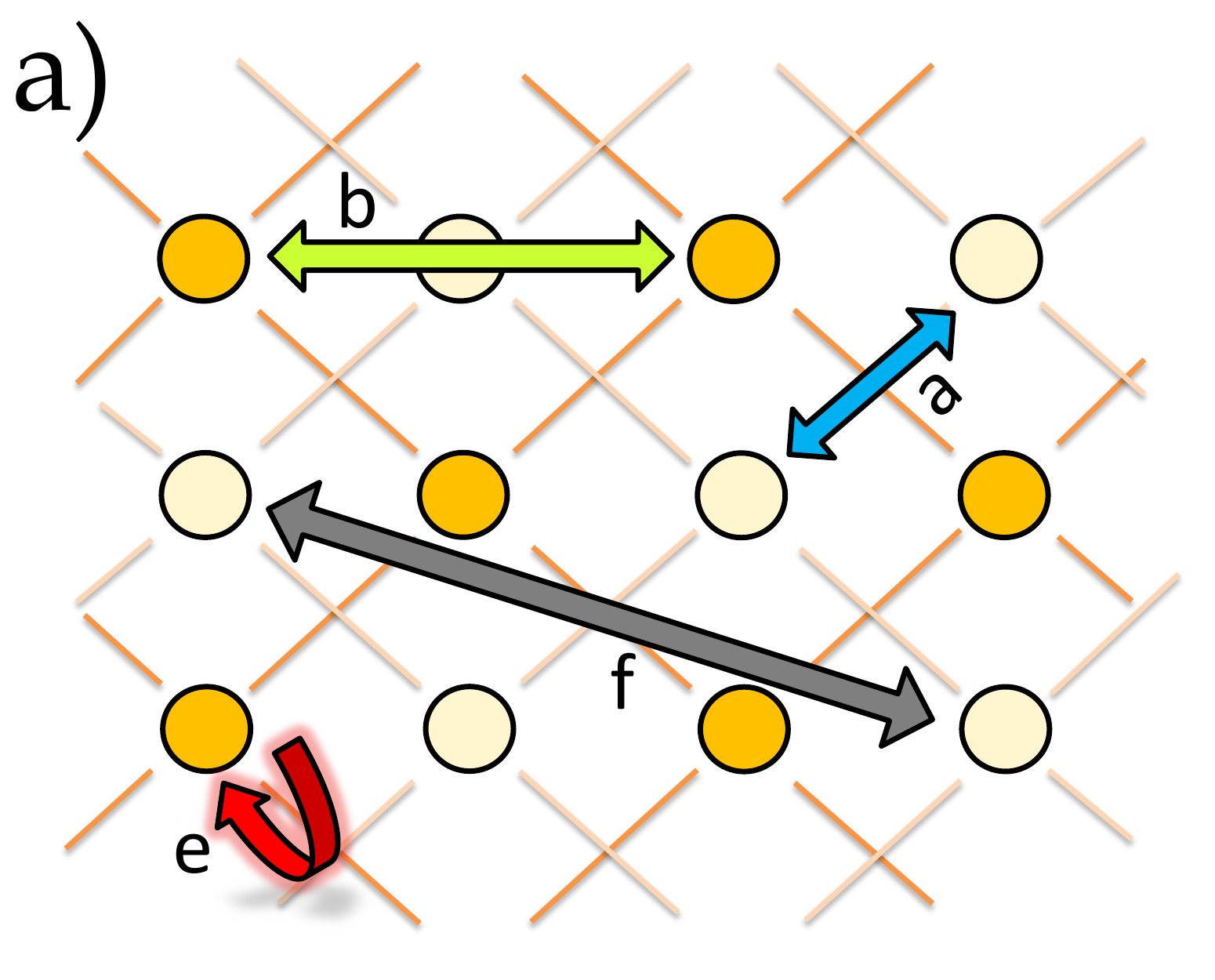}
\includegraphics[width=0.49\linewidth]{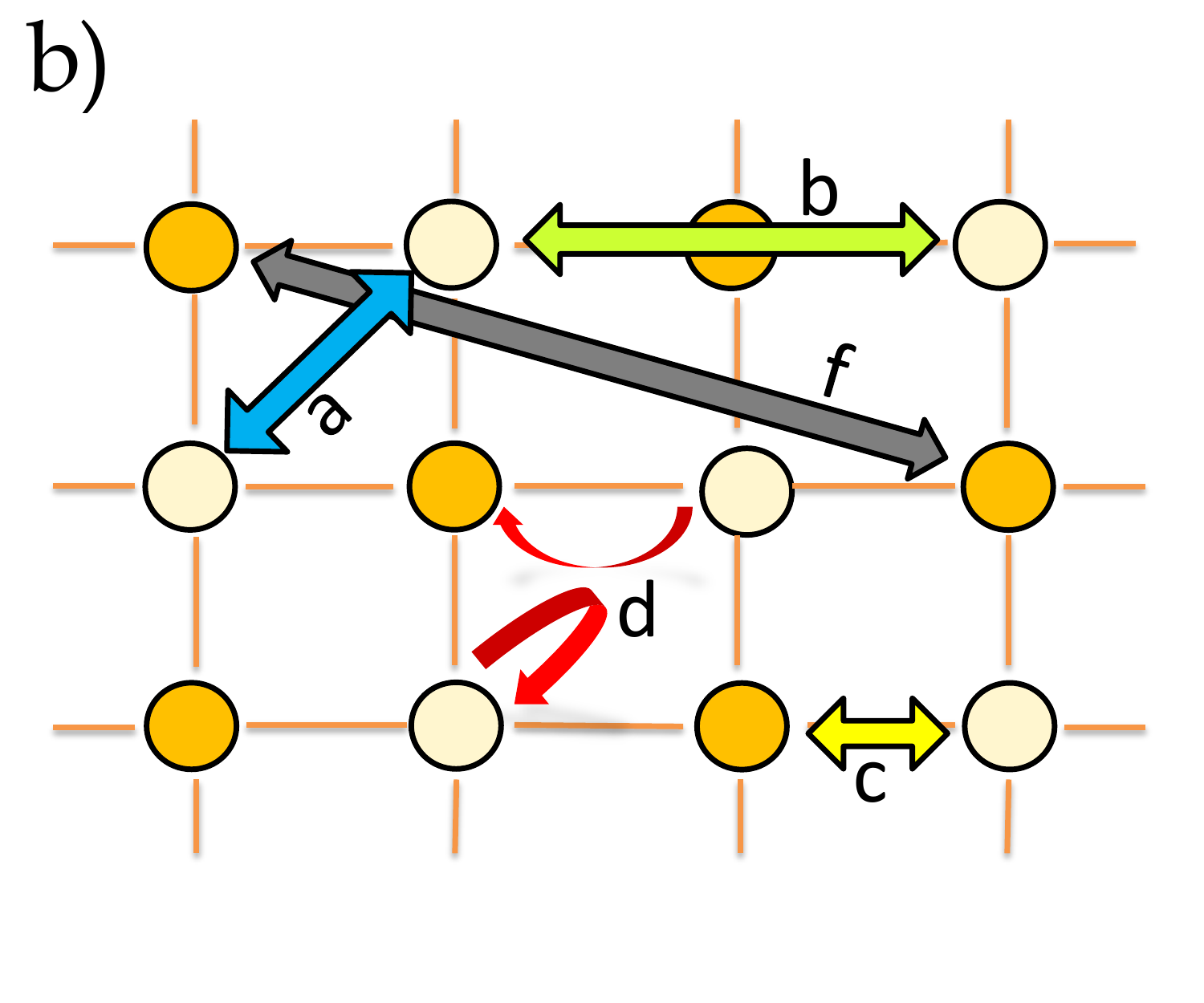}
\end{minipage}
\caption{  a) The truncated $U^1$ PP for the decoupled Dirac limit $H'_{\lambda=0}$. It has the same interactions as that of the original Dirac model (Fig. \ref{diracmodel}), but is rescaled by a factor of $\sqrt{2}$ and rotated by $\pi/4$. For instance, the NN interaction in the original Dirac model becomes the NNN interaction. Note that all interactions connect only sites in the same sublattice, as indicated by the diagonal bonds illustrated above.
b) The truncated $U^1$ PP for the d-wave limit $H'_{\lambda=1}$. It consists of four density-density interaction terms (blue, yellow light green and grey of relative strengths $1$, $0.76$, $0.44$ and $0.26$) and a pair hopping term (red of relative strength $0.57$), as indicated in Eq. \ref{dwaveeq}. The red interaction involves one self-hopping (i.e. density term) and one hopping among sites arranged in a triangle as shown. There is no decoupling between the sublattices in this case. All interactions are labeled by letters which will appear again in Fig. \ref{strengthgraph}.}
\label{interpol}
\end{figure}

The diagonally directed Wannier states and coherent states can be constructed through a redefinition of the unit cell $x'=x-y,y'=y$ or $k'_x=k_x,k'_y=k_x+k_y$. The Wannier states defined after the unit cell redefinition are $k_y'$ eigenstates. A flux that couples to $k'_y$ will now cause a spectral flow of the WFs in the $x'$ direction which thus stays in the same sublattice. In this definition, the decomposition of the $C_1=2$ band into two layers automatically reduce to the decomposition to the Dirac model on the two sublattices in the $\lambda=0$ limit. In this new coordinates, the Hamiltonian reads
\begin{eqnarray}
H'_\lambda(k') &=&\sigma_x \left[(1-\lambda)[\sin k'_y-\sin (2k'_x-k'_y)]\right.\nonumber\\
 & &\left.+\lambda (\sin k'_x -\sin (k'_y-k'_x))\right]\notag\\
 && +\sigma_y \left[(1-\lambda)[\sin k'_y+\sin(2k'_x-k'_y)]\right.\nonumber\\
 & &\left.+\lambda (\sin k'_x +\sin (k'_y-k'_x))\right] \notag\\
 &&+ \sigma_z\left[(1-\lambda)+\cos k'_y+\cos(2k'_x-k'_y)\right]
\label{interpol2}
\end{eqnarray}
This is the expression of the Hamiltonian which we will use for calculating $U^1$. For $\lambda=0$ 0the Hamiltonian reduces to that of the $C_1=1$ Dirac Hamiltonian Eq. \ref{dirac01} if we transform its coordinates to $r''=(x'',y'')$ on each sublattice:
\begin{eqnarray}
x''&=&\frac{1}{2}x'=\frac{x-y}{2},~
y''=\frac{1}{2}x'+y'=\frac{x+y}{2}\nonumber\\
%\end{eqnarray}
%or
%\begin{eqnarray}
k''_x&=&2k'_x-k'_y,~
k''_y=k'_y
\end{eqnarray}
This unit cell redefinition belongs to the type described in Eqs. \ref{abredef1} and \ref{abredef2} with $\alpha=\frac{1}{2}$ and $\beta = -\frac{1}{2}$. As elucidated in the sentences preceding them, the coherent states in the new coordinates will be identical to those from the old ones as long as the periodic part of the Bloch state $\phi(k')$ remain invariant under such a transformation. Indeed, we mathematically see why the PP of the decoupled Dirac model ($\lambda=0$) is identical to that of the $C_1=1$ Dirac model up to rescaling and rotation.

\begin{figure}[H]
\begin{minipage}{\linewidth}
\includegraphics[width=.99\linewidth]{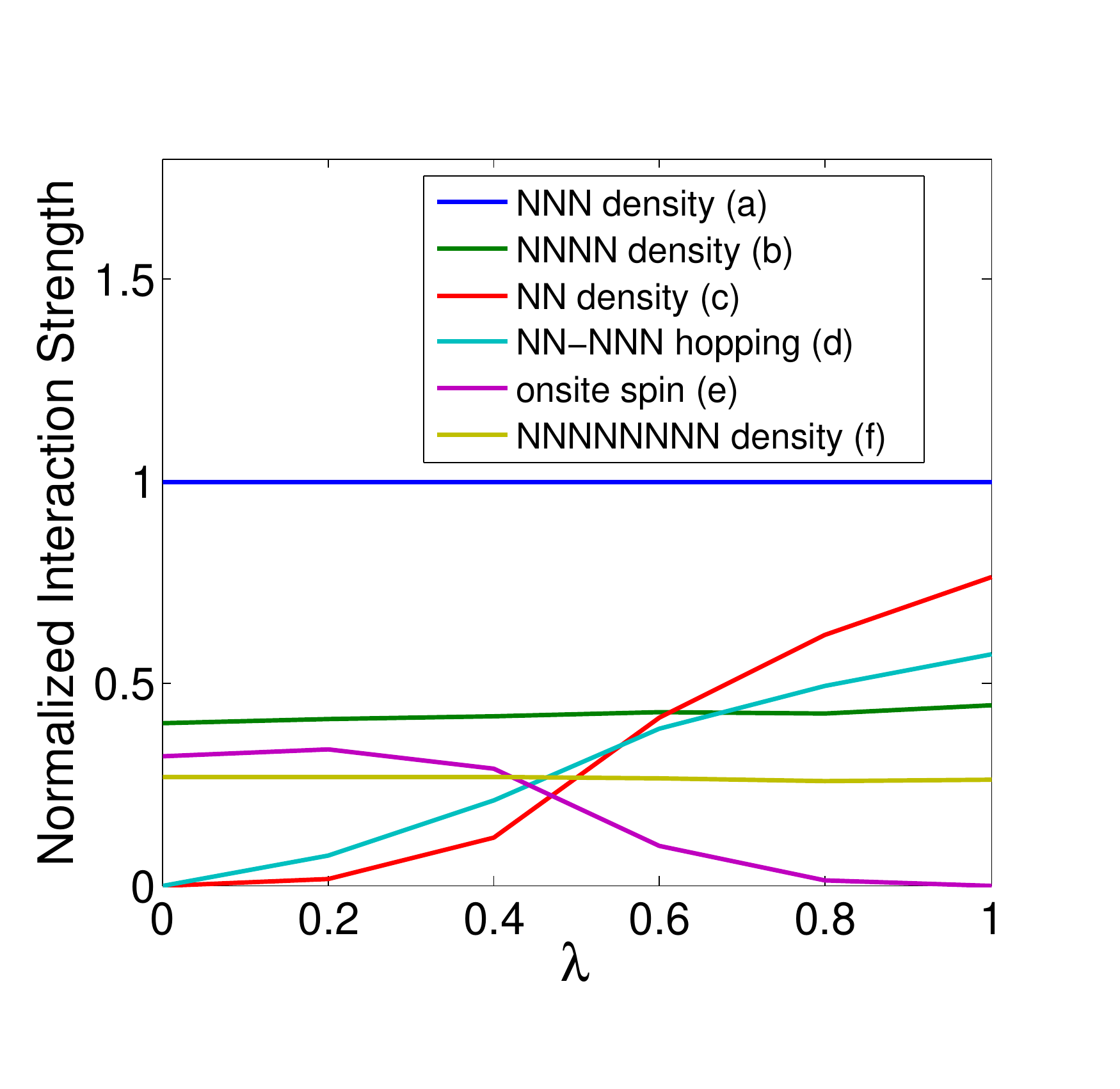}
\end{minipage}
\caption{The relative magnitudes of the various interaction terms in $U^1$ of $H'_\lambda$. The amplitude of the leading term has been normalized to $1$. Terms which are present in both models remain stable throughout the interpolation. Their exact definitions can be found in Fig. \ref{interpol}, where they are labelled graphically from a) to f). }
\label{strengthgraph}\end{figure}

The numerical results of the truncated PP Hamiltonian is shown in Fig. \ref{interpol} and \ref{strengthgraph}. For $\lambda=0$, the dominant terms of $U^1$ is identical to that in Eq. \ref{diraceq} after rescaling and $\frac{\pi}4$ rotation. For the $d$-wave model at $\lambda=1$, the PP Hamiltonian is given by
\begin{eqnarray}
H&=&\displaystyle\sum_{<<ij>>}\rho_{i1}\rho_{j2} +0.76 \sum_{<ij>} \rho_{i1}\rho_{j2}\notag\\
&& +[0.57e^{2.5i}\sum_{ijk\in \Delta}\rho_{i2}c^\dagger_{j1}c_{k2}+h.c.]\notag\\
&&+0.44 \sum_{<<<ij>>>}\rho_{i2}\rho_{j2} -0.26\sum_{[ij]} \rho_{i2}\rho_{j2}
\label{dwaveeq}
\end{eqnarray}
As shown in Fig. \ref{strengthgraph}, the leading term remains the same as $\lambda=0$, while the relative magnitudes of NNN, NNNN and NNNNNNNN (eighth-nearest neighbor) density-density interaction terms of $U^1$ remain stable across the interpolation. This attests to the robustness of our PP construction w.r.t. deformations that do not change the topology of the model. The other three terms vary significantly because they exist only in $\lambda\neq 0$ models. For instance, the NN density-density and NN-NNN hopping terms do not exist in the decoupled Dirac limit because the NN site does not belong to the same sublattice.

\subsubsection{Triangular lattice model}

\begin{figure}[H]
\begin{minipage}{0.99\linewidth}
\includegraphics[width=.5\linewidth]{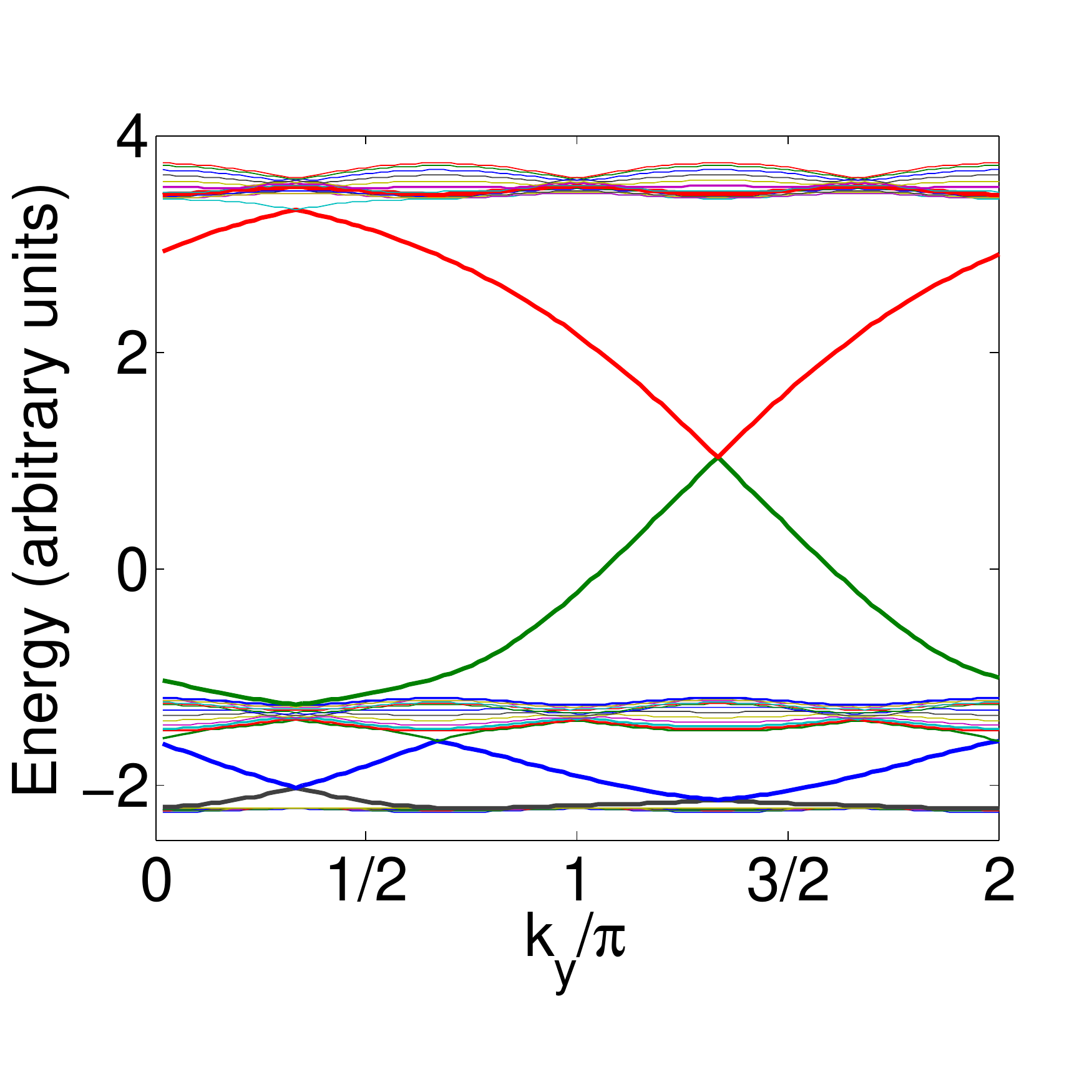}
\includegraphics[width=.5\linewidth]{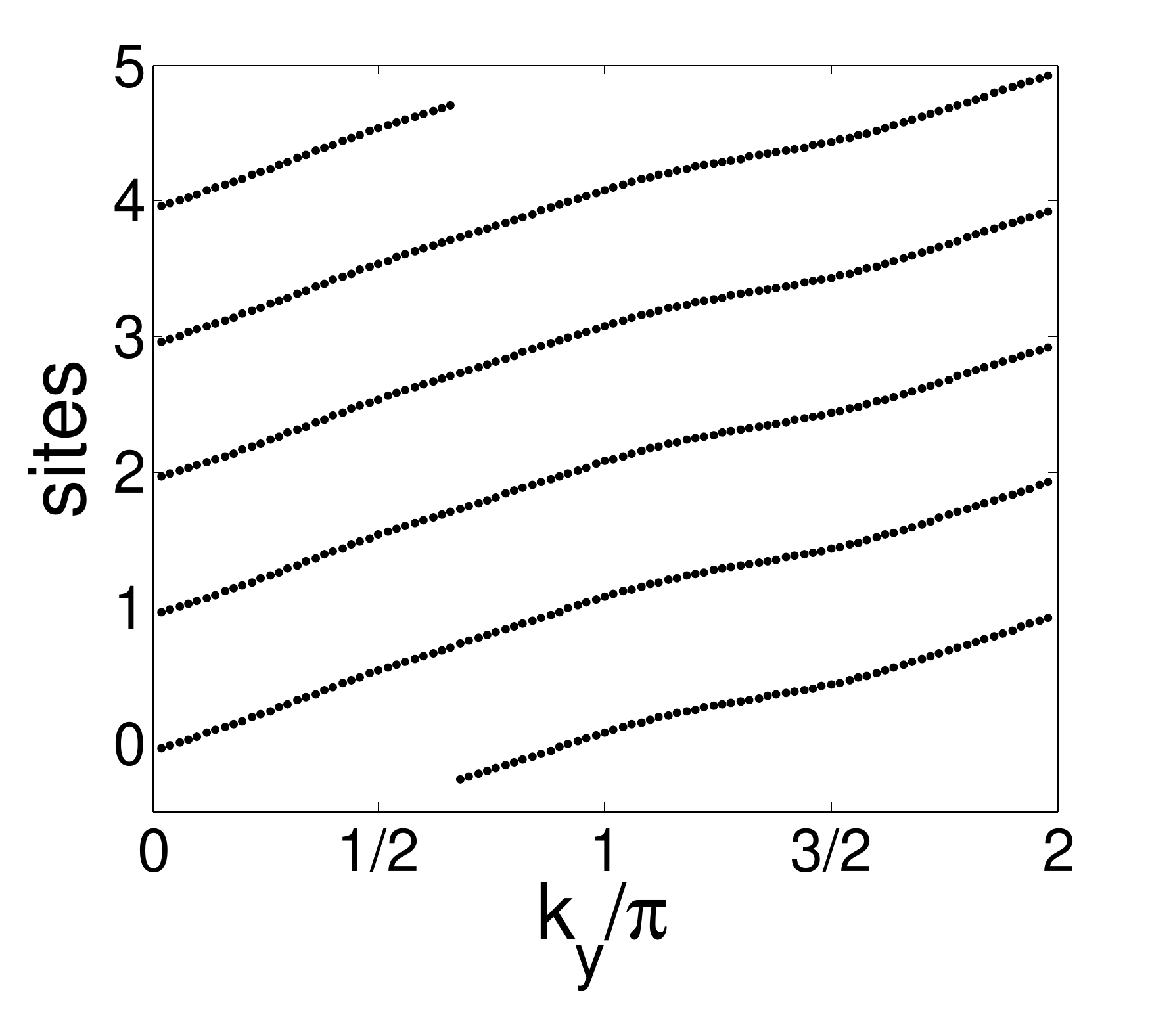}
\end{minipage}
\caption{Left) Band structure of the triangle lattice model, which is the same as in Ref. \onlinecite{wang2011}. We can see the flatness of the lowest occupied band. Right) The Wannier polarization evolves by two sites as the flux $k_y\rightarrow k_y+A_y$ cycles over $2\pi$. }
\label{triangularband}
\end{figure}

As a last example, we consider another $C_1=2$ model, the 3-band triangular lattice flatband model introduced in Ref. \onlinecite{wang2011}. Its lowest (occupied) topological flat band carries a Chern number of $2$, as evidenced in the presence of its two edge states on Fig. \ref{triangularband}. Each unit cell contains three inequivalent spinless sites, leading to three bands with asymmetrical dispersions. From its real-space description detailed in Ref. \onlinecite{wang2011}, we obtain its momentum-space Hamiltonian $h^{Tri}_{ij}(k)$ with
\begin{eqnarray}
h_{11}&=&t'(\cos(k_y+\phi)+\cos(k_x-\phi)+\cos(k_y-k_x-\phi))\notag\\
h_{12}&=& t(1+e^{i(k_x-k_y)}+e^{-ik_y})\notag\\
h_{13}&=& t(e^{i(k_x-2k_y)}+e^{i(k_x-k_y+2\phi)}+e^{-i(k_y+2\phi)})\notag\\
h_{22}&=&t'(\cos(k_y-\phi)+\cos(k_x+\phi)+\cos(k_y-k_x+\phi))\notag\\
h_{23}&=&-t(1+e^{i(k_x-k_y-2\phi)}+e^{i(2\phi-k_y)})\notag\\
h_{33}&=& -t'(\cos k_y+\cos k_x +\cos(k_y-k_x))
\end{eqnarray}
with $h_{ij}=h^*_{ji}$. $t$ and $t'$ parameterizes the magnitudes of the NN and NNN hoppings respectively, and $\phi$ provides the necessary time-reversal symmetry breaking. These parameters are chosen to take values of $t=t'=1/4$ and $\phi=\pi/3$ for maximum flatness of the occupied band.

\begin{figure}
\begin{minipage}{0.99\linewidth}
\includegraphics[width=0.8\linewidth]{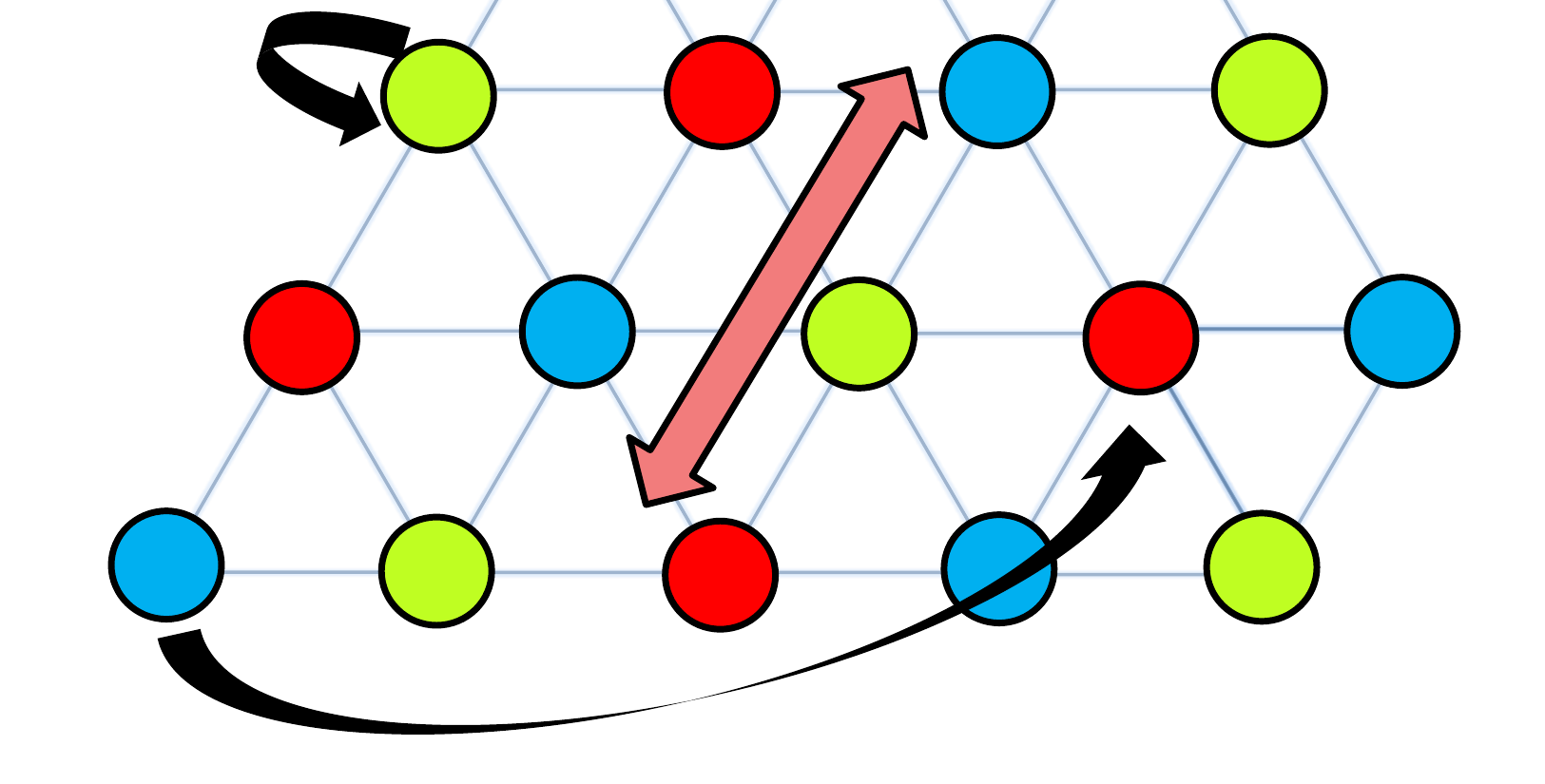}
\end{minipage}
\caption{  The truncated $U^1$ PP of the triangular model consists of a leading density-density interaction term (pink) followed by a pair  hopping term (black of relative strength $0.8$), as indicated in Eq. \ref{triangulareq}. As in previous figures, the density-density term is represented by a double-headed arrow. The black hoppings involve one self-hopping (density interaction) and a rather far jump. Note that the pink interaction between sites of type A,B is not equivalent to a similar-looking one between NNN A,C and B,C pairs. }
\label{triangular}
\end{figure}

For this model, the dominant terms in $U^1$ are given by
\begin{eqnarray}
H&=&\displaystyle\sum_{<<AB>>}\rho_A\rho_B +0.8 \sum_{ABC\in \Gamma}\rho_C c^\dagger_A c_B
\label{triangulareq}
\end{eqnarray}
where A,B and C refers to the inequivalent sites in the unit cell shown in Fig. \ref{triangular}. The notation $<<AB>>$, for instance, refers to the NNN A and B pairs, and not the NNN pairs among all three types.

In this model, the subleading interaction (of magnitude $0.8$) is still almost as large as the leading one. This is because the coherent states in this model are less localized than other models we studied. As captured in Eq. \ref{j}, the spread of the coherent states increases with the quantity $J$ which is a measure of the total complexity of the Bloch states. This complexity increases with the number of bands $N$. In addition, it becomes more difficult to find a distinctive dominant term when $N$ is large because the number of possible interaction terms scales like $N^4$.

\section{Conclusion for mapping I: WSR}

We have defined an exact mapping between the Hilbert spaces of the occupied band of a Fractional Chern Insulator, and the lowest Landau level of a Quantum Hall system. This mapping has allowed us to compare the FCI interactions with the QH Pseudopotentials at the operator level. 

We have employed the comparison from both directions: FCI to FQH and FQH to FCI. The mapping in the former direction has allowed us to decompose bare FCI interactions 
into their PP components, as well as generically PP-nondecomposable parts. The appearance of the latter arises from the partial breaking of the magnetic
translation group, which appears to be reemergent at low energies in
the FCI models. 

At the same time, we have also developed an elegant approach for generalizing the second-quantized PPs to arbitrarily many bodies on the cylinder as well as the torus. This helped establish a basis for further studying unconventional FCI phases via appropriately designed trial Hamiltonians. We believe that it will be interesting to
employ the PP perspective as a complementary tool to entanglement measures studies that further develop our understanding of
fractional Chern phases, as well as to extend its applicability to Chern bands
of higher Chern number~\cite{wang2011c,maissam,maxemil,Yang2012,liulauch} and two-dimensional fractional topological insulator models~\cite{levin2009,scharfenberger2011,Neupert2011a,lu2012}. Indeed, the use of many-body pseudopotential both as trial Hamiltonians and as effective Hamiltonians opens up new branches of research. For example, it will be interesting to apply the pseudopotential formalism to the model discussed in the work Ref.~\onlinecite{wangsheng}, where convincing numerical evidence for a stable $\nu=1$ non-Abelian state is presented. Other states that can be realized with our many-body PPS include fractional Abelian states such as the Laughlin
series, fractional non-Abelian states such as the Read-Rezayi series, and
also even more exotic states such as a FCI Gaffnian state~\cite{yoshioka1989,simon-07prb075317}. For the Pfaffian FCI state, for example, it is interesting to study the interpolation of the exact FCI Hamiltonian to a generic FCI
model and see whether adiabatic connectivity can be reached both at
the level of Hamiltonians and entanglement spectra~\cite{thomale-10prl180502}. Some results along these exciting directions have already materialized in the recent work~\cite{lee2015pp}.

Applying the WSR mapping in the latter FQH to FCI direction allows one to ``reverse engineer" a FQH Pseudopotential onto a generic FCI systems. Doing so enables us to once and for all write down the exact interaction Hamiltonian that admits certain desired groundstates while precluding the onset of magnetic translation symmetry breaking contributions. However, there is an inevitable proliferation of terms and a truncation in the range of interaction will be necessary for any practical implementation. The truncation error will be minimized if we use the Coulomb Gauge and choose an FQH basis corresponding to a special coherent state aspect ratio $A_{opt}=J/\pi+\pi/3$, where $J$ depends only on the geometric properties of the FCI system, namely its Berry curvature and Fubini-Study metric. To illustrate the robustness of our approach, we numerically obtained the first PP in various models with Chern number $1$ and $2$, including the CB model, Dirac model, d-wave model and triangular lattice model. The PPs are found to be reasonably well-approximated by a few dominating terms shown in Table I, especially for that of the flattened Dirac model where only one term dominates. Hence one prediction of our work is that topological nematic states are likely to be realized with the interaction Hamiltonians shown in Table I (or perhaps even further truncated versions thereof). One concomitant challenge will be to identify FCI systems where the proliferation of terms is sufficiently benign for a sensible truncation to still support their coveted exotic states.

While generic FCI models will only contain two-body interactions at the bare level,
our many-body pseudopotentials will be useful for studying
interactions beyond the effective single-band level. For instance, they are generically generated when interband scattering is considered. Moreover, following the construction introduced in Ref.~\onlinecite{lee2004} for the hardcore potential, we can map all these states to effectively one-dimensional models (see also Appendix~\ref{nbody} and~\ref{3body}), and even higher dimensional generalizations~\cite{chern2010}, of featureless Mott insulators with exotic ground state and quasiparticle properties, which are already worth studying in their own right.

\chapter{Mapping II: Exact Holographic mapping}

\section{Introduction}
In the recent years, holographic duality, also known as the Anti-de-Sitter space/Conformal Field Theory (AdS/CFT) correspondence\cite{maldacena1998,witten1998,gubser1998}, has attracted tremendous research interest in both high energy and condensed matter physics. This correspondence is defined as a duality between a $D+1$-dimensional field theory on a fixed background geometry and a $D+2$-dimensional quantum gravity theory. 
The best understood example of holographic duality is the correspondence between $4$-dimensional super-Yang-Mills theory and $5$-dimensional supergravity. There, the large-$N$ limit of the super-Yang-Mills theory corresponds to the classical limit of the dual gravity theory, which provides a helpful description of strongly coupled gauge theories. What makes holographic duality particularly interesting is its generality. When the boundary theory is not a conformal field theory, a dual theory with a different space-time geometry may still be well-defined.\cite{witten1998b} Physically, holographic duality can be understood as a generalization of the renormalization group (RG) flow of the boundary theory\cite{akhmedov1998,boer2000,skenderis2002}, where bulk gravitational dynamics generalize the RG flow equations and the emergent dimension perpendicular to the boundary has the physical interpretation of energy scale\cite{heemskerk2011,lee2010}. 
Indeed, holographic duality has been applied to condensed matter physics as a new tool to characterize strongly correlated systems\cite{hartnoll2009,horowitz2009,mcgreevy2010,sachdev2012}.

More recently, holographic duality has been proposed to be related to another approach developed in condensed matter physics, namely tensor networks\cite{swingle2012,swingle2012b,evenbly2011,nozaki2012,hartman2013,czech2014,miyaji2014,miyaji2015,pastawski2015}. 
In its most general form, tensor networks refer to a description of many-body wavefunctions and operators (i.e. linear maps) by contracting tensors defined on vertices of a graph.\cite{white1992,klumper1993,verstraete2004,vidal2007,vidal2008,gu2009} %\red{should we mention and cite PEPs at some point?}%projected entangled pair states (PEPS)\cite{verstraete2004,murg2007,verstraete2008,aguado2008,gu2008,gu2009}, and the
More specifically, the tensor network state proposed to be related to holographic duality is the multiscale entanglement renormalization ansatz (MERA)\cite{vidal2007,vidal2008}, which is defined on a graph with hyperbolic structure, with external indices (corresponding to the physical degrees of freedom) at the boundary and internal indices contracted in the bulk. An important feature of states described by tensor networks is that the entanglement entropy of a given region is bounded by the number of links between the region and its complement. This property motivated its relation to holographic duality\cite{swingle2012}, where the entanglement entropy of a given region is determined by the area of the minimal surface bounding it, in accordance to the Ryu-Takayanagi formula\cite{ryu2006}. 

There are many open questions in the proposed tensor network interpretation of the holographic duality. One important question is how to describe space-time geometry rather than spatial geometry. Another (related) question is how to understand excitations (quantum fields) living in the bulk. Motivated by these questions, one of us\cite{qi2013} proposed a tensor network which defines not a many-body state but a unitary mapping between the boundary and bulk systems, known as the exact holographic mapping （EHM）. The EHM is a tensor network very similar to MERA, except that it is a one-to-one unitary mapping between boundary and bulk degrees of freedom. Each boundary state $|\psi\rangle$ is mapped to a bulk state $|\tilde{\psi}\rangle=M|\psi\rangle$, and each boundary operator $O$ is mapped to a bulk operator $\tilde{O}=MOM^{-1}$. Physically, the EHM is a ``lossless" version of real space renormalization group. Denoting a site in the bulk as ${\bf x}$, a local operator at that site $\tilde{O}_{{\bf x}}$ is dual to a generically nonlocal operator on the boundary $O_{\bf x}=M^{-1}\tilde{O}_{{\bf x}}M$. Different bulk sites ${\bf x}$ correspond to operators $O_{\bf x}$ on the boundary with different energy scales and different center-of-mass locations of their support. Once a mapping $M$ is chosen, bulk correlation functions can in principle be calculated. Motivated by the general principle of relativity, the bulk geometry was proposed to be determined by the bulk correlation functions. More specifically, the distance between two points was proposed to depend logarithmically on the connected two point correlation functions. Compared to previous tensor network proposals, the EHM is different in two aspects: i) The bulk geometry is not determined by the structure of the tensor network but by the correlation structure of the bulk state; ii) The bulk geometry can be studied in both the spatial and temporal direction by studying the bulk correlation functions. In Ref. \onlinecite{qi2013}, an explicit choice of the mapping $M$ for $(1+1)$-dim lattice fermions was proposed, and the consequent dual geometries corresponding to different boundary states were studied. They included the ground state of massless and massive fermions, the nonzero temperature thermal ensemble of massless fermions, and a thermal double state which is a purification of the thermal ensemble. Dynamics after a quantum quench was also studied in the thermal double system, motivated by a comparison with geometrical properties of a two-sided black hole space-time\cite{hartman2013}.

The results in Ref. \onlinecite{qi2013} for the abovementioned free fermion systems were obtained numerically. This limits the extent of analysis, due especially to the exponential growth of boundary system size. To have a well-defined bulk geometry with $N$ layers in the emergent direction of the bulk perpendicular to the boundary, the boundary system has to have $2^N$ sites. In this paper, we shall obtain analytic results on the free fermion EHM, which will enable us to rigorously determine asymptotic properties of the dual geometry, and also to discuss more general boundary systems. For instance, the existence of a black hole horizon in the geometry dual to a nonzero temperature state at the boundary can be studied more explicitly from the \emph{asympotic} infrared behavior of correlation functions in both spatial and temporal directions. In addition to reproducing the results of Ref. \onlinecite{qi2013} analytically, we shall also explore a few other interesting emergent bulk geometries, such as that corresponding to a critical fermion with nontrivial dynamic critical exponent. Our analytic framework also allows us to generalize the EHM to boundary theories with dimension $2+1$ or higher\footnote{The higher dimensional generalization of EHM is also independently investigated by Xueda Wen, Gil Y. Cho and Shinsei Ryu.}, in which case the analytic approach is more essential due to the increasing difficulty of numerical calculations\footnote{In $(1+1)$-d, at least $2^{15}$ sites are needed for analyzing the dual geometry with reasonable precision. In $(2+1)$-d the same number of layers in the bulk will require $2^{30}$ sites.}. An added advantage of an analytic approach is that it allows one to identify properties of the bulk geometry that are insensitive to details of the choice of the mapping which thus reflects intrinsic properties of the boundary state. 

This part on the EHM will be structured as follows. In Section 3.2, we first review the EHM construction by describing its general principles and the definition of bulk geometry. These ideas will be elaborated in Section 3.3 for free lattice fermions, where an explicit Haar wavelet representation of the EHM will be presented. In Section 3.4, we provide detailed descriptions of the asymptotic correlator behavior and corresponding bulk geometries for the prototypical $1+1$-dim Dirac model at various combinations of zero and nonzero temperature and mass. These developments will be further extended to higher dimensions and generic energy dispersions in Section 3.5, where we discuss the emergence of interesting geometries like anisotropic black hole horizons with nontrivial topology. Next, in Section 3.6, the EHM will be applied on the $2+1$-dimensional Chern Insulator introduced in the previous part on the Wannier state representation. While the WSR exactly maps the Chern Insulator onto a Quantum Hall system, the EHM maps the Chern Insulator on the holographic boundary to a $3+1$-dimensional topological insulator in the bulk. Finally, in Section 3.7, some new ideas for generalizing the EHM will be discussed.

\section{Review of the Exact Holographic Mapping}\label{sec:review}

In this section, we shall review the motivation and construction of the EHM proposed in Ref. \onlinecite{qi2013} in a formalism that will be helpful for the later part of this paper. We will also include some new insights that are not discussed in the original proposal. The EHM approach is defined by the following two principles:
\begin{enumerate}
\item The bulk theory and boundary theory are defined in the same Hilbert space. The bulk local operators are determined by a unitary mapping acting on the boundary local operators.
\item The bulk geometry is determined by physical correlation functions. More specifically, the distance between two space-time points ${\bf x}, {\bf y}$ in the bulk is determined by the connected correlation functions between the two points.
\end{enumerate}
Although the abovementioned unitary transformation can be very generic in principle, the types of transformations that are relevant for holographic duality are those which are physically analogous to the renormalization group\cite{wilson1974,wilson1975}. The bulk operators at different locations should represent boundary degrees of freedom with different energy scales. The key difference from the conventional RG approach is that the high energy degrees of freedom are spatially separated from low energy ones, instead of being integrated out. This enables us to concretely answer many new questions, such as how the high and low energy degrees of freedom (DOFs) are entangled/correlated. In the following, we will elaborate on the two abovementioned principles in the context of free fermion systems, and discuss the transformation of free fermion Hamiltonians under EHM.

\subsection{General construction of EHM}

\begin{figure}[H]
\centering
\includegraphics[scale=.5]{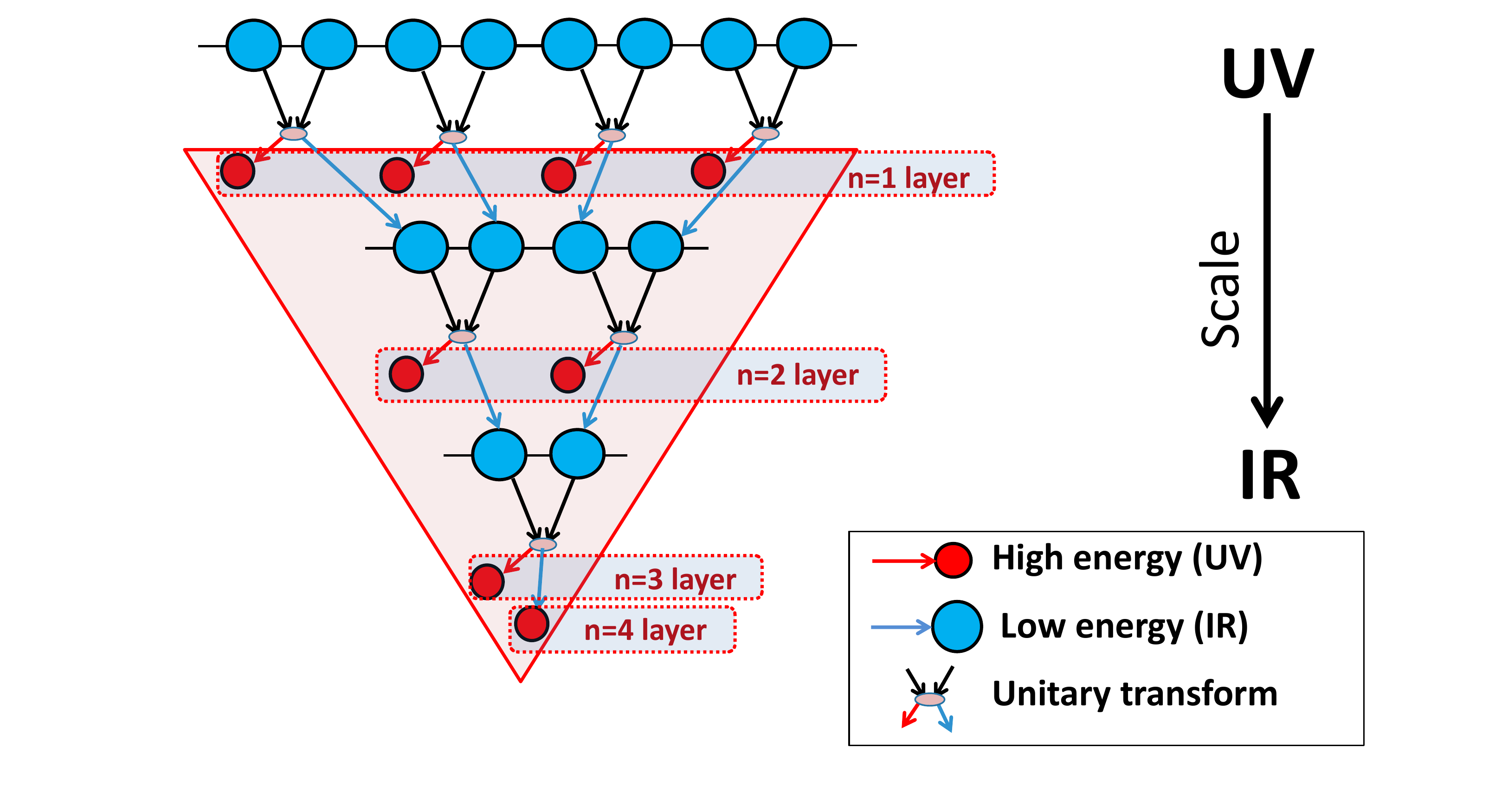}
\caption{Illustration of the EHM on $2^3=8$ sites. At each iteration, two auxiliary sites (blue) is fed into a unitary transform that produces a UV (red) DOF that defines a bulk site, and a IR (blue) DOF that becomes the auxiliary site for the next iteration. The bulk sites are arranged in a tree-like structure (red triangle) with $4$ layers, inclusive of the last (lowest energy) IR that forms the last "layer". }
\label{ehmtree}
\end{figure}

The Exact Holographic Mapping is a unitary transformation defined by a tensor network or, equivalently, a quantum circuit consisting of local unitary operators. As proposed in Ref. \onlinecite{qi2013}, a simple construction of the EHM is given by a tree-shaped tensor network depicted in Fig. \ref{ehmtree}, where bulk (red) sites at the same level belong to the same \textsc{\char13}layer\textsc{\char13}. To construct it, we first take a $D+1$-dimensional boundary system to be the zeroth bulk layer with $L^D=2^{ND}$ sites. To construct the first bulk layer, one performs a unitary transform $U$ on every set of $2^D$ adjacent sites such that the UV and IR (high and low momentum, assuming a monotonic energy dispersion) degrees of freedom are separated out. For $D=1$, this can be written as
\begin{equation}
U_{12}|\psi_1\psi_2\rangle =\sum_{\alpha,\beta} U^{\alpha\beta}_{\psi_1\psi_2}|\alpha\rangle_{IR} |\beta\rangle_{UV}
\label{unitary}
\end{equation}
where $U_{12}$ only acts on states $\psi_1,\psi_2$ on sites $1$ and $2$ respectively, and $|\alpha\rangle $ and $|\beta\rangle$ capture the higher and lower momentum (shorter and longer scale) degrees of freedom respectively. The construction of these $|\alpha\rangle $ and $|\beta\rangle$ states will be shown in detail in the next section. The full transformation on the zeroth layer is given by
\begin{equation}
U= U_{12}\otimes U_{34}\otimes ...\otimes U_{2^N-1,2^N},
\label{U}
\end{equation}
which is a unitary transform on the Hilbert space of the whole layer. For $D>1$ dimensions, $U$ will be given by the direct product of $D$ copies of the expression in Eq. \ref{U}.

We construct the first bulk layer from the component $|\beta\rangle=|\beta\rangle_{UV}$ in Eq. \ref{unitary}, which has the UV half of the degrees of freedom in the original layer. The other lower energy half $|\alpha\rangle$, which we shall call the \emph{auxillary sites} in deference to Ref. \onlinecite{qi2013}, are fed into another copy of $U$ with half the number of sites. This process is iterated for $N$ times, each time producing a new layer in the bulk that has $1/2^D$ the number of sites as the preceding layer, until only one site is left. The resultant (bulk) tree\footnote{In the continuum limit, the bulk system is topologically half the suspension (cone) of the boundary system, i.e. the latter with successively smaller copies of itself connected in a prism-like manner. Loosely speaking, the bulk system can be visualized as the solid 'interior' of the boundary manifold.} is unitary equivalent to the original (boundary) system, and is illustrated in Fig. \ref{ehmtree}. Note that in this construction, the bulk is made up of discrete layers; in Sect. \ref{EHMfurther}, an alternative construction with a continuous emergent direction will be proposed, although that is not strictly an exact mapping.

\subsection{Emergent bulk geometry through boundary correlators}

The key motivation behind the EHM approach is to uncover the relationship between space-time geometry and the quantum entanglement properties of a quantum many-body system. The unitary mapping defined by the tensor network defines a new direct-product decomposition of the Hilbert space , and is chosen to make physical correlation functions more local in this new basis. To be more precise, we assume that the two-point connected correlation functions in the bulk \emph{always} decay exponentially, according to the geodesic distance of certain emergent geometry:
\begin{eqnarray}
C_{({\bf x}_1,t_1)({\bf x}_2,t_2)}&\equiv&\left\langle O_{\bf x_1}(t_1)O_{\bf x_2}(t_2)\right\rangle-\left\langle O_{\bf x_1}(t_1)\right\rangle\left\langle O_{\bf x_2}(t_2)\right\rangle\nonumber\\
&\propto &\exp\left(-d_{({\bf x}_1,t_1),({\bf x}_2,t_2)}/\xi\right)
\end{eqnarray}
This assumption can conversely be used as a definition of the distance\cite{qi2013}:
\begin{equation}
d_{(\bold x_1,t_1),(\bold x_2,t_2)}= -\xi \log \frac{C_{({\bf x}_1,t_1)({\bf x}_2,t_2)}}{C_0}
\label{d}
\end{equation}
where $C_0$ and $\xi$ control the overall offset and scaling respectively. $\xi$ can be physically interpreted as the inverse mass of the emergent bulk theory, which may depend slightly on how we perform the EHM. The logarithmic dependence is physically motivated by the observation that for a massive system, $d_{(\bold x_1,t_1),(\bold x_2,t_2)}$ should recover the Euclidean distance in the original system.

In this work, we shall for simplicity focus on systems that are translationally-invariant in space and time, and study only correlators with purely space or time intervals, i.e. $\Delta t=t_2-t_1=0$ or $\Delta x =x _2-x_1=0$.

In the former case with purely spatial interval, all two-point connected correlators are bounded above\cite{wolf2008} by the \emph{mutual information}
\begin{equation}
I_{\bold x \bold y}=S_{\bold x} + S_{\bold y} - S_{\bold x \bold y}
\label{mutual}
\end{equation}
where $S_{\bold x }$ and $S_{\bold x \bold y}$ are the entanglement entropies (EE) of a single site and two sites respectively. Roughly speaking, the mutual information between two sites measures how much the entanglement entropy of two sites will be reduced if the correlation between the two sites are known. Hence a basis-independent definition of the spatial geometry is given by the mutual information:
%\inf d_{\Delta \bold x}\
\begin{equation}
d_{\Delta \bold x}
=-\xi \log \frac{I_{\bold x \bold y}}{I_0}
\label{dx}
\end{equation}
where $\Delta \bold x=|\bold x -\bold y|$ and $I_0$ is a reference value for the mutual information. One reasonable choice of $I_0$ is $I_0=2\aleph\log 2 $, which is the maximal mutual information between two sites, each with $\aleph$ internal DOFs (spins, bands, etc). This bound is saturated in the (hypothetical) situation when $S_{\bold x \bold y}=0$ but $S_{\bold x}=S_{\bold y}=\aleph\log 2$, i.e. when the DOFs of the two sites are maximally entangled with each other but not with those of the other sites. A more detailed explanation for the mutual information is given in Appendix \ref{app:mutual}.

In the latter case with purely temporal interval, we specialize Eq. \ref{d} to
\begin{equation}
d_{\tau}= -\xi_\tau \log \frac{C(\tau)}{ C(0)}
\label{dt}
\end{equation}
where $C(\tau)$ represents chosen component/s of $\langle O_{\bold x}(0) O_{\bold x}(\tau)\rangle_{bulk}$, $\tau$ being the imaginary (Wick-rotated) time interval $\tau=-i\Delta t$. The imaginary time direction is preferred over real time as the latter typically exhibits oscillatory behavior that makes an asymptotic comparison difficult. Further discussion on the relationship between the real and imaginary time correlators will be deferred to Appendix \ref{sec:time}. Note that unlike the case with spatial intervals, there is no known operator that yields the upper bound of correlators across temporal intervals.

Henceforth, we shall use Eq. \ref{dx} and \ref{dt} as the expressions for the distance between two points in the bulk, and compare them with the geodesics of classical geometries.

\section{Exact holographic mapping for free lattice fermions}

We now specialize the above developments to free lattice fermions, for which the correlators and mutual information possess nice analytic behavior, at least asymptotically. First, we recall the following well-known result for the entanglement entropy of free fermions\cite{peschel2002,klich2006, lee2014exact}:
\begin{equation}
S_X=-\text{Tr}~(C_X\log C_X + (\mathbb{I}-C_X)\log(\mathbb{I}-C_X))
\label{Sx}
\end{equation}
where $S_X=-\text{Tr}~(\rho_X\log\rho_X)$ is the entanglement entropy for the region $X$, and $C_X$ is the projector (correlator) onto region $X$. With the help of Eq. \ref{Sx}, it is shown in Appendix \ref{app:mutualcorr} that the Mutual Information is approximately
\begin{eqnarray}
I_{\bold x \bold y  }&=&S_\bold x +S_\bold y  -S_{\bold x \bold y  }\notag\\
&\approx& \frac{1}{2}\text{Tr}~\left[C_{\bold x -\bold y  }\frac{1}{C_\bold y  (\mathbb{I}-C_\bold y  )}C_{\bold y  -\bold x }+(\bold x \leftrightarrow \bold y  )\right]\notag\\
&\sim& \text{Tr}~ [C^\dagger_{\bold y  -\bold x }C_{\bold y  -\bold x }]
\label{Ixy00}
\end{eqnarray}
where $C_\bold x ,C_\bold y  $ are the single-particle onsite correlators, and $C_{\bold x -\bold y  }$ is the \emph{single-particle} propagator between the \emph{two different sites} $\bold x $ and $\bold y  $. This result is completely general, and implies that
\begin{equation}
d_{\Delta \bold x }=-\xi_{\Delta \bold x }\log \frac{I_{\bold x \bold y  }}{I_0}\sim 2\xi_{\Delta \bold x } \log \text{Tr}~ C_{\bold y  -\bold x }
\end{equation}
in the limit of large spatial separation $|\bold x -\bold y  |$. That $C_\bold x ,C_\bold y  $ drops out is hardly surprising, as they each depend only on one site, and have no knowledge about their separation. Indeed, most of the information transfer in the asymptotic limit is dominated by the single-particle propagator.

We also define the temporal distance via
\begin{equation}
d_{\tau}= -\xi_\tau \log \frac{\text{Tr}~ C(\tau)}{\text{Tr}~ C(0)}
\label{dt2}
\end{equation}
where a trace of the fermion states have been taken. This is the simplest possible basis-independent combination of the components of $C(\tau)$.

In the next two subsections, we shall introduce prototypical fermionic models as the boundary systems in $1+1$ and higher dimensions, and show how their corresponding bulk distances and hence geometries can be computed via suitable holographic unitary mappings.

\subsection{EHM for (1+1)-dimensional lattice Dirac fermions}
\label{subsec:kspace}

The (1+1)-dimensional lattice Dirac model is among the simplest models with a single critical point. In this subsection, we will summarize the explicit construction of the EHM for this system. Its simplicity allows us to study its multitude of entanglement and geometric properties analytically with minimal complication.

The $(1+1)$-dim Dirac Hamiltonian is a 2-band Hamiltonian given by
\begin{equation}
H_{Dirac}(k)=v_F[\sin k \sigma_1 + M(m+1-\cos k)\sigma_2]
\label{dirac1}
\end{equation}
where $\sigma_1,\sigma_2$ are the Pauli matrices and $v_F$, the Fermi velocity, controls the overall scale of the dispersion. $M$ is controls the relative weight between the $\sin k$ and $m+1-\cos k$ terms, and will be set to unity here. A discussion for generic $M$ will be given in Appendix \ref{genericdirac}. When $m=0$ or $\pm 2$, its gap closes at $k=0$ and it becomes critical with two crossing bands with linear dispersion. To explore or ''zoom into'' the low energy (IR) degrees of freedom (DOFs), we utilize a unitary transform that maps states $|\psi_1^{s_1}\rangle,|\psi_2^{s_2}\rangle$ on neighboring sites into symmetric (low energy) and antisymmetric (high energy) linear combinations $\frac1{\sqrt{2}}\left(|\psi_1^{s_1}\rangle\pm|\psi_2^{s_2}\right)$. Note that the unitary transform does not rotate the spin labels $s_1,s_2$, which we shall suppress in the following. In matrix form, the unitary transform is written as
\begin{equation}
U_{12}=\frac{1}{\sqrt{2}} \left(\begin{matrix}
 & 1 & 1 \\
 & 1 & -1\\
\end{matrix}\right)
\label{haarU}
\end{equation}
The symmetric combination has a Fourier peak at $k=0$, which is exactly the gapless point of the critical ($m=0$) Dirac model. The discerning reader will notice that $U_{12}$ is nothing other than the defining expression for the Haar transform. Indeed, the construction of the EHM basis is mathematically identical to performing a wavelet decomposition\cite{meyer1989,daubechies1992,strang1996}. A systematic study of all possible wavelet descriptions of the EHM will be deferred to future work, since for this work we will be primarily concerned about the behavior of the bulk geometries due to qualitatively different boundary systems, not the details of the wavelet mapping. The transform given by Eq. \ref{haarU} possess the virtue of simplicity and, most importantly, fixes the archetypal Dirac Hamiltonian, a property we shall prove in the next subsection.

\begin{figure}[H]
\begin{minipage}{0.99\linewidth}
\includegraphics[width=0.99\linewidth]{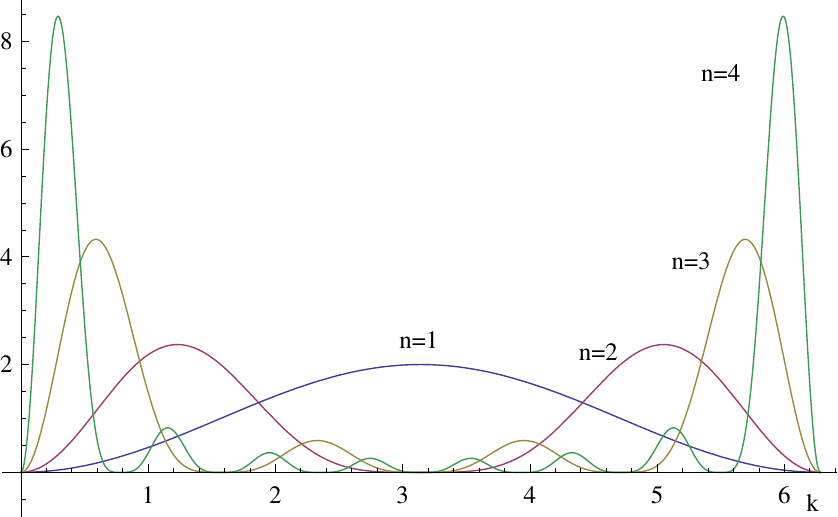}
\end{minipage}
\caption{(Color Online) Plot of the normalized spectral weight $|W_n(e^{ik})|^2$ of the bulk basis for $n=1,2,3,4$. We see that as $n$ increases, the dominant spectral peak approaches the unique 
IR point at $k=0$ (and its periodic image at $k=2\pi$)
exponentially viz. $k_0\approx \frac{2\pi}{2^n}\rightarrow 0$. Also, it becomes narrower since its spectral width also goes like $\sim 2^{-n}$. $W_1$ is peaked at the highest momentum $k=\pi$, attesting to the fact that it contains all the UV DOFs.   } 
\label{wavelets}
\end{figure}

More insight into the EHM can be gleaned in momentum space, where one can directly see how the Hilbert space is decomposed into layers with different momentum spectral distributions. Fourier transforming the action of Eq. \ref{haarU} on the single particle states, we obtain $|\alpha_k\rangle = \sum_{2k} C(e^{ik})|\psi_{2k}\rangle$ and $|\beta_k\rangle = \sum_{2k} D(e^{ik})|\psi_{2k}\rangle$ for the auxiliary and bulk states respectively, where $|\psi_k\rangle$ is the periodic part of the Bloch state and
\begin{equation}C(e^{ik})=\frac{1}{\sqrt{2}}\left(1+e^{ik}\right),\label{C}\end{equation}
\begin{equation}D(e^{ik})=\frac{1}{\sqrt{2}}\left(1-e^{ik}\right)\label{D}\end{equation}
We shall call $C,D$ the IR and UV (low energy and high energy) projectors. Physically, they represent the spectral weight projected to the auxiliary and bulk DOFs at each iteration. Through these iterations, we obtain successive basis projectors for each bulk layer that are increasingly sharply peaked in the IR. To understand this, note that the basis projector of the $n^{th}$ layer $W_n(z)=W_n(e^{ik})$ is obtained from $n-1$ consecutive IR outputs $|\alpha\rangle$ and one final UV output $|\beta\rangle$. Hence the first bulk layer should contain the DOFs projected from the UV projector $D(e^{ik})$, while the second layer should contain an IR projector $C(e^{ik})$ followed by an UV projector $D(e^{2ik})$ that peaks at half the momentum. This reasoning generalizes to all the $N$ layers, so the normalized projector for the $n^{th}$ layer is given in momentum space by (writing $z=e^{ik}$)
\begin{eqnarray}
W_n(z)&=&\frac{1}{\sqrt{2 \pi}}D\left(z^{2^{n-1}}\right)\prod_{j=1}^{n-1}C\left(z^{2^{j-1}}\right)\notag\\
&=&\frac{1}{\sqrt{2 \pi}}\frac{1-z^{2^n}}{\sqrt{2^n}} \prod_{j=1}^{n-1}\left(1+z^{2^{j-1}}\right)\notag\\
&=&\frac{1}{\sqrt{2 \pi}}\frac{1}{\sqrt{2^n}}  \frac{\left(1-z^{2^{n-1}}\right)^2}{1-z}.\notag\\
\label{Wn}
\end{eqnarray}
$W_n(e^{ik})$ contains a series of peaks interspersed by valleys at $e^{i2^{n-1}k}=1$. The dominant peaks occur at $k=\pm k_0\approx \frac{2\pi}{2^n}$ where the denominator is most singular, as shown in Fig. \ref{wavelets}, and has magnitude $|W_n(e^{ik_0})|=\sqrt{\frac{2^{n+1}}{\pi^3}}$. This means that as $n$ increases, the spectral weight of the $n^{th}$ bulk layer exponentially approach the IR point at $k=0$. One can further show that the $W_n$'s form a complete an orthonormal basis, i.e. $\int_{-\pi}^{\pi} W_m^*(e^{ik})W_n(e^{ik}) dk = \frac1{2\pi i}\oint_{|z|=1}W^*_m(z^{-1})W_n(z)\frac{dz}{z}=\delta_{mn}$, where the conjugation symbol in $W^*$ denotes that only the coefficients of $W(z)$, not the argument $z$, are complex conjugated. Indeed, that the $W_n$'s are orthonormal with peaks $k_0\sim 2^{-n}$ is testimony to the fact that the EHM is a unitary mapping that separates the momentum (or energy) scale.

Note that the auxiliary projector, i.e. projection to auxiliary sites with the IR (low energy) half of the DOF, is just the orthogonal complement of $W_n(z)$ in Eq. \ref{Wn}: It is given by $\frac{1}{\sqrt{2\pi}}\prod_{j=1}^n C(z^{2^{j-1}})$, comprising IR projectors $C(z)$ only.

From Fig. \ref{wavelets}, we see that the bulk basis from Eq. \ref{Wn} become more effective in separating different momentum (and hence energy) scales as $n$ increases, since they become more sharply peaked. This property is due to the recursive nature of the definition Eq. \ref{Wn}, which also implies that physical quantities, i.e. correlators \emph{must scale universally with $n$} in the IR limit. However, departures from universal scaling may occur at small $n$ (UV regime) due to the non-universal high energy characteristics of the boundary system.

Note that the abovementioned unitary EHM transform successively ''zooms into'' the low energy DOFs of any model with a critical point at $k=0$, and not just that of the Dirac model.

%\subsubsection{Expression for the bulk correlator}

Having defined the mapping explicitly, we now write down the explicit expression of the bulk correlators. Previously, we have seen how the bulk distance can be expressed in terms of the bulk correlators $C_x, C_{x-y}$ and $C(\tau)$, which are the onsite, spatial and temporal propagators respectively. 
In the free fermion system discussed here, the bulk correlators are determined by the boundary two-point correlators
\begin{equation}
G_\bold q(\tau)=e^{\tau(H(\bold q)-\mu )}(\mathbb{I}+e^{\beta(H(\bold q)-\mu)})^{-1}
\label{gq0}
\end{equation}
where $H(\bold q)$ is the boundary single particle Hamiltonian matrix, $\mu$ the chemical potential and $T=1/\beta$ the temperature. Explicit expressions for $G_\bold q(\tau)$ as well as their resultant mutual information $I_{\bold x\bold y}$ are derived in Appendix \ref{app:twoband} For cases with particle-hole symmetry, which includes the Dirac model. In the zero temperature limit, $G_\bold q(0)$ reduces to a projector onto the occupied bands below the chemical potential. In the extremely high temperature limit, it becomes nearly the identity operator, which just means that almost every state is equally accessible. In the remainder of the paper, we will always focus on imaginary time correlator unless otherwise stated, since it is still unclear how to define time-drection distance from the rapidly oscillating real time correlators (as discussed further in Appendix \ref{sec:time}).

The bulk correlators are most easily expressed as a sum in momentum space, since we have already found projections to the various bulk layers in terms of the spectral weight. Taking the thermodynamic limit where the boundary system is infinitely large, i.e. $N\rightarrow \infty$, the sum over momenta can be replaced with an integral. The bulk correlator between two bulk points $(x_1,n_1,0)$ and $(x_2,n_2,\tau)$ is given by
\begin{eqnarray}
&&C(n_1,n_2, \Delta (2^n  x),\tau)\notag\\&=&\sum_{  q} W^*_{n_1}( e^{i q})W_{n_2}(  e^{iq})e^{i    q  \Delta (2^n  x) }G_{  q }(\tau)
\label{bulkcorrelator1d}
\end{eqnarray}
where $\Delta (2^n x)=2^{n_2}x_2-2^{n_1}x_1$ is the bulk angular interval and $G_{ q}(\tau)$ is the (matrix-valued) $(1+1)$-dim boundary correlator. The indices $n_1,n_2$ specify the coordinates ('layers') in the emergent 'radial' momentum-scale direction in the bulk. There is no translational symmetry in this momentum(or energy)-scale direction, unlike the original spatial and temporal directions.

We see that $C(n_1,n_2, \Delta   x,\tau)$ is just a Fourier transform of the boundary correlator $G_{  q}$ weighted by the spectral contributions $W_{n_1}^{*}W_{n_2}$ of the respective bulk layers. One important observation is that the spatial coordinate in the exponential factor is $\Delta (2^n   x)$, not $\Delta   x$. This is because the correlation comes from the holographic projection of the bulk sites onto the boundary, which depends on the angle subtended by the bulk displacement: In the simplest case of a circular boundary, the angle subtended by $\Delta x$ at the $n^{th}$ layer is $\Delta \theta = \frac{2\pi \Delta x}{2^{N-n}}= \frac{2\pi}{L} (2^n\Delta x)$. The bulk correlator still possesses translation invariance, but of $\Delta(2^n  x)$, the angular interval \emph{projected on to the boundary}, not of the bulk interval $\Delta x$ itself. Mathematically, we find that the $2^n$ rescaling of angular distance is also required for the orthogonality of the basis $\{W_{n}(  q)e^{i 2^n   q \cdot  x }\}$, $x=1,2,..., \frac{L}{2^n}$.

\subsection{Transformation of the Hamiltonian under the EHM and its fixed points}
The EHM is an exact version of renormalization group (RG) transformation. In each step, the DOFs of the system are split into the high energy (UV) and the low energy (IR) parts, with the procedure iterated on the low energy part. If we write the Hamiltonian in the new basis after an EHM step and ignore the coupling between the IR and UV degrees of freedom, we can write down a "low energy effective Hamiltonian" of the IR states. This resembles the renormalization group flow of the effective Hamiltonian in ordinary RG. For a given choice of the EHM tensor network, there are certain boundary systems for which the IR Hamiltonian is at an RG fixed point. In the following, we will show that the massless Dirac Hamiltonians are RG fixed points of the EHM transformation we defined earlier.

We start by writing down the effective IR Hamiltonian in the momentum basis. In $D=1$ spatial dimensions, the EHM for each iteration is given by the change of basis in Eqs. \ref{C} and \ref{D}, so the single particle Hamiltonian matrix $h^n$ of the $n^{th}$ layer is transformed according to
 \begin{equation} h^{n+1}(e^{ik/2}) = \left[V^\dagger \left(\begin{matrix}
 & h^n(e^{ik/2}) & 0 \\
 & 0 & h^n(e^{i(k/2+\pi)}) \\
\end{matrix}\right)V \right]_{11}
\label{RG0}
\end{equation}
with the two components of the matrix representing the IR and UV DOFs.  Here \[V(w)=\frac{1}{\sqrt{2}}\left(\begin{matrix}
 & C(w) & D(w) \\
 & C(-w) & D(-w) \\
\end{matrix}\right)=\frac{1}{2}\left(\begin{matrix}
 & 1+w & 1-w \\
 & 1-w & 1+w \\
\end{matrix}\right),\]where $w=e^{ik/2}$. To obtain the Hamiltonian at the $(n+1)^{th}$ layer, one projects onto the upper left or IR component of the RHS. With $D>1$ spatial dimensions, $V$ will be given by $V(w_1)\otimes ... \otimes V(w_D)$. If the Hamiltonian is to remain invariant under the RG, the $11$ (IR) component in Eq. \ref{RG0} gives $2h^{n+1}(w)=2\lambda h^n(w^2)$ or, in detail:
\begin{eqnarray} %\nonumber\\
%&&\notag\\
%\text{or~}
&&2\lambda h^n(w^2)\notag\\
&=&h^n(w)C(w)C^*(w^{-1}) + h^n(-w) C(-w)C^*(-w^{-1}) \notag\\
%&=& \sum_{\pm} h^n(\pm w)\left(1\pm\frac{w+w^{-1}}{2}\right)\notag\\
&=& (h^n(w)+h^n(-w))+ \frac{w+w^{-1}}{2}(h^n(w)-h^n(-w))\notag\\
\label{RG}
\end{eqnarray}
where $\lambda$ is a constant scale factor for each EHM step. Upon setting $w=1$, we obtain
\begin{equation}
\lambda h(1) = h(1)
\end{equation}
which implies that $\lambda=1$ unless $h(1)=0$, i.e. that the rescaling $\lambda$ for each step can be nontrivial ($\lambda \neq 1$) \emph{only if} the Hamiltonian is gapless \footnote{$k=0$ is special because this is where the UV projector has zero weight.} at the IR point $k=0$ or $w=1$. In other words, only gapless Hamiltonians can have nontrivial scale invariance, as is expected. This has very important implications in spatial dimensions $D>1$, since it implies that if the Hamiltonian contribution $h$ does not depend explicitly on $k_j$, the Hamiltonian will not be scale invariant by any EHM transformation in the $j^{th}$ dimension. 

One can derive solutions of Eq. \ref{RG} by comparing terms power by power. For instance, the second-order terms in $h(w^2)$ on the LHS force $h(w)$ to be a linear function of $w$ that is either symmetric or antisymmetric under $w\leftrightarrow w^{-1}$, i.e. a function of $w+w^{-1}$ or $w-w^{-1}$. Hence we find the two linearly independent solutions to Eq. \ref{RG} to be $h(z)= \frac{z-z^{-1}}{2i}=\sin k$ and
$h(z)=\frac{2-z-z^{-1}}{4}=\frac{1-\cos k}{2} $, both with the EHM rescaling $\lambda =\frac{1}{2}$. They are both gapless at $k=0$, as they should be, and can be combined to form the massless Dirac Hamiltonian Eq. \ref{dirac01} in $1+1$ dimensions:
\begin{equation}
H_{Dirac}(k)=v_F[\sin k \sigma_1 + M(1-\cos k)\sigma_2]
\end{equation}
where $\sigma_i$ are the Pauli matrices and $M$ controls the relative weight of the two terms. 

Note that the two terms $\sin k$ and $1-\cos k$ do not have the same scaling dimension if one takes the continuum limit $\sin k\sim k,~1-\cos k\sim k^2/2$ in ordinary RG. This illustrates the distinction of real space EHM transformation from simple momentum rescaling, due to the nontrivial influence of lattice regularization that replaces functions in $k$-space with periodic trigonometrical functions.

\subsubsection{Example: Transformation of the 2-dimensional massive Dirac Hamiltonian}

Here let's examine the 2-dimensional analogue of Eq. \ref{RG0} in more detail. For each dimension, the $11$ and $22$ components of the matrix on the RHS correspond to the $UV$ and $IR$ sectors respectively. In two dimensions, these will be generalized to the four sectors $UV,UV$, $UV,IR$, $IR,UV$ and $IR,IR$. The first three form the bulk sectors. Call their respective Hamiltonians, at layer $n$, as $H^{UV,UV}_n$, etc. We have
\begin{equation} H^{UV,UV}_{n+1}(z^2,w^2) = \frac1{2}(|D(z)|^2 |D(w)|^2 H^{IR,IR}_n(z,w)+|D(-w)|^2|D(-z)|^2 H^{IR,IR}_n(-z,-w)) 
\end{equation}

 \begin{equation} H^{IR,IR}_{n+1}(z^2,w^2) = \frac1{2}(|C(z)|^2|C(w)|^2 H^{IR,IR}_n(z,w)+|C(-w)|^2|C(-z)|^2 H^{IR,IR}_n(-z,-w)) 
\end{equation}

 \begin{equation} H^{UV,IR}_{n+1}(z^2,w^2) = \frac1{2}(|D(z)|^2|C(w)|^2 H^{IR,IR}_n(z,w)+|C(-w)|^2|D(-z)|^2 H^{IR,IR}_n(-z,-w)) 
\end{equation}
where $z=e^{ik_x},w=e^{ik_y}$.%, $C(z),D(z)=\frac{1\pm z}{\sqrt{2}}$ are the IR and UV filters for the Haar wavelet. 
The $\cos k$ and $\sin k$ terms transform simply as
 \begin{equation} \sin k |C(e^{ik}|^2= \frac1{2}\sin 2k \end{equation}
 \begin{equation} \cos k |C(e^{ik}|^2= \frac{\cos 2k +1}{2}\end{equation}
and 
 \begin{equation} \sin k |D(e^{ik}|^2= -\frac1{2}\sin 2k \end{equation} 
 \begin{equation} \cos k |D(e^{ik}|^2= -\frac{\cos 2k +1}{2} \end{equation} 
Constant terms remain unchanged. Starting from the Dirac Hamiltonian $H_0(k_x,k_y)$ with $d$-vector $(\sin k_x, \sin k_y, m+\cos k_x + \cos k_y)$ (Eq. \ref{dirac1}), we have the bulk layer Hamiltonians
\begin{equation}
H^{IR,UV}_n= \frac1{2^n}\sigma \cdot (\sin k_x, \sin k_y, 2^n m + \cos k_x -\cos k_y -1)
\end{equation}
\begin{equation}
H^{UV,UV}_n= \frac1{2^n}\sigma \cdot (\sin k_x, \sin k_y, 2^n m - \cos k_x -\cos k_y -2)
\end{equation}
While these are 2-dimensional Hamiltonians in their own right, they do not exist in isolation. Notably, we have neglected the \emph{interlayer} coupling Hamiltonians. The fractional chern number from layer $n$, which should be quantized as an integer for an isolated 2-dimensional lattice system, arises from the (unnormalized) part of the boundary eigenstate at layer $n$, and not the layer Hamiltonians derived above. 

In general, the gaps of these Hamiltonians drop to a minimum near the layer where the entanglement gap closes (see Sect. \ref{sec:holotopo}), and then increases to $m$ deep in the IR, where the d-vector $\rightarrow (0,0,m)$ and the topology is trivial. The minimal gap is usually a few orders of magnitude smaller than $m$. 

\begin{figure}[H]
\centering
\includegraphics[scale=1.8]{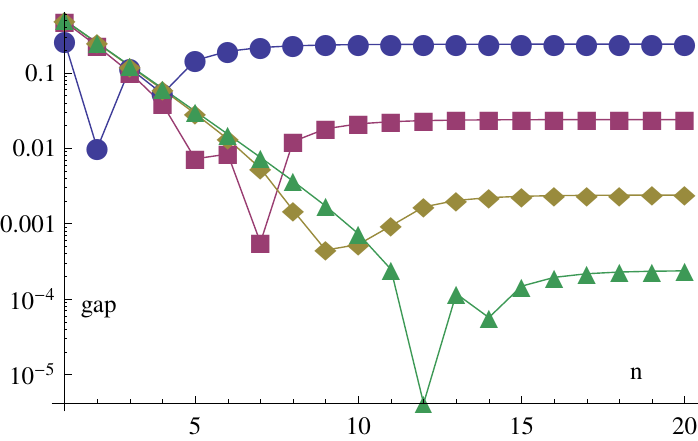}
\caption{Logplots of the gap of $H^{UV,IR}$ as a function of $n$ for masses $m=0.24,0.024,0.0024$ and $0.00024$ in the order (blue,purple,brown,green). We see that the min. gap is a few order of magnitude smaller than $m$. }
\label{fig:mingap}
\end{figure}

\section{Analytic results for $(1+1)$-dimensional boundary systems}

In Ref. \onlinecite{qi2013}, the behaviors of correlation functions and their associated bulk distances were studied numerically. In the following, we will obtain analytic results of the asymptotic behavior of bulk correlators for $(1+1)$-dim translationally-invariant boundary systems, which in turn determine the asymptotic large scale behavior of the bulk geometry we define. We will compare them with the geodesic distances between analogous points in candidate classical geometries. The higher-dimensional extensions of these results will be discussed in the next section. 

\subsection{General setup}
\label{gensetup}
We shall consider four distinct physical scenarios, all at zero chemical potential, with the representative Hamiltonian for the first three cases taken to be the Dirac Hamiltonian given in Eq. \ref{dirac1}. The results obtained should also be valid for more generic Hamiltonians, since the qualitative bulk geometry properties remain robust as long as the long distance behavior of correlators remain the same. In approximately increasing levels of sophistication, the four scenarios are:
\begin{enumerate}
\item Critical boundary Dirac model at $T=0$, corresponding to a bulk AdS (Anti-de-Sitter) geometry.
\item Massive boundary Dirac model at $T= 0$, corresponding to a ``confined geometry" with an IR termination surface.
\item Critical boundary Dirac model at $T\neq 0$, corresponding to a bulk BTZ (Ba\~nados, Teitelboim, and Zanelli) black hole geometry.
\item Critical boundary model with nonlinear dispersion at $T\neq 0$, corresponding to a bulk Lifshifz black hole geometry.
\end{enumerate}

The first two cases were already explored numerically in Ref. \onlinecite{qi2013}, with results in excellent agreement with our analytical results below.

As previously explained, the fundamental quantity to be calculated is the bulk correlator $C(n_1,n_2,\Delta (2^n x),\tau)$ given in Eq. \ref{bulkcorrelator1d}. In the thermodynamic limit $L\rightarrow \infty$, all momentum sums can be replaced by integrals:
\begin{eqnarray}
&&C(n_1,n_2,\Delta (2^nx), \tau)\notag\\
&=&\sum_q  W_{n_1}^*(q)W_{n_2}(q) e^{iq\Delta (2^nx)}G_q(\tau)\notag\\
&=&\oint_{|z|=1}\frac{dz}{z}W^*_{n_1}(z^{-1})W_{n_2}(z)z^{\Delta (2^nx)}G_{z}(\tau)\notag\\
\label{correlator1d}
\end{eqnarray}
where, as before, the conjugation symbol $^*$ indicates that only the coefficients of the polynomial $W(z)$ are complex conjugated. It is insightful to analytically continue the momentum $q$ into the \emph{complex} $z=e^{iq}$ plane, where the decay properties of the correlators can be directly read from the properties of the complex poles and branch cuts.

For comparison with the geodesic distances, we shall specialize to 2-point correlators of the following three directions in the $(2+1)$-dim bulk:
\begin{itemize}
\item Equal time, same layer ``angular'' correlator
\begin{eqnarray}
C_n(\Delta x)&=&C(n,n,\Delta x,0)\notag\\
&=&\oint_{|z|=1}\frac{dz}{z}W^*_n(z^{-1})W_n(z)z^{2^n\Delta x}G_{z}(0)
\label{correlatorx}
\end{eqnarray}
\item Equal time, different layer ``radial'' correlator
\begin{eqnarray}
C(n_1,n_2)&=&C(n_1,n_2,0,0)\notag\\
&=&\oint_{|z|=1}\frac{dz}{z}W^*_{n_1}(z^{-1})W_{n_2}(z)G_{z}(0)
\label{correlatorr}
\end{eqnarray}
\item Same site imaginary-time correlator
\begin{equation}
C_n(\tau)=C(n,n,0,\tau)=\oint_{|z|=1}\frac{dz}{z}W^*_n(z^{-1})W_n(z)G_{z}(\tau)\\
\label{correlatort}
\end{equation}
\end{itemize}
Here, we have assumed translational invariance in $x$, which is necessary for defining the correlator in terms of a Fourier integral in the angular direction.

Next we specify the boundary Hamiltonian. We shall use the Dirac Hamiltonian
\[ H_{Dirac}(k)=v_F[\sin k \sigma_1 +M(m+1-\cos k)\sigma_2]\]
from Eq. \ref{dirac1} for cases $(1)$ to $(3)$.  For the sake of conciseness in the already sundry results, we shall henceforth set $v_F$ and $M$ to unity unless otherwise stated, and consider only cases at zero chemical potential $\mu$. Indeed, $v_F$, which couples to $\tau$ under imaginary time evolution $e^{-H\tau}$ merely leads to a trivial rescaling $\tau\rightarrow v_F\tau$ in the results. The value of $M$ does not affect the leading asymptotic behavior of the correlators in general, and its study is relegated to Appendix \ref{genericdirac}. For case $(4)$, we shall simply base our calculations on the non-linear dispersion relation 
\begin{equation}E_k=k^\gamma,\end{equation}
since we will be primarily interested in the effect of setting $\gamma\neq 1$. 

We next elaborate on the complex analytic structure of the Dirac Hamiltonian and correlator. The positions of the complex singularities play a crucial role in determining the \emph{asymptotic} decay behavior of the correlators, typically with power-law decay when all singularities lie on the unit circle and exponential decay otherwise. The correlator in the spatial ''angular'' direction, in particular, is a Fourier transform for which there exist results that relate the decay of Fourier coefficients with the location of singularities. For a meromorphic function $f(z)$, the Fourier coefficients $f_l = \oint_{|z|=1} \frac{dz}{z} f(z) z^l$ decay like
\begin{equation}
f_l\sim l^{-(1+B)}|z_0|^l
\label{decay}
\end{equation}
for $|z_0|<1$, $l\gg 1$, with $z_0$ the branch point of $f(z)$ closest to the unit circle and $B$ is its corresponding branching number: 
\begin{equation} f(z_0+ \Delta z) \sim f(z_0) + \left(\frac{\Delta z}{z_0}\right)^B
\label{decay2}
\end{equation}
for $z$ near $z_0$. Note that $B$ cannot be a non-negative integer, since otherwise the Riemann surface will not be ramified or even divergent at $z_0$. Proofs, together with other physical applications of this result, can be found in Refs. \onlinecite{kohn1959,leeandy2014,leeye2015} and especially \onlinecite{he2001}.

In our correlators of interest, $f(z)$ takes the explicit forms $h_z$ or $h_z/E_z$ from the expressions to follow. $h_z$ is the Dirac Hamiltonian with $M=1,v_F=1$:
\begin{equation}h_z=\left(\begin{matrix}
 & 0 & i(\frac{1}{z}-(1+m)) \\
 & -i(z-(1+m)) & 0 \\
\end{matrix}\right)\label{hz}\end{equation}
with eigenenergies
\begin{eqnarray}
E_z&=&\sqrt{1+(m+1)^2-(m+1)\left(z+\frac{1}{z}\right)}\notag\\
%& \stackrel{\rightarrow}{m=0}& -i(z^\frac{1}{2}-z^{-\frac{1}{2}})
&  \rightarrow_{m=0}&\; \; -i(z^\frac{1}{2}-z^{-\frac{1}{2}})
\label{hz2}
\end{eqnarray}
Due to the square root, $z$ is an analytic function on a 2-sheeted Riemann surface with ramification (branch points) at $z=\infty,m+1,\frac{1}{m+1}$ and $0$. When $m=0$, the two points $z=(1+m)^{\pm 1}$ coincide and annihilate, leaving a single branch cut from $0$ to $\infty$. These branch points also appear in the flattened Hamiltonian $\frac{h_z}{E_z}$ that appears in the correlators Eqs. \ref{gq}, \ref{mueq} and \ref{gqtemp}:
\begin{eqnarray}\frac{h_z}{E_z}&=&\left(\begin{matrix}
 & 0 & \sqrt{\frac{m+1}{z}}\sqrt{\frac{z-\frac{1}{m+1}}{z-(m+1)}} \\
 & \sqrt{\frac{z}{m+1}}\sqrt{\frac{z-(m+1)}{z-\frac{1}{m+1}}} & 0 \\
\end{matrix}\right)\notag\\
&  =_{m\rightarrow 0}&\; \left(\begin{matrix}
 & 0 & \frac{1}{\sqrt{z}} \\
 & \sqrt{z} & 0 \\
\end{matrix}\right)
\end{eqnarray}
When $m=0$ at criticality, $\frac{h_z}{E_z}$ takes a particularly simple form that does not contain any nonzero pole in the unit circle. This still holds true for much more generic critical systems, being a necessary condition for power-law decay as required by Eq. \ref{decay}. Note that all the nontrivial branch points in the integrand of the bulk correlator must come from $G(z)$, since the wavelet basis functions $W_n(z)$ or $W_n(z^{-1})$ are polynomials in $z$ or $\frac{1}{z}$.

Interestingly, this simple form of $\frac{h_z}{E_z}$ admit a real space correlator $G(\Delta x)\propto \int z^{\Delta x}\frac{h_z}{E_z}\frac{dz}{z}$ with a nontrivial phase winding. Indeed, a short calculation reveals that $G(\Delta x)\propto \frac{e^{i\pi(\Delta x \pm 1/2)/L}}{\sin\left[\frac{\pi}{L}(\Delta x + 1/2)\right]}$, which exactly agrees with numerical results for different models with a Dirac point\cite{hermanns2014}.

Although we have only explicitly studied the complex topology of the Dirac model, the important point is that more generic models, i.e. multi-band models with arbitrary dispersion also have analogous topologies that lead to similar asymptotic correlator behavior. This will be further elaborated in the last part of Appendix \ref{genericdirac}. 

Table \ref{table1ehm} contains a summary of the asymptotic behavior of the mutual information, correlators, entanglement entropy and bulk geometry parameters for the different boundary systems. More details about each case are discussed in the remainder of this section.

%\begin{onecolumn}
\begin{table*}%[H]
\centering
\renewcommand{\arraystretch}{2}
\begin{tabular}{|l|l|l|l|}\hline
 &\ $m=T=0 $  &\ $m\neq0,T=0 $   &\ $T\neq0,m=0 $ \\ \hline%&\ $\mu\neq 0,m=T=0$ \\ \hline
 %$|A|$  &$\sim 2^{-n}\rightarrow 0$, Eq. \ref{entropy1} &\ $\frac{1}{2}-\frac{4^{-n}}{3m^2}$ &\ $\rightarrow 0$, $\approx$ Eq. \ref{entropy1} \\ \hline%&\ $\sim 2^{-n}$ for $n>n^*$ \\ \hline
%a  & $\frac{1}{2}$ &\ $\frac{1}{2}$ &\ $\frac{1}{2}$ &\ $\rightarrow \frac{1+sgn(\mu)}{2}$ for  $n\rightarrow \infty$  \\ \hline
 $S_x$  &\ $\rightarrow \log 4$, Eq. \ref{entropy8}  &\ $\propto\frac{n}{4^{n}}\rightarrow 0$,   &\  $\rightarrow \log 4$ \\ %&\ Eq. \ref{entropy3}, $\rightarrow 0$ as $n\rightarrow \infty$\\ \hline
 &\ &\ Eq. \ref{massentropy} &\  \\ \hline
$I_n(\Delta x)$  &\ $\sim \frac{1}{x^6}$, Eq. \ref{Ixyangular} &\ $\sim \frac{e^{-2^{n+1}m\Delta x}2^n}{\Delta x}$,  &\ $\sim e^{-2\pi T 2^n \Delta x}$ for $\gamma=1$; and in general \\%&\ Eq \ref{6mu}; $\sim \frac{16^n\nu^4}{x^2}$, Eq. \ref{Ixy3mu}, $n<n^*$ \\ %\hline
 &\ &\ Eq. \ref{mass2} &\ $e^{-[2(\pi T)^{1/\gamma} (2^n\Delta x)]\sin \frac{\pi}{2\gamma}}$, Eq. \ref{nonlinearIxy} \\ \hline%&\ $\sim \frac{1}{x^28^n\nu^4}$, Eqs. \ref{8mu}, \ref{9mu}, $n\gg n^*$ \\ \hline
$I(n_1,n_2)$  &\ $\sim 2^{-\Delta n}$, Eq. \ref{Ixyradial} &\ unexplored &\ unexplored \\ \hline%&\ Eq. \ref{radialmu} \\ \hline
$\text{Tr} C_n(\tau)$  &\ $\sim \frac{2^{3n}}{v_F^3\tau^3}$, Eq. \ref{time1} &\ $\sim e^{-m\tau}$ &\ $\text{Tr}~[C_n(\tau)+C_n(\beta-\tau)]_{T=0}$, $2^n\ll 2\pi\beta$;\\ % &\ Eqs. \ref{time5}, \ref{time4}; $\sim \frac{8^n\nu^2}{v_F\tau}$, Eq. \ref{timeUV}, $n<n^*$ \\ %\hline
  &\  &\ &\ lower bound $e^{-\frac{\beta }{2^{(2\gamma-1)n}}}$, Eqs. \ref{rhoBTZ1},\ref{rhononlinear} \\
&\ &\ &\ for $2^n> 2\pi\beta$ \\ \hline
%&\ $\sim \frac{1}{2^n\nu^2v_F\tau}$, Eq. \ref{timeIR}, $n\gg n^*$  \\ \hline
$\frac{\xi_{\theta}}{R }$  &\ $\frac{1}{3}$ &\ n.a. &\ $1$ for $\gamma=1$ \\ \hline%&\ $1$ \\ \hline
%$\frac{R}{\xi}|_{I_{(x,1),(x,n)}}$  &\ $-\frac{\log C(0)^2}{\log 2}$ &\ $-\frac{\log C(0)^2}{\log 2}+\text{const.}$ &\ unexplored &\ exact for Haar \\ \hline
$\frac{\xi_\rho}{R }$  &\ $1$ &\ n.a. &\ $1$ for $2^n\ll 2\pi \beta$; n.a. otherwise \\ \hline%&\ $\frac{1}{3}$ for $n<n^*$, $1$ for $n\gg n^*$ \\ \hline
$\frac{\xi_\tau}{R_\tau}$  &\ $\frac{2}{3}$ &\ n.a. &\ $\frac{2}{3}$ \\ \hline%&\ $2$\\ \hline
$b$  &\ n.a. &\ n.a. &\ $L\xi_\theta T$ for $\gamma=1$; $\propto T^{1/\gamma}$ for $\gamma>1$  \\ \hline%&\ n.a. \\ \hline
$R_{\theta}$  &\ $\frac{1}{2}\left(\frac{1}{\pi}\sqrt{\frac{2}{I_0}}\right)^{\frac{1}{3}}$ &\ n.a. &\ not uniquely determined  \\ \hline%&\ $\propto \nu^{2}$ \\ \hline
$R_{\tau}$  &\ $(2v_F \pi^{1/3})^{-1}$, Eq. \ref{timeR} &\ n.a. &\ $\sqrt[3]{\frac{2}{\pi^4}}$ for $\gamma=1$  \\ \hline%&\ $\propto \frac{\nu^{2}}{v_F}$ \\ \hline

%$Q$  &\ n.a. &\ n.a. &\ n.a.  &\ $\frac{L}{2\pi}\mu$?? \\ \hline

$\rho_n$  &\ $\frac{L}{2^{n+1}\pi}$   &\ n.a. &\ $\frac{L}{2^{n+1}\pi}$ for $2^n\ll 2\pi\beta$ \\ %&\ $\frac{L}{8^{n}2\pi}$ for $n<n^*  $  \\
  &\  &\ &\ $b\left(1+\frac{2\beta^4}{9\pi^2 4^n}\right)$ for $2^n>2\pi\beta$  \\ \hline%&\ $\frac{L2^n\nu^4}{(2\pi)^5}$ for $n\gg n^*$\\ \hline
\end{tabular}

\caption{Summary of results for various physical quantities in all of the $D=1$ scenarios considered, where $L$ is the size of the boundary system. Case $(1)$ ( $m=T=0$) from Sect. \ref{gensetup} is fitted onto the AdS metric with radius being either $R$ or $R_\tau$ (which are numerically close) depending on whether the fit is done for the spatial direction or imaginary time direction. The massive ($m\neq 0$) case $(2)$ is not fitted with any classical geometry. For cases $(3)$ and $(4)$ with temperature $T\neq 0$, $\gamma$ controls the dispersion via $E_q=q^\gamma$, and results are given for the BTZ case ($\gamma=1$) and the general Lifshitz case ($\gamma>1$), when applicable. The energy scale of $T$ divides the bulk into two regions demarcated by $2^n=2\pi \beta$, each with different qualitative properties. }
\label{table1ehm}
\end{table*}
%\end{onecolumn}

\subsection{Critical boundary Dirac model vs. bulk AdS space}

For a start, let us consider the massless $(1+1)$-dim massless Dirac model as the boundary theory. We show that the correlators in all three directions (Eqs. \ref{correlatorx} to \ref{correlatort}) all suggest a bulk geometry of a (2+1)-dim Anti de-Sitter space (further detailed in Appendix \ref{app:geodesics}) given by the metric (Eq. \ref{adsmetric}):
\begin{equation}
\frac{1+\frac{\rho^2}{R^2}}{1+\frac{L^2}{4\pi^2R^2}}d\tau^2+\frac{d\rho^2}{1+\frac{\rho^2}{R^2}}+\rho^2d\theta^2
\end{equation}
where $L$ is the boundary system size and $\tau$ is scaled such that the metric is $O(2)$-invariant at the critical boundary where $\rho=\frac{L}{2\pi}\gg R$. In the following, we shall only consider asymptotically large angular and temporal intervals with $\Delta x,\tau\rightarrow \infty$, and/or not necessarily large radial intervals between layers $1$ and $n$.  Since we have already taken the thermodynamic limit $L\rightarrow \infty$, $\Delta x<L$ can still be satisfied for arbitrarily large $\Delta x$.

\subsubsection{Spatial directions}

As is detailed in Appendix \ref{app:twoband}, the spatial decay of the Mutual Information is given by (see Eq. (\ref{Ixy2})):
\begin{equation}
I_{xy}=\frac{4(|u|^2+|v|^2)}{1-4A^2}
\end{equation}
where $|u|,|v|$ are the off-diagonal (unequal spin) part of the propagator $C_{x-y}$, and $A$ is the off-diagonal part of the onsite propagator $C_x$. From Eq. \ref{hz}, they are explicitly: 
\begin{equation}
A=-iC_n(\Delta x=0)=i\oint_{|z|=1}\frac{dz}{z}W^*_n(z^{-1})W_n(z)\sqrt{z}
\label{A}
\end{equation}
which, being onsite, does not depend on the displacement between $x$ and $y$. For the angular direction where $x,y$ are on the same layer,
\begin{equation}
u,v=-iC_n(\Delta x)=-i\oint_{|z|=1}\frac{dz}{z}W^*_n(z^{-1})W_n(z)z^{\mp \frac{1}{2}}z^{2^n\Delta x}
\label{uv}
\end{equation}
while for the radial direction where $n_1\neq n_2$ but $\Delta x=0$,
\begin{equation}
u,v=-iC(n_1,n_2)=-i\oint_{|z|=1}\frac{dz}{z}W^*_{n_1}(z^{-1})W_{n_2}(z)z^{\mp \frac{1}{2}}
\label{uv2}
\end{equation}
In the above equations, the integrands do not contain poles. However, the integrals are nonzero due to the branch cut from $z=0$ to $z=\infty$ from the square root factor. They can be evaluated by standard deformations of the contour, as demonstrated in detail in Appendix \ref{app:critical}.

After some computation, we obtain for the angular direction
\begin{equation}
|u|,|v|\sim \frac{1}{16 \pi}\frac{1}{(\Delta x)^3}
\label{uv3}
\end{equation}
As elaborated in Appendix \ref{app:critical1}, such a power-law decay of the single-particle propagator is generally expected in the presence of a branch cut. Physically, it is a signature of criticality, with a power of $3$ instead of $1$ due to the additional 'destructive interference' from the antisymmetric combinations of adjacent sites in the Haar wavelet basis. There is a striking absence of the layer index $n$ in Eq. \ref{uv3}, which reflects the scale-invariance of the boundary theory. Eq. \ref{uv3} also holds for general values of $M$ in the Dirac Model Eq. \ref{dirac1}, where $M$ controls the ratio between the quadratically dispersive $1-\cos k$ and linearly dispersive $\sin k$ terms near the IR point. While $M$ can affect the details of the branch cut, it cannot change the decay exponent, as is shown in Appendix \ref{genericdirac}.

For the radial direction with $n_1=1$ and $n_2=n$, Appendix \ref{app:critical2} also tells us that

\begin{equation}
|u|,|v|\sim \left(\frac{1}{\sqrt{2}}\right)^{n-1}
\label{uv4}
\end{equation}

Hence we have exponential decay of the single-particle propagator in the radial direction, which is consistent with scale invariance\footnote{ Scale invariance entail the multiplicative property $C(n_1,n_3)\sim C(n_1,n_2)C(n_2,n_3)$ for all $n_1<n_2<n_3$. This is only satisfied by an exponential dependence on the internal $\Delta n$.}.

Strictly speaking, the mutual information across the radial direction involves both $A_1$ and $A_n$, the unequal spins onsite propagators $A$ at layers $1$ and $n$, and a more general (and complicated) version of Eq. (\ref{Ixy2}) should be used. However, $A_n \rightarrow 0$ rapidly as $n$ increases, effectively leading to no asymptotic correction. Mathematically, this is because as $n$ increases, the peaks of $W_n(e^{iq})$ approaches a delta function at $q_0=\frac{2\pi}{2^n}\rightarrow 0$ which gets exponentially closer to the IR point, where contributions to the integral are penalized by the momentum correlator $h_q/E_q$. Explicitly, 

\begin{eqnarray}
|A_n|&=&-\frac{i}{2}\int_{-\pi}^\pi |W_n(e^{iq})|^2 \text{sgn}(q)e^{iq/2}dq\notag\\
&\rightarrow &  \frac{1}{2}\int_{-q_0}^{q_0} |W_n(e^{iq})|^2 \frac{|q|}{2}dq\notag\\
&\sim &  \frac{1}{4}|W_n(e^{iq_0})|^2\int_{-q_0}^{q_0}  |q|dq\notag\\
&=& \frac{2^{n-1}}{\pi^3}q_0^2\notag\\
&=& \frac{2}{\pi}\frac{1}{2^n}\rightarrow 0
\label{entropy1}
\end{eqnarray}
in the large $n$ limit. Physically, this means that unequal spins become totally decoupled in the IR regime. According to Eq. \ref{entropy3}, the sites in the IR layers are hence maximally entangled with the rest of the bulk:
\begin{equation}
S_x\rightarrow -4\frac{1}{2}\log \frac{1}{2}=\log{4}
\label{entropy8}
\end{equation}
This maximal entanglement in the IR also exists in generic critical systems, since the IR DOFs become harder and harder to isolate unless an energy scale (i.e. mass) exists.

Putting it all together, the mutual information behaves like
\begin{equation}
-\log I_n(\Delta x)\sim 6\log \Delta x +\log (32\pi^2)
\label{Ixyangular}
\end{equation}
for angular intervals and
\begin{equation}
-\log I(1,n)\sim (n-1)\log 2 + \text{small const.}\sim \Delta n\log 2
\label{Ixyradial}
\end{equation}
for radial intervals. These asymptotic behaviors are in excellent agreement with those of the geodesic distances on AdS space, if one uses the proposed correspondence given by $\frac{d^{min}_{\Delta \bold x}}{\xi}=- \log \frac{I_{\bold x \bold y}}{I_0}$ in Eq. \ref{dx}. The parameters $\xi$ and $I_0$ respectively set the scales of bulk distance and mutual information, and are related in a precise way discussed later. In principle, $\xi$ can be different in different independent directions.

We first study the correspondence in the angular direction. The geodesic distance between two equal-time points $(\rho,\theta_1)$ and $(\rho,\theta_2)$ with angular interval $\Delta \theta = |\theta_2-\theta_1|=\frac{\Delta x}{\rho}$ is given by Eq. \ref{ads1}:

\begin{equation}
d^{min}_{\Delta x}= R\cosh^{-1}\left(1+\frac{2\rho^2}{R^2}\sin^2\frac{\Delta \theta}{2}\right)\sim 2R \log \frac{\Delta x }{R}\label{ads1prime}
%\label{ads1}
\end{equation}
where $R$ is the AdS radius that determines the length scale below which we expect significant deviations from logarithmic behavior. Comparing Eqs. \ref{Ixyangular} and \ref{ads1prime}, we see that 
\begin{equation}
\frac{R }{\xi_\theta}=3
\label{Ixy4}
\end{equation}
and
\begin{equation}
R =\frac{1}{2}\left(\frac{1}{\pi}\sqrt{\frac{2}{I_0}}\right)^{\frac{1}{3}}
\label{Ixy5}
\end{equation}
We see that $R $ decreases weakly with increasing $I_0$. For the Dirac model we take $I_0=2\log 4$, the theoretical maximal mutual information mentioned below Eq. \ref{dx}. With it, we obtain $R =0.3233$ and $\xi_\theta=0.1078$, which is in excellent agreement with the numerical values obtained in Ref. \onlinecite{qi2013}.

For the radial direction, we compare Eq. \ref{Ixyradial} with the AdS geodesic distance (Eq. \ref{ads2}):
\begin{eqnarray}
d^{min}_{\Delta \rho}&= &R |\Delta(\log \rho)|\notag\\
&=& R (\log 2) |\Delta n|
\end{eqnarray}
Here we have taken $\rho=\rho_n= \frac{L}{2\pi 2^n}$, as required by the scale-invariance of radius $\rho$ and the circumference $\frac{L}{2^n}$ at any layer $n$. We easily obtain
\begin{equation}
\xi_\rho = R 
\label{Rradial}
\end{equation}
which means that the AdS radius is nothing but the radial length scale $\xi_\rho$ of radial geodesics. This is also in agreement with the numerical results in Ref. \onlinecite{qi2013}. 

It should be noted that the ratio $R/\xi$ is different for the angular and radial directions, which is a manifestation of the fact that the mapping does not preserve the entire conformal symmetry of AdS space (since conformal symmetry is only emergent in long wavelength limit and does not exist rigorously in a lattice model). The bulk theory has scale invariance and a reduced translation symmetry with unit cell size varying with the layer index $n$, but the correlations are anisotropic between radial and angle directions.

\subsubsection{Imaginary time direction}

As previously postulated by Eq. \ref{dt}, we can deduce a classical bulk geometry from the decay properties of the imaginary time correlator $C_n(\tau)$ given by
\begin{equation}
C_{n}(\tau)=\frac{1}{2}\int_{-\pi}^\pi dq |W_n(e^{iq})|^2 e^{-\tau E_q}\left(\mathbb{I}-\frac{h_q}{E_q}\right)
\label{time0}
\end{equation}
When $\tau$ is large, the value of $C_n(\tau)$ arises from competing contributions from the IR regime and the momentum scale set by the layer index $n$. While most of the spectral weight of $W_n(e^{iq})$ is concentrated around the momentum $q_0=\frac{2\pi}{2^n}$, the exponential factor exponentially suppresses contributions above the IR  regime set by $E_q=v_F q \sim \tau^{-1}$. Hence we expect $C_n(\tau)$ to increase as $n$ goes deeper into the IR. Indeed, this is exactly contained in the analytical result Eq. \ref{time3} derived in Appendix \ref{app:time}:
\begin{equation}
\text{Tr}~ C_n(\tau)\sim \frac{1}{8\pi} \left(\frac{2^n}{v_F \tau}\right)^3 = \frac{1}{(8\pi^2)^2} \left(\frac{L}{v_F \rho\tau}\right)^3
\label{time1}
\end{equation}
where $\rho=\frac{L}{2\pi 2^n}$ is the bulk radial AdS coordinate of layer $n$. Eq. \ref{time1} also holds for generic $M$ in the Dirac model (Eq. \ref{dirac1}), except for the degenerate case where $M=\infty$ and there is no linearly dispersive term $\sin k\sigma_1$.

Comparing the logarithm of Eq. \ref{time1} with the AdS timelike geodesic $d_\tau \sim 2R\log \frac{2\pi\rho \tau}{LR}$ from Eq. \ref{ads3}, we obtain
\begin{equation} \frac{R}{\xi_\tau}=\frac{3}{2} \end{equation}
and
\begin{equation}
 R_\tau=\frac{1}{2v_F\pi^{1/3}}\approx \frac{0.3414}{v_F}
\label{timeR}
\end{equation}
These results are valid in the regime where $\rho\gg R_\tau$ and $R_\tau\ll \tau \ll L$, i.e. when the geodesics do not come close to circumnavigating the AdS space. The AdS radius $R_\tau$ obtained here is slightly different from $R $ obtained through spatial geodesics in Eq. \ref{Ixy5}, with the latter containing a weak dependence on the reference mutual information $I_0^{-1/6}$. 

We note that $\frac{R}{\xi_\tau}=\frac{3}{2}$ obtained from the imaginary time correlator is exactly half of that of $\frac{R}{\xi_\theta}=3$ obtained from the (spatial) mutual information. This is not an inconsistency, but a manifestation of the anisotropy in the definitions: the mutual information is asymptotically quadratic, not linear, in the two-point correlator, and its associated geodesic distance must be doubled.

\subsection{Boundary Dirac model at nonzero mass $m$, $T=0$ }

This is our first and simplest example of a non-critical system. A nonzero mass scale $m$ introduces a pole in the correlator, which leads to the exponential decay of correlation functions. As we will discuss in this subsection, the dual geometry has a spatial "termination surface" which makes the spatial geometry topologically different from that of critical fermions. The temporal geodesics, however, exhibit no unusual behavior at the termination surface, so the surface is not a black hole horizon but a purely spatial cutoff.

\subsubsection{Angular direction}

As previously discussed, the geodesics in the spatial angular direction depend on the mutual information $I_{xy}=\frac{|u|^2+|v|^2}{1-4A^2}$ where $u$ and $v$ are the unequal-spin propagators between sites separated by an angular distance of $\Delta x$ within layer $n$ , and $A$ is the unequal-spin onsite propagator. 

We first look at how $u$ and $v$ differ from those of the critical case. They are given by the off-diagonal components of
\begin{equation}
-i\oint_{|z|=1} \frac{dz}{z}W^*_n(z^{-1})W_n(z)z^{2^n\Delta x}\frac{h_z}{E_z}
\label{mass1}
\end{equation}
where $\frac{h_z}{E_z}$ is the $2\times 2$ matrix given by Eq. \ref{hz}. For the Dirac model with $m\neq 0$, $\frac{h_z}{E_z}$ and hence the integrand of Eq. \ref{mass1} now has square root branch points away from the origin, namely at $z_0=1+m$ and $z_0=\frac{1}{1+m}$. Unlike in the critical case, the singularity $z_0=\frac1{1+m}$ now satisfies $|z_0|<1$ and by Eq. \ref{decay} determines the asymptotic exponential correlator decay with $f(z)=W_n^*(z^{-1})W_n(z)\frac{h_z}{zE_z}$. The branching numbers $B$ (Eq. \ref{decay2}) at $z_0$ are given by $B=\pm \frac{1}{2}$ for $u$ and $v$, which shall thus decay (for $2^n\Delta x \gg 1$ and small $m$) like
\begin{equation}
u,v\sim \frac{(2^n\Delta x)^{-1\mp 1/2}}{(1+m)^{2^n\Delta x}}\approx \frac{e^{-m 2^n \Delta x}}{(2^n\Delta x)^{1\pm 1/2}}
\label{udecay}
\end{equation}
We see that $v$ dominates with a factor of $2^n\Delta x$, and is itself exponentially decaying with a subleading power-law decay. Hence $-\log I_{xy}$ will define a rescaled Euclidean distance with length scale given by $\frac{1}{m}$. For a Dirac model with a generic value of $M$ in Eq. \ref{dirac1}, $z_0=\frac{1}{1+m}$ should be replaced by the pole \begin{equation}z_0=\frac{M(1+m)-\sqrt{1+2mM^2+m^2M^2}}{M\pm 1},\end{equation}
whose corresponding effective mass is given by $|z_0|^{-1}-1$.

As contrasted with the critical case, the mutual information in the massive case also depends nontrivially on $A$, the unequal-spin onsite propagator, as we go deep into the IR regime below the energy scale of layer $n$. Recall that at layer $n$, $A$ is given by
\begin{equation}  A_n =  \frac{-i}{2}\int_{-\pi}^{\pi}|W_n(e^{iq})|^2 \frac{h_q^{offdiag}}{E_q} dq \label{entropy5}
\end{equation}
For large $n$, the spectral weight is mainly concentrated around $q_0=\frac{2\pi}{2^n}$, where
\begin{eqnarray} \frac{h^{offdiag}_q}{E_q} &=& \frac{\sin q \pm i(m+1-\cos q)} {E_q}\notag\\
&\rightarrow& \frac{i(m+\frac{q^2}{2})}{\sqrt{m^2+q^2}}\notag\\
& \approx& i\left(1+\frac{m-1}{2m^2}q^2 \right)
\label{hE}
\end{eqnarray}
where the $\sin q$ term had been dropped because it is odd. Hence Eq. \ref{entropy5} can be rewritten as
\begin{eqnarray}
|A_n|&=&\frac{1}{2}\int_{-\pi}^\pi |W_n(e^{iq})|^2 \left(1+\frac{m-1}{2m^2}q^2 \right)dq\notag\\
&\rightarrow &  \frac{1}{2}+\frac{m-1}{4m^2}\int_{-q_0}^{q_0} |W_n(e^{iq})|^2q^2 dq\notag\\
&\sim &  \frac{1}{2}-\frac{1}{4m^2}|W_n(e^{iq_0})|^2\int_{-q_0}^{q_0}  q^2 dq\notag\\
&=& \frac{1}{2}-\frac{2^{n}}{8m^2\pi^3}\frac{q_0^3}{3}\notag\\
&=& \frac{1}{2}-\frac{1}{3m^2}\frac{1}{4^n}
\label{massa}
\end{eqnarray}
This leads to an overall enhancement of $\frac{1}{1-4A_n^2}\sim 4^{n}$ in the mutual information, which thus behaves like
\begin{equation}
-\log I_n(\Delta x) \sim 2^{n+1}m\Delta x + \log \Delta x - n\log 2
\label{mass2}
\end{equation}
Its leading contribution is proportional to $2^n\Delta x$, which leads to a  key difference of its dual geometry from that of the critical fermion. Since the corresponding geodesic distance $d_{\Delta x}=\xi_{\Delta x}I_n(\Delta x)$ is linear in $\Delta x$, i.e. like an Euclidean distance, we can obtain the circumference $\alpha_n$ of layer $n$ simply by taking $\Delta x=2^{N-n}$, the number of sites in the whole layer. More rigorously, the circumference $\alpha_n$ should be measured by first taking equally spaced points with $2^{N'}$ sites between them, and then taking the $N\rightarrow \infty$ limit before taking the $N'\rightarrow \infty$ limit. The circumference will be given by the product of the geodesic distance between neighboring points $2m\xi_{\Delta x}\cdot 2^{n+N'}$ and the number of points $2^{N-n-N'}$, i.e. $\alpha_n=2m\xi_{\Delta x}\cdot 2^N$ which is a finite portion of the boundary circumference $2^N$.

\begin{figure}[H]
\centering
\includegraphics[scale=.74]{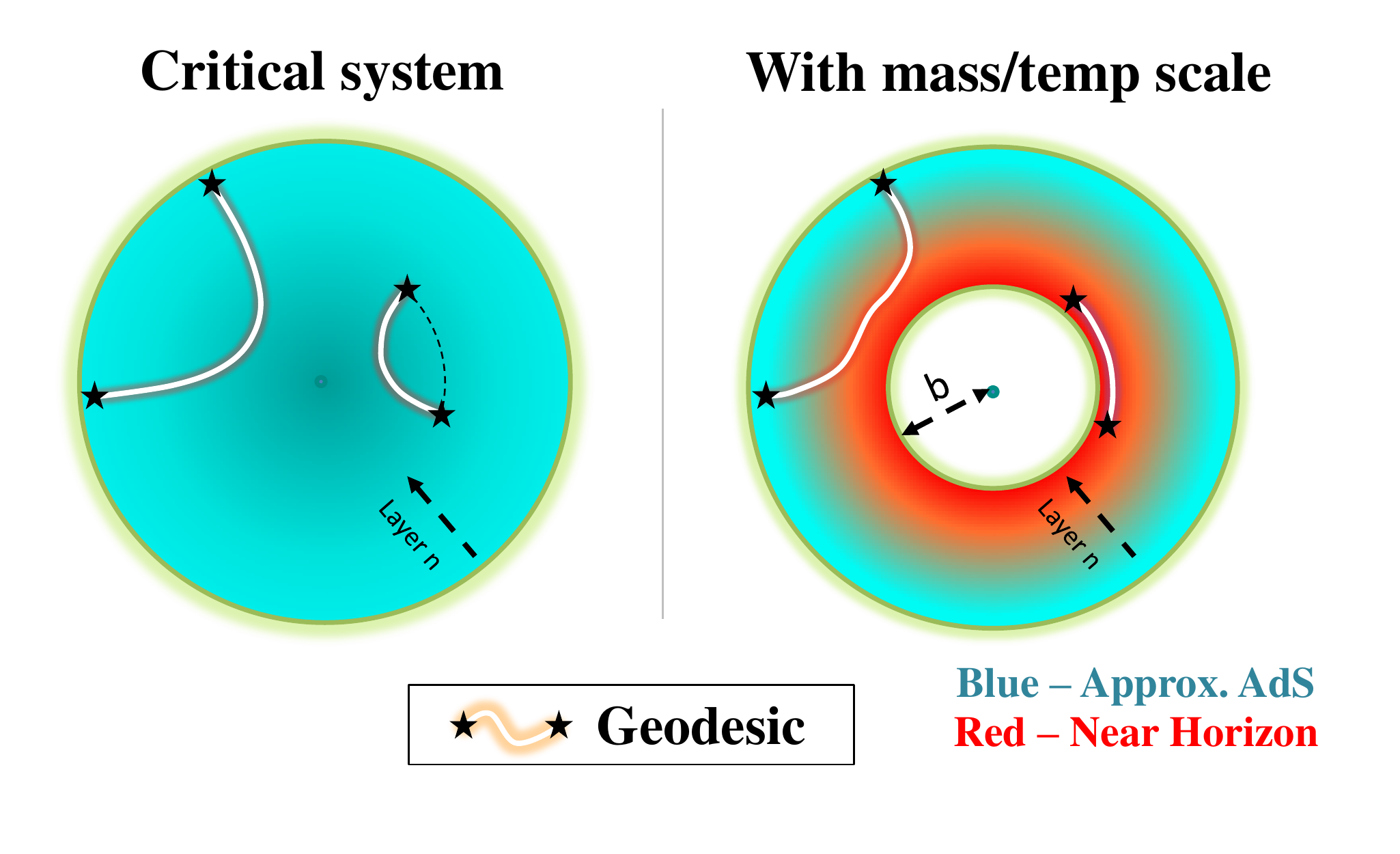}
\caption{(Color Online) The spatial bulk geometries of critical (Left) and non-critical (Right) systems and examples of their geodesics. In the AdS bulk (Left) corresponding to a critical boundary system, geodesics extend inwards towards the center, where there is far less 'space'. With an energy scale introduced by nonzero mass or temperature, the bulk geodesics develops a spatial 'termination' of space (Right). Geodesics wrap around a 'prohibited' region of radius $\propto m$ or $T$, acquiring an Euclidean character is that regime. As explained further in the main text, only the case at nonzero temperature corresponds to a true horizon. }
\label{fig:ADS_BTZ}
\end{figure}

The fact that $\alpha_n$ is finite in the large $n$ (infrared or IR) limit tells us that when $n\rightarrow \infty$, we are not approaching the center of a hyperbolic disk, but rather approaching a surface with finite area which acts as a ``termination surface" of the space. This is illustrated in Fig. \ref{fig:ADS_BTZ}. The IR surface shrinks with decreasing mass $m$. By comparison, in the dual geometry corresponding to the critical fermion, the distance between points $\Delta x$ sites apart does not depend on the layer index, so that the circumference $\alpha_n\propto 2^{-n}$ decays exponentially in the IR limit.

It is important to note that this surface is not a black-hole horizon, which distinguishes it from the case of nonzero  temperature state we will discuss in subsequent sections. One evidence for this conclusion is that each site carries a vanishing entanglement entropy $S_x$ in the IR limit $n\rightarrow \infty$, since
\begin{eqnarray}
S_x%&=& -2\left[(1-\frac{1}{3m^2}\frac{1}{4^n})\log\left(1-\frac{1}{3m^2}\frac{1}{4^n}\right)+\frac{1}{3m^2}\frac{1}{4^n}\log\left(\frac{1}{3m^2}\frac{1}{4^n}\right)\right]\notag\\
&\approx& -\frac{2}{3m^2}\frac{1}{4^n}\left(\frac{1}{3m^2}\frac{1}{4^n}+\log\left(\frac{1}{3m^2}\frac{1}{4^n}\right)-1\right)\notag\\
&\sim& \frac{ \log 16}{3m^2}\frac{n}{4^n}
\label{massentropy}
\end{eqnarray}
which is obtained by substituting Eq. \ref{massa} into Eq. \ref{entropy3}. Since the entropy decays exponentially at large $n$, the IR sites actually for direct product states unentangled with one  another. This is consistent with the physical picture that mass is renormalized to exponentially larger values in the IR limit (with respect to the kinetic energy scale), which forces the IR ground state to be simply a direct product of the single-site ground states with a large mass term. 

\subsubsection{Imaginary time direction}

The correlator in the imaginary time direction can be obtained pretty straightforwardly. Substituting the expression for the energy dispersion \[E_q=\sqrt{\sin^2q+(1+m-\cos q)^2}\approx m+\left(\frac{1+m}{2m^2}\right)q^2\] into Eq. \ref{time0}, we obtain
\begin{eqnarray}
\text{Tr}~ C_{n}(\tau)&=&e^{-m\tau}\int_{-\pi}^\pi dq |W_n(e^{iq})|^2 e^{-\tau \frac{1+m}{2m^2}q^2}\notag\\
&\propto& e^{-m\tau}
\end{eqnarray}
which is an exponentially decaying term multiplied by a nonuniversal Gaussian Integral. Hence $-\log \text{Tr}~ C_{n}(\tau)\sim m \tau$ defines an Euclidean bulk geodesic in the imaginary time direction. Together with the  previous results on spatial geometry, we see that although there is a spatial termination surface in the IR limit, the temporal direction still extends as usual. This is consistent with our earlier statement that the IR termination surface is not a black hole horizon, since the time direction is not infinitely redshifted. Topologically, the space-time we obtain for the massive Dirac model is $\mathcal{R}\times \mathcal{M}$, where $\mathcal{R}$ is the line in time direction, and $\mathcal{M}$ is a spatial annulus. 

\subsection{Critical linear boundary Dirac model at nonzero temperature $T$ vs. bulk BTZ black hole}
\label{sec:temp}

Now we study the effect of nonzero temperature in the massless Dirac model, the bulk geometry dual to which will be interpreted as a BTZ (Ba$\tilde{n}$ados, Teitelboim and Zanelli) black hole\cite{banados1992}. The BTZ black hole is a black hole solution for $(2+1)$-dim gravity with a negative cosmological constant, whose metric is
\begin{equation}
ds^2=\frac{V(\rho)}{V(L/(2\pi))}d\tau^2 + \frac{d\rho^2}{V(\rho)} +\rho^2 d\theta^2
\label{BTZmetric}
\end{equation}
where $V(\rho)=\frac{\rho^2-b^2}{R^2}$ is the Lapse function, $b$ is the horizon radius and $R$ is an overall length scale. We have rescaled $\tau$ by a factor of $\left[V\left(\frac{L}{2\pi}\right)\right]^{-1/2}$ so that $ds^2\rightarrow d\tau^2+\rho^2d\theta^2$ possesses $O(2)$ symmmetry at the boundary where $2\pi\rho =L$. The rescaling makes $\tau$ the same imaginary time variable as that in the boundary system (the canonical conjugate coordinate to the boundary Hamiltonian). 

Most notably, we find that the layers deep in the IR accumulate at the bulk horizon radius $\rho\rightarrow b$ from above, where $b$ is proportional to the temperature $T=\beta^{-1}$. This result is rigorously accurate in the $m\sim O(1)$ range of the Dirac model given by Eq. \ref{dirac1}, where the critical dispersion is essentially linear. 

\subsubsection{Angular direction}

Like a nonzero mass, a nonzero temperature also introduces an energy scale into the system. This energy scale is manifested as an imaginary gap $-log |z_0|>0$ where $z_0$ is the singularity of the momentum-space correlator $z_0$ closest to the unit circle. In this case, the singularities originate from the $\tanh\frac{\beta E_q}{2}$ term\footnote{The $\tanh\frac{\beta E_q}{2}$ term also cancels the singularity from $\frac{1}{E_q}$ on the unit circle, hence opening up the possibility of faster-than-power-law decay.} given in Eq. \ref{gqtemp}.

Above the energy scale of $T$, all correlators and hence the mutual information do not feel the effect of thermal excitations, and define an approximate AdS geometry just as in the $T=0$ critical case.

As one goes below the energy scale of $T$, Eq. \ref{decay} states that the unequal-spin propagators $u,v$ decay exponentially as $|z_0|^{2^n\Delta x}$, with a subleading power-law term $(2^n\Delta x)^{-(1+B)}$. $B$ and $|z_0|$ can be found as follows. 

Expanding $\cosh\frac{\beta E_z}{2}$, the denominator of the singular term, about $z_0$, we find that
\begin{equation}
 \cosh\frac{\beta E_z}{2}=0+\frac{\beta(z-z_0)}{2}\sinh\frac{\beta E_z}{2} \frac{dE_z}{dz} \propto z-z_0
\label{powerdecay}
\end{equation}
so the branching number $B=-1$. This value of branching number holds universally for generic systems at nonzero temperature\footnote{Except in the rare case where the energy scales associated with the mass and temperature exactly coincide. Singularities due to massive branch points diverge at zeros of $E_z$, unlike in the case of nonzero  temperature, and may consequently possess a fractional $B$. }, since Eq. \ref{powerdecay} does not depend on the form of $E_z$. Therefore, there is \emph{rigorously no} subleading power-law term in the decay of correlators in the nonzero  temperature case.

We now proceed to find $|z_0|$. $\tanh\frac{\beta E_q}{2}$ is singular when $\cosh \frac{\beta E_q}{2}=0$, i.e. $i\pi=\beta E_q =-i\beta(z^{1/2}-z^{-1/2})$ (see Eq. \ref{hz2}). This is satisfied by
$z+\frac{1}{z}-2=\left( \frac{2\pi(2l+1)}{\beta}\right)^2$, where $l\in \mathbb{Z}$. Hence the poles occur at
\begin{eqnarray}
 z_0^\pm&=&1+\frac{\pi^2 (2l+1)^2}{2\beta^2}\pm \frac{\pi(1+2l)\sqrt{4\beta^2+(2l+1)^2\pi^2}}{2\beta^2}\notag\\
&\rightarrow& 1\pm \pi\frac{2l+1}{\beta}
\end{eqnarray}
in the limit of small $T=\frac{1}{\beta}$. The pole with $l=0$ is closest to the unit circle, and we thus have the asymptotic decay
\begin{equation}
u\sim v\sim |z_0^-(l=0)|^{2^n\Delta x}\rightarrow(1-\pi T)^{2^n\Delta x}\approx e^{-\pi T 2^n \Delta x}
\label{udecay2}
\end{equation}
so that
\begin{equation}
I_n(\Delta x)\sim \frac{8e^{-\pi T 2^{n+1} \Delta x}}{1-4A_n^2}
\label{udecay3}
\end{equation}
Since $A_n\rightarrow \frac{1}{2}$ after the first few $n$, as explained in the subsection on the zero temperature critical case $(1)$, the mutual information behaves like (recalling that $2^n\Delta x = \frac{L\Delta \theta}{2\pi}$)
\begin{equation}
-\log I_n(\Delta x)\sim 	2\pi T 2^n\Delta x =LT \Delta \theta
\label{temp1}
\end{equation}
with no logarithmic subleading term. This asymptotic form for $I_n\left(\Delta x\right)$ is, to leading order, the same as that of Eq. (\ref{mass2}) for the massive zero temperature case, if we replace $2m$ by $2\pi T$. Following along the same lines as the previous section, we conclude that the circumference of each circle at layer $n$ is asymptotically $\alpha_n\simeq 2\pi T\cdot 2^N$. Thus there is a termination surface with this circumference (1-dim area) in the IR (low energy) limit. However, as we will verify by the single-site entropy and imaginary time direction distance later in this section, this surface is not just a termination surface of space, but a black hole horizon. Before discussing that, we shall first make a detailed comparison of the angular direction distance defined by Eq. \ref{temp1} with the angular geodesic distance given by a BTZ black hole metric. 

We can obtain a precise relationship between the temperature $T$ and the black hole radius $b$. Since Eq. \ref{temp1} holds in the IR regime, we compare it with the BTZ geodesic distance\footnote{Intuitively, the geodesic distance is strictly linear in $\Delta x$ infinitesimally near the horizon because the geodesics are not allowed to have any radial extent. This occurs when $V(\rho)\rightarrow 0$, which results in a radial displacement $\Delta s= \frac{\Delta \rho}{V(\rho)}$ becoming infinitely costly. } $d^{min}_{\Delta x}=2R \sinh^{-1}\left[\frac{\rho}{b}\sinh\frac{b\Delta \theta}{2R}\right]$ from Eq. 23 of Ref. \onlinecite{qi2013} in the near horizon limit $0<\rho-b\ll b$. 
\begin{eqnarray}
-\log I_n(\Delta x)= \frac{d}{\xi_\theta}&=& \frac{2R}{\xi_\theta} \sinh^{-1}\left[\frac{\rho}{b}\sinh\frac{b\Delta \theta}{2R}\right]\notag\\
&\approx& \frac{2R}{\xi_\theta} \log\left(\frac{\rho}{b}e^{b\Delta \theta/2R}\right)\notag\\
&=&\frac{ 2R}{\xi_\theta} \log\frac{\rho}{b}+\frac{2^{n+1}b \pi \Delta x}{\xi L}\notag\\
&\approx &\frac{2 \pi b(2^n \Delta x)}{\xi_\theta L}
\label{BTZcompare1}
\end{eqnarray}
which implies that
\begin{equation} b=L\xi_\theta T
\label{T1}
\end{equation}
Here $\xi_\theta$ is a yet-undertermined length scale. Our bulk geometry has agreed remarkably well with that of an actual BTZ black hole, with the LHS and RHS of Eq. \ref{BTZcompare1} agreeing on not just the leading term linear in $\Delta x$, but also the \emph{vanishing} logarithmic subleading term. Further discussions of the near-horizon geometry can be found in Appendix \ref{app:rindler}.

$\xi_\theta$ can be determined by compairing Eq. (\ref{T1}) with the relation between $T$ and $b$ in classical gravity. The requirement that the geometry is smooth in imaginary time at $\rho=b$, i.e. without a conical singularity, requires\cite{gibbons1977,charmousis}
\begin{eqnarray}
T&=& \left(\frac{2\pi R^2}{b}\frac{L}{2\pi R}\right)^{-1}= \frac{b}{RL}
\label{T2}
\end{eqnarray}
More details about this formula are given in Appendix \ref{app:rindler}. Comparing Eqs. (\ref{T1}) and (\ref{T2}) we obtain
\begin{equation} R=\xi_\theta
 \end{equation}
so that $\xi_\theta$ and $R$ are actually one and the same length scale parameter.

\subsubsection{Imaginary time direction}

The nonzero temperature features prominently in the imaginary time correlator because a finite $\beta$ corresponds to a finite periodicity in imaginary time. We shall show that the correlator $\text{Tr}~ C_n(\tau)$ defines a bulk geometry that is qualitatively similar to that of BTZ black hole, and deduce the effective radius $\rho=\rho_n$ of layer $n$ by looking at the maximal value of $-\log \text{Tr}~ C_n(\tau)$ achieved at half-period $\tau=\frac{\beta}{2}$.

The quantity to be computed is
\begin{equation}
\text{Tr}~ C_n(\tau)= \frac{1}{2}\int^{\pi}_{-\pi} dq |W_n(e^{iq})|^2 \text{Tr}~ G_q(\tau)
\end{equation}
where, from Eq. \ref{gq0},
\begin{equation}
%G_q= \frac{1}{2}\sum_{\lambda_q}e^{\lambda_q \tau}\left(1-sgn(\lambda_q)\tanh\frac{\beta \lambda_q}{2}\right)\rightarrow \sum_{\lambda_q<0}e^{\lambda_q \tau}=\sum_{E_q>0}e^{-E_q \tau}
\text{Tr}~ G_q(\tau) = \sum_{\lambda_q}\frac{e^{\lambda_q \tau}}{1+e^{\beta \lambda_q}}=\frac{1}{2}\sum_{\lambda_q}\frac{e^{\eta \lambda_q }}{\cosh \frac{\beta \lambda_q}{2}}=\frac{\cosh\eta E_q }{\cosh \frac{\beta E_q}{2}}.
\label{gqD}
\end{equation}
where $\eta=\tau-\frac{\beta}{2}$ and $\lambda_q$ denotes an eigenenergy of the system. The intermediate steps of Eq. \ref{gqD} are valid for any number of bands, but we have specialized to $\lambda_q=\pm E_q$ for the particle-hole symmetric 2-band case at the last step. $\text{Tr}~G_q(\tau)$ is manifestly even in $\eta$, which guarantees that $\text{Tr}~ G_q(0)=\text{Tr}~ G_q(\beta)$. It can also be obtained from Eq. \ref{gq} through direct simplification.

Due to the presence of the energy scale set by $T$, there are two distinct regimes. In the (high energy) UV limit $2^n\ll 2\pi \beta$ the kinetic energy dominates the temperature, and we have $\frac{\cosh\eta E_q }{\cosh \frac{\beta E_q}{2}}\approx e^{-\tau E_q}+e^{(\tau-\beta)E_q}$, yielding the nice relation
\begin{equation}
\text{Tr}~ C_n(\tau)\approx \text{Tr}~ C_n(\tau)|_{T=0}+\text{Tr}~ C_n(\beta-\tau)|_{T=0}.
\label{timetemp}
\end{equation}
This equation tells us that the nonzero  temperature correlator above the energy scale of $T$ is just the superposition of two copies of the zero temperature correlator reflected about $\tau=\frac{\beta}{2}$. This is consistent with how correlation functions in BTZ geometry are obtained by a periodic quotient of those in AdS space\cite{lifschytz1994}. Eq. \ref{timetemp} is completely general, since we haven't used any particular form for the Hamiltonian.

\begin{figure}[H]
\centering
\includegraphics[scale=1.05]{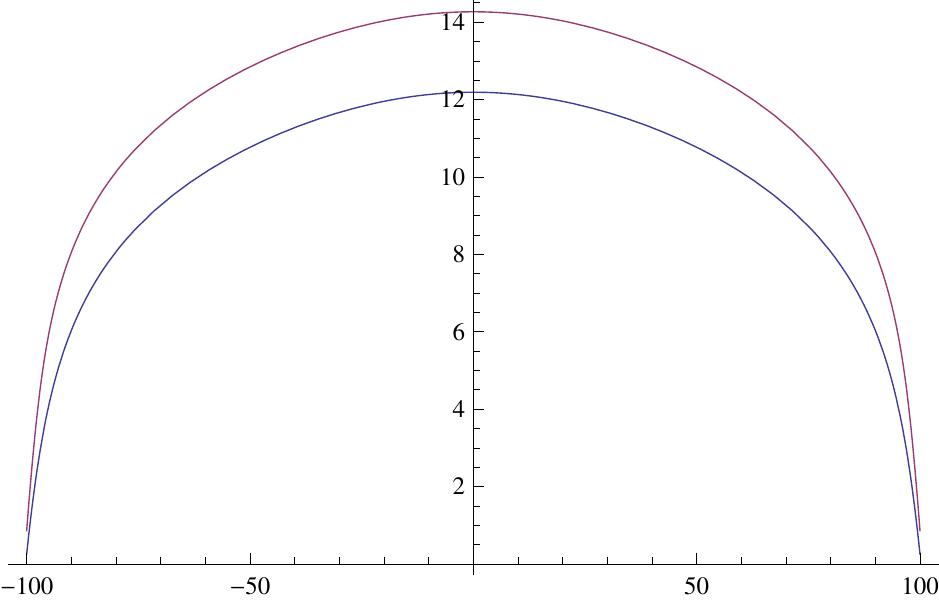}
\caption{$-\log\text{Tr}~C_n(\tau)$ according to Eq. \ref{timetempD}. Plotted are the $n=1,2$ curves for  $T=0.005$, which corresponds to an energy scale of $n\approx 10$. }
\label{periodic_time}
\end{figure}

Inserting the result of $\text{Tr}~ C_n(\tau)|_{T=0}$ from Eq. \ref{time1}, we find that
\begin{eqnarray}
&&-log  \;\text{Tr}~ C_n(\tau)\notag\\
&=& -\log\left(\tau^{-3}+(\beta-\tau)^{-3}\right)-3(n-1)\log2 +\log (\pi )\notag\\
\label{timetempD}
\end{eqnarray}
As suggested in Fig. \ref{periodic_time}, the first term gives a curve with geodesic distance qualitatively similar to that outside a BTZ black hole where $\rho\gg b$:
\begin{eqnarray}
d_\tau&=& R\cosh^{-1}\left[\frac{2\rho^2}{b^2}\sin^2\frac{\pi\tau}{\beta}\right]
\label{dtau}
\end{eqnarray}
which is derived in the Appendix of Ref. \onlinecite{qi2013}.

We can find the length scales $R$ and $\xi_\tau$ by comparing the maximal value of $-\log \text{Tr}~ C_n(\tau)$ with $\frac{d_\tau}{\xi_\tau}$, $\tau=\frac{\beta}{2}$. We have
\begin{eqnarray}
\frac{d_{\beta/2}}{\xi_\tau}&=& \frac{R}{\xi_\tau}\cosh^{-1}\frac{2\rho^2}{b^2}\notag\\
&\approx & \frac{2R}{\xi_\tau}\log\frac{2\rho}{b}\notag\\
&=& \frac{2R}{\xi_\tau}\log\frac{2\rho\beta}{LR}\notag\\
&\approx& \frac{2R}{\xi_\tau}\log\frac{\beta}{R\pi 2^n}
\end{eqnarray}
where we have used $\rho=\rho_n\approx \frac{L}{2\pi2^n}$ in the approximate AdS geometry far from the horizon. Comparing this with $-\log \text{Tr}~ C_n(\beta/2)= 3\log\frac{\beta }{2^n}+\log\frac{\pi}{2}$, we obtain
\begin{equation}
\frac{R}{\xi_\tau}= \frac{3}{2}
\end{equation}
which is exactly the same as in the $T=0$ case. This must be true because the bulk geometry is still asymptotically AdS above the energy of $T$. However, the value of $R$ is somewhat different due to the different functional forms of Eqs. \ref{timetempD} and \ref{dtau}, and takes the value
\begin{equation}
R=\sqrt[3]{\frac{2}{\pi^4}}\approx 0.274
\end{equation}
In the IR limit below the energy scale of $T$, i.e. $2^n> 2\pi\beta$, Eqs. \ref{timetemp} and hence \ref{timetempD} no longer hold because we must use the small $\beta E_q$ approximation in the denominator of $\text{Tr}~G_q(\tau)=\frac{\cosh{\eta E_q}}{\cosh \frac{\beta E_q}{2}} $. At $\tau=\frac{\beta}{2}$ or $\eta=0$, we have, to 2nd order in $\beta E_q \ll 1$,
\begin{eqnarray}
\text{Tr}~ G_q\;(\tau=\beta/2)&\approx & \sum_{\lambda_q}\left(1-\frac{\beta^2 \lambda_q^2}{8}\right)\approx  1-\frac{\beta^2}{8}q^2
\label{rhoBTZ0}
\end{eqnarray}
Hence,
\begin{eqnarray}
&&-\log \text{Tr}~ C(\beta/2)\notag\\
&=& -\log \text{Tr}~\int dq  G_q(\beta/2) |W_n(e^{iq})|^2\notag\\
&=& -\log\left(\int dq  |W_n(e^{iq})|^2 - \frac{\beta^2 }{8}\int dq q^2 |W_n(e^{iq})|^2\right)\notag\\
&\approx & -\log\left(1 - \frac{\beta^2 }{8}q_0^2  |W_n(e^{iq_0})|^2\right)\notag\\
&\approx & \frac{\beta^2 }{2^n\pi}
\label{rhoBTZ1}
\end{eqnarray}
where we have used the fact that $W_n(e^{iq})$ is sharply peaked at $q_0=\frac{2\pi}{2^n}$ with magnitude $\sqrt{\frac{2^{n+1}}{\pi^3}}$. Comparing Eq. \ref{rhoBTZ1} to the BTZ geodesic distance in Eq. \ref{dtau}:
\begin{equation} -\log \text{Tr}~ C(\beta/2)\sim \frac{2R_\tau}{\xi_\tau}\cosh^{-1}\frac{\rho}{b}\approx \frac{3}{2}\sqrt{2\left(\frac{\rho}{b}-1\right)}
\end{equation}
we arrive at
\begin{equation}
\rho_n\propto b\left(1+\frac{2}{9\pi^2}\frac{\beta^4 }{2^{2n}}\right)
\label{rhoBTZ}
\end{equation}
Indeed, $\rho\rightarrow b$ exponentially from above. This justifies the previous assumption that $\frac{\rho}{b}-1\ll 1$ which is, among other implications, consistent with the fact that no non-negligible logarithmic subleading terms exist in Eq. \ref{BTZcompare1}. Physically, the agreement between the imaginary time distance of the bulk geometry and the BTZ black hole means that the rate of imaginary time correlator decay slows down exponentially as one goes deeper into the IR, which translates into the infinite redshift an outside observer sees for any physical process near the BTZ horizon. %just like those near an actual BTZ horizon.

\subsection{Critical boundary model with nonlinear dispersion vs. bulk Lifshitz black hole}

As a sequel to the previous subsection, we now consider a nonzero  temperature boundary system with nonlinear dispersion in the long wavelength limit. As we have seen in the previous subsection, the energy-momentum dispersion is sufficient for determining the decay properties of the correlators. Here we consider the simplest nonlinear critical dispersion
\begin{equation}
E_q\simeq q^\gamma,\text{~for~}q\approx  0
\label{eqgamma}
\end{equation}
For higher $q$, $E_q$ should be regularized to a periodic function. However, the details of the regularization do not affect the critical behavior as we have seen time and again, and Eq. \ref{eqgamma} is sufficient as it stands. In this sense, Eq. \ref{eqgamma} subsumes the Dirac model (Eq. \ref{dirac1}) with general $M$, which is merely an interpolation between a $\gamma=1$ and a $\gamma=2$ Hamiltonian.

The asymptotic behavior of the mutual information depends solely on the position of the singularities of $G_q(\tau)\propto \text{sech}\frac{\beta E_q}{2}$. They occur when $\beta E_q=i(2l+1)\pi$, $l\in\mathbb{Z}$, i.e. when
\begin{equation}
E_q^2=q^{2\gamma}=-\pi^2 T^2(2l+1)^2
\end{equation}
In terms of $z=e^{iq}$, they occur at $z_0=e^{i\left(i\pi T (2l+1)\right)^{1/\gamma}}$. Clearly, the $l=0$ singularity has the largest magnitude within the unit circle, which is given by
\begin{equation}
|z_0|=e^{-(\pi T)^{1/\gamma}\sin \frac{\pi}{2\gamma}}
\end{equation}
From Eq. \ref{decay} and discussions surrounding Eqs. \ref{udecay3}, we conclude that the mutual information between two points with angular separation of $\Delta x$ sites decay like
\begin{eqnarray}
-\log I_n(\Delta x) &\sim& 2(\pi T)^{1/\gamma}(2^n\Delta x)\sin \frac{\pi}{2\gamma} \notag\\
&=&\frac{L\Delta \theta}{\pi}(\pi T)^{1/\gamma}\sin \frac{\pi}{2\gamma}
\label{nonlinearIxy}
\end{eqnarray}
Similar to case $(3)$ with linear dispersion, the $\Delta \theta$ dependence in the mutual information suggests that the circumference approaches a finite value in the IR limit, so that there is an event horizon. However, the nonlinear dispersion leads to a different $T$ dependence. Since the bulk geodesic distance $\propto - \log I_n(\Delta x)\propto T^{1/\gamma}\Delta \theta $, the black hole radius is $b\propto T^{1/\gamma}$, different from the $b\propto T$ behavior of the BTZ black hole\footnote{This is incompatible with metrics with Galilean symmetry, i.e the BTZ metric in Eq. \ref{BTZmetric}, because they always require $b\propto T$ to avoid a conical singularity in Rindler space.}.

There are indeed black hole solutions to classical gravity with the property $b\propto T^{1/\gamma}$, if one considers spacetimes with anisotropic scale invariance, i.e. with metric\cite{ayon2009}
\begin{equation}
ds^2=-\frac{r^{2\gamma}}{R^{2\gamma}}dt^2 + \frac{R^2}{r^2}dr^2 + \frac{r^2}{R^2}d\vec x^2
\end{equation}
invariant under the rescaling $(t,\vec x,r)\rightarrow (\lambda^\gamma t, \lambda \vec x, \lambda^{-1}r)$, with $\gamma$ the dynamical critical exponent and $R$ a length scale. 

If we include quadratic curvature tensor terms like $\Omega^2$, $\Omega_{\alpha\beta}\Omega^{\alpha\beta}$ or $\Omega_{\alpha\beta\mu\nu}\Omega^{\alpha\beta\mu\nu}$ to the gravitational action\cite{ayon2009,ayon2010}, the resultant Einstein's equations in the non Galilean-invariant spacetime will possess black hole solutions for certain ranges of parameters. Such solutions are known as \emph{Lifshitz Black Holes}, which are well-studied\cite{ayon2009,ayon2010, balasubramanian2009, cai2009} and proposed as possible gravity duals to Lifshitz fixed points in condensed matter physics. The explicit solutions of these black holes are known for certain values of $\gamma$, especially in the $2+1$ dimensions relevant to our current context.  For instance, the black hole metric for $\gamma=3$ and the gravitational action $S=\frac1{16\pi G}\int d^3x \sqrt{-g}\left[\Omega -2 \Gamma +2R^2\left(\Omega_{\alpha\beta}\Omega^{\alpha\beta}-\frac{3}{8}\Omega^2\right)\right]$, where $\Gamma$ is the cosmological constant, is given by\cite{ayon2009}
\begin{equation}
ds^2=-\frac{\rho^6}{R^6}\left(1-\frac{b^2}{\rho^2}\right)dt^2+\frac{R^2d\rho^2}{\rho^2-b^2}+r^2d\theta^2
\label{lifshitzmetric}
\end{equation}
By examining the near-horizon geometry in Euclidean time, we explicitly find its Hawking temperature to be $T=\frac{b^3}{2\pi R^4}$, which agree with the horizon area we obtained from a boundary theory with cubic dispersion.

The $T\propto b^\gamma$ dependence can also be expected from a simple counting argument. A system at a temperature $T$ can be physically understood as one with states randomly distributed in a energy width of $T$. This randomness is quantified by the entanglement entropy $S$ of the system with the thermal bath, most of which is carried by the IR region of the system. In the bulk system obtained through the EHM, the IR states carry maximal entropy per each site, so that the thermal entropy is proportional to the number of sites in the ``stretched horizon", which is proportional to the horizon area $b$. For a system with energy dispersion $E\propto q^\gamma$, the momentum range that has energy below $T$ is $\Delta k\propto T^{1/\gamma}$, so that the entropy $S\propto \Delta k\propto T^{1/\gamma}$. Consequently, $b\propto T^{1/\gamma}$. 

The imaginary time bulk correlator behaves in a similar way as that with linear dispersion (Case $(3)$). For the layers with energy scale above $T$, we still have, of course,
\[\text{Tr}~ C_n(\tau)=\text{Tr}~[C_n(\tau)+C_n(\beta-\tau)]_{T=0}\]
which holds independently of the dispersion. In the IR limit with energy scale below $T$, the minimal value for $\text{Tr}~ G_q(\tau)$ at $\tau=\frac{\beta}{2}$ still follows from Eq. \ref{rhoBTZ0} and \ref{rhoBTZ1}, except that the energy eigenvalues are now $\lambda=q^\gamma$. Hence we obtain a nontrivial (but still simple) nonlinear correction
\begin{equation}
-\log \text{Tr}~ C(\beta/2)\sim \frac{\beta^2}{2^{(2\gamma-1)n}\pi}
\label{rhononlinear}
\end{equation}
which is consistent with the metric in Eq. \ref{lifshitzmetric}, which has the geodesic distance at $\tau=\frac{\beta}{2}$ vanishing as a power of $\frac{\rho-b}{b}$. We still have $\rho\rightarrow b$ exponentially as $n$ increases, but at a different rate compared to the BTZ (Galilean-invariant) case.

\section{Generalization of EHM to higher dimensions}
\label{sec:higherdim}

\subsection{General setup}

\begin{figure}[H]
\centering
\includegraphics[scale=.7]{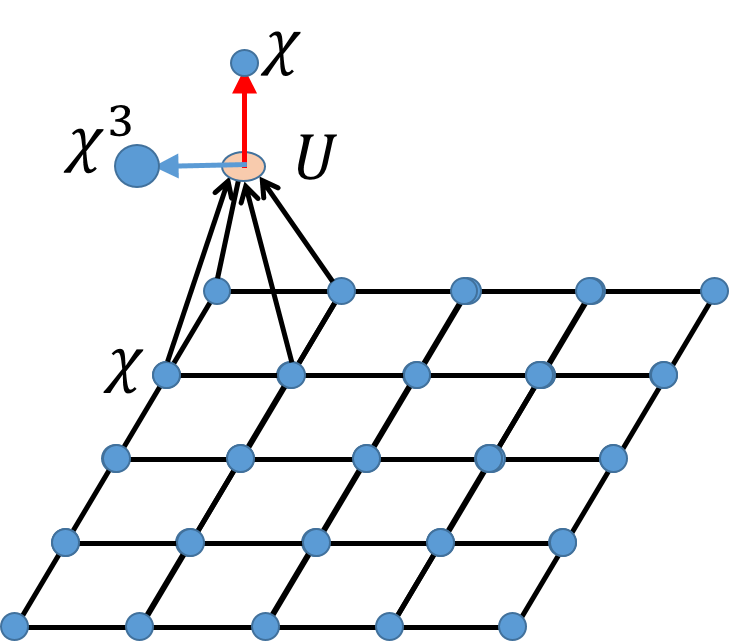}
\caption{Schematic picture of the EHM for two-dimensional boundary states. A unitary mapping (pink oval) is defined for four sites, each with a $\chi$-dimensional Hilbert space. It maps the DOFs from the four sites to two output sites, the IR site (red arrow) with dimension $\chi$ and the UV site (blue arrow) with dimension $\chi^3$. }
\label{fig:EHM2d} \end{figure}

When we generalize the boundary system to $D$ spatial dimensions, the bulk system will contain $D+1$ spatial dimensions, with a new emergent direction representing the energy scale. The boundary theory and bulk theory can be related by EHM in the same way as in the $D=1$ case. For example, with a boundary theory defined on the two-dimensional square lattice, a unitary mapping can be defined on four sites around a plaquette, which maps it to two sites representing the high energy and low energy degrees of freedom of the four sites. The mapping is illustrated in Fig. \ref{fig:EHM2d}. If the Hilbert space dimension is $\chi$ on each site, the output IR site should have dimension $\chi$ while the UV site now has a higher dimension $\chi^3$, corresponding to the $UV,IR$, $IR,UV$ and $UV,UV$ sectors of the 1-dim case. More generally in $D$ dimensions with a square lattice, one can map the $2^D$ sites in a cube to one IR site with Hilbert space dimension $\chi$ and one UV site with dimension $\chi^{2^D-1}$.

For free fermions systems, the mapping is equivalent to a wavelet transformation on the single-particle wavefunctions, just like in the case with one spatial dimension. The simplest higher-dimensional wavelet basis can be obtained via direct products of $1$-dim wavelet bases. To define them, we first label the 1-dim wavelet functions as% $W_n(z)$ with
\begin{eqnarray}
W^\upsilon_n(z)=\left\{\begin{array}{cc}C(z^{2^{n-1}})\prod_{j=1}^{n-1}C(z^{2^{j-1}}),&~\upsilon=1\\
D(z^{2^{n-1}})\prod_{j=1}^{n-1}C(z^{2^{j-1}}),&~\upsilon=2\end{array}\right.
\label{wavelet1D}
\end{eqnarray}
so that $\upsilon=1,2$ corresponding to the IR and UV wavelets in layer $n$, respectively. The $D$-dimensional wavelet functions can then be defined by
\begin{equation}
W^{\upsilon_1\upsilon_2...\upsilon_D}_n(z_1,...,z_D)=\prod_{j=1}^D W^{\upsilon_j}_n(z_j)
\label{WD}
\end{equation}
These $2^D$ wavefunctions for $\upsilon_j=1,2$ include one IR wavelet defined by $\upsilon_j=1,~\forall j$ and $2^D-1$ other wavelets that are regarded as UV degrees of freedom. The bulk correlators $C^{\mu\nu}$ can be obtained from the boundary correlator $G_{\vec q}(\tau)$ via this basis transform:
\begin{eqnarray}
&&C^{\mu\nu}(n_1,n_2, \Delta(2^n\bold x),\tau)\notag\\&=&\sum_{\bold q} W^{\mu*}_{n_1}(\bold q)W^{\nu}_{n_2}(\bold q)e^{i \bold q \cdot \Delta (2^n\bold x) }G_{\bold q }(\tau)\label{bulkcorrelator}
\end{eqnarray}
where we have denote the $2^D-1$ dimensional label of UV states $\upsilon_1\upsilon_2...\upsilon_D$ (with at least one $\upsilon_j=2$) by $\mu$ or $\nu$ for simplicity. 

When there is $\aleph$ number of orbitals at each site, the correlation matrix is $(2^D-1)\aleph \times (2^D-1)\aleph$. For analyzing the asymptotic bulk geometry, it suffices to consider only the slowest decaying elements of $C^{\mu\nu}$, which is determined by the lowest power of ${\bf q}$ in $W^\mu_n({\bf q})$. Using the long wavelength asymptotic behavior $C(e^{iq_j})\approx \sqrt{2}$ and $D(e^{iq_j})\approx \frac{-iq_j}{\sqrt{2}}$, we see that among the $2^D-1$ UV wavelets $W^{\upsilon_1\upsilon_2...\upsilon_D}_n({\bf q})$, the ones that play a leading role in the long wavelength correlation functions are those with only one $\upsilon_j=2$ and all other $\upsilon_k=1,~k\neq j$. Therefore the asymptotic behavior of the bulk correlator is given by  
\begin{eqnarray}
C^{\mu\nu}(n,n, \Delta\bold X,\tau)\sim 2^{D(n-2)}\partial_{X_{j}}\partial_{X_{k}}\sum_{\bold q} e^{i \bold q \cdot \Delta \bold X}G_{\bold q }(\tau)
\label{multidim1}
\end{eqnarray}
where $X_j=2^nx_j$, and $j,k$ are the directions where a 1-dim UV wavelet function $D$ is taken, {\it i.e.}, $\upsilon_{j,k}=2$. 

In the following, we shall examine the bulk geometries of various critical boundary systems at both zero and nonzero temperature, and highlight how they are different from those of $(1+1)$-dim boundary systems.

\subsection{Critical boundary model at zero $T$}

Here, we shall examine in detail the decay properties of the bulk mutual information and bulk imaginary time correlator corresponding to a critical $(D+1)$-d boundary system. We will find that they describe a bulk geometry of a higher-dimensional AdS space, in close analogy to the $(1+1)$-dim case described previously. Similar to one-dimensional case, we consider the Dirac model in (D+1)-dimensions\cite{golterman1993,creutz2001,qi2008topological}:
\begin{eqnarray}
%&&H\notag\\
H&=&\sum_kc_k^\dagger\left[\sum_{i=1}^D\Gamma^i\sin k_i+\left(M+D-\sum_{i=1}^D\cos k_i\right)\Gamma^0\right]c_k\notag\\
\label{dirac}
\end{eqnarray}
where $\Gamma^0,\Gamma^i$ are Hermitian Dirac matrices satisfying $\left\{\Gamma^\mu,\Gamma^\nu\right\}=\delta^{\mu\nu}$ for $\mu,\nu=0,1,...,D$. 
For $M$ close to $0$, the lowest energy excitations of this system are centered around $k=0$, where the EHM defined by wavelets in Eq. (\ref{wavelet1D}) and (\ref{WD}) correctly separates low energy and high energy degrees of freedom. In the following, we study the behaviors of the correlation function and dual geometry along different directions. For simplicity, we shall only explicitly study the $(2+1)$-dim case.
\\
\noindent{\bf Spatial (angular) directions.} 

We build on the results of the decay of unequal spin propagators $u,v$ for $(1+1)$-dim, with two obvious extensions mandated by Eq. \ref{multidim1}: Firstly, we now need to perform a multi-dimensional sum over $\bold q$ and secondly, we need to decide which sequence of derivatives $\partial^2_{X_{j_i}}$ produce the slowest decaying correlator.

We recall that in the absence of a EHM transform, the critical correlator behaves like the inverse first power of distance, i.e. $\sim \frac{1}{\sqrt{x^2+y^2}}$. Under the EHM transform, the correlators $u$ or $v$ decay faster due to derivatives introduced by UV projectors $D(e^{iq})$. Their slowest-decaying elements involve $D=2$ derivatives, i.e. $\partial^2_X$, $\partial^2_Y$ or $\partial^2_{XY}$. Hence  
\begin{equation} |u|\sim |v|\sim \frac{1}{(x^2+y^2)^{3/2}} \end{equation}
i.e.
\begin{equation} I_n(\Delta \bold x)\sim 6 \log |\Delta \bold x|+\text{const.} \label{Ixyangular2d}\end{equation}
which is an almost trivial generalization of the result in $(1+1)$-dim (Eq. \ref{Ixyangular}). The undetermined constant defines the AdS radius of the corresponding AdS geometry, and is a complicated function of the full correlator involving $u\sim v$. Here, we have not been careful in keeping track of the powers of $2^n$, and readers interested in doing so are invited to generalize the more rigorous derivation in Appendix \ref{app:critical}. The apparent isotropy of Eq. \ref{Ixyangular2d} may not be exact due to the numerous approximations made. However, any angular dependency should only manifest itself as a form factor in the correlators, with the leading $\log$ term in the mutual information remaining unaffected.
\\
\noindent{\bf Imaginary time direction.}

A critical $D+1$-d boundary system also has power-law decaying imaginary time correlators, consistent with the interpretation of the bulk as a higher-dimensional AdS spacetime. Explicitly,
\begin{equation}
\text{Tr}~ C_n(\tau)\sim \left(\frac{2^{n-1}}{v_F \tau}\right)^{D+2}
\label{multidim4}
\end{equation}
with dimension-dependent critical exponent of $D+2$. This is unlike that of the spatial correlators, which do not depend on $D$. Interpreted as a bulk geodesic distance $d_\tau$, we have
\begin{equation}
\frac{d_\tau}{\xi_\tau}=-\log \text{Tr}~ C_n(\tau)\sim (D+2) \left[\log(v_F\tau)-n\log 2\right] + \text{const.}
\end{equation}
which is proportional to the dimensionality. Physically, we can understand the origin of the $D+2$ exponent as follows. Each spatial direction provides an additional dimension for the decay, and contributes a power of $\frac{2^n}{v_F \tau}\propto \frac{L}{v_F \rho \tau}$. There has to be at least one direction where only the UV half of the degrees of freedom are selected, since a separation of energy scales is necessary for the EHM network. This direction contributes an additional power of $2$ due to the gradient-like property of the UV projector $D(z)$. The above statements are justified with more mathematical rigor in Appendix \ref{app:multidimtime}, where the subleading powers in the decay are also explicitly evaluated.

\subsection{Boundary model with generic dispersion at nonzero temperature}

When an energy scale is introduced by the temperature $T$, the decay of correlators is dominated by the energy scale which is independent of the EHM basis. Hence we can calculate the bulk correlators in a way similar to that of the $(1+1)$-dim case, taking note only of the multidimensionality of $\bold q$.

Instead of just analyzing the Dirac model, we make our discussion more general by allowing our energy dispersion to take the following generic asymptotic form:
\begin{equation}
E_{\bold q}= \sqrt{\sum_j^D v_j^2 q_j^{2\gamma_j}}
\label{criticaldispersion}
\end{equation}
This form encompasses various physical scenarios, and reduces to the linear Dirac model in the simplest case of $\gamma_j=1$. When $D=1$ and $\gamma>1$, Eq. \ref{criticaldispersion} represents the nonlinear dispersions discussed previously. More interestingly, it can also describe semi-Dirac points characterized by anisotropic dispersions, i.e. with $D=2$, $\gamma_1=1$ and $\gamma_2=2$. Such dispersions have been observed in realistic systems involving ultrathin (001) $VO_2$ layers embedded in $TiO_2$, which exhibit unusual electromagnetic properties\cite{pardo2009, banerjee2009,delplace2010}.

According to Eq. \ref{decay} and subsection \ref{sec:temp}, the correlators and hence mutual information decay exponentially according to the complex root of $\cosh\frac{\beta E_{\bold q}}{2}=0$ closest to the real axis. The roots are given by the values of $\bf{q}$ satisfying
\[E_{\bold q}^2 = -\pi^2 T^2 (2l+1)^2,\]
with $l\in \mathbb{Z}$, with the temperature $T$ functioning as an imaginary gap. To find the exponential decay rate in direction $j$, we have to find 
$|Im(q_j)|$, the imaginary part of the complex root $q=q_j$ of
\begin{equation}
v_j^2 q^{2\gamma_j}=-(m^2+\pi^2 T^2 (2l+1)^2)
\label{criticalTD}
\end{equation}
where $m^2=\sum_{i\neq j}^D v_i^2 q_i^{2\gamma_i}$ denotes an effective mass from the momentum contributions from all the other directions. Since $T^2$ and $m^2$ are both positive, we clearly choose $l=0$ for $q_j$ to have the smallest imaginary part, i.e. slowest decay. While the decay rate also depends on $m$, the combination of momentum components giving $m=0$ yields the slowest decay rate $|Im(q_j)|$. We can take $m=0$ to be the dominant contribution to the overall decay rate $h_j$, and the $m>0$ contributions as the subleading corrections. This will be discussed explicitly for the two cases below, with calculational details relegated to Appendix \ref{app:temp2}.

\subsubsection{Finite temperature Dirac fermions}

We first discuss the massless Dirac case with
\begin{equation}
E_{\bold q}= v|\bold q| = v\sqrt{\sum_j^D  q_j^{2}},
\label{multidirac}
\end{equation}
i.e. with $v_j=v$ and $\gamma_j=1$ for $j=1,2,...,D$. Let $\Delta \vec x$ be the displacement between two distant points within the same layer in the bulk. The mutual information decays like $I_{xy}\sim 8|u|^2\sim 8|v|^2$ where, as shown in Appendix \ref{app:temp2},
\begin{eqnarray}
u\sim v &\sim& \int e^{i2^n \bold q \cdot \Delta \vec x}\tanh\frac{\beta E_{\bold q}}{2}d^D \bold q \notag\\
%&\sim & \prod_{j=2}^D \int dq_j e^{-2^n h_1(q_2,...,q_D)\Delta x_1 }e^{i\sum_{j\geq 2}^D \Delta x_jq_j}
&\sim& e^{-\frac{2^n\pi T}{v}|\Delta \vec x|}
\label{multidirac2}
\end{eqnarray}
Hence
\begin{equation}
-\log I_n( \Delta \vec x)\sim 	\frac{2^{n+1}\pi T}{v}|\Delta \vec x| =\frac{LT}{v} \sqrt{\sum_j^D (\Delta \theta_j)^2}\\
\label{multidirac3}
\end{equation}
This is a direct generalization of Eq. \ref{temp1} for the $(1+1)$-dim  critical Dirac model at temperature $T$, whose bulk geometry corresponds to that of a BTZ black hole horizon in the IR limit. Here, we have exactly the same asymptotic behavior, with the horizon having the same topology as the boundary system. When the latter is defined on a D-dimensional lattice with periodic boundary condition along all directions, the horizon is a D-dimensional torus $T^D$.%, where $T=S^1$.

\subsubsection{$(2+1)$-dim anisotropic dispersion}

We now consider a generic anisotropic dispersion in $D=2$, and show that the event horizon can also become anisotropic. The dispersion is given by
\begin{equation}
E_{\bold q}=  \sqrt{ v_1^2 q_1^{2\gamma_1}+ v_2^2 q_2^{2\gamma_2}},
\label{multianisotropic1}
\end{equation}
with correlator decay rates $\left(\frac{\pi^2 T^2 +v_{j'}^2q_{j'}^2}{v_j^2}\right)^{\frac{1}{2\gamma_j}}\sin \frac{\pi}{2\gamma_j}$ where $j'=1,2$ for $j=2,1$. It is mathematically tricky to obtain the asymptotic behavior of $I_{xy}$ for arbitrary $\Delta \vec x$, when all components of $x$ are not small. For our current purpose, it suffices to expand the asymptotic behavior about the limiting directions $\Delta \vec x = x \hat e_x$ and $ y\hat e_y$. After a Gaussian integral computation detailed in Appendix \ref{app:temp2}, the mutual information at $\Delta \vec x = |\Delta \vec x|(\cos\phi  \hat e_x + \sin\phi  \hat e_y)$ for $\phi$ near $0$ is approximately given by
\begin{eqnarray}
&&-\log I_n(\vec \Delta x )|_{\phi  \approx 0}\notag\\
&\sim& 	2^{n+1}\left (\sin\frac{\pi}{2\gamma_1}\left(\frac{\pi T}{v_1}\right)^{1/\gamma_1}x+\frac{\gamma_1(\pi T)^{2-1/\gamma_1}}{2\sin\frac{\pi}{2\gamma_1}}\frac{v_1^{1/\gamma_1}}{v_2^2}\frac{y^2}{x}\right)\notag\\
&=& 2^{n+1}|\Delta \vec x|\sin\frac{\pi}{2\gamma_1}\left(\frac{\pi T}{v_1}\right)^{1/\gamma_1}\left(1+\frac{\left(\alpha_1^2-1\right)}{2}\phi ^2 + O\left(\phi ^4\right)\right)\notag\\
\label{multianisotropic2}
\end{eqnarray}
where $\alpha_j=\sqrt{\gamma_j}\frac{(\pi T)^{1-1/\gamma_j}v_1^{1/\gamma_j}}{v_{\bar j}\sin \frac{\pi}{2\gamma_j}}$. An exactly analogous result holds near $\phi  \approx \frac{\pi}{2}$, with $\gamma_1,\alpha_1$ and $v_1$ replaced by $\gamma_2,\alpha_2$ and $v_2$. 

Notably, in the isotropic linear Dirac case where $\gamma_j=1$ and $v_1=v_2$, $\alpha_j=1$, the mutual information is manifestly asymptotically isotropic to third order by Eq. \ref{multianisotropic2}. This is despite the fact that the  wavelet basis was constructed via tensor products of those of each direction and hence only possess four-fold rotation symmetry. 
\begin{figure}[H]
\centering
\includegraphics[scale=1.55]{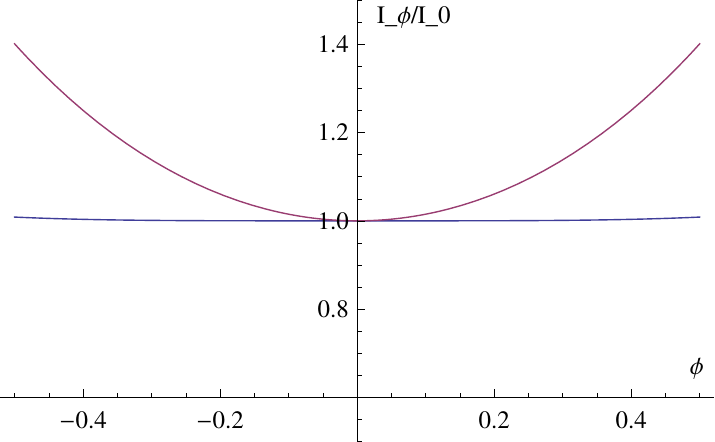}
\caption{Plots of $I_{xy}(\phi)/I_{xy}(\phi=0)$ for $\alpha=1$ (almost horizontal line) and $\alpha=2$ (upper curved line) according to Eq. \ref{multianisotropic2}, with higher order terms in $\phi^2$ kept. We see that in the isotropic linear Dirac case with $\alpha=1$, $I_{xy} \propto |\Delta \vec x|$ to a high degree of accuracy even away from $\phi\ll 1$. }\label{horizon_angle} \end{figure}

By contrast, when the dispersion acquires some nonlinearity, $\gamma_j>1$ and $\alpha_j\neq 1$ for some $j$ and we expect the mutual information to be significantly anisotropic in a temperature dependent way. This is illustrated in Fig. \ref{horizon_angle}, where the angular dependence of the mutual information is compared for $\alpha=1$ and $2$. In any case, the factor $2^{n+1}\Delta |\vec{x}|$ in Eq. (\ref{multianisotropic2}) suggests that there is still a finite area (anisotropic) horizon in the IR limit, since the circumference of a closed circle around any periodic direction approaches a finite value while $n\rightarrow \infty$.

\section{Holographic Topological Insulator from Chern Insulator }
\label{sec:holotopo}

In this section, we apply the EHM to a topological state of matter, and study the new insights that emerge. Topological states of matter are gapped states which can be distinguished from trivial insulators (defined by a system in which electrons are completely localized on each atom) by topological properties, such as robust gapless edge states, topological response properties and/or ground state degeneracy dependent on the system manifold.  A proptotypical candidate for a topological state is the ground state of the Chern Insulator (also known as a Quantum Anomalous Hall insulator) introduced in the previous part~\cite{klitzing1980, haldane88prl2015, qi2006topological, liu2008quantum, yu2010quantized}, for which a nonzero Chern number gives rise to a nonvanishing quantized Hall conductivity.  The integer quantum Hall state is the first electronic topological state of matter. The past decade saw a surge in interest in topological states of matter since the discovery of time-reversal invariant topological insulators and topological superconductors\cite{qi2010quantum,qi2011topological,hasan2010colloquium,moore2010birth}.

Naively, one may expect the holographic dual theory of a topological state to be trivial, since there is no low energy excitation in the infrared limit. Indeed, as we have previously seen, the dual geometry of such a gapped system terminates at a ``end-of-the-world brane" at the infrared end, with its position in the emergent direction determined by the energy gap. When the topologically ordered state has a fixed point wavefunction description\cite{levin2005string}, the multiscale entanglement renormalization ansatz (MERA) approach can be applied to construct a dual theory that has the topological information completely encoded at the infrared brane\cite{aguado2008entanglement,vidal2007,vidal2008,gu2009,matsueda2013tensor}. However, as we will show below, the situation is even more nontrivial for fermionic topological states such as Chern Insulator states, since a fixed point wavefunction with zero correlation length does not exist. 

By applying the EHM approach to a $2+1$-dimensional Chern insulator, we obtain a dual bulk $3+1$-dimensional system containing a  time-reversal invariant topological insulator (TI) within a slab of finite thickness in the emergent direction. This slab is demarcated by surfaces representing the length scales characterizing the time-reversal breaking of the original (boundary) Chern insulator, i.e. the Berry flux which make up the nonzero Chern number. The fact that the bulk is a TI is a direct manifestation of the nontrivial quantum entanglement between degrees of freedom at different length scales. As an explicit verification of the bulk topological property, we study the entanglement spectrum of the system with an entanglement cut in the emergent bulk direction. This corresponds to studying the entanglement between long-wavelength and short-wavelength degrees of freedom in the Chern insulator ground state, something that is also well-defined in real-space only after carrying the EHM transformation. For such an entanglement cut we obtain a gapless entanglement spectrum with odd number of massless Dirac cones, a fact that directly verifies the TI nature of the bulk theory\cite{li2008, fidkowski2010, turner2010, qi2012}. %\cite{thomale2010entanglement}\cite{lauchli2010disentangling}%\cite{regnault2009topological}

\subsection{Mapping of topological invariants: Berry curvature to Chern number density}

We use as the Chern insulator system the $2+1$-dimensional Dirac model
\begin{eqnarray}
\hat{H}&=&\sum_{k} c^\dagger_{  k} H(k) c_{  k}=\sum_{k} c^\dagger_{{  k}} \left[ \vec{d}({  k}) \cdot \vec{\sigma} \right] c_{  k} \\
\vec{d}({ k})&=&\left( \sin k_x, \sin k_y,m+2-\cos k_x-\cos k_y\right)\nonumber
\label{model}
\end{eqnarray} 
This simple two-band model has topological phases with quantized Hall conductance\cite{qi2006topological} of $\frac{C e^2}{h}=\pm \frac{e^2}{h}$ for  $m\in(-2,0)$ and $m\in(-4,-2)$ respectively. %Such a quantum Hall phase without orbital magnetic field is known as quantum anomalous Hall (QAH) state or Chern insulator\cite{haldane1988,qi2006topological}. 

Recall that the band topology of the Chern insulator is characterized by the integral of the Berry curvature $F_{xy}=\partial_x A_y-\partial_y A_x$, $A_j=-i\langle \phi |\partial_{k_j}\phi\rangle$ over the Brillouin zone:
\[ C= \frac1{2\pi}\int_{k\in BZ} d^2k F_{xy} \]
This BZ is decomposed by the EHM into layers $n$ via $c^\dagger_{k_j}\rightarrow [W_n(e^{ik_j})]^* c^\dagger_{k_j}$, where $W_n(z)=D(z^{2^{n-1}})\prod_{j=0}^{n-2}C(z^{2^j})$, $C(z)=\frac{1+z}{2}$, $D(z)=\frac{1-z}{2}$ defines the wavelet basis. Note the different normalization of $C,D$ from Eqs. \ref{C} and \ref{D} in Sect. \ref{subsec:kspace}: Here, an extra normalization factor of $\frac1{\sqrt{2}}$ is introduced to rescale the wavelet bases of each layer according to their number of degrees of freedom, such that $\sum_{j=1}^\infty |W_j(z)|^2=1$ for all $z=e^{ik}$.
Since the Berry curvature operator is diagonal in $k$, the EHM simply projects the Berry curvature into each layer by multiplying it with an appropriate spectral weight. The contributions from all three bulk sectors of each layer $n$ can be most easily expressed in terms of $Y_n(z)=\prod_{j=0}^{n-1}C(z^{2^j})$, where $|Y_n(e^{ik_x})|^2|Y_n(e^{ik_y})|^2$ is the combined spectral weight of \emph{all} layers $n+1$ and deeper \footnote{In one dimensional EHM, we simply recover $|Y_{n-1}(e^{ik})|^2-|Y_{n}(e^{ik})|^2=|Y_{n-1}(e^{ik})|^2|D(e^{i 2^{n-1}k })|^2=|W_n(e^{ik})|^2$, the spectral weight of layer $n$. Note that $\int_0^{2\pi} 2^n|W_n(e^{ik})|^2dk=2\pi$.}: 
\begin{eqnarray}
F_{xy}(k_x,k_y)\rightarrow 
&=&|Y_{n-1}(e^{ik_x})|^2|Y_{n-1}(e^{ik_y})|^2\left[1-|C_n(e^{ik_x})|^2|C_n(e^{ik_y})|^2\right] F_{xy}(k_x,k_y)\notag\\
&=&\left[|Y_{n-1}(e^{ik_x})Y_{n-1}(e^{ik_y})|^2-|Y_{n}(e^{ik_x})Y_{n}(e^{ik_y})|^2\right]F_{xy}(k_x,k_y)\notag\\
\end{eqnarray}
Integrated over the BZ, we obtain the \emph{Chern number density} $C(n)$ which is the contribution to the Chern number $C$ from layer $n$:
\begin{eqnarray}
C(n)%&=& \frac1{2\pi 4^n}\int d^2k |Y_{n-1}(e^{ik_x})|^2|Y_{n-1}(e^{ik_y})|^2\left[|D_n(e^{ik_x})|^2|D_n(e^{ik_y})|^2+|C_n(e^{ik_x})|^2|D_n(e^{ik_y})|^2+|D_n(e^{ik_x})|^2|C_n(e^{ik_y})|^2\right] F_{xy}(k_x,k_y)\notag\\
%&=& \frac1{2\pi }\int d^2k |Y_{n-1}(e^{ik_x})|^2|Y_{n-1}(e^{ik_y})|^2\left[1-|C_n(e^{ik_x})|^2|C_n(e^{ik_y})|^2\right] F_{xy}(k_x,k_y)
&=& \frac1{2\pi }\int d^2k \left[|Y_{n-1}(e^{ik_x})Y_{n-1}(e^{ik_y})|^2-|Y_{n}(e^{ik_x})Y_{n}(e^{ik_y})|^2\right] F_{xy}(k_x,k_y)\notag\\
\label{Cn}
\end{eqnarray}
This expression can be trivially extended to \emph{any} number of dimensions by including the $Y$ factors in the other directions. By the telescoping property, it is easily seen that
\begin{eqnarray}
\sum_{n=1}^\infty C(n) &=& \frac1{2\pi}\int d^2k |Y_0(e^{ik_x})Y_0(e^{ik_y})|^2 F_{xy} \notag\\
&=& C,
\end{eqnarray}
the Chern number of the original system. Since the Chern number physically corresponds to the Hall conductance, the Chern number density $C(n)$ corresponds to the (non-quantized) layer Hall conductance\footnote{A non-quantized Hall conductance is only possible for 2-dimensional layers that do not exist in isolation, but which forms a higher-dimensional system.}. 

\subsection{Identifying bulk holographic topological insulator I: Axion field theory}

The layer Hall conductance interpretation above naturally leads us to treat the $(3+1)$-dimensional bulk region as a topological insulator, where the noninteger layer Chern number is the difference of the topological order parameter (Axion angle) $\theta$ between the two surfaces of a finite region. This $\theta$ appears in the field theoretic description of a time reversal invariant $(3+1)$d topological insulator Lagrangian constructed from the seminal works~\cite{qi2008topological, essin2009magnetoelectric}:
\begin{eqnarray}
\mathcal{L} &=& \frac{\theta}{8\pi^2} \epsilon^{\mu\nu\sigma\tau} \partial_\mu A_\nu \partial_\sigma A_\tau
\label{axion}
\end{eqnarray}
Time reversal invariance restricts $\theta$, which is periodic in $2\pi$, to take values of only $0$ or $\pi$. These values represent the trivial and nontrivial topological insulator states respectively. In this sense, the Lagrangian in Eq. \ref{axion} offers a local field theoretic understanding of the $\mathbb{Z}_2$ topological invariant.

The spatial transition between states with $\theta=0$ and $\pi$ mod $2\pi$ necessarily interpolates between them, and thus involve time reversal symmetry breaking, i.e. through magnetic doping. Suppose $\theta$ changes gradually in the $\mu$ direction. Then %on surface of a nontrivial topological insulator, we will see a gradually changing theta angle from $\pi$ to $0$ (mod $2\pi$), and the direction that angle winds depends on the time reversal breaking field you choose. Moreover, one important physical property of topological insulator in this scenario is the nontrivial surface state carrying half QH effect: 
\begin{eqnarray}
S&=&\int d^4 x \mathcal{L}\notag\\
&=& -\frac{1}{8\pi^2} \int d^4 x  \partial_{\mu}\theta \epsilon^{\mu\nu\sigma\tau } A_\nu \partial_\sigma A_\tau\notag\\ 
&=& - \frac{\Delta \theta}{8 \pi^2} \int_{\partial \Sigma} d^3 x \epsilon^{\nu \sigma \tau } A_\nu \partial_ \sigma A_\tau
\end{eqnarray}
where the last line was obtained via integrating by parts. This is just a Chern-Simons term proportional to $\Delta \theta$, the difference in $\theta$ between the two bounding regions which must take values of $\pm \pi$ between two time-reversal invariant topological states. The sign of the jump in $\theta$ depends on the time-reversal breaking direction chosen. From \cite{qi2008}, a nontrivial $\mathbb{Z}_2$  region has $\theta=\pi$, and is sandwiched between vacua taking $\theta=0$ and $\theta=2\pi$ on either side. The boundaries between these regions each carry a Hall conductance of $\sigma_H=\pm \frac{e^2}{2h}$~\cite{stone1992quantum}, corresponding to the half QH effect on the surfaces of a TI.

Let us now look at how the $\theta$ angle varies in the holographic bulk. In the continuum limit, it corresponds to the integral of the Chern number density $C(n)$ in the emergent layer direction, which we say call the $z$ direction. Assuming that $\theta=\theta(z)$ we have $C(z)=\frac{1}{2\pi} \partial_z \theta(z)$ or $\theta(z_2)-\theta(z_1)= 2\pi \int_{z_1}^{z_2} dz^\prime c(z^\prime)$. In the discrete case, $\theta=\theta(n)$ is simply given by a summation over the layers $n$: $\theta(n)=2\pi \sum_{n'=1}C(n')$, assuming that $\theta=0$ on the UV surface.

\begin{figure}[htb]
\includegraphics[scale=0.45]{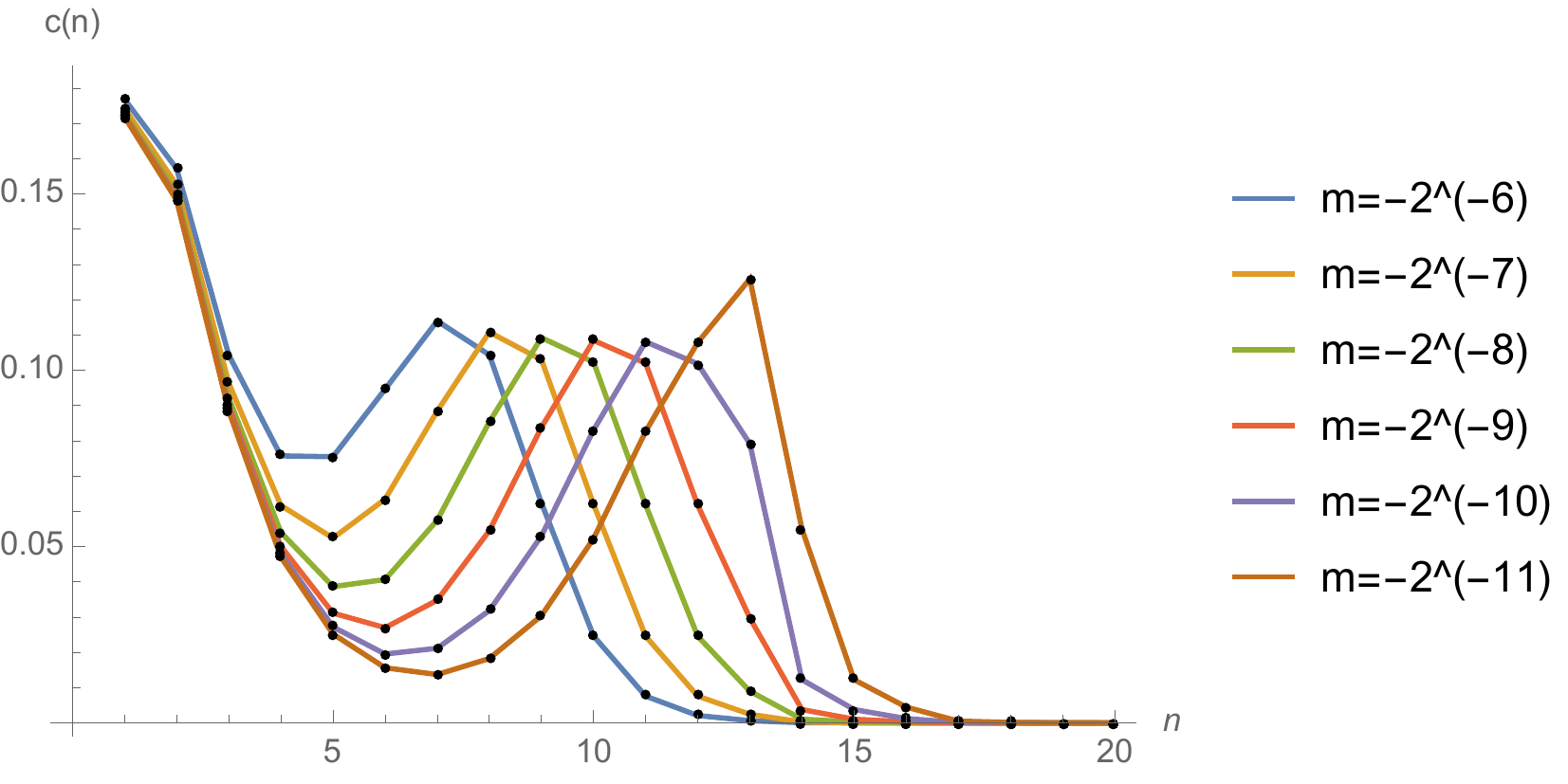}
\includegraphics[scale=0.45]{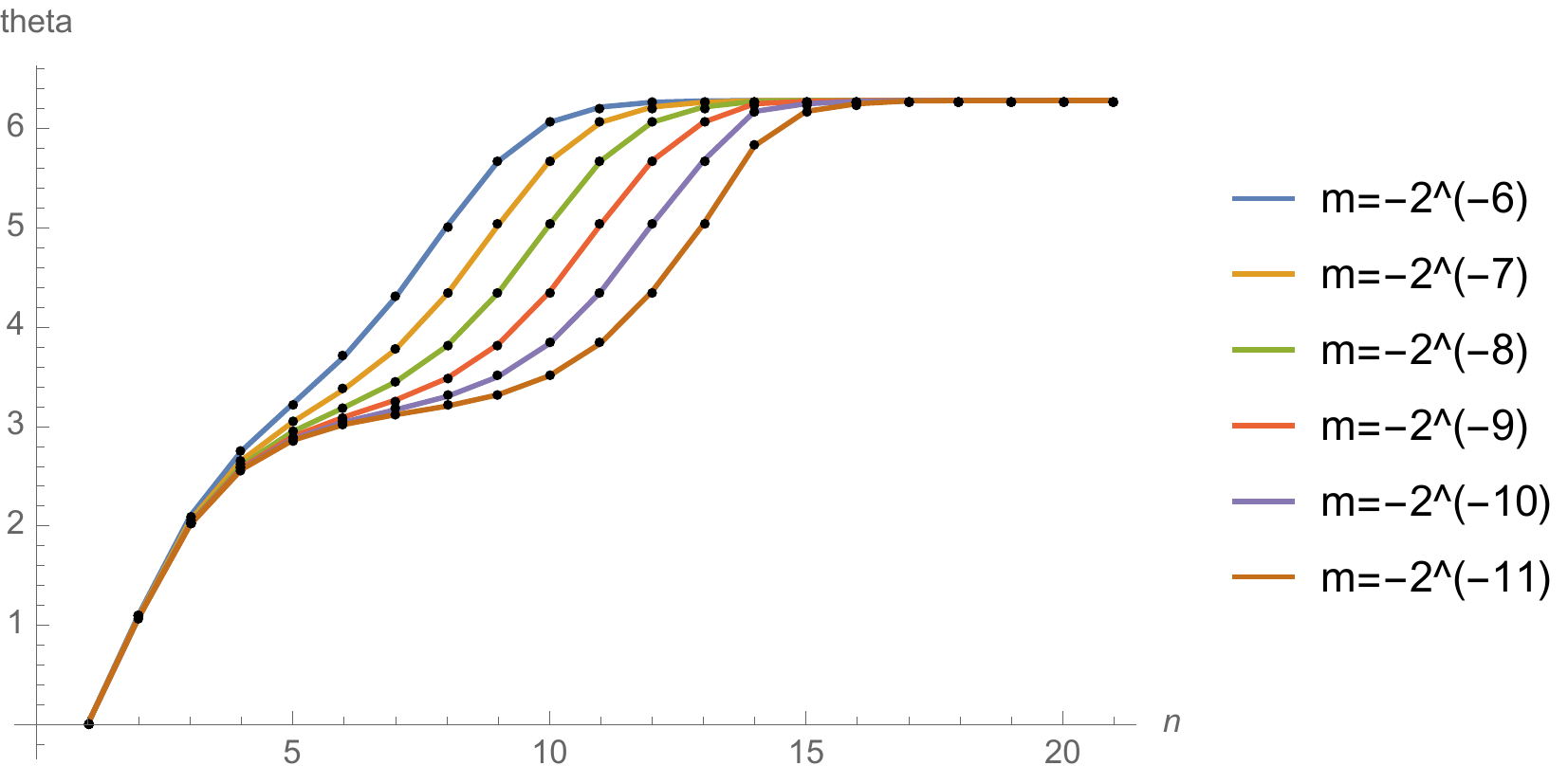}
\caption{Numerical results on the Chern number density $C(n)$ and theta angle $\theta(n)=2\pi \sum_{n'=1}C(n')$ for the Dirac model with different $m$. There are two peaks in $C(n)$, one situated in the UV (Left) which is not sensitive to the smallness of the gap $|m|$, and the other in the IR (Right) which moves deeper into the IR as $|m|$ decreases. Hence the two peaks are well separated for very small $|m|\leq 2^{-11}$, and a plateau at $\theta=\pi$ corresponding to a nontrivial $\mathbb{Z}_2$ phase appears.}
\label{numerical}
\end{figure} 

Fig(\ref{numerical}) contains the numerical results for the Chern number density $C(n)$ and theta angle $\theta(n)$ of the Dirac model Eq. \ref{model} with different $m$. There are two peaks in $C(n)$, one in the UV (small $n$) and the other in the IR (large $n$). The UV peak peaks at $n=1$, and does not change appreciably as $m$ is made exponentially small. It physically arises from the Berry curvature contributions away from the $\Gamma$ point $\vec k=(0,0)$, and depends non-universally on the UV regulator chosen for the lattice model. The IR peak corresponds to the Berry curvature accumulated near the $\Gamma$ (IR) point. As $|m|$ decreases, the Berry curvature profile becomes sharper, i.e. more localized around $k=(0,0)$. The IR peak consequently moves deeper into the IR with peak position $n_{IR}\sim -\log_2 (m)$. This relation, together with a more detailed analysis of $C(n)$, will be rigorously presented in Appendix \ref{Canalytic}.

As $|m|$ becomes sufficiently small, the UV and IR peaks become well separated, with a region of vanishing $C(n)$ between them . This region corresponds to a plateau of $\theta\approx \pi$ in the theta angle, which should just be the region with nontrivial $\mathbb{Z}_2$ phase. This region is demarcated by two transition regions with spatially nonuniform $\theta$ with a total $\Delta \theta =\pi$ each. They are just the two surfaces of a $\mathbb{Z}_2$ topological insulator which possesses a half Hall conductivity each.

In general, the topology of the bulk depends on the distribution of the Berry curvature in energy scale, and a topologically nontrivial $\mathbb{Z}_2$ phase may or may not exist depending on whether there exists a region with $\theta =\pi$ mod $2\pi$ in the middle of a nontrivial winding of $\theta$. For instance, Eq. \ref{model} with $m>0$, which leads to a zero Chern number, will still lead to two peaks in $C(n)$, but with \emph{opposite} signs. The total winding of $\theta$ is zero, as it should be, and the $\theta=\pi$ region between the two peaks is topologically trivial.

%In summary, we are making an identification between the ground state obtained via the EHM of a Chern Insulator (Dirac model), with the ground state of a $(3+1)$d topological insulator in network shaped lattice. The bulk layers between the two peaks of $c(n)$ corresponds to the topologically nontrivial region of the TI, as can been seen by $\theta$ taking the value of $\pi$. Outside of that region is the time reversal breaking surfaces, where $\theta \neq 0$ or $\pi$.

%There is one remark about the topological property of the 2d model we start with. Although we had started from model(\ref{model}) which has Chern number $1$, the resulting bulk topological insulator does not really depends on this choice. If we perform the EHM of a Dirac model with positive mass, the resulting bulk topological insulator is essentially the same except that the time reversal breaking direction of IR surface will change. Actually, the important ingredient of the holographic topological insulator is a Dirac cone, which carries half Chern number. The origin of the non trivial topology will be discussed later. 

\subsection{Identifying the bulk theory (II): entanglement spectrum}

A defining feature of topological insulators is the appearance of gapless surface states. However, the surface we have in the (3+1)-dimensional bulk here are time reversal symmetry breaking, so there are no gapless surfaces at hand. In order to probe the gapless state, we shall perform an entanglement cut in the bulk region, where time reversal symmetry is preserved, and study the surface states via the entanglement spectrum.

\begin{figure}[htb]
\centering
\includegraphics[scale=0.6]{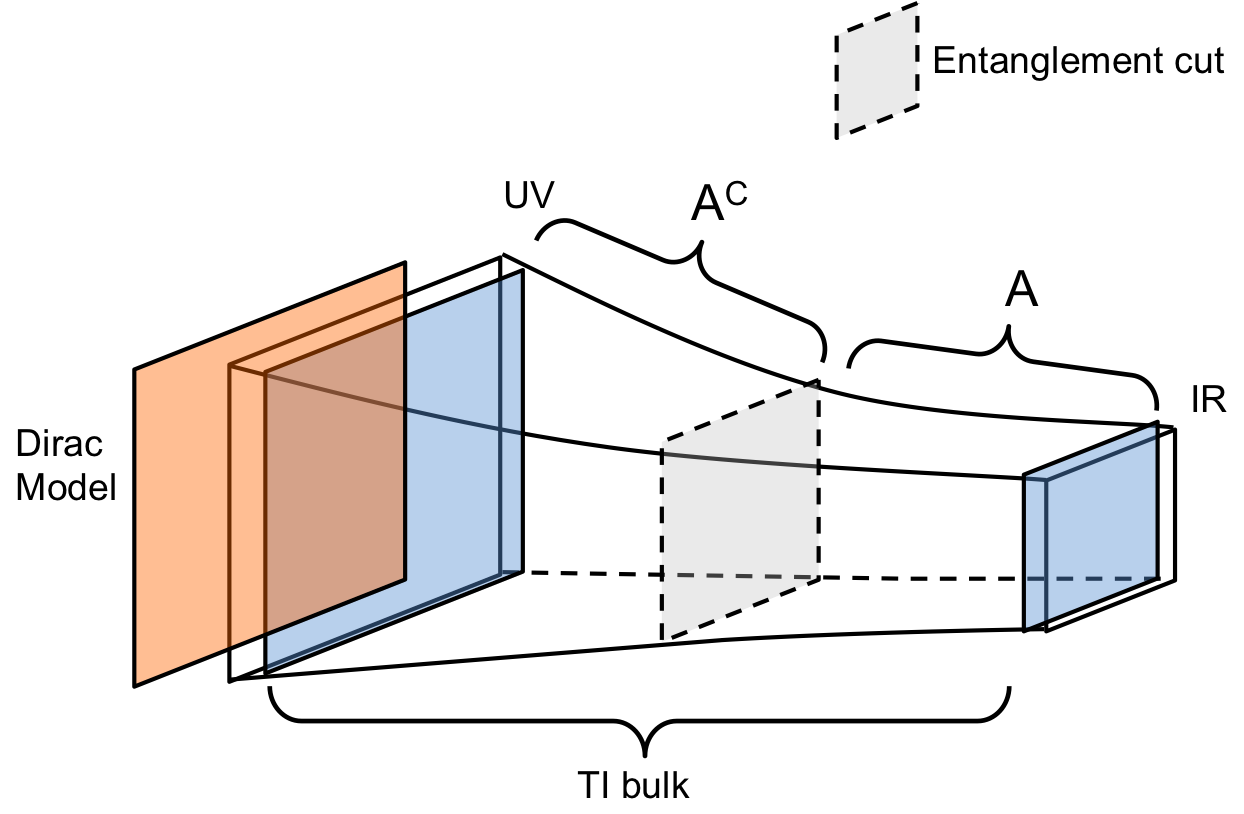}
\caption{An entanglement cut in the holographic topological insulator bulk: The whole system is divided into two parts $A$ and $A^C$(complimentary), and $A^C$ is traced out to obtain the reduced density matrix on subsystem A. In our context, $A$ contains the low energy (IR) degrees of freedom.}
\label{EC1}
\end{figure}

Physically, quantum entanglement characterizes the amount by which classically independent regions in the system are correlated. It has been extensively studied in the context of novel phases and critical phenomena \cite{holzhey1994,PhysRevLett.96.110404,ryu2006,wolf2006,verstraete2006,plenio2005,li2008,regnault2009,laeuchli2010,yao2010,turner2010,matsueda2014comment,fidkowski2010,chandran2011,qi2012,alba2012,alba2012xxz,alba2013, hermanns2014,lai2014,puspus2014,lee2016random}, particularly of exotic topological states\cite{kitaev2006,levin2006,zozulya2009}. Information about the extent of entanglement between two subsystems is contained in the reduced density matrix (RDM) $\rho$, which is obtained by tracing the density matrix over one of the subsystems. The entanglement entropy (EE) is simply defined by $S=-{\rm tr}\left(\rho\log\rho\right)$, while the entanglement spectrum (ES), which we are going to use here, consists of all the eigenvalues of $\rho$, and contains more precise information on the topological nature of the system~\cite{li2008}.

As shown in Figure \ref{EC1}, we divide the entire bulk system into two parts $A$ and $A^C$. The density matrix of the whole system is $\rho=|\psi\rangle \langle \psi |$, where $|\psi \rangle$ is the ground state wavefunction in the entire bulk. The reduced density matrix $\rho_A$ and entanglement Hamiltonian $H_A$ for subsystem A are defined as:
\begin{eqnarray}
\rho_A : = Tr_{A^C}(|\psi\rangle \langle \psi |) \equiv e^{-H_A}
\end{eqnarray}
Since we have residual translational symmetry in the bulk layers, we can define the entanglement Hamiltonian $H_A$ in momentum space $\vec{k}$. Furthermore, for our fermion model, there is a simple relation\cite{peschel2002} between the entanglement Hamiltonian $H_A(\vec{k})$ and the single particle projector $C(\vec{k})$ (not to be confused with the Chern number) onto subsystem A:
\begin{equation} 
H_A(\vec k)=\log\frac{1-C(\vec k)}{C(\vec k)}
\end{equation}
These notions will be elaborated in Appendices \ref{wannES}, \ref{criticalEE} and \ref{sec:criticalspacing}, where their relation with the Wannier polarization and conformal field theory (CFT) will be drawn.

\begin{figure}[htb]
\includegraphics[scale=0.55]{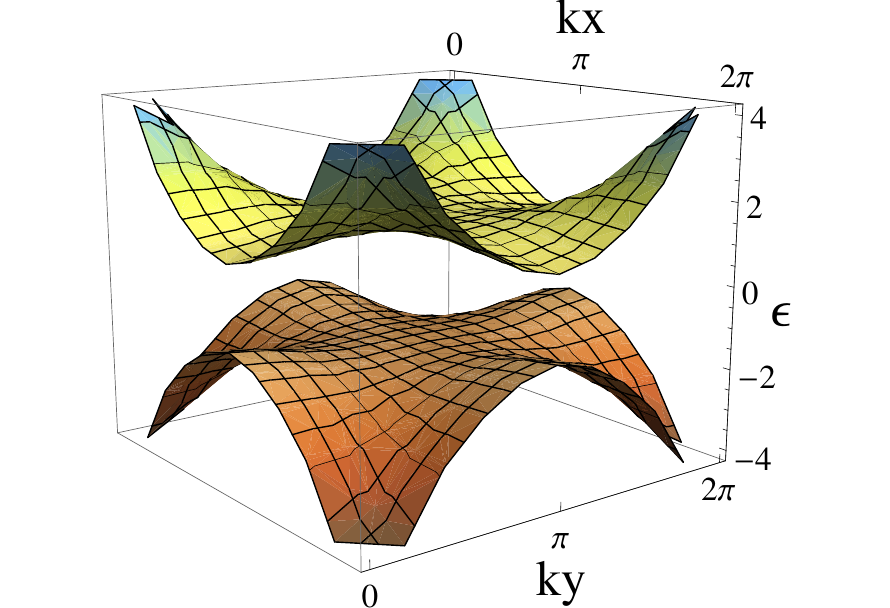}
\includegraphics[scale=0.55]{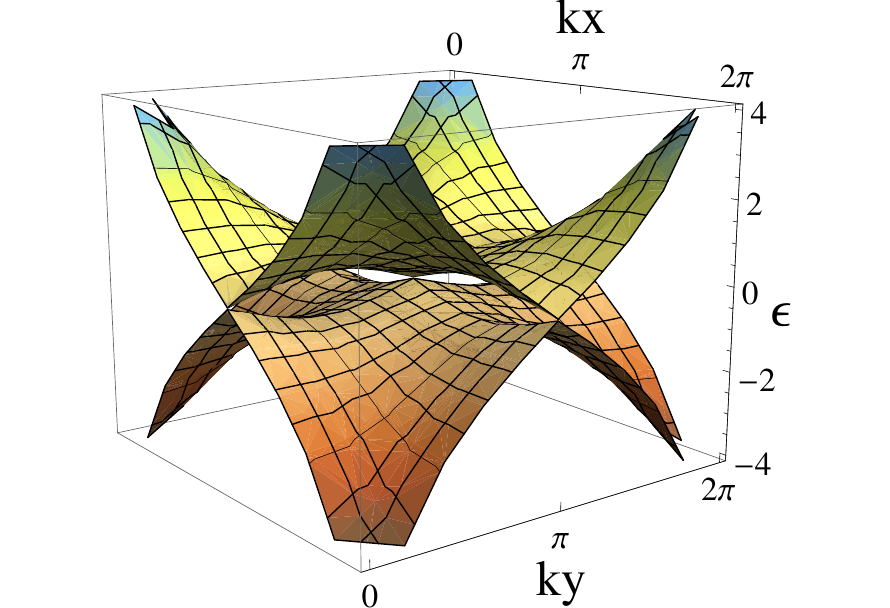}
\includegraphics[scale=0.55]{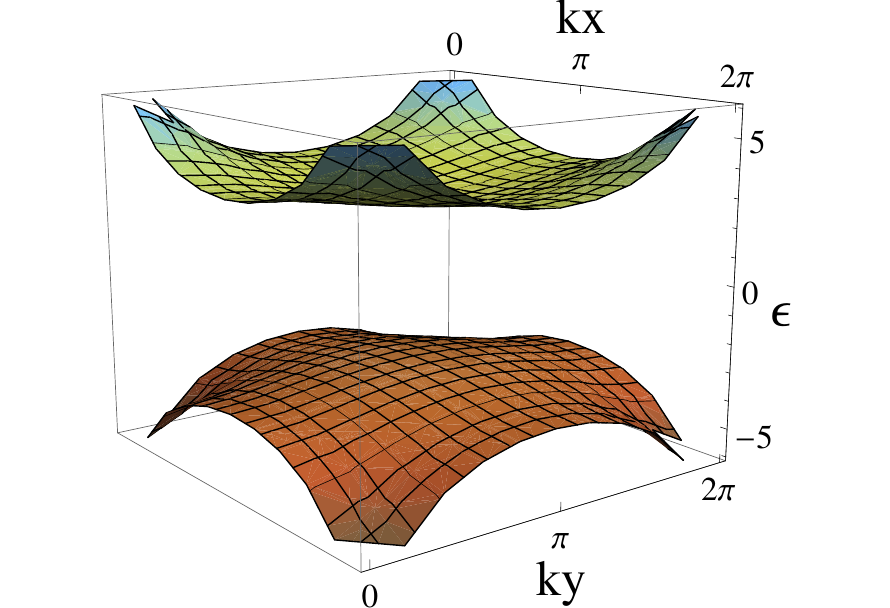}
\caption{Left to Right: The entanglement spectra for the Dirac model with $m=-2^{-20}$, with entanglement cuts located at layers $n=2$, $n=14$ and $n=30$ respectively. The spectrum is gapped for cuts in both in the UV ($n=2$) and extreme IR ($n=30$) regimes, but closes when the cut is made in the bulk TI region. The three gap closure points are situated at $\vec k = (\pi,0)$, $(0,\pi)$ and $(\pi,\pi)$, each with linear Dirac dispersion. $k_x$ and $k_y$ denote the good momentum quantum number associated with the enlarged unit cell in the particular layer.}
\label{ES}
\end{figure}

\begin{figure}[htb]
\includegraphics[scale=0.33]{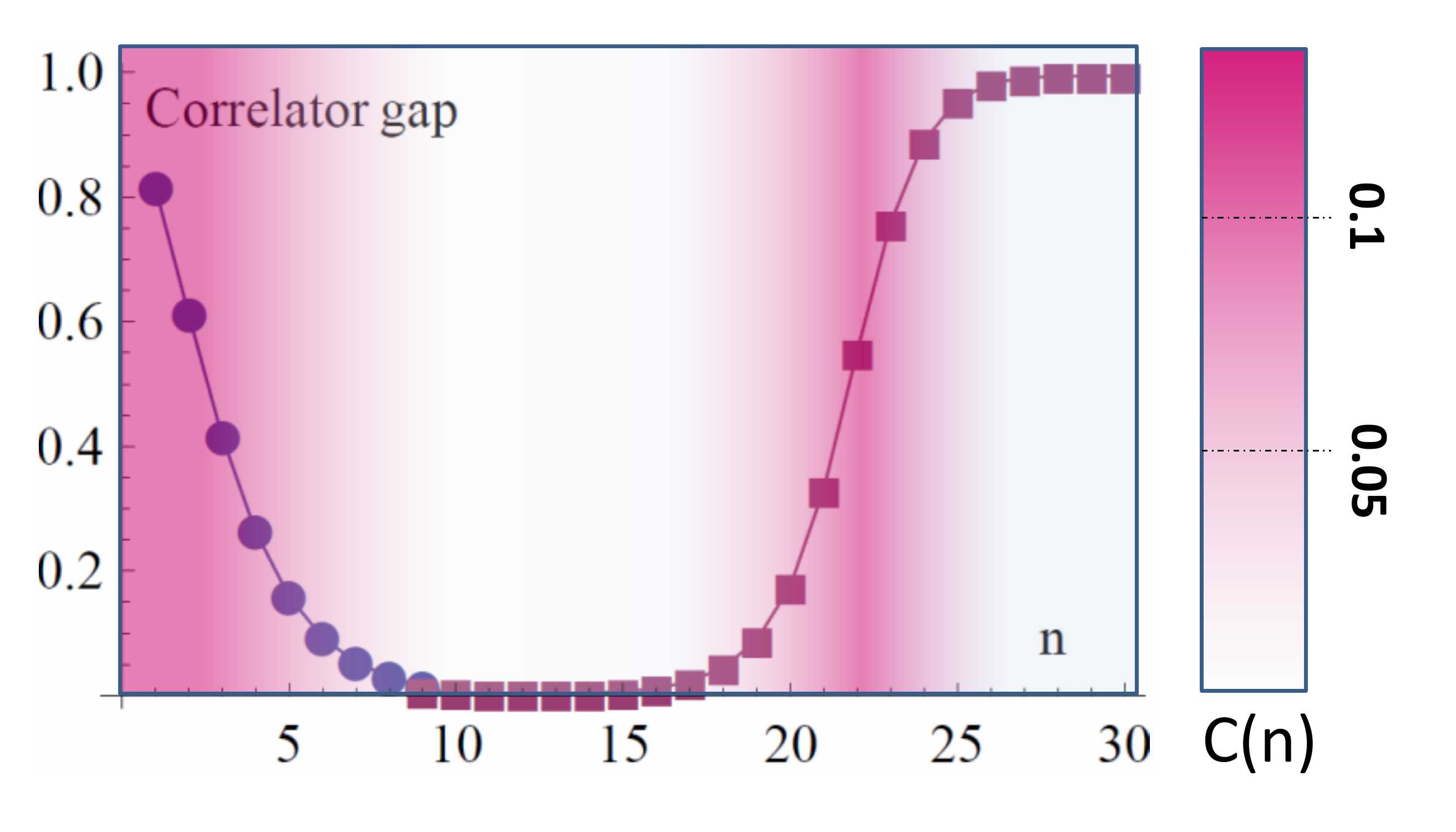}
\includegraphics[scale=1.4]{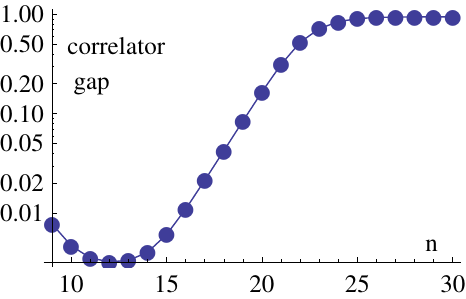}
\caption{Left) Minimal gap in the correlator $C_n(q_x,q_y)$ for $m=-2^{-20}$ as a function of position of layer cut $n$. Plotted on the blue curve from $n=1$ to $9$ are exact numerical results, while those on the purple curve from $n=9$ to $30$ are obtained analytically. The background color density represents the $C(n)$ profile. Indeed, the minimal gap vanishes only between the two $C(n)$ peaks where time reversal symmetry is broken. Right) The same minimal gap, but plotted on a logarithmic scale. It decreases exponentially into the region between the two $C(n)$ peaks, reflecting the exponentially decaying profile of the IR peak (as shown in Appendix \ref{Canalytic}.) }
\label{ESgap}
\end{figure}

We shall demonstrate the nontrivial topology of the bulk region between the two $C(n)$ peaks by performing a layer entanglement cut on it. We choose a very small $m=-2^{-20}$ so that the two peaks are clearly separated. From Fig. \ref{ESgap}, the region between them with $\theta\approx \pi$ occurs between layers $n=7$ and $n=18$. As evident in Fig. \ref{ES}, the entanglement gap closes at $n=14$, where $\theta=\pi$, but opens up when the cut is performed on layers with nonzero Chern number density. Furthermore, at $n=14$ (or other layers where $\theta\approx \pi$) the gap closes at the three points $(\pi,0)$, $(0,\pi)$ and $(\pi,\pi)$, each with a linear dispersion. That there is an odd number of Dirac cones on this entanglement cut further attests to the nontrivial $\mathbb{Z}_2$ topological nature of the region between the $C(n)$ peaks.

%The single particle projector to subsystem A could be labeled by the layer index of cut n. Here $C_n$ stands for the single particle projector with a cut at the link between layer $(n-1)$ and $n$. The numerical data of figure(\ref{ES}) shows the entanglement spectrum at three representative position $n=2, n=14, n=30$, in which, only the $n=14$ cut is lying in the bulk region of holographic topological insulator. Other two cut are inside the trivial insulator region. So we can see from the entanglement spectrum, the spectrum inside the TI bulk is gapless at three point $(\pi,0)$, $(0,\pi)$ and $(\pi,\pi)$ which indicates three gapless point on the physical surface energy spectrum when we have a physical cut at layer $n=14$(inside the bulk).

\subsubsection{Geometric picture with the $\hat d$-vector}
In this simple model used, the position of entanglement gap closure also reveals an interesting observation pertaining to the topology of the $d$-vector characterizing the Dirac Hamiltonian $H=d\cdot \sigma$. First, note that the single-particle boundary correlator $G$, which projects onto the occupied band, is given by $G_{q_x, q_y}= \left( I-\hat d(q_x,q_y)\cdot \sigma\right)/2$ where $\hat d(\vec q)= \frac{d(\vec q)}{|d(\vec q)|}$. Essentially, it is (up to a constant) the "flattened" Hamiltonian with eigenenergies $\pm \frac1{2}$. Just like the Chern number density $C(n)$, the layer $n$ single-particle correlator $C_n(k_x,k_y)$ is given by 
\begin{eqnarray}
 C_n(k_x,k_y)\simeq \sum_{m=-\chi}^{\chi-1} \sum_{l=-\chi}^{\chi-1} |Y(q_x,q_y)|^2 G_{q_x,q_y}
\label{CnG}
\end{eqnarray}
where $q_x$ and $q_y$ take values at  $q_x=k_x/2^n+2\pi m/2^n$ and $q_y=k_y/2^n+2\pi l/2^n$, $m,n$ integers taking values on $2\chi\times 2\chi$ points where $\chi$ is a cutoff parameter. While $\chi$ can in principle be as large as $2^{n-1}$, $Y(\vec q)$ is sharply peaked at $|\vec q|\sim \frac{2\pi}{2^n}$. In other words, we only need to include $\chi\sim O(1)$ points. This peak becomes more pronounced as $n$ increases. As such, in the IR limit, most of $C_n(k_x,k_y)$ is contributed by $G_{\vec q}$ with $\vec q\approx (\pm \frac{2\pi}{2^n},\pm \frac{2\pi}{2^n})$.  

This observation allows us to understand why the gap of $C_n$ should close between the two $C(n)$ peaks. Around the two peaks, which corresponds to contributions of half a Chern number each, the $\hat d$-vector traces out nearly half a Bloch sphere. This implies that away from them, the $\hat d$-vector lie in the equatorial directions. From the arguments below Eq. \ref{CnG}, the main contributions to $C_n$ occur around the four points $\vec q\approx (\pm \frac{2\pi}{2^n},\pm \frac{2\pi}{2^n})$, above which the equatorial-lying $\hat d$-vector point in mutually destructive directions. Hence $\sum_{m=-\chi}^{\chi-1} \sum_{l=-\chi}^{\chi-1} |Y(q_x,q_y)|^2 \hat d(q_x,q_y)\approx 0$ and the entanglement gap must close.

%The $m$ sum serves to 'distribute' the contributions to $H^{UV}(q)$ across the whole BZ, for each $q$. In particular, the relative contributions are dictated by the weight $|W(e^{i\frac{q+2\pi m }{2^n}})|^2$, and not the position of $q$ per se. Indeed, for large $n$ the contributions tend towards an integral independent of $q$, i.e. an average peaked in the IR.
That this cancellation of the contributions to $\hat d$ must occur is fundamentally due to topology: each Dirac cone has already taken up half a Chern number, forcing $\hat d$ on the region in between to take values on a ring of zero solid angle. Indeed, it is through the virtue of the EHM that this (topologically protected) cancellation effect acquires the physical interpretation of a gapless entanglement spectrum.

\subsubsection{Physical origins of the nontrivial topology}

The physical origin of the nontrivial $\mathbb{Z}_2$ topology lies in the notion of an anomaly. A single massless Dirac cone cannot be realized in lattice model without a UV regulator. This can be understood via the distribution of its Berry curvature in momentum space. A single Dirac cone carries half a Chern number, but another half must be present since the total Chern number is quantized as an integer in a lattice system. Hence there must be another half a Chern number contributed by the UV part of the Berry curvature. This leads to an unavoidable entanglement between the UV and IR degrees of freedom of the lattice Dirac model. 

This UV-IR entanglement is usually very hard to show explicitly, However, the EHM, as a holographic mapping, is able to separate the physics of different energy scale without introducing nonlocality. %In fact, as mentioned in the introduction, the emergent direction of the EHM can be interpreted as RG flow direction of the physical system.
It spreads a Dirac cone (and the accompanying UV contributions) onto a region in the emergent energy scale direction, where an entanglement cut may be made. With that, the topological nature of a Chern Insulator (Dirac model) can be understood in terms of the entanglement properties of a Topological Insulator one dimension higher.

\section{Further extensions to the EHM}
\label{EHMfurther}
\subsection{``Zooming in'' onto arbitrary Fermi points}

The EHM basis construction described in Sect. \ref{subsec:kspace} does \emph{not} require that the UV and IR projectors $C(z), D(z)$ be the same at each layer iteration. Recall that at each iteration, we keep one set of basis states as the bulk layer, and pass its complement to the next iteration. In particular, this exact mapping process is still well-defined if we interchange the $C(z),D(z)$ projectors at certain layers. Doing so will allow us to ``zoom in'' onto points other than the zero momentum point (long wavelength limit) as $n$ increases. To understand why, let's first study this alternative definition of the EHM in detail.

Define an $n$-length vector $k=(\pm,\pm,...)$ that keeps track of the orders of the $C,D$ projectors which we will relabel as $B_j^\pm$. For instance, if $k(j)=1$, $B_j^+=D$ and $B_j^-=C$. If $k(j)=-1$, $B_j^+=C$ and $B_j^-=D$. Hence the EHM construction previously introduced, which involve the sequence $(C,C,C,...,D)$, can be encoded by the sequence $k=(1,1,1,1,...,-1)$. 

The EHM wavelet basis is now modified to become
\begin{eqnarray}
%W_n(z)&=& D(z^{2^{n-1}})\prod_{b\in B}^{n-2}C(z^{2^b})\prod_{b\in [0,...,n-2]/B}^{n-2}D(z^{2^b})
W_n(z)&=& B^+_n(z^{2^{n-1}})\prod_j^{n-1}B^-_j (z^{2^{j-1}})
\label{wavelet3}
\end{eqnarray} 

\subsubsection{Orthogonality of the new basis}
The $W_n$s above are still orthogonal, as shown below. Given $m>n$,
\begin{eqnarray}
(W_n,W_m)&\propto &\oint_{|z|=1}\frac{dz}{z} W^*_n(z^{-1})W_m(z)\notag\\
&=&\oint_{|z|=1}\frac{dz}{z} B^{+*}(z^{-2^{n-1}})B^+(z^{2^{m-1}})\prod_{a=0}^{n-2}B^{-*}(z^{-2^a})\prod_{b=0}^{m-2}B^-(z^{2^b})\notag\\
%&=&\oint_{|z|=1}\frac{dz}{z} \left[B^{+*}(z^{-2^{n-2}})B^-(z^{2^{n-2}})\right]\prod_{a=0}^{n-3} \left[B^{-*}(z^{-2^a})B^-(z^{2^a})\right]\left(1+O(z^{-2^{n-1}})+\text{higher orders...}\right)\notag\\
&=&\oint_{|z|=1}\frac{dz}{z} \left[B^{+*}(z^{-2^{n-1}})B^-(z^{2^{n-1}})\right]\prod_{a=0}^{n-2} \left[B^{-*}(z^{-2^a})B^-(z^{2^a})\right]\left(1+O(z^{2^{n}})\right)\notag\\
&=& \oint_{|z|=1}\frac{dz}{z} (\text{nonconstant})\notag\\
&=&0
\label{ortho3}
\end{eqnarray} 
%The first term in line 3, which is equal to $C^{*}(z^{-2^{n-2}})D^-(z^{2^{n-2}})$ or $D^{*}(z^{-2^{n-2}})C^-(z^{2^{n-2}})$, has no constant term by Eq. \ref{ortho1}, and has a smallest power of $\pm 2^{n-2}r$, $r\in \mathbb{Z}^+$. This power cannot be canceled by any combination of terms in the product in the second term, since each term is equal to $C^{*}(z^{-2^a})C^-(z^{2^a})$ or $D^{*}(z^{-2^a})D^-(z^{2^a})$ and has no even power of $z^{\pm 2^a}$. Explicitly, each postive power term in the product has the form
The first term in line 3, which is equal to $C^{*}(z^{-2^{n-1}})D^-(z^{2^{n-1}})$ or $D^{*}(z^{-2^{n-1}})C^-(z^{2^{n-1}})$, has no constant term since $C,D$ are orthogonal projectors. It has a smallest power of $\pm 2^{n-1}r$, $r$ an odd positive integer. This power cannot be canceled by any combination of terms in the product in the second term, since each term is equal to $C^{*}(z^{-2^a})C^-(z^{2^a})$ or $D^{*}(z^{-2^a})D^-(z^{2^a})$ and has no even power of $z^{\pm 2^a}$. Explicitly, each postive power term in the product has the form
%\[ z^{\sum_{j=0}^{n-3}2^j(2m_j+1)}=z^{\sum_{j=1}^{n-2}m_{j-1}2^j+2^{n-2}-1}\]
\begin{equation}
z^{\sum_{j=0}^{n-2}2^j(2m_j+1)}=z^{\sum_{j=1}^{n-1}m_{j-1}2^j+2^{n-1}-1}
\end{equation}
where $m_j$ is either a non-negative integer or $-\frac{1}{2}$, the latter corresponding to the case when $z^{2^j}$ is not used. The exponent is thus odd and unable to cancel the power in $z^{-2^{n-1}r}$. This holds for the negative power terms too. The remaining terms from $W_m$ are either constant or have degree exceeding $\pm 2^{n-1}$, and so cannot form a constant term. Hence the integral is zero by the residue theorem.

If $W_n$ or $W_m$ were to be displaced from each other by a distance $x$, there will be an addition factor of $z^{\pm2^mx}$ or $z^{\pm2^nx}$ in the integral. However, it is clear from the above argument that such a term also cannot be combined with an other term to produce a constant term. Hence the displaced wavelet bases are also orthogonal, as required earlier on. 

%\textbf{Note that this above proof does not require $C$ and $D$ to be the same for each layer $j$, only that they must all satisfy the conditions mentioned. }

\subsubsection{Peak positions of new basis}

\begin{figure}[H]
\includegraphics[scale=1.2]{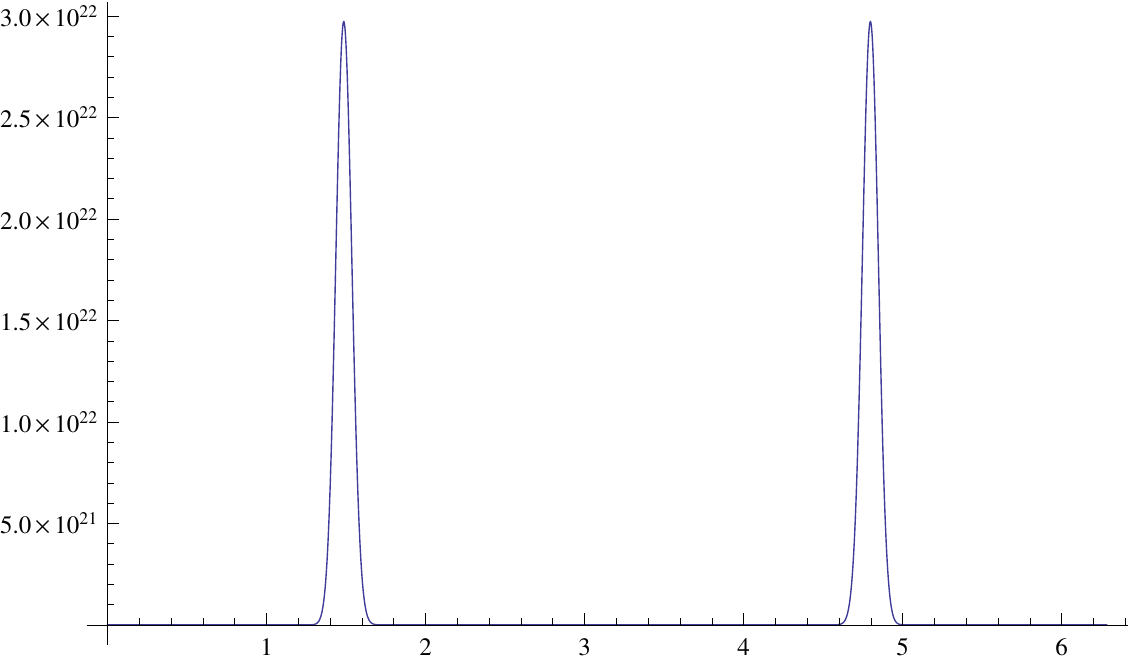}
\caption{Un-normalized momentum space spectral weight of the EHM basis with k-vector $\vec k=(1,-1,1)$ and $C,D$ chosen to be the order $74$ Debauchies' wavelet filters. Note that it also have the same peak position at $\pm \frac{\pi}{2}$ as the basis defined by $\vec k=(-1,-1,1)$. A certain linear combination of these two basis states will lead to an extinction of one of the peaks.  } \label{D74}\end{figure}

At each iteration, a $C$ (IR) projector makes the spectral weight of the basis more strongly peaked around the zero momentum point, while a $D$ (UV) projector creates a peak near the highest momentum point in the complement subspace of all the previous layers. Let $\pm p_m$ be the position of the peaks for the $m^{th}$ level wavelet involving the first $m$ coefficients of the k-vector. the $p_m$'s satisfies the relation
\begin{eqnarray}
p_{m+2}= \frac{1}{2}p_{m+1} +  \frac{p_{m+1}}{2}\left(\frac{1+k_{m+2}}{2}\right)+ \frac{p_{m}}{2}\left(\frac{1-k_{m+2}}{2}\right)
\end{eqnarray}
In other words, $p_{m+2}=p_{m+1}$ if $k_{m+2}=+1$, and $p_{m+2}=\frac{p_{m+1}+p_m}{2}$ if $k_{m+2}=-1$, as expected form the behavior of the IR and UV projectors respectively. One example of this construction is illustrated in Fig. \ref{D74}.

In general, there are small contributions to such EHM basis away from the intended peaks, i.e. the secondary lobes in Fig. \ref{wavelets}. Choices of $C,D$ corresponding to more sophisticated wavelets\footnote{Will be further discussed in a work in progress.} can help exponentially suppress these contributions, leading to well-defined peaks like those in Fig. \ref{D74}. 

However, the important point is that single particle correlation functions under the EHM only depends on the basis and boundary system near the critical point, even if the EHM basis has nonvanishing spectral weight elsewhere. This is most easily seen from the large $\tau$ limit of the factor $e^{\tau E_q}$ in the imaginary time correlator Eq. \ref{time} or \ref{timeimag}, although the spatial correlators actually behave similarly as elaborated below.
%Even for interacting systems, we can control the amount of spectral weight away from the Fermi surface via $\kappa$, the no. of vanishing moments of $C(z)$ at $z=1$ or $k=0$. \textbf{The weight of the other peaks decay exponentially with $\kappa$}, which for the optimal choice by Daubechies is proportional to the real-space locality of the basis.

\subsubsection{Angular direction}
In the angular direction, contour integration still gives a power law correlator behavior $\sim \frac{1}{x^{2\kappa+1}}$, where $\kappa$ is the first nonzero derivative of the EHM basis $W_n(e^{ik})$ near $E_k=\mu$. However, $\kappa=0$ if the wavelet basis does not zoom in onto the IR point. Mathematically, this is because the asymptotic contributions only comes from the branch cut introduced by the critical point. 

When there is inversion symmetry, the leading $\kappa$ terms from the Fermi points $\pm \nu$ may cancel, resulting in a correlator decay of $\sim \frac{1}{x^{2\kappa+2}}$. In all of these case, the decay is always power-law, corresponding to a hyperbolic geometry.

\subsubsection{Radial direction}

The radial decay is due to the 'halving' of the peak distances at each successive $n$ about the critical point. As shown in Appendix \ref{app:critical}, the correlation function decays like $\sim \left(\frac1{\sqrt{2}}\right)^n$. The same result actually holds true for arbitrary Fermi points, as shown in the numerical results in Fig. \ref{muplots} below.
%Intuitively, $C(0)$ measures the amount of 'spread' of the IR filter $1\rightarrow C(z)$: a large $C(0)$ means that most of the DOFs are retained onsite, i.e. not smoothened out onto neighboring sites. Hence there is more interlayer overlap, and radial decay will be slower. 

\begin{figure}[H]
\includegraphics[scale=.9]{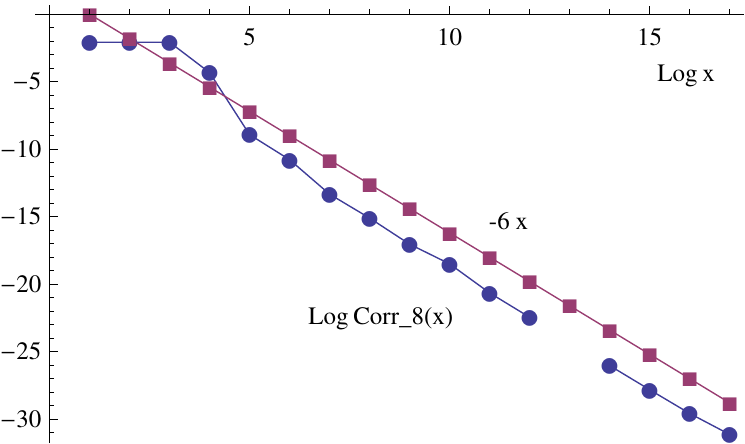}
\includegraphics[scale=.9]{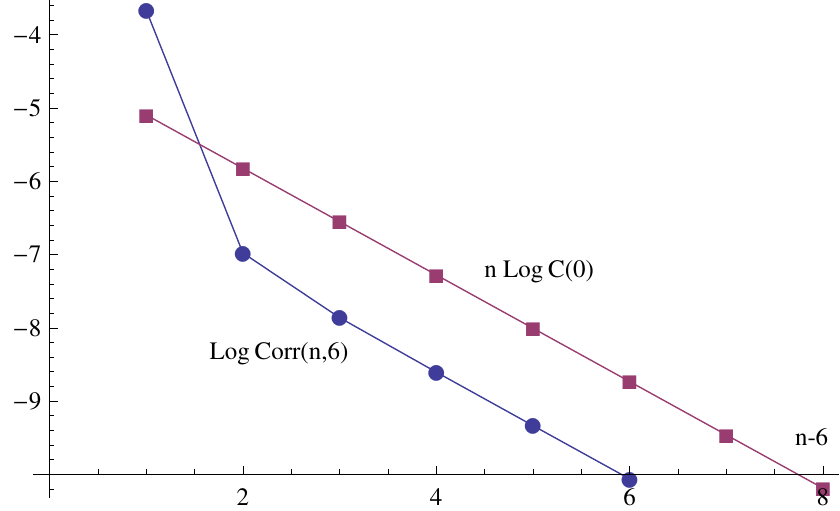}
\caption{Numerics and analytic fits for the off-diagonal correlator in the a) angular and b) radial directions. The Fermi points are at $\pm \frac{\pi}{16}$, so that the wavelet bases are of the form $CCCCDCC...CCD$ with $C,D$ corresponding to $\kappa=2$ and $C(0)=\frac{1+\sqrt{3}}{4\sqrt{2}}$. Due to inversion symmetry, the decay exponent is $2\kappa+2=6$, which is seen to hold very accurately. %The 2nd Daubechies' wavelet  has been used. There is excellent numerical agreement in both directions, which cannot be merely coincidental as $C(0)=\frac{1+\sqrt{3}}{4\sqrt{2}}$ has such a strange value.
} \label{muplots}\end{figure}

%\textbf{The angular decay depends crucially on $D(e^{i\nu})$, where $\nu$ are the Fermi momenta. By contrast, the radial decay depends explicitly on $C(0)$.}

\subsubsection{Imaginary time direction}

The imaginary time correlator takes the form
\begin{eqnarray}
Tr G_q(\tau)&=& Tr \frac{e^{\tau(h_q-\mu)}}{\mathbb{I}+e^{\beta(h_q-\mu)}}\notag\\
&\rightarrow & Tr e^{\tau(h_q-\mu)}\theta(\mu-h_q)\notag\\
&=& \sum_{E_q<\mu} e^{\tau(E_q-\mu)}
\label{timeimag}
\end{eqnarray}
so that the dominant contributions at large $\tau$ always arise from the eigenvalues $E_q$ that are closest to $\mu$, i.e. where $q$ is closest to the critical point. Since contributions away from criticality are exponentially suppressed, wavelet peaks elsewhere, which at most grow like a power, will still be suppressed. 

\subsubsection{Nonzero temperature cases}

From previous sections, the asymptotic behavior at nonzero temperature only depends on the complex singularities associated with the nonzero temperature Green's function. It is wavelet-independent. This holds for any other kind of mass-scale type of modification. Hence the bulk geometry at nonzero temperature is not affected by the order of the UV and IR layer projectors.

\subsection{Towards a continuous bulk with the continuous Wavelet Transform (CWT)}

%\subsection{Critical bulk propagators under Haar CWT}

Till now, we have considered the bulk to be made up of discrete layers. This is necessary due to the discreteness of the scaling parameter in the discrete wavelet transform. The continuous Wavelet transform offers an approach to generalize this scale dimension to a continuum. However, the wavelet mapping will not be an injective map anymore, as there will be more degrees of freedom in the continuous bulk. But this can be remedied by a "gauge-fixing" step where only certain types of linear combinations of the bulk states are chosen to reproduce the boundary state. Mathematically, it is realized by the inverse Wavelet transform.% formula Eq. \ref{inv_wavelet}. 

To understand it concretely, let us first define $C(a,a',b,b')=C(a,a',\Delta b)$ to be the bulk correlator, with $a,a'$ the scale of the two points, and $\Delta b=b'-b$ their displacement coordinate. They are related to the discrete case via $a\propto 2^n$, and $\frac{b}{a} \propto \frac{b}{2^n} = X$, where $X$ is the angular site position in the discrete tree.

The bulk correlator is given by
\begin{equation}
C(a,a',b,b')=\frac{1}{|a|}\int\int dx dx' w^*\left(\frac{x'-b'}{a'}\right)w\left(\frac{x-b}{a}\right)C(x-x')
\label{inv_wavelet0}
\end{equation}
%which has the inverse\red{???????????????????????}
%\label{inv_wavelet}

For an illustration, we input the Haar mother wavelet $w(x)=\theta(1-x^2)sgn(x)$ in Eq. \ref{inv_wavelet0}, and assume the form of the correlator $C(\Delta x)=\frac{1}{\sqrt{(\Delta x)^2 + c^2}}$, where $c$ is a small regulator. In momentum space, it corresponds to an exponential decay envelope. After some algebra, the asymptotic behaviors in the angular and radial (scale) directions are found to be
\begin{equation}
C(a,a,\Delta b)= \frac{2a^3}{(\Delta b)^3}+\frac{2\left(2- 3 \frac{c^2}{a^2}\right)a^5}{(\Delta b)^5}+O(b^{-7}) \sim \frac{2}{X^3}
\end{equation}
and 
\begin{eqnarray}
C(a,a',0)&=& \sqrt{\frac{a}{a'}}\left [ 4(1+\log \frac{c}{a})+2\frac{a}{a'}+\frac{a^2}{a'^2}\left(\frac{1}{3}-\frac{c^2}{a^2}\right)\right]+O\left(\frac{a^3}{a'^3}\right)\notag\\
&\sim& 4(1+\log \frac{c}{a})\sqrt{\frac{a}{a'}}\sim 2^{-\Delta n/2}
\end{eqnarray}
respectively. In the radial direction, the two scales $a\ll a'$ are assumed to be well-separated. Both of these correlator scales in exactly the same way as those of the discrete EHM. The cutoff scale $c$ enters logarithmically in the radial direction, where it provides a small correction to the overall proportionality constant.

\section{Conclusion for mapping II: EHM}
\label{sec:ehmconcl}

The EHM affords a way to separate the physics of different energy scales in an emergent dimension of a ``bulk'' system. We have analytically studied the emergent bulk geometries of several different boundary systems through the EHM approach. In general, critical boundary systems at zero temperature correspond to scale invariant bulk geometries. A spatial boundary appears in the infrared region when a mass scale is introduced by a nonzero mass. At nonzero temperature, a horizon appears in the infrared region, which is distinguished from the spatial boundary by the infinite red-shift that can be observed in the behavior of correlation functions along the imaginary time direction. For critical boundary theories with different dynamical exponents in time spatial and temporal directions, the spatial geometry is similar but the space-time geometry depends on the dynamical exponent, at both zero temperature and nonzero temperature. Most of the above results qualitatively still hold true when the EHM is generalized to higher dimensions.

As an exact mapping of the Hilbert space, the EHM also has applications beyond probing the bulk geometry. As we have also seen, it can be used to decompose the Chern number of a boundary Chern Insulator into the Chern number density spread across the layers of the corresponding bulk system. The former corresponds to the Hall conductance of single layer in the bulk, which is reminiscent of the hall conductance of a time-reversal symmetry breaking slab in (3+1)-dimensional topological insulator described by Axion field theory. We rigorously justified this analogy by showing that an entanglement cut within the purported nontrivial $\mathbb{Z}_2$ region indeed gives rise to the signature odd number of gapless entanglement modes. Indeed, the above application of EHM has lead to another realization of a bulk/edge correspondence involving topological systems, distinct from the usual bulk/edge correspondence relating topological edge states with the bulk. 

A major open question concerns the dual geometry for a Fermi gas with finite charge density. The existence of a finite Fermi momentum makes it inappropriate to use the same EHM mapping defined here, since the long wavelength limit will no longer correspond to the low energy limit. A modified tensor network is required in order to describe the correct infrared physics near the Fermi surface. 

Our result can be generalized straightforwardly to free fermion topological states in other dimensions and symmetry classes, which suggests that nontrivial quantum entanglement between different energy scales and the duality between states in different dimensions are generic features of fermionic topological states of matter. Some hints for possible generalizations of the basis were described in Sect. \ref{EHMfurther}, where the bulk basis was made continuous with the layers ``zooming in'' onto a point not representing the extreme long wavelength regime.

%\appendix
\chapter{Appendices}

\section{Second-quantized pseudopotentials vs. real-space projection Hamiltonians}\label{sec:TK}

Here we provide a brief comparison between the second-quantized PPs that have been derived in this thesis and the first-quantized real-space projection Hamiltonians elsewhere in the literature, e.g. Trugman-Kivelson type Hamiltonians on the infinite plane. We show that our geometric PP construction avoids certain ambiguities that plague the Trugman-Kivelson type PPs.

Neglecting internal DOFs for simplicity, a generic real-space PP living in total relative angular momentum sectors up to $m$ takes the form
\begin{equation} H(\vec r_1,...,\vec r_N)=\left(\prod^N_{j=1}\nabla_j^{2d_j}\right)\delta^2(\vec r_1-\vec r_2)...\delta^2(\vec r_{N-1}-\vec r_N)\end{equation}
such that $\sum_j d_j=m$. The various $d_j$'s refer to how the derivatives are distributed among the particles. In this form, there is \emph{no} simple one-to-one correspondence between the $d_j$ values and the relative weight of $H$ in the various $m$ sectors. To find the relative weights, one has to project $H$ onto the sector spanned by a state with angular momentum $m$. It is most convenient to adopt the symmetric gauge, where such a state takes the form $\ket{\Psi_m}\propto p(\tilde z_1,...,\tilde z_N)e^{-\frac{1}{4}\sum_j |z_j|^2}|z_1,...,z_N\rangle$ where $p$ is a symmetric or antisymmetric polynomial and $z_j=x_j+iy_j$ refers to the particle position and $\tilde z_j=z_j-\frac1{N}\sum_k z_k$:
\begin{eqnarray} 
&&\langle \Psi_m|H\ket{\Psi_m} = \nonumber \\
&& \left(\prod_{j=1}^N \int d^2\vec r_j\right)\Psi_m^\dagger(\vec r_1,...,\vec r_N)H(\vec r_1,...,\vec r_N) \Psi_m(\vec r_1,...,\vec r_N). \nonumber \\
\end{eqnarray}
Upon substituting $\Psi_m$ and integrating by parts, we can express $\langle \Psi_m|H|\Psi_m\rangle$ as 
\begin{eqnarray}
&& \langle \Psi_m|H\ket{\Psi_m} = \notag \\
&&  \left(\prod_{j=1}^N \int d^2\vec r_j\right)\delta^2(\vec r_1-\vec r_2)...\delta^2(\vec r_{N-1}-\vec r_N) \notag \\
&\times&  \left(\prod^N_{j=1}\nabla_j^{2d_j}\right)[\Psi_m^\dagger(\vec r_1,...,\vec r_N)\Psi_m(\vec r_1,...,\vec r_N)],
\end{eqnarray}
which can be evaluated as
\begin{eqnarray}
&& \langle \Psi_m|H\ket{\Psi_m} = \notag \\
&=& 4^m\left(\prod_{j=1}^N \int dz_jd z^*_j\right)\delta^2( z_1- z_2)...\delta^2( z_{N-1}- z_N)  \notag \\
&& \left(\prod^N_{j=1}\partial_{ z^*_j}^{2d_j}\Psi_m^\dagger(  z_1^*,...,z_N^*)\right)\left(\prod^N_{j=1}\partial_{z_j}^{2d_j}\Psi_m( z_,...,z_N)\right),
\label{realproject}
\end{eqnarray}
where $\nabla^2 = 4 \partial_z\partial_{\bar z}$ has been used. The delta functions, which enforce that all $z_j$ be equal, also enforce $\tilde z_j=z_j-\frac{1}{N}\sum_k z_k=0 \; \forall \; j$. Hence we only get a nonzero contribution from constant integrands. Specific examples were worked out in the appendices of Ref. \onlinecite{simon2007}; in general, an operator $H$ has an overlap with various $m$ sectors, and a PP that lies \emph{purely} in one $m$ sector must be a complicated linear combination of the the Trugman-Kivelson $H$'s with various sets $d_j$'s. Some progress was made in Ref.~\onlinecite{lee2013}, where operators containing $L_m\left(\frac{N\kappa^2}{2(N-1})\nabla_1^{-2}\right)$ were shown to be contained purely in one $m$ sector. That, however, does not always work for fermions as will be explained below.

The mathematical complications above are further aggravated when the space of PPs in the sector $m$ is degenerate. In particular, the matrix $\langle \Psi_{ma}|H\ket{\Psi_{mb}}$ is not of full rank and some combinations of bona-fide PPs of degree $m$ will still not give a positive energy penalty to states in the $m$ sector~\cite{simon2007}. 
In particular, it is possible for certain Trugman-Kivelson Hamiltonians with $2m$ derivatives to evaluate to zero upon taking into account the fermionic antisymmetry. Consider for instance the $N=3$ fermionic state $\ket{\Psi_3} \propto  (\tilde z_1 - \tilde z_2)(\tilde z_2 - \tilde z_3)(\tilde z_3 - \tilde z_1)|z_1,z_2,z_3\rangle$. The following operators have a zero projection on it:

\[ \langle \Psi_3|\nabla^6_j\delta^2(r_1-r_2)\delta^2(r_2-r_3)|\Psi_3\rangle =0\] and
\[ \langle \Psi_3|\nabla^2_1\nabla^2_2\nabla^2_3\delta^2(r_1-r_2)\delta^2(r_2-r_3)|\Psi_3\rangle =0.\]  

Indeed, the only nonvanishing Trugman-Kivelson type fermionic PP in the $U^{N=3}_{m=3}$ sector is $H=\nabla^4_i\nabla^2_j\delta^2(r_1-r_2)\delta^2(r_2-r_3)$. As $|\Psi_3\rangle$ is of degree $\le 2$ in each variable, $H$ must contain a derivative of degrees $2d_j\leq 4$ in each variable for a nonzero projection. Since only constant terms in the integrand of Eq. \ref{realproject} contribute, we also need derivatives in all three particle coordinates. 

By contrast, our PP construction appeals to the orthogonality structure of the PPs {\it from the outset}, and avoids all of the above problems that generically arise when one attempts to use the Trugman-Kivelson approach in more complicated many-body cases. 

\section{Pseudopotential parent Hamiltonian from a given wave function}
\label{sec:power}

Here we briefly summarize the heuristic procedure how to determine the PPs whose null spaces contain a given QH state. These are the PPs that should be included in its parent Hamiltonian, since they penalize denser or ``simpler" states that will otherwise be realized at the same filling fraction. The method consists of two steps: (i) identify the thin-torus root pattern from the first-quantized wave function (Sect.~\ref{sec:root}), and (ii) use the root pattern and power counting to obtain the PPs (Sect.~\ref{sec:parent}).

\subsection{Wave function to root pattern}\label{sec:root}

Suppose we are given a QH wave function with polynomial part $\psi(z_1,z_2,...)$, where the $z_i$'s denote the positions of particles. In the infinite plane, $z_j$ is the usual complex coordinate $z_j=x_j+iy_j$. To represent a cylindrical geometry, however, we need to perform a stereographic mapping: $z_j \to e^{\kappa z_j}$, where $\kappa=2\pi/L_y$. For simplicity, assume that there are no internal degrees of freedom here. 

%\begin{figure}[htb]
%\includegraphics[width=0.92\linewidth]{thincylinder.pdf}
%\caption{  In the limit of thin cylinder ($\kappa\to\infty$), the single-particle orbitals become well-separated from one another. As a consequence, the dominant configurations (``root patterns") in the expansion of a many-body state $\Psi$ are those Fock states where particles minimize their classical electrostatic energy.}
%\label{fig:thincylinder}
%\end{figure}
To each wave function defined on a cylinder, we can assign one or several thin-cylinder \emph{root patterns}. Those are the Fock states with non-zero weight as the circumference of the cylinder is taken to zero ($\kappa\to\infty$). Typically, any given FQH state $\ket{\psi}$ in the thermodynamic limit decomposes onto many Fock states, as $\ket{\psi}$ represents a strongly correlated state. However, when the cylinder is stretched, the weight of most of the Fock states in the decomposition vanishes. The Fock states surviving in the limit $\kappa\to\infty$ are called ``root patterns" and are useful for finding the parent Hamiltonian (Laplacian) for $\ket{\psi}$ as its kernel.

In order to find the root pattern, we must examine the dominant terms in the decomposition of $\psi$ as $\kappa\to\infty$. Usually, this amounts to solving a classical electrostatic problem. Here we consider the case of an open cylinder, where the root pattern will typically be unique. To get the remaining root patterns, that arise due to topological degeneracy or non-Abelian statistics, one must consider the system on a torus. Due to the complicated form of torus wave functions, it is more convenient to work on a cylinder but perform a flux insertion which allows one to access other topological sectors. Details of this are given e.g. in Ref.~\onlinecite{Seidel-PhysRevLett.101.036804}. 

Fixing the total number of particles, we study the dominating term in $\psi$. It is the term $z_1^{m_1}z_2^{m_2}z_3^{m_3}...$ such that \emph{no} other term (denoted by $'$) has $m'_1>m_1$ or, $m'_2>m_2$ if $m'_1=m_1$ or, $m'_3>m_3$ if $m'_1=m_1$ and $m'_2=m_2$, etc. The root pattern can then be written down as the partition defined by $m_1,m_2,...$:
\begin{equation}
(m_1,m_2,m_3,...)\rightarrow 10...010...01...
\end{equation}
where the $1$'s are at position $m_1,m_2,...$. If two $m_i$'s coincide, as can occur for bosons or multicomponent fermions, we shall label that position by '$2$' and so forth.
For instance, consider the $1/m$ Laughlin state with $2$ particles. We have $\psi(z_1,z_2)=(z_1-z_2)^m$ which, when expanded out, contains monomials
 \begin{equation}
\{ z_1^m, z_1^{m-1}z_2, z_1^{m-2}z_2^2,...\}
\end{equation}
Evidently, the term $z_1^mz_2^0$ is in the root pattern. Had we considered a droplet of $N$ particles, this term would have become $z_1^{m(N-1)}z_2^{m(N-2)}...z_{N-1}^mz_N^0$, with total degree $\frac{mN(n-1)}{2}$ and the root
\begin{equation}10^{m-1}10^{m-1}1...,\end{equation}
where $0^{m-1}$ represents a string of $m-1$ consecutive zeroes. For this simplest case, a droplet of $N=2$ particles is sufficient to determine the root configuration. Because we deal with translationally invariant states, a root configuration for larger $N$ is obtained by repeating the minimal pattern.

Let us now consider a more complicated example such as the generalized Pfaffian state introduced in the Examples section of~\cite{lee2015pp}. By using the identity $[Pf(A)]^2=Det(A)$, one can show by induction starting from $n=4$, the most compact particle droplet, that the dominating term is $z_1^{3q-1}z_2^{2q}z_3^{q-1}$. This defines the root configuration $10^{q-2}10^{q}10^{q-2}10^q...$. It is also easy to read off the filling fraction from the above: we have $2$ particles per period, and each period has length $2q$. Hence the filling fraction is $\frac{2}{2q}=1/q$.

\subsection{Root pattern to parent Hamiltonian}\label{sec:parent}

Given a root pattern, we can easily infer what relative angular momentum states should be penalized to construct a projector onto the state. For instance, if the closest pair of particles are $...10^l1...$, no two particles have relative angular momentum less than $l+1$. Hence the state should lie in the null space of any two-body PPs with $U^2_m$, $m\leq l$. Consequently, the parent Hamiltonian should be a linear combination of these PPs, so that it penalizes all states denser than the given one. This is discussed at length in Ref. \onlinecite{simon2007generalized}.

To figure out the required PPs with more than $N>2$ bodies, we apply the above procedure analogously to clusters of $N$ particles with minimal relative angular momentum.  Note that some many-body PPs need not be included if the interaction is already precluded by particle (anti)symmetry, as we will discuss next:
Consider the $1/q$ Pfaffian state described above, with a root given by $10^{q-2}10^{q}10^{q-2}10^q...$. Obviously, no two particles have relative angular momentum less than $q-1$. Naively, we will then need to include all $U^2_{m}$ with $m\leq q-2$. The $1/q$ Pfaffian state, however, is fermionic (bosonic) when $q$ is even(odd), and $U^2_{q-2}$ exactly vanishes in both cases. Hence the parent Hamiltonian only contains $U^2_{m< q-2}$. 

Let us analyze the 3-body interactions. From the root pattern, the minimal total relative angular momentum is $l=(q-1)+2q=3q-1$. However, configurations with $l=3q-2=2(q-1)+q$ automatically vanish since they require pairs of particles with relative angular momenta $q$ and $q-1$ simultaneously, which are forbidden by either fermionic or bosonic statistics. With lower 3-body relative angular momentum already excluded by the 2-body PPs, the only 3-body PP we need to include in the parent Hamiltonian is $U^3_{3q-3}$.  As the root pattern repeats from the third particle, there is no need to consider interactions with $4$ or more bodies.  

\section{Equivalence of the Landau-level-projected  and Cartesian description on
  the cylinder}
\label{equi}
The Landau level on a cylinder can be seen as a hybrid version
of the spherical and planar scenario. Starting from a Cartesian $(x,y)$
plane, we impose periodic boundary conditions along $y$ and, in analogy
to the sphere, quantize the pseudo-momentum along $y$ according to the total
magnetic flux $N_\phi$, constraining the available area for
the guiding center motion. We assume an infinite cylinder along the $x$-direction. On the cylinder geometry, the PP Hamiltonian takes
the explicit form~\cite{rezayi-94prb17199}
\begin{eqnarray}
H&=&\gamma \sum_{i<j}\sum_{n,m} \int  d q \; l_B V^m L_m(q^2l_B^2
+\gamma^2n^2) e^{-q^2 l_B^2/2} \nonumber \\
&& \times e^{-\gamma^2 n^2/2} e^{i\gamma n \hat{x}_i/l_B} e^{iq
  (\hat{y}_i-\hat{y}_j)} e^{-i\gamma n \hat{x}_j /l_B}, \label{cyl}
\end{eqnarray}
where $\gamma$ is the aspect ratio of the cylinder, $l_B$ is the
magnetic length, $q$ denotes the momentum variable along $x$, $V^m$ is the PP energy of a state with
relative angular momentum $m$, and $L_m$ denotes the $m$th Laguerre
polynomial~\cite{Abra}. (Note that we have explicity included all
units in \eqref{cyl} as compared to Ref.~\onlinecite{rezayi-94prb17199}.) Eq.~\ref{cyl}
gives a pair interaction energy for each
$m$. In the plane limit where $\gamma n \rightarrow k_xl$, the sum over
$n$ reduces to a momentum integral along $y$ analogous to the $q$ integration
along $x$, resulting in a two-dimensional momentum integral which reduces to the pair
interaction
\begin{eqnarray}
V(\bs{r}_i-\bs{r}_j)&=&\sum_m V^mP^m(\bs{r}_i-\bs{r}_j) \nonumber \\
%&=& \sum_m l \int d^2\bs{p} V_m L_m(\bs{p}^2l^2)
%e^{-\bs{p}^2 l^2/2} e^{i\bs{p}(\bs{r}_i-\bs{r}_j)} \nonumber \\
&=& \sum_m V^m L_m (-l_B^2 \nabla^2) \delta^2 (\bs{r}_i -\bs{r}_j).
%e^{-\vert \bs{r}_i -
%\bs{r}_j \vert^2/2l^2}.
%\label{lag}
\end{eqnarray}
We reconcile the two expressions for the delta function potentials on
the cylinder appearing in Refs.~\onlinecite{rezayi-94prb17199,lee2004}. The following arguments are valid for generic $l_B$, which has been set to unity in
the following. We show that
\begin{equation} \langle 0 | \delta^2(r-r')|0\rangle =
  e^{-(r_p-r'_p)^2/2},
\end{equation}
where $|0\rangle\sim f(z) e^{-|z|^2/4}$ denotes the LLL wavefunction and $r_p=(x_p,y_p)$ are the guiding-center coordinates defined by
\[ x_p = :\partial_z +\frac{z}{2}:, \]
\[ y_p = :-i\partial_z+\frac{iz}{2}: \]
with $z=x-iy$. These are the LLL-projected coordinates since $z_p =
z$, $\bar z_p = :2\partial_z:$(see Ref.~\onlinecite{jainbook}).
Here, the normal-ordering symbols ":'' indicate that any derivative
within them does not act on the Gaussian factor $e^{-|z|^2/4}$ present
in LL wavefunctions. Note that $[x_p,y_p]=i$, i.e. the guiding-center coordinates do not commute.
Consider a generic interaction \begin{equation} V(r-r')= \int \frac{d^2k}{(2\pi)^2}V(k)e^{ik\cdot (r-r')}.\end{equation}
Let us project it onto the LLL $|0\rangle = f(z)e^{-|z|^2/4}$, i.e.
\begin{equation} \langle 0 |V(r-r')|0\rangle = \int \frac{d^2k}{(2\pi)^2}V(k)\langle 0 |e^{ik\cdot (r-r')}|0\rangle. \end{equation}
The quantity in the angle brackets is evaluated as
\begin{eqnarray}
\langle 0 |e^{ik\cdot r}|0\rangle &=&\langle 0 |e^{i(k\bar z + \bar{k}
  z)/2}|0\rangle \nonumber \\
&=& \langle 0 | e^{\frac{i}{\sqrt{2}}(\bar{k} a + k a^\dagger)}
e^{\frac{i}{\sqrt{2}}(\bar{k} b^\dagger + k b)}|0\rangle \nonumber \\
&=& e^{-|k|^2/2}\langle 0| e^{\frac{i}{\sqrt{2}}(k
  a^\dagger)}e^{\frac{i}{\sqrt{2}}(\bar{k} a )}
e^{\frac{i}{\sqrt{2}}(\bar{k} b^\dagger + k b)}|0\rangle \nonumber \\
&=& e^{-|k|^2/2}\langle 0| e^{\frac{i}{\sqrt{2}}(\bar{k} b^\dagger + k b)}|0\rangle,
\end{eqnarray}
where $k=k_x-ik_y$ and $a,b,a^\dagger,b^\dagger$ are lowering and
raising operators of angular momentum and LL (see also Appendix~\ref{nbody}). Due to the specific form of $|0\rangle$, we also have
\begin{eqnarray}
b |0\rangle &=& \frac{1}{\sqrt{2}}(\frac{\bar
  z}{2}+2\partial_z)f(z)e^{-|z|^2/4} =
\frac{1}{\sqrt{2}}:2\partial_z:f(z)e^{-|z|^2/4} =\frac{1}{\sqrt{2}}z_p
|0\rangle, \nonumber \\
b^\dagger |0\rangle &=& \frac{1}{\sqrt{2}}(\frac{ z}{2}-2\partial_{\bar z})f(z)e^{-|z|^2/4} =   \frac{1}{\sqrt{2}}:z:f(z)e^{-|z|^2/4} = \frac{1}{\sqrt{2}}\bar z_p |0\rangle.
\end{eqnarray}
Hence,
\begin{eqnarray}
\langle 0 |e^{ik\cdot r}|0\rangle &=& e^{-|k|^2/2}\langle 0|
e^{\frac{i}{2}(\bar{k} z_p + k \bar z)}|0\rangle \nonumber \\
&=& e^{-|k|^2/2}\langle 0| e^{ik\cdot r_p}|0\rangle.
\end{eqnarray}
If we start from a delta function in real space, $V(k)=1$, we obtain
\begin{equation} \langle 0 |\delta^2(r-r')|0\rangle = \int \frac{d^2k}{(2\pi)^2} e^{-|k|^2/2}e^{ik\cdot r_p} = e^{-(r_p-r'_p)^2/2}.\end{equation}
Higher PPs in the LLL will thus be of the form
$L_m(\nabla^2_p) e^{-(r_p-r'_p)^2/2}$. To summarize, the LLL projection leads to two types of modifications. Firstly, the delta function is replaced by a Gaussian
in guiding-center coordinates. Secondly, the derivative will also be
taken with respect to guiding-center coordinates.

\section{First-principle derivation of Generalized Haldane Pseudopotentials}
\label{nbody}

We generalize the Haldane pseudopotentials to \emph{General} Haldane Pseudopotentials (GHPs) involving $N>2$ bodies. Such interactions include the most general first-quantized $N$-body interactions, although we will be primarily concerned with those that conserve the center-of-mass (CM). The main technical steps in
this appendix include (i)  a derivation of the GHP starting from
an original interaction Hamiltonian involving arbitrarily many bodies
 (ii) a clarification on how the total relative angular
momentum can be defined through an appropriate change of coordinates. The results of this
appendix will be directly utilized in Appendix~\ref{3body} for explicit calculations of the first $3$-body PPs , which
always take the factorized form $\sim \sum_n \hat{b}^\dagger_n \hat{b}_n$.

We start with the Hamiltonian involving $N$ electrons in a magnetic field interaction via the potential $V$: 
\begin{eqnarray}
H&=& \frac{1}{2m}\sum_i^N (\vec{p}_i + \frac{e}{c}\vec{A}_i)^2 +
V(\vec{x}_1,\vec{x}_2,...,\vec{x}_N) \nonumber \\
&=& \frac{1}{2m}\sum_i^N \left(-i\frac{\partial}{\partial
    x_i}-\frac{eB}{c}\frac{y_i}{2}\right)^2
+\left(-i\frac{\partial}{\partial
    y_i}+\frac{eB}{c}\frac{x_i}{2}\right)^2 \nonumber \\
&&+ V(\vec{x}_1,\vec{x}_2,...,\vec{x}_N) \nonumber \\
&=&\frac{1}{2m}\sum_i^N \left[ -4l_B^2\frac{\partial^2}{\partial
    z_i \partial \bar{z}_i}+\frac{1}{4l^2}|z_i|^2 +
  \left(\bar{z}_i\frac{\partial}{\partial
    \bar{z}_i}-z_i\frac{\partial}{\partial z_i}\right)\right ]\nonumber \\
&&+V(\vec{x}_1,\vec{x}_2,...,\vec{x}_N),
\label{symmham}
\end{eqnarray}
where $z_i=x_i-iy_i$, $\vec{A}_i=B(-y_i,x_i)/2$ and
$l_B=\sqrt{\frac{\hbar c}{eB}}$. The symmetric gauge has been used so
that the Hamiltonian eigenstates will be conveniently labelled by
angular momentum. However, the results that follow will be
gauge invariant. For now, we will not make any assumption about the
form of the interaction potential $V$.

Haldane's original procedure was to first seperate this Hamiltonian into a CM part and relative part, and then project the relative part into different angular momentum sectors~\cite{haldane1983}. The same will be done here, but with $N$ particles instead of two. Define a change of coordinate
\[w_i = R_{ij} z_j, \]
with $w_1=(z_1+z_2+\dots + z_N)/N$, i.e. the CM coordinate. Any part of the resultant Hamiltonian depending only on $w_1$ will not affect our PP expansion.

\subsection{Allowed coordinate transforms $R_{ij}$}
It turns out that we are not free to choose the rest of $R_{ij}$ arbitrarily. If we want to have a well-defined angular momentum decomposition in terms of the new variables $\{ w_i \}$, we will need to ensure that the resultant Hamiltonian is of the same form as the last line of  Eq.~\ref{symmham}. This is to ensure that the angular momentum operator can still be defined in a similar way with the new coordinates. To see why, first write the kinetic term in~\eqref{symmham} as
\begin{eqnarray}
H_{kin}&=&\frac{1}{2m}\sum_i^N \left[ -4l_B^2\frac{\partial^2}{\partial
    z_i \partial \bar{z}_i}+\frac{1}{4l^2}|z_i|^2 +
  \left(\bar{z}_i\frac{\partial}{\partial
    \bar{z}_i}-z_i\frac{\partial}{\partial z_i}\right)\right ] \nonumber \\
&=& \hbar \omega \sum_i( b^\dagger_i b ^{\phantom{\dagger}}_i+\frac{1}{2})+\hbar
\omega(a^\dagger_i a ^{\phantom{\dagger}}_i -b^\dagger_i b ^{\phantom{\dagger}}_i) \nonumber \\
&=&\hbar\omega\sum_i \left(a^\dagger_i a ^{\phantom{\dagger}}_i+\frac{1}{2}\right),
\end{eqnarray}
with second-quantized operators defined by (particle index $i$ suppressed) \[a=\frac{1}{\sqrt{2}}\left(\frac{z}{2l_B}+2l_B\frac{\partial}{\partial\bar{z}}\right),\]
\[b=\frac{1}{\sqrt{2}}\left(\frac{\bar{z}}{2l_B}+2l_B\frac{\partial}{\partial z}\right)\]
so that its eigenstates are labeled by angular momentum $m$ and LL index $n$ :
\begin{equation}
|n,m\rangle = \frac{(b^\dagger)^{m+n}}{\sqrt{(m+n)!}}\frac{(a^\dagger)^n}{\sqrt{n!}}|0,0\label{nm}\rangle.
\end{equation}
with the angular momentum operator given by \begin{equation} L=\hbar \left(\bar{z}\frac{\partial}{\partial\bar{z}}-z\frac{\partial}{\partial z}\right ) = \hbar(a^\dagger a - b^\dagger b).\end{equation}
We will still desire the eigenstates to take to form Eq. \ref{nm} after the coordinate transform $R_{ij}$. Hence the latter must leave the form of each term of the last line of Eq.~\ref{symmham} invariant, i.e. $|z_i|^2$ must transform into a sum of similar quadratic terms, etc.

Denoting $z=(z_1,z_2,...,z_N)^T$, $\partial z=(\frac{\partial}{\partial z_1},\frac{\partial}{\partial z_2},...,\frac{\partial}{\partial z_N})^T$ and likewise for the $w^i$s, the various terms transform as
\begin{equation} \sum_i \frac{\partial^2}{\partial z_i \partial \bar{z}_i}=\partial \bar{z}^T \partial z = \partial \bar{w}^T[ R R^T] \partial w,
\end{equation}
\begin{equation} \sum_i |z_i|^2= \bar{z}^T z =  \bar{w}^T[ (R^{-1})^T(R^{-1})]  w=\bar{w}^T[R R^T]^{-1} w,
\end{equation}
\begin{equation} \sum_i z_i\frac{\partial}{\partial z_i }=z^T \partial z = w^T[ (R^{-1})^T R^T] \partial w = w^T \partial w.
\end{equation}
The Hamiltonian retains the same form if $R R^T$ is diagonal. If we regard $R$ as a rotation matrix, we see that this condition is satisfied whenever $R$ maps an orthogonal basis to another orthogonal basis. Hence an allowed $R$ consists of mutually orthogonal rows. As a simple example, the $R$ matrix for 2 particles takes the valid form of
\[
\left( \begin{array}{cc}
1/2& 1/2  \\
1 & -1\end{array} \right),
\]

according to the CM coordinate $w_1=(z_1+z_2)/2$ and the relative coordinate $w_2=z_2-z_1$.

\subsection{The explicit form of the QH Hamiltonian in total relative coordinates}

The next step is to explicitly find the coefficients of the transformed Hamiltonian. Since we are interested in a PP expansion in the total angular momentum, we define the total relative coordinate
\begin{equation}
w_2=\frac{1}{N-1}\sum_{n=1}^{N-1}(z_n-z_N)=\frac{\sum_i^{N-1}z_i}{N-1}-z_N
.
\end{equation}
The other coordinates can be arbitrarily defined as long as they are orthogonal to $w_2$ and $w_1=\frac{1}{N}\sum_i^{N}z_i$.
With this choice, the diagonal elements of $RR^T$ are
\[ \lambda_1 = N, \lambda_2 = \frac{N}{N-1} , \dots \]
Hence the kinetic part of the Hamiltonian becomes
\begin{eqnarray}
H_{\text{kin}}&=&\frac{1}{2m}\sum_i^N \left[
  -4l_B^2\lambda_i\frac{\partial^2}{\partial w_i \partial
    \bar{w}_i}+\frac{1}{4l_B^2\lambda_i}|w_i|^2 \right. \nonumber \\
&& \left. +
  \left(\bar{w}_i\frac{\partial}{\partial
    \bar{w}_i}-w_i\frac{\partial}{\partial w_i}\right)\right ] \nonumber \\
&=&\frac{1}{2m} \left[ -4l^2_{\text{rel}}\frac{\partial^2}{\partial
    w_2 \partial \bar{w}_2}+\frac{1}{4l^2_{\text{rel}}}|w_2|^2 \right. \nonumber \\
&& \left. + \left(\bar{w}_2\frac{\partial}{\partial \bar{w}_2}-w_2\frac{\partial}{\partial w_2}\right)\right ]+\dots,
\end{eqnarray}
where $l_{\text{rel}}=l_B\sqrt{\lambda_2}=l_B\sqrt{\frac{N}{N-1}}$ is the effective "magnetic length" for the total relative coordinate. Only the terms corresponding to the total relative coordinate are shown in the second line.
In general, the diagonal elements of $RR^T$, i.e. $\lambda_1,\lambda_2,\lambda_3,...,\lambda_N$ define a set of effective magnetic lengths $l_i=l_B\sqrt{\lambda_i}$.

\subsection{Derivation of the N-body pseudopotential}

We are now set up to find $\langle m_1,...,m_N|V| m_1,...,m_N
\rangle$, the projection of an interaction potential
$V(w_1,w_2,...,w_N)$ onto the angular momentum sectors $m_1,m_2,...,m_N$ in the LLL. This projection is of course dependent on $w=Rz$. For the $w_1$ and $w_2$
previously defined as the CM and total relative coordinates, $m_1$ and
$m_2$ correspond to the CM angular momentum and total relative angular
momentum, respectively. To evaluate this matrix element, we Fourier transform to shift the coordinate dependencies onto an universal exponential factor:
\begin{eqnarray}
V^{m_1m_2\dots m_N}
&=&(4\pi)^N\langle m_1,...,m_N|V(w_1,...,w_N)|m_1,...,m_N\rangle \nonumber \\
&=&(4\pi)^N \langle
m_1,...,m_N|\left(\prod_j^N\int\frac{d^2(k_jl_j)}{(2\pi)^2}\right)V(k_1,...,k_N)\prod_j^N
e^{i k_j\cdot w_j}|m_1,...,m_N\rangle \nonumber \\
&=& (4\pi )^N\left(\prod_j^N\int\frac{d^2(k_jl_j)}{(2\pi)^2}\right)\prod_j^N \langle
m_j|V(k_1,...,k_N) e^{i k_j\cdot w_j}|m_j\rangle \nonumber \\
&=& (4\pi )^N\left(\prod_j^N\int\frac{d^2(k_jl_j)}{(2\pi)^2}\right)V(k_1,...,k_N)\prod_j^N \langle m_j| e^{i k_j\cdot w_j}|m_j\rangle.
\end{eqnarray}
The momentum-space potential $V$ in the last line can be taken out of
the expectation value since the momenta labeled by $k_j$s are regarded as
complex numbers. For each $j$,
\begin{eqnarray}
\langle m_j| e^{i k\cdot w_j}|m_j\rangle &=&\langle m_j| e^{i
  \frac{l_j}{2}(k\bar{z}+\bar{k}z)}|m_j\rangle =\langle m_j| e^{\frac{il_j}{\sqrt{2}}(\bar{k}a_j +k a^\dagger_j)
}e^{\frac{il_j}{\sqrt{2}}(\bar{k}b^\dagger_j +k b_j) }|m_j\rangle
\nonumber \\
&=&e^{-|k|^2l_j^2/4}\langle m_j| e^{i\frac{l_j}{\sqrt{2}}ka^\dagger_j
}e^{i\frac{l_j}{\sqrt{2}}\bar{k} a_j
}e^{i\frac{l_j}{\sqrt{2}}(\bar{k}b^\dagger_j +k b_j) }|m_j\rangle\notag\\
&=&e^{-|k|^2l_j^2/2}\langle
m_j|e^{\frac{il_j}{\sqrt{2}}\bar{k}b^\dagger_j
}e^{\frac{il_j}{\sqrt{2}}k b_j } |m_j\rangle \nonumber \\
&=&e^{-|k|^2l_j^2/2}\sum_{s=0}\frac{1}{(s!)^2}\langle m_j|
\left(\frac{il_j\bar{k}b^\dagger}{\sqrt{2}}\right)^s\left(\frac{il_j k
 b}{\sqrt{2}}\right)^s|m_j\rangle \notag\\
&=& e^{-|k|^2l_j^2/2}\sum_{s=0}\frac{m!}{(s!)^2(m-s)!}\left(\frac{-l_j^2
    |k|^2}{2}\right)^s \nonumber \\
&=&e^{-|k|^2l_j^2/2}L_m\left(\frac{l_j^2|k|^2}{2}\right).
\label{mvm}
\end{eqnarray}
The terms containing the $a_j$ and $a_j^\dagger$ operators in the
third line reduce to unity because the states are already defined to
be in the LLL. Use has been made of the Baker-Campbell-Hausdorff
formula in producing the factors of $e^{-|k|^2l_j^2/4}$.

The LLL projected pseudopotential component reads
\begin{eqnarray}
&&V^{m_1m_2\dots m_N} \nonumber \\
&=&\langle m_1,...,m_N|V(w_1,w_2,...,w_N)|m_1,...,m_N\rangle \nonumber
\\
&=& \prod_j^N\int\frac{d^2(k_jl_j)}{\pi}e^{-|k_j|^2l_j^2/2}L_{m_j}\left(\frac{l_j^2|k_j|^2}{2}\right)V(k_1,...,k_N),
\label{vmm}
\end{eqnarray}
where $V(k_1,...,k_N)$ is the Fourier transform of $V(w_1,...,w_N)$.
We focus on the $w_2$ degree of freedom. Let $|m\rangle$ denote the
state where $m_2=m$ and all other $m_i=0$, i.e. the state with total
relative angular momentum $m$. Then the $m$th PP component for a translationally invariant interaction is
\begin{eqnarray}
V^m&=&
\left(\prod_{j=2}^N\int\frac{d^2(k_jl_j)}{\pi}e^{-|k_j|^2l_j^2/2}\right)\nonumber
\\
&&\times L_m\left(\frac{N l_B^2|k_j|^2}{2(N-1)}\right)V(k_2,k_3,...,k_N).
\label{vm}
\end{eqnarray}
The integral over $k_1$ has been omitted since $V$ does not depend on the CM coordinate $w_1$. Also, the rest of the Laguerre polynomial factors have disappeared since $L_0=1$.
We still need to define $V^{m_2,...m_N}(k_2,k_3,...,k_N)$ such that it
corresponds to a PP component \[\langle
m_1,...,m_N|V(w_1,w_2,...,w_N)|m_1,...,m_N\rangle \] that is nonzero
only at the simultaneous set of angular momenta $m_2,...,m_N$. This is
done by exploiting the orthonormality relation
$\int_0^{\infty}2qe^{-q^2} L_s(q^2)L_t(q^2)dq=\delta_{st}$. Switching
to polar coordinates, we see that the functional form of the
pseudopotential is given by
\begin{equation}
U^{m_2,...m_N}=V_0\prod_{j=2}^N L_{m_j}\left(\frac{k_j^2 l_j^2}{2}\right),
\end{equation}
where $V_0$ is again a constant with units of energy. If we want a PP that has no angular momentum on the spurious degrees of freedom $w_3,w_4,...$, we find
\begin{equation}
U^m(k)= V_0L_m\left(\frac{k^2 l_B^2N}{2 (N-1)}\right).
\label{l2}
\end{equation}
This reproduces the familiar result $U^m(k)= L_m\left(k^2
  l_B^2\right)$ for $N=2$. For $N=3$, $U^m(k)= L_m\left(\frac{3}{4}k^2
  l_B^2\right)$ where $m$ is the total relative angular momentum
characterized by $w_2=\frac{1}{2}((z_1-z_3) + (z_2-z_3))$. The latter
will be used extensively in the calculations of the next section.

\subsection{Discussion and Generalizations}
We can determine how the effective magnetic length $l_j$ should be
rescaled without assuming the explicit form $w_j$ defined in terms of the $z_i$s. By examining the diagonal elements of $R R^T$, we find that
\begin{equation}
l_i= l_B\sqrt{\sum_j^{N-1}R_{ij}^2+\sum_{j<k}^{N-1}R_{ij}R_{ik}} 
\end{equation}
for $i>1$. For $i=1$, $l_1=\sqrt{N}l_B$. Here, $w_1$ is the CM position
and $l_1$ is the effective magnetic length for the CM angular
momentum.

Eq. \eqref{vmm} can also be extended to cases beyond the LLL. There, the
$a$ and $a^\dagger$ terms in the third line of Eq.~\ref{mvm} will
not yield unity. If we consider the case where each particle occupies
a specific Landau Level, we will have to first calculate expressions
such as $e^{i\frac {l_j}{\sqrt{2}}ka^\dagger_j }e^{i\frac{l_j}{\sqrt{2}}\bar{k} a_j}$ before making the change of coordinates from $z$ to $w$. This is because the positions of the particles are indexed by $z$, not $w$.

To begin with, we rearrange the exponential factor in the Fourier transform $e^{i k_j\cdot w_j}=e^{i(k_jR_{ji})z_i}$ so that it depends explicitly on the $z_i$s, albeit with modified $k=(k_jR_{ji})$, summation implied. After some algebra, we find
\begin{equation}\langle n'| e^{\frac{il_j}{\sqrt{2}}ka^\dagger_j }e^{\frac{il_j}{\sqrt{2}}\bar{k} a_j}|n\rangle=
\left( - \frac{i l_Bk}{\sqrt{2}}\right)^{n'-n}\sqrt{\frac{n!}{n'!}}L_n^{n'-n}\left(\frac{l_B^2|k^2|}{2}\right),\end{equation}
where $n$ and $n'$ denote the initial and final Landau levels of the
particle. When $n=n'$, we just have $L_n(k^2 l_B^2/2)$. With the
states being reduced to the LLL, we proceed as in~\eqref{vmm}, arriving
at the general formula for the PP between $N$ particles
initially at LLs $n_1,n_2,\dots,n_N$ mapping to LLs
$n'_1,n'_2,\dots,n'_N$, with angular momenta $m_1,m_2,\dots,m_N$
associated with the coordinates $w_i=R_{ij}z_j$:
\begin{eqnarray}
V_{n'_1\dots n'_N;n_1\dots n_N}^{m_1m_2\dots m_N}&=&
\langle  n'_1,...,n'_N;m_1,...,m_N|V(w_1,w_2,...,w_N)| n_1,...,n_N;m_1,...,m_N\rangle\notag\\
&=&\left(\prod_j^N\int\frac{d^2(k_jl_j)}{\pi}e^{-|k_j|^2l_j^2/2}L_{m_j}\left(\frac{l_j^2|k_j|^2}{2}\right)\right)\prod_i^N
\left( \frac{-i l_B(R_{li}k_l)}{\sqrt{2}}\right)^{n'_i-n_i}\notag\\
&&\times\sqrt{\frac{n_i!}{n'_i!}}L_{n_i}^{n'_i-n_i}\left(\frac{l_B^2|(R_{li}k_l)^2|}{2}\right) V(k_1,...,k_N),
\label{grand}
\end{eqnarray}
with the effective magnetic lengths $l_j$ defined as before. This is the most general result in this section.

If $V$ is translationally invariant, i.e. it does not depend on $w_1=(z_1+\dots+z_N)/N$, the $k_1$ integral produces a delta function $\delta(k_1)$. Hence, as is usually the case, $k_1$ should be excluded from all sums in Eq.~\ref{grand}.
As an illustration for $N=2$ bodies with interaction independent of
the CM, only the $k_2$ integration survives. We have $R_{li}k_l=\pm k_2$
and $V=V(k_2)$. $m_1$, the CM angular momentum, is irrelevant for
the interaction, so $V_{m_1}=\delta_{m_1,0}$. We can also
deduce this result from the orthogonality of the Laguerre polynomials:
if we impose the further restriction $n_i=n'_i$ for all $i$, i.e. particles stay in their respective LLs, Eq.~\ref{grand} reduces to
\begin{equation}
V_{n_1n_2;n_1n_2}^{m_2}\propto\int d^2k e^{-l_B^2k_2^2}L_m(l_B^2k_2^2)L_{n_1}\left(\frac{l_B^2k_2^2}{2}\right)L_{n_2}\left(\frac{l_B^2k_2^2}{2}\right)V(k_2).
\label{simple}
\end{equation}

\section{Derivation of 2-body Pseudopotential Hamiltonians via explicit integration}
\label{2body}

Here we switch back to the second-quantized form of the PPs, and present the details of the derivation of the two-body PPs $U^m$. Here the details of the brute-force integration approach will be presented; an alternative geometric approach has been explained in the main text around Tables \ref{N2} and \ref{N3}. We shall explicitly work through only the fermionic case, since the bosonic case can be analogously derived.
From Eq.~\ref{lag}, a special case of Eq.~\ref{simple}, the potential $U^m$ that is nonzero only for the relative angular momentum sector $m$ is given by $U^m(r-r')= V_0 L_m(-l_B^2\nabla^2)\delta^2(r-r')$, where $V_0$ is a constant with units of energy.  We find its LLL Landau gauge basis matrix elements $U^m_{n_1n_2n_3n_4}$ by projecting onto the basis wavefunctions $\psi_n(r)=\frac{1}{\sqrt{\sqrt{\pi}L_yl_B}}e^{i\frac{\kappa}{l_B}ny}e^{-\frac{(x-\kappa nl_B)^2}{2l_B^2}}$, where $L_y$ is the circumference of the cylinder:
\begin{eqnarray}
&&U^m_{n_1n_2n_3n_4}\notag\\
&=&\frac{V_0}{4}\int d^2r
d^2r'\psi^\dagger_{n_1}(r)\psi^\dagger_{n_2}(r') L_m(-l_B^2
\nabla^2)\delta^2(r-r')\psi ^{\phantom{\dagger}}_{n_3}(r')\psi ^{\phantom{\dagger}}_{n_4}(r) + \text{antisymm} \nonumber \\
&=&\frac{V_0}{4}\int \frac{l_B^2 d^2q}{(2\pi)^2}\int d^2r d^2r' L_m(q^2l_B^2)e^{i q\cdot (r-r')}\psi^\dagger_{n_1}(r)\psi^\dagger_{n_2}(r')\psi ^{\phantom{\dagger}}_{n_3}(r')\psi ^{\phantom{\dagger}}_{n_4}(r) + \text{antisymm}.\notag\\
\label{pseudopotmain}
\end{eqnarray}
Recall that $\kappa=\frac{2\pi l_B}{L_y}$ is a dimensionless ratio that
is small in the limit of large magnetic fields. The two types of MT
symmetry constraints mentioned in Section V are manifest in the above
expression as (i) the CM conservation which
corresponds to $n_1+n_2=n_3+n_4$, and (ii) one-dimensional translation
symmetry, which is the invariance of $U^m$ under $n_i\rightarrow n_i
+a$, where $a$ is an integer and $i=1,2,3,4$. (A similar observation
has been independently made in Ref.~\onlinecite{gunnar}.)
CM conservation must be present because the $\int dy$ and $\int dy'$ integrals produce delta functions of the form
\[ \delta\left(\frac{2\pi}{L_y}(n_4-n_1)+q_y\right)\] and
\[ \delta\left(\frac{2\pi}{L_y}(n_3-n_2)-q_y\right).\]

The CM conservation condition $n_1+n_2=n_3+n_4$, i.e. $n=n'$, appears after we combine these two delta functions. Since the CM of each LLL wavefunction occurs along $x=\kappa l_Bn\propto n$, we see that the CM of the particles must indeed be equal before and after a two-body hopping.

\begin{figure}[t]
\begin{minipage}{0.89\linewidth}
\includegraphics[width=0.47\linewidth]{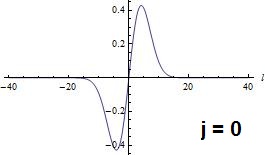}
\includegraphics[width=0.47\linewidth]{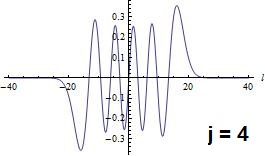}
\includegraphics[width=0.47\linewidth]{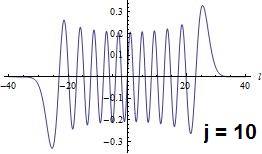}
\includegraphics[width=0.47\linewidth]{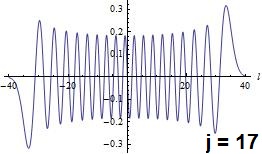}
\end{minipage}
\caption{Profiles of $b^{2j+1}$ from Eq.~\ref{bj} for $j=0,4,10,17$ with $\kappa=\frac{1}{6}$. As $j$ increases, the
  main region of contribution of $b^{2j+1}$ shifts in the direction of
  larger $|l|$.}
\label{bpic}
\end{figure}

We explicitly resolve the MT symmetry via the translation
$n_i\rightarrow n_i+1$, as Fourier terms cancel upon
$\psi_i(r)\rightarrow \psi_i\left(r-\frac{2\pi l_B^2}{L_y}\right)$. This
is exactly a magnetic translation under which the system in a magnetic
field $B=\frac{\hbar c}{el_B^2}$ is expected to be invariant.
To continue the calculation, reduce $U^m$ to the $\int d^2q$ integral
\begin{eqnarray}
&&U^m_{l_1l_2}\notag\\
&=& \frac{V^0\pi l_B^2}{4\pi L_y^2l_B^4}\frac{L_y}{2\pi}\int d^2q
\delta\left(q_y+\frac{2\pi}{L_y}(l_1-l_2)\right)L_m(l_B^2q^2)e^{-\frac{l_B^2q^2_x}{2}}e^{i\kappa
  l_B(l_1+l_2)q_x}e^{-\frac{\kappa^2 (l_1-l_2)^2}{2}}\notag\\ &&+\text{antisymm}
\nonumber \\
&=&\frac{V^0\pi l_B^2}{4\pi L_y^2l_B^2}\frac{L_y}{2\pi}\int d^2q
\delta\left(q_y+\frac{2\pi}{L_y}(l_1-l_2)\right)L_m(l_B^2q^2)e^{-\frac{l_B^2q^2}{2}}e^{i\kappa
  l_B(l_1+l_2)q_x}+\text{antisymm}.\notag\\
\label{qint}
\end{eqnarray}
The second line follows from the fact that $q_y$ is constrained to be
$q_y=\frac{2\pi}{L_y}(l_2-l_1)=\frac{\kappa}{l_B}(l_2-l_1)$. Note that
$n$ completely disappears from the expression, as expected from MT
symmetry. A closed form expression for $U^m$ is given by
\begin{eqnarray}
U^m_{l_1l_2}
&=&g\kappa^3\sum_{p=0}^m\frac{(-1)^{p+1}m!}{p!(m-p)!}\sum_{j=0}^p\sum_{r=0}^{j/2}\frac{\Gamma(p-j+r+1/2)2^{m+j-r-3}(i\kappa)^{2(j-r-1)}}{(p-j)!(2r)!(j-2r)!\sqrt{\pi}(l_1l_2)^{2r-j}}\notag\\
&&\times\left[(l_1+l_2)^{2r}-(-1)^j(l_1-l_2)^{2r}
\right]e^{-\kappa^2(l_1^2+l_2^2)}.\nonumber \\
\end{eqnarray}
The lengthy expression above can be factorized into the form
$U^m_{l_1l_2}=g\kappa^3 b^m_{l_1}b^m_{l_2}$, $g={4V_0}{(2\pi)^{3/2}}$, where
\begin{equation}
b^{2j+1}_l = le^{-\kappa^2l^2}\sum_{p=0}^j \frac{(-2)^{3p-j}(\kappa l)^{2p}\sqrt{(2j+1)!}}{(j-p)!(2p+1)!}.
\label{bj}
\end{equation}
This result can be proven by induction. The first few $b^{2j+1}$s are
\[ b^1_l=le^{-\kappa^2l^2},\]
\[ b^3_l=\frac{1}{\sqrt{3!}}(-3+4\kappa^2l^2)le^{-\kappa^2l^2},\]
\[ b^5_l=\frac{1}{\sqrt{5!}}(15-40\kappa^2l^2+16\kappa^4l^4)le^{-\kappa^2l^2},\]
\[...\]
\[ b^{2j}_l=0.\]
which agree exactly with the expressions in Table \ref{N2} equivalently obtained via orthogonalization (with $p$ the polynomial part of $b$). The profiles of a few chosen $b$'s are depicted in Fig.~\ref{bpic}. Note that the $U^m$ operators can always be decomposed into the
product of two $b^m$ operators that are $m$ degree polynomials of
$l_1$ and $l_2$. These polynomials have the physically relevant
property that (i) the $b^m_l$ of higher $m$ are "localized" at larger
values of $l$ and (ii)
that they are orthogonal in the limit of $\kappa\rightarrow 0$ before
we enforce the $L_xL_y$ periodicity in the $x$-direction of the cylinder. This is
further explained in Appendix~\ref{ortho}.

We can similarly calculate the bosonic PPs through Eq. \ref{qint}, but with terms symmetrized instead of antisymmetrized over $l_1$ and $l_2$. As before, the PPs can be written as $U^m_{l_1l_2}=g\kappa^3 c^m_{l_1}c^m_{l_2}$, $g={4V_0}{(2\pi)^{3/2}}$. The results are also given in Table \ref{N2}. 

\section{Orthogonality of 2-body PPs $U^m$ on the torus}
\label{ortho}

The FQH PPs $U^m$ which we have obtained by integrating over a continuum are not strictly orthogonal once we place them on a finite cylinder or torus. The physical reason for this is that relative angular momentum is not even well-defined when its corresponding spatial separation is comparable to the dimensions of the system. While we are only concerned with the $L_x=L_y$ case in this appendix, it will be interesting to investigate how deviations from orthogonality depend on the aspect ratio set by $L_x$ and $L_y$. 

\subsection{From plane to cylinder}
When we compactify the plane into a cylinder, the notion of relative
angular momentum is no longer well-defined. In a fixed LL, the
relative angular momentum is proportional to the interparticle
distance which can only be meaningfully defined when the latter is much smaller than $L_y$. Recall that the effective $\kappa$ relation $\kappa=\frac{2\pi l_B}{L_y}=\frac{1}{L_y}$ after mapping to the FCI system. We thus expect orthogonality to occur only in the limit of large $L_y$, or, equivalently, small $\kappa$.
To proceed, we evaluate the overlap elements of two PPs $V_{\text{cyl}}^m$ and $V_{\text{cyl}}^n$, and show that the latter are orthogonal for sufficiently small $\kappa$ ratios.
\begin{eqnarray}
&&\langle {U}_{\text{cyl}}^m,U_{\text{cyl}}^n \rangle\notag\\
&\propto & \sum_{ l_1
  l_2}U^m_{\text{cyl }  l_1, l_2}U^n_{\text{cyl }  l_1, l_2}
\nonumber \\
&\propto&\sum_{ l_1 l_2}\int d^2q\int d^2q'
\delta(q_y+\frac{2\pi}{L_y}(l_1-l_2))\delta(q'_y-q_y)L_m(l_B^2q^2)L_n(l_B^2q'^2)\notag\\ && \times e^{-\frac{l_B^2(q^2+q'^2)}{2}}e^{i\kappa
  l_B(l_1+l_2)(q_x+q'_x)}
  \nonumber \\
&\propto &\sum_{ l_1+l_2, l_1-l_2}\int dq_y\int dq_x \int dq'_x
\delta(q_y+\frac{2\pi}{L_y}(l_1-l_2))L_m(l_B^2q^2)L_n(l_B^2q^2)e^{-\frac{l_B^2(q^2+q'^2)}{2}} \notag\\ && \times e^{i\kappa
  l_B(l_1+l_2)(q_x+q'_x)}
 \nonumber \\
&\propto&\sum_{l_1-l_2}\int dq_y\int dq_x \int dq'_x
\delta(q_y+\frac{2\pi}{L_y}(l_1-l_2))L_m(l_B^2q^2)L_n(l_B^2q^2)e^{-l_B^2q^2}\delta(q_x+q'_x)
\nonumber \\
&\propto&\int d^2q
\left(\sum_{l_1-l_2}\delta(q_y+\frac{2\pi}{L_y}(l_1-l_2))\right)L_m(l_B^2q^2)L_n(l_B^2q^2)e^{-l_B^2q^2}
\nonumber \\
&\approx&\pi\int
2(ql_B)d(ql_B)\left(\sum_{l_1-l_2}\delta((q_yl_B)+\kappa(l_1-l_2))\right)L_m(l_B^2q^2)L_n(l_B^2q^2)e^{-l_B^2q^2}
\nonumber \\
&\rightarrow & \pi \delta_{mn}.
\end{eqnarray}
Two approximations have been made above. From the third last to second
last line, we replace the non-rotationally invariant integral over
$\int dq_x dq_y$ by the rotationally invariant integral $\pi\int
2qdq$. From the second last to the last line, the delta function sum
in the large parentheses was replaced by unity, i.e. taking the limit where the range of $l_1-l_2$ (and hence $q$) tends to infinity. These approximations become exact when the discrete $q_yl_B$ becomes a continuum, which occurs precisely when $\kappa\rightarrow 0$. Indeed, this agrees with the physical intuition that the relative angular momentum becomes a well-defined quantity when $L_y$ is large.
From the viewpoint of polynomial orthogonality, we see that the
orthogonality of the $U_{\text{cyl}}^m$ is respected as much as as the
integral over Laguerre polynomials is allowed to be made
continuous. Specifically, the orthogonality of the Laguerre
polynomials is exact only for the continuum $q^2=q^2_x+q^2_y$ and not
for discrete points specified by $\Delta l =l_1-l_2$ whose separations do not vanish unless $\kappa= 0$.

\subsection{From cylinder to torus}

The compactification of the cylinder into a torus introduces a periodicity in the $L_x$-direction.
This introduces periodic copies of $U^m_{\text{cyl}}$ in
$U^m_{\text{tor}}$, each displaced from another by $(L_xL_y,L_xL_y)$
or $(L_xL_y,-L_xL_y)$ sites in $l_1-l_2$ space. Nonorthogonality is
expected when there is significant overlap between these images. Let us obtain a bound by finding the conditions where $|U_{\text{cyl}}^m|$, whose explicit form is given by~\eqref{bj}, is not negligible, with $l_1,l_2$ being of the order of $L_xL_y$. At such values,
\begin{eqnarray}
 U^{m}_{\text{cyl} l_1 l_2} &\sim& e^{-2(\kappa L_xL_y)^2}(2^{3/2}\kappa L_xL_y)^{2m}\nonumber \\
 & =& e^{-2w^2}(2^{3/2}w)^{2m}\nonumber \\
 &=&f(w),
\end{eqnarray}
where $w=\kappa L_xL_y$. The function $f(w)$ exhibits a rapid Gaussian decay beyond $w\approx 2\sqrt{m}$. Hence we expect the PP $U^m$ to be nonorthogonal when
\begin{equation}  < \frac{2\sqrt{m}}{L_xL_y}, \end{equation}
which implies that
\begin{equation} L_x <2\sqrt{m},
\label{orthobound}
\end{equation}
a bound well-verified by calculations in the following subsection.

\subsection{Numerical results for the orthonormality of the fermionic PPs}
\begin{figure}
\centering
\begin{minipage}{0.7\linewidth}
\includegraphics[width=0.99\linewidth]{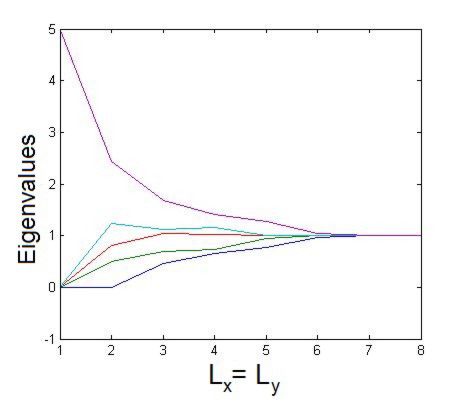}
\end{minipage}
\caption{  The eigenvalues of the overlap matrix of
  $U^1,U^3,U^5,U^7$ and $U^{9}$ as a function of the system size
  $L=L_x=L_y$. Indeed, we observe orthogonality when $L\geq 6$, in
  excellent agreement with the bound $2\sqrt{9}= 6$ from
  Eq.~\ref{orthobound}.}
\label{fig:phaseKM}
\end{figure}

The orthonormality of the $U^m$s on a finite system can be studied quantitatively through
their overlap matrix defined by
\begin{equation}
M_{mn}=\langle U^m, U^n \rangle
\end{equation}
according to Eq.~\ref{over}. When the $U^m$s are orthonormal, it holds
$M_{mn}=\mathbb{I}$. If the $U^m$s are not orthonormal while they span an orthonormal basis, the
eigenvalues of $M_{mn}$ will still be unity since we can find a
unitary transformation where $M_{mn}$ is diagonal. When $U^m$s are
overcomplete, however, the spectrum of $M$ broadens and yields
eigenvalues deviating from this limit.
In Fig.~\ref{fig:phaseKM}, we plot the eigenvalues of the overlap
matrix for the first few PPs as a function of
$L$. Orthogonality is hence broken when the eigenvalues differ
from unity. Indeed, we observe that orthogonality improves with system
size, in agreement with the conclusions of the preceding
subsections.
\\

\section{Derivation of 3-body bosonic Pseudopotentials via explicit integration}
\label{3body}

Let $U^m$ be the bosonic 3-body CM-invariant potential which is
nonzero only in the sector of total relative angular momentum
$m$. Such $3$-body $U^m$s will be useful as a basis through which arbitrary CM conserving potentials can be expanded. In this appendix, we will show the detailed derivation of $U^m$ in the basis of LLL Landau gauge eigenfunctions, starting from Eq.~\ref{pseudopotmain}.

According to Eq.~\ref{l2}, we have to replace the magnetic length
$l_B^2$ by $\frac{3}{4}l_B^2$. This, however, causes no difference when
$m=0$, since $L_0=1$, so the results in Refs. \onlinecite{lee2004} and
\onlinecite{dunghai2006} for $U^0$ could have been obtained without
this modification along our GHPs.

After performing the $\int d^2r_i$ integrations according to Eq.~\ref{pseudopotmain},
\begin{equation}
U^m_{n_1n_2n_3n_4n_5n_6}\propto\sum_\sigma\int \frac{d^2qd^2p}{(2\pi)^4} I(q,n_1,n_6)I(p-q,n_2,n_5)I(-p,n_3,n_4)L_m\left(\frac{3}{4}(p-q)^2 l_B^2\right),
\end{equation}
where
\begin{equation}
I(q,n,n')=2\pi\delta\left(q_y+\frac{\kappa}{l_B}(n-n')\right)e^{-l_B^2q^2_x/4}e^{i\kappa l_B(n+n')q_x/2}e^{-\kappa^2(n-n')^2/4}.
\end{equation}
Here, $\sum_\sigma=\frac{1}{(3!)^2}\sum_{perm(n_1n_2n_3)}\text{sgn}(perm)^f\sum_{perm(n_6n_5n_4)}\text{sgn}(perm)^f$ where $f=1$ for fermions and $f=0$ for bosons. $perm$ refers to the permutations of the $n_i$s while $sgn(perm)=\pm 1$ depending on whether the permutation is even or odd.
Up to now, the expression is easily generalizable to any number of fermions or bosons by increasing the number of integrals and $I(q,n,n')$s. If we further restrict ourselves to the 3-body bosonic case, we can simplify it to
\begin{eqnarray}
&&U^m_{n_1n_2n_3n_4n_5n_6} \nonumber \\
&\propto& e^{-\kappa ^2([(n_1-n_6)^2+(n_2-n_5)^2+(n_3-n_4)^2]/4}\sum_\sigma
\int \frac{d^2q_xd^2p_x}{(2\pi)^4}L_m\left(\frac{3}{4}(p-q)^2
  l_B^2\right)\notag\\
	&&e^{-l_B^2(p^2+q^2+(p-q)^2)/4}e^{i\kappa l_B N_1 q/2}e^{i\kappa l_B N_2 p/2}
\nonumber \\
&\propto& \sum_\sigma P_m(n_1,n_2,n_3,n_4,n_5,n_6;\kappa ^2)
\frac{4\pi}{\sqrt{3}}e^{-\frac{\kappa ^2}{6}(N_1^2+N_1N_2+N_2^2)}, \nonumber \\
\end{eqnarray}
where $N_1=n_1+n_6-n_2-n_5$,
$N_2=n_2+n_5-n_3-n_4$. $P_m(n_1,n_2,n_3,n_4,n_5,n_6;\kappa ^2)$ is a
potentially complicated polynomial whose form depends on $m$. We
define the conserved CM "Landau level wavefunction index" $R$ via $3 \text{R \text{mod} N}=n_1+n_2+n_3=n_6+n_5+n_4$. The existence of $R$ is a conseqence of the total CM conservation of $U^m$.

Below we shall present some calculation details of the first few $U^m$s of the bosonic 3-body PPs, although the same results which appear in Table \ref{N3} can also be obtained via a geometric approach.

\subsection{m=0 for bosons on a torus}
In this case, $L_0=1$, and $P_0(n_1,n_2,n_3,n_4,n_5,n_6;\kappa^2)=1$. Since each of the $n_i$'s is also defined modulo $N$ (but constrained to sum to $3R$ as shown above), we make the replacement
\begin{eqnarray}
&& e^{-\frac{\kappa ^2}{2}((R-n_1)^2+(R-n_2)^2+(R-n_3)^2)}\rightarrow
\nonumber \\
&& \sum_{s,t} e^{-\frac{\kappa ^2}{2}((R-(n_1+Ns))^2+(R-(n_2+Nt))^2+(R-(n_3-N(t+s)))^2)}, 
\end{eqnarray}
and likewise for the identical factor involving $n_4,n_5,n_6$. Hence the $U^m$ factorizes into a product of nonlocal operators $\hat{b}_R$:
\[ U^0 \propto \sum_R \hat{b}_R^\dagger \hat{b}_R^{\phantom{\dagger}}, \]
where
\begin{eqnarray}
b_R&=& \sum_{\sum_i n_i=3 R \text{mod} N}\left[\sum_{\sum s_i=0}
  e^{-\frac{\kappa ^2}{2}\sum_i(R-(n_i+Ns_i))^2} \right]
c_{n_1}c_{n_2}c_{n_3} \nonumber \\
&=&\sum_{n_1+n_2+n_3=3 R \text{mod} N}\left[\sum_{s,t} e^{-\frac{\kappa ^2}{3}W_{st}} \right] c_{n_1}c_{n_2}c_{n_3},
\end{eqnarray}
with $W_{st}=\sum_i {n'}_i^2 -\sum_{i<j}n'_i n'_j$, $n'_1=n_1+sN$, $n'_2=n_2+tN$ and $n'_3=n_3-N(s+t)$.
This result is identical to that in Ref. \onlinecite{dunghai2006}.

\subsection{m=1 for bosons on a torus}

In this case, the 1st Laguerre polynomial gives a factor $1-\frac{3}{4}(\kappa ^2(n_2-n_5)^2+p_x^2)$, and $\sum_\sigma P_1$ evaluates to
\begin{eqnarray}
\sum_\sigma P_1&=& \sum_\sigma 3\kappa ^2(n_2-R)(n_5-R) \nonumber \\
&=&
3\kappa ^2\left(\frac{n_2+n_1+n_3}{3}-R\right)\left(\frac{n_4+n_5+n_6}{3}-R\right)
\nonumber \\
&=&0
\end{eqnarray}
Hence we have
\begin{equation}
U^1=0.
\end{equation}
This agrees with the result from Ref. \onlinecite{simon2007}, in that
there is no PP of total relative angular momentum $m=1$
for bosons. This is because the CM invariance of the interaction
precludes any symmetric wave function of total degree 1.

\subsection{m=2 for bosons on a torus}
After some algebra,
\[P_2=\frac{1}{2}(1-3\kappa ^2(n_2-R)^2)(1-3\kappa ^2(n_5-R)^2).\]
Since
\begin{eqnarray}
\sum_\sigma (n_2-R)^2=&=& \sum_\sigma( n_2^2-2n_2 R + R^2) \nonumber \\
&=& \frac{1}{3}\sum_{i=1}^3 (n_i^2 - 2R^2 + R^2)\nonumber \\
&=& \frac{2}{9}(\sum_i n_i^2 -\sum_{i<j}n_in_j),
\end{eqnarray}
we have
\[ U^2 \propto \sum_R \hat{b}_R^\dagger \hat{b}_R ^{\phantom{\dagger}}, \]
where
\begin{equation}
\hat{b}_R=\sum_{\sum_i n_i=3 R \text{mod} N}\left[\sum_{s,t} \left(1-\frac{2\kappa ^2}{3}W_{st}\right)e^{-\frac{\kappa ^2}{3}W_{st}} \right] c_{n_1}c_{n_2}c_{n_3}
\end{equation}
with $W_{st}=\sum_i {n'}_i^2 -\sum_{i<j}n'_i n'_j$, $n'_1=n_1+sN$, $n'_2=n_2+tN$, and $n'_3=n_3-N(s+t)$ as before.

\subsection{m=3 for bosons on a torus}

This is zero for 2 bosons, but not for 3 bosons~\cite{simon2007}. Indeed,
\[\sum_\sigma P_3= \sum_\sigma \frac{9}{2}\kappa ^6(n_2-R)^3(n_5-R)^3 = \frac{9}{2}\kappa ^6\prod_{i=1}^6(n_i-R).\]
Hence
\[ U^3 \propto \sum_R \hat{b}_R^\dagger \hat{b}_R ^{\phantom{\dagger}}, \]
where
\begin{eqnarray}
\hat{b}_R&=&\frac{-3\kappa ^3}{\sqrt{2}}\sum_{\sum_n n_i=3 R \text{mod} N}\notag\\
&&\times\left[\sum_{s,t} (n_1+sN-R)(n_2+tN-R)(n3-(s+t)N-R)e^{-\frac{\kappa ^2}{3}W_{st}} \right] c_{n_1}c_{n_2}c_{n_3}.\notag\\
\end{eqnarray}

\subsection{m=4 for bosons on a torus}
After the smoke clears, we find
\[ U^4 \propto \sum_R \hat{b}_R^\dagger \hat{b}_R^{\phantom{\dagger}}, \]
where
\begin{eqnarray}
&&\hat{b}_R=\nonumber \\
&&\sum_{\sum_i n_i=3 R \text{mod} N}\left[\sum_{s,t}
  \left(1-\frac{4\kappa ^2W_{st}}{3}+\frac{\kappa
      ^4W^2_{st}}{9}\right)e^{-\frac{\kappa ^2}{3}W_{st}} \right]
c_{n_1}c_{n_2}c_{n_3}, \nonumber \\
\end{eqnarray}
with $W_{st}=\sum_i {n'}_i^2 -\sum_{i<j} n'_i n'_j$, $n'_1=n_1+sN$,
$n'_2=n_2+tN$, and $n'_3=n_3-N(s+t)$ as before.

The above results form the first few entries of Table \ref{N3}, the rest of which are obtained via geometric orthogonalization. 

\section{Barycentric coordinates for many-body pseudopotentials}
\label{sec:barycentric}

Here we describe how to find a manifestly $S_N$-symmetric embedding of the tuple $(\bar n_1,...,\bar n_N)$ onto the $(N-1)$-dimensional simplex in $\mathbb{R}^{N-1}$. This can be achieved in barycentric coordinates, which is useful for diverse applications involving permutation symmetry with a linear constraint~\cite{leeandy2014}. The most straightforward construction is to write 
\begin{equation}
\vec x = \kappa\sum_{k=1}^N \bar n_k \vec\beta_k,
\label{xbary}
\end{equation}where $\vec x\in \mathbb{R}^{N-1}$ and $\{\vec\beta_k\}$, $k=1,...,N$ forms a linearly dependent set of basis vectors, also in $\mathbb{R}^{N-1}$, normalized so that 
\begin{equation}
\vec\beta_j \cdot \vec\beta_k = \frac{N}{N-1}\delta_{jk} - \frac{1}{N-1}.
\label{betadot}\end{equation}
Geometrically, the $\beta_k$'s define the vertices of a simplex, and are at angle of $\cos^{-1}(-1/(N-1))$ from one another. The vertex $k$ corresponds to the least isotropic configuration with $\bar n_k\propto \frac{N-1}{N}$ and $\bar n_{j}\propto -\frac{1}{N}$, $j\neq k$. A basis consistent with the above requirements is \begin{subequations}\begin{align}
\vec\beta_1 &= (1,0,\ldots,0), \\
\vec\beta_2 &= (C_1, S_1,\ldots,0), \\
\vec\beta_3 &= (C_1, S_1C_2, S_1S_2,\ldots,0), \\
\vec\beta_4 &= (C_1, S_1C_2, S_1S_2C_3,S_1S_2S_3,\ldots,0), \\
&\vdots \notag \\
\vec\beta_{N-1} &= (C_1,S_1C_2,S_1S_2C_3,\ldots, S_1\cdots S_{N-2}), \\
\vec\beta_N &= (C_1,S_1C_2,S_1S_2C_3,\ldots, -S_1\cdots S_{N-2}),
\end{align}\label{barybasis}\end{subequations}
with $S^2_k+C^2_k=1$, $1\leq k\leq N-2$. Upon enforcing the scalar product constraint in Eq. \ref{betadot}, we require that $S_{k+1}=1-(1+1/N)/(\prod_{j=1}^k S_k)^2)$, which also implies that $C_k^2+S_k^2C_{k+1}=C_k$.  
A simple solution fortunately exists: 
\begin{equation}
C_k = -\frac{1}{N-k}, %\;\;\;\; S_k^2+C_k^2 = 1,
\label{recur2}\end{equation}
which implies that the $k$-th projected component of the relative angles between the position vectors of the vertices approaches $\pi$ as $k$ and $N$ increase. We proceed by substituting Eq.~\ref{recur2} into the explicit form of vertex positions $\vec\beta_k$, and expressing the latter in terms of the spherical coordinates.
From Eq. \ref{xbary}, we can easily check \begin{equation}
\bar n_k =  \frac{N-1}{N\kappa} \vec x \cdot \vec\beta_k,  \label{eqg10}
\end{equation}
and that 
\begin{equation}
W^2=|\vec x|^2 = \frac{N\kappa^2}{N-1}\sum_k^N\left(n_k-\frac{n}{N}\right)^2=\frac{N\kappa^2}{N-1}\sum_k^N \bar n_k^2.
\label{eqg11}
\end{equation}
For the purpose of orthogonalizing the PPs over the inner product measure in Eq.~\ref{innerproduct}, we will also need to express $\vec x$ explicitly in terms of angles in origin-centered spherical coordinates: \begin{subequations}\begin{align}
x_1 &= W \cos \varphi_1, \\
x_2 &= W \sin\varphi_1 \cos\varphi_2, \\
&\vdots \notag \\
x_{N-2} &= W  \cos \varphi_k \prod_{k=1}^{N-3}\sin\varphi_k, \\
x_{N-1} &= W \prod_{k=1}^{N-2}\sin\varphi_k.
\end{align}\label{spherical}\end{subequations}

Substituting the explicit expressions from Eqs. \ref{barybasis} and \ref{spherical} into Eq. \ref{eqg10}, we obtain
\begin{align}
\bar n_1&=  \frac{N-1}{N\kappa }W\cos\varphi_1,\\
\bar n_2&=  \frac{W}{N\kappa }\left(-\cos\varphi_1+\sqrt{N(N-2)}\sin\varphi_1\cos\varphi_2\right),\\
\bar n_k&= \frac{W}{N\kappa }[-\cos\varphi_1-\sqrt{\frac{N}{N-2}}\sin\varphi_1\cos\varphi_2 \notag \\
& \;\;\;\;\; - \sum_{j=2}^{k-2}\sqrt{\frac{N!}{(N-j-2)!}}\frac{\left(\prod_{i=1}^{j} \sin\varphi_i\right )\cos\varphi_{j+1}}{(N-j)(N-j-1)}\notag\\
& \;\;\;\;\; + \sqrt{\frac{N!}{(N-k-1)!}}\frac{\left(\prod_{i=1}^{k-1} \sin\varphi_i\right )\cos\varphi_{k}}{N-k+1}],\label{thirdline}\\
&\vdots \notag \\
\bar n_{N-1}&= \frac{W}{N\kappa }[-\cos\varphi_1-\sqrt{\frac{N}{N-2}}\sin\varphi_1\cos\varphi_2 \notag \\
& \;\;\;\;\; - \sum_{j=2}^{N-3}\sqrt{\frac{N!}{(N-j-2)!}}\frac{\left(\prod_{i=1}^{j} \sin\varphi_i\right )\cos\varphi_{j+1}}{(N-j)(N-j-1)}\notag\\
& \;\;\;\;\; + \frac{\sqrt{N!}}{2}\left(\prod_{i=1}^{N-2} \sin\varphi_i\right)],\\
\bar n_{N}&= \frac{W}{N\kappa }[-\cos\varphi_1-\sqrt{\frac{N}{N-2}}\sin\varphi_1\cos\varphi_2 \notag \\
& \;\;\;\;\; - \sum_{j=2}^{N-3}\sqrt{\frac{N!}{(N-j-2)!}}\frac{\left(\prod_{i=1}^{j} \sin\varphi_i\right )\cos\varphi_{j+1}}{(N-j)(N-j-1)}\notag\\
& \;\;\;\;\; - \frac{\sqrt{N!}}{2}\left(\prod_{i=1}^{N-2} \sin\varphi_i\right)]
\label{qs}
\end{align}
These are the explicit expressions for transcribing the $\bar n_j$, $1\leq j\leq N$ indices directly into spherical coordinates. In~\eqref{thirdline}, $k$ ranges from  for $3$ to $N-2$.

\section{Geometric construction of pseudopotentials: spinful case}
\label{sec:spinful}

In the presence of ``internal degrees of freedom" (DOFs) which we also refer to as ``spins" or ``components'' for simplicity, there is a considerably greater diversity PP possibilities. This is because the PPs consist of products of spatial and spin parts, and either part can have many possible symmetry types, as long as they conspire to produce an overall (anti)symmetric PP in the case of (fermions) bosons.

A generic multicomponent PP takes the form
\begin{equation} 
U^N_m=\sum_\lambda\sum_\sigma U^N_{m,\lambda} \ket{\sigma^\lambda }\bra{\sigma^\lambda }. 
\end{equation}
The notation here requires some explanation. $\lambda=[\lambda_1,\lambda_2,...]$ refers to the symmetry type and $U^m_\lambda$ is a spatial term (in $(\bar n_1,...,\bar n_N)$-space) possessing that symmetry. $\ket{\sigma^\lambda} =\ket{\sigma_1\sigma_2...}$ refers to an (internal) spin basis that is consistent with the symmetry type $\lambda$. Each symmetry type corresponds to a partition of $N$, with $\sum_j \lambda_j=N$. For $N$ bosons, $\lambda$ represents the situation where there is permutation symmetry among the first $\lambda_1$ particles, among the next $\lambda_2$ particles, etc. but \emph{no additional symmetry} between the $\lambda_i$ subsets. This is often represented by the Young Tableau with $\lambda_i$ boxes in the $i^{th}$ row. For fermions, we use the conjugate representation $\bar\lambda$, with the rows replaced by columns and symmetry conditions replaced by antisymmetry ones. For instance, the totally (anti)symmetric types are $[1,1,...,1]$ and $[N]$, respectively.

Hence the space of PPs is specified by $3$ important parameters: $N$ -- the number of particles that  interact, $m$ -- the total relative angular momentum or the total polynomial degree in $\bar n_i$, and $b$ -- the number of internal DOFs (spin). While the symmetry type $\lambda$ and hence $U^N_{m,\lambda}$ depends only on $N$ and $m$, the set of possible $\ket{\sigma^\lambda }$ also depends on $b$.
To further illustrate our notation, we specify $N,m,b$ parameters for the interactions relevant to some commonly known states. In the archetypal single-layer FQH states, we have $b=1$ components, and the $N=2$ PP interactions for the Laughlin state penalize pairs of particles with relative angular momentum $m$, where $\frac1{m+2}$ is the filling fraction. For bilayer FQH states, we have $b=2$ and $N=2$-body interactions. PPs as energy penalties in the $[2]$ sectors with angular momentum $<m$ and $[1,1]$ sectors with angular momentum $<n$ give rise to the Halperin $(mmn)$ states. Here, the $[2]$ sector is also known as the triplet channel, as it is spanned by the following three basis vectors: $\{\ket{\uparrow\uparrow},\ket{\uparrow\downarrow}+\ket{\downarrow\uparrow},\ket{\downarrow\downarrow}\}$. By contrast, the $[1,1]$ sector only contains $\{\ket{\uparrow\downarrow}-\ket{\downarrow\uparrow}\}$ as dictated by antisymmetry.

\subsection{Multicomponent pseudopotentials for a given symmetry}

The construction of multicomponent PPs here parallels that of multicomponent wave functions described in Ref.~\onlinecite{davenport2012}. For completeness, we first review this construction, and proceed to show how an orthonormal multicomponent PP basis, adapted to the cylinder or torus, can be explicitly found through the geometric approach. We first describe how to find the spatial part of the PP $U^N_{m,\lambda}$, and then the spin basis $\ket{\sigma^\lambda}$. 

\subsubsection{Spatial part}

For each symmetry type $\lambda$, we can construct the spatial part $U^N_{m,\lambda}$ with elementary symmetric polynomials in \emph{subsets} of the particle indices $\bar n_i=n_i-n/N$. They are, for instance, $S_{1,12}= \bar n_1+\bar n_2$, $S_{2,234}=\bar n_2\bar n_3+\bar n_3 \bar n_4 +\bar n_2\bar n_4$, etc. Of course, we must have $S_{1,123...N}=\sum_i \bar n_i =0$.

Like in the single-component case, the spatial part $U^N_{m,\lambda}$ consists of a primitive polynomial which enforces the symmetry, and a totally symmetric factor that does not change the symmetry. Here, the main step in the multicomponent generalization is the \emph{replacement} of primitive polynomials $1$ and $A$ by primitive polynomials consistent with the symmetry type $\lambda$. 
As the simplest example, the primitive polynomial in 
$U^{N=3}_{m=1,[2,1]}$ is $S_{1,12}=-\bar n_3$ (and cyclic permutations). It is the only possible degree $m=1$ expression symmetric in two (but not all three) of the indices. 

In general, there can be more than one candidate monomial obeying a symmetry consistent with $\lambda$. 
For instance, for $U^{N=3}_{m=2,[2,1]}$ they are $S_{1,12}^2=(\bar n_1+\bar n_2)^2$ and $S_{2,12}=\bar n_1\bar n_2$ (and cyclic permutations thereof). To find the primitive polynomials, we will have to construct one or more linear combinations of these terms which do not have any higher symmetry other than $[2,1]$ (i.e. in this case, this higher symmetry channel could be $[3]$). Elementary computation reveals that the only primitive polynomial should be $S_{1,12}^2+2S_{2,12}$, because it is the only linear combination that is manifestly symmetric in indices $1,2$ {\it and} disappears upon symmetrization over all three particles.  

As a more involved example, we demonstrate how to find the primitive polynomial corresponding to the symmetry type $[2,2]$ for $m=2$, $N=4$. Independent monomials that satisfy this symmetry include $S_{1,12}^2$, $S_{2,12}$ and $S_{2,34}$. The primitive polynomial is then given by the linear combination 
\begin{equation}
S_{1,12}^2+\mu_1 S_{2,12}+\mu_2 S_{2,34}
\end{equation}
with $\mu_1$ and $\mu_2$ to be determined by demanding that the linear combination disappears upon symmetrizing over permutations under $[3,1]$ and $[4]$. The symmetrized sums are
\begin{eqnarray}
&&2((n_2+n_3)^2+(n_2+n_4)^2+(n_3+n_4)^2+\notag\\
&&\mu_1(n_1n_2+n_1n_3+n_2n_3)+\mu_2(n_1n_4+n_2n_4+n_3n_4))
\end{eqnarray} and 
\begin{eqnarray}
4\left(n_2^2+n_3^2+n_4^2+n_2n_3+n_3n_4+n_2n_4\right)(4-\mu_1-\mu_2).
\end{eqnarray}
Setting them both to zero, we find $\mu_1=\mu_2=2$, so the primitive polynomial is $S_{1,12}^2+2 S_{2,12}+2 S_{2,34}$.

The above procedure works for arbitrarily complicated cases, but quickly becomes cumbersome. This is when our geometric approach again becomes useful. %Below, we shall show how one can elegantly read out the the primitive polynomials through graphical inspection. 
We first write down the relevant monomials in barycentric coordinates given by Eqs.~\ref{n3def} for $N=3$ bodies (or Eq.~\ref{qs} for general $N$). The coefficients in the primitive polynomial can then be elegantly determined through graphical inspection. 

\begin{figure}[ttt]
\centering
\includegraphics[width=0.75\linewidth]{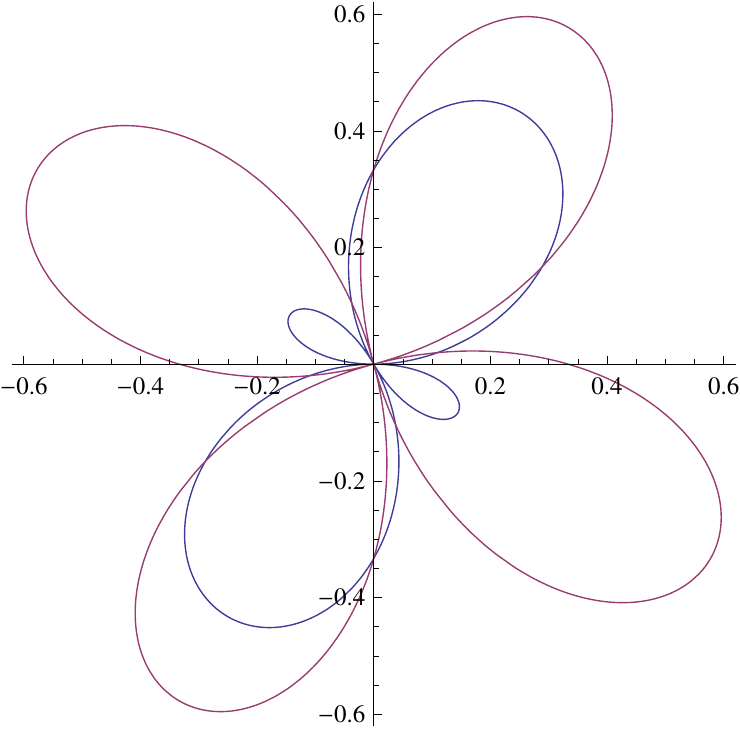}
%\captionsetup{justification=centerlast}
\caption{(Color online) Polar plots of $\beta_2(\theta)$ as a function of $\theta$, with $\mu=0.5$ (blue curve) and $\mu=2$ (purple curve). $\beta_2(\theta)+\beta_2(\theta+2\pi/3)+\beta_2(\theta+4\pi/3)$ sum to zero only when the lobes are of equal size, which is the case for $\mu=2$ only.   }
\label{fig:beta2}
\end{figure}

We demonstrate this explicitly by revisiting the example on $\lambda=[2,1]$. Recall that the primitive polynomial (call it $\beta_2$) is a linear combination of $S_{1,12}^2$ and $S_{2,12}$, i.e. 
\begin{eqnarray}
\beta_2 &=& S_{1,12}^2+\mu S_{2,12}\notag\\
&=& -\frac{W^2}{9\kappa}\left(\mu-2+(1+\mu)(\cos 2 \theta -\sqrt{3}\sin 2 \theta)\right)
\end{eqnarray}
where $\theta=0,2\pi/3,4\pi/3$ points towards the vertices favoring $\bar n_1,\bar n_2,\bar n_3$ respectively. The correct value of $\mu$ will cause $\beta_2$ to disappear under symmetrization of the $3$ particles. Graphically, it means that the lobes of three copies of the plot of $\beta_2$, each rotated an angle $2\pi/3$ from each other, must cancel upon addition. This is illustrated in Fig. \ref{fig:beta2}, where $\mu=2$ is readily identified as the correct value. 

All in all, we have the $m=2$ primitive polynomial for $[2,1]$: 
\begin{equation}\beta_2= (\bar n_3)^2+2 \bar n_1 \bar n_2 =\frac{W^2}{3}\left(-\cos 2\theta+\sqrt{3}\sin 2\theta\right).
\label{beta2}
\end{equation}
The reader is also invited to visualize the simpler case for $m=1$, $[2,1]$, which takes the form
\begin{equation}\beta_1= S_{1,12} =-\bar n_3=\frac{W}{3\kappa}\left(\cos\theta+\sqrt{3}\sin\theta\right).
\label{beta1}
\end{equation}
Compared to the antisymmetric case for fermions, we have two primitive polynomials $\beta_1,\beta_2$, instead of just $A$. The primitive polynomials for a wide variety of cases are also listed in Table II and III of Ref. \onlinecite{davenport2012}, and can alternatively be systematically derived via the theory of Matric units explained in the appendix of the same reference.

\subsubsection{Admissible spin bases}

In general, not all possible $N$-particle spin bases survive under symmetrization under a given symmetry type $\lambda$. Only those that survive should be included, i.e. are admissible, in the set of basis in $\ket{\sigma^\lambda}$. For instance, the basis $\ket{\alpha\alpha\alpha }$ does not appear in $\ket{\sigma^{[2,1]}}$. To see this, try symmetrizing the combination $p(\bar n_1,\bar n_2,\bar n_3)\ket{\alpha\alpha\alpha }$ subject to the condition that $p(\bar n_1,\bar n_2,\bar n_3)$ has no higher symmetry than $\lambda=[2,1]$, i.e.  the totally symmetrized sum $\mathcal{S}\left[ p(\bar n_1,\bar n_2,\bar n_3)\right]=0$. Obviously, $p(\bar n_1,\bar n_2,\bar n_3)\ket{\alpha\alpha\alpha }$ has to symmetrize to zero too. 

There is a nice way to write down the set of admissible bases by looking at the labelings of semistandard Young Tableaux. From Ref. \onlinecite{davenport2012}, the admissible bases in $\ket{\sigma^\lambda}$, up to permutation, are in one-to-one correspondence with the labelings of semistandard Young Tableaux with numbers $1$ to $N$. (A semistandard Young Tableau has non-decreasing entries along each row and strictly increasing entries along each column.) 
For instance, a symmetry type of $[3,1]$ with $b=3$ corresponds to labelings (with labels $a,b,c$ defined to be in increasing order): 
\begin{eqnarray}
\nonumber && [aaa,b], \; [aab,b], \; [aac,b], \; [abb,b], \; [abc,b], \\
\nonumber && [acc,b], \; [aaa,c], \; [aab,c], \; [aac,c], [abb,c], \\
 \nonumber && [abc,c], \; [acc,c], \; [bbb,c], \; [bbc,c], \; [bcc,c] \; {\rm and} \; [ccc,c],
\end{eqnarray}
where rows are separated by commas. From these, the admissible states for $\lambda=[3,1]$ can be written down by copying the labelings verbatim:
\begin{eqnarray} 
\nonumber && \ket{aaab}, \; \ket{aabb}, \; \ket{aacb}, \; \ket{abbb}, \; \ket{abcb}, \\
\nonumber && \ket{accb}, \; \ket{aaac}, \; \ket{aabc}, \; \ket{aacc}, \; \ket{abbc}, \\
\nonumber && \ket{abcc}, \; \ket{accc}, \; \ket{bbbc}, \; \ket{bbcc}, \; \ket{bccc} \; {\rm and} \; \ket{cccc}.
\end{eqnarray}
In the Young Tableau corresponding to $\lambda$, each row represents a set of particles that are symmetric with each other in the $\lambda$ representation. The requirement that labels increase monotonically within each row defines an ordering, and prevents the repeated listing of bases related by permutation. The requirement of strictly increasing labelings down each column also prevent that, and also avoids the listing of bases that do not survive. 

With these preliminary considerations, we are now in position to formulate the general recipe for constructing PPs in the multicomponent case, which is given in Table~\ref{table:summary}. In the following section we apply this recipe to the case of SU(2) spins with 3-body interactions.
\begin{table}[htb]
\centering
\renewcommand{\arraystretch}{2}
\begin{tabular}{|p{10.5cm}|}\hline
(1) Given $N$,$m$ and $b$, specify which symmetry type
the PP is associated with. For example, we specify the $[2]$ or the $[1,1]$ channel when computing PPs realizing the fermionic Halperin states in bilayer QH systems ($b=2,N=2$). \\ \hline
(2) Next, determine the appropriate primitive polynomials by finding the coefficients multiplying the allowed monomials. Then multiply the primitive polynomial by symmetric polynomials, and orthogonalize to obtain the spatial parts $U^N_{m,\lambda}$; \\ \hline
(3) Finally, choose the desired admissible spin channels, and (anti)symmetrize the resultant product of the spatial and spin parts depending on whether we want a bosonic (fermionic) PP. \\ \hline
\end{tabular}
%\captionsetup{justification=centerlast}
\caption{Summary of the PP construction procedure for the multicomponent case.}
\label{table:summary}
\end{table}

\subsection{Example: $3$-body case for SU(2) spins }

Here we explicitly work out the simplest case with $b=2$ components which is arguably also the most common. We focus on $N=3$-body interactions to illustrate the nontrivial aspects of our approach.

The symmetry type is given by $[\lambda_1,\lambda_2]=[N/2+S,N/2-S]$, where $S$ is the total spin of the particles. Let us denote the spins by $\uparrow,\downarrow$. There are only $N=3$ boxes in the Young Tableaux, with the following possible symmetry types: $[3]$, $[2,1]$, $[1,1,1]$. 

For type $\lambda=[3]$, the possible spin bases are $\ket{\uparrow \uparrow \uparrow }$, $\ket{\uparrow \uparrow \downarrow }$, $\ket{\uparrow \downarrow \downarrow }$ and $\ket{\downarrow \downarrow \downarrow }$. For $\lambda=[2,1]$, the possible spin bases are $\ket{\uparrow \downarrow \downarrow }$, $\ket{\uparrow \uparrow \downarrow }$, corresponding to Tableau labellings $[\uparrow\downarrow,\downarrow]$ and $[\uparrow\uparrow,\downarrow]$ respectively. For $\lambda=[1,1,1]$, there is actually \emph{no} admissible spin basis: total (internal DOF) antisymmetry is impossible for 3 particles, if there are only 2 different spin states to choose from.

Consider bosonic particles in the following. The symmetry type $[3]$ case corresponds to the primitive polynomial $1$, so the resultant PP takes the same form as in the single-component case. After symmetrizing the spin part, we have the following spin channels for $[3]$: 
\begin{eqnarray}
\nonumber && \{\ket{\uparrow \uparrow \uparrow }\}, \\
\nonumber && \{\ket{\uparrow \uparrow \downarrow },\ket{\uparrow \downarrow \uparrow },\ket{\downarrow \uparrow \uparrow }\}, \\
\nonumber && \{\ket{\downarrow \downarrow \uparrow },\ket{\downarrow \uparrow \downarrow },\ket{\uparrow \downarrow \downarrow }\}, \\
\nonumber && \{\ket{\downarrow \downarrow \downarrow }\}.
\end{eqnarray}
This corresponds to $U^3_{m,[3]}\propto \kappa^3\sum_n b^{m\dagger}_nb^m_n$, where
\begin{equation}
b^{m\dagger }_n\ket{0} = \sum_{\sum_j n_j=n}p_m e^{-\frac1{3}W^2}\ket{\uparrow \uparrow \uparrow }\
\label{bS32}
\end{equation}
for $S=3/2$, and
\begin{equation}
b^{\dagger m}_n\ket{0} =\mathcal{S} \left[ \sum_{\sum_j n_j=n}p_m e^{-\frac1{3}W^2}\ket{\uparrow \uparrow \downarrow }\right]
\label{bS12}
\end{equation}
for $S=1/2$. Other contributions with $S\rightarrow -S$ are obtained with the identification $\ket{\uparrow} \leftrightarrow \ket{\downarrow}$. Here $\ket{\uparrow \uparrow \downarrow }$ is the shorthand for $c^\dagger_{n_1\uparrow}c^\dagger_{n_2\uparrow}c^\dagger_{n_3\downarrow}\ket{0}$, etc. $p_m=p_{m,[3]}$ refers to the \emph{same} polynomial as in the single-component case.

To construct the orthonormal PP basis for the first few $m$, we orthogonalize the set $\{\beta_1,\beta_2,\beta_1 W^2, \beta_2W^2,\beta_1 Y,\beta_2Y,\beta_1W^4,... \}$. 
The results are shown in Table \ref{N21}. Notice that the first PP for symmetry type $[2,1]$ occurs at $m=1$, whereas that of single-component bosons/fermions occur at $m=0$ and $m=3$ respectively. Indeed, there are more ways of constructing PP polynomials when only a subgroup of the full symmetric/alternating group is involved. The onset of PPs degenerate in $m$ also occurs earlier, at $m=4$.

\begin{table}[ttt]
\centering
\renewcommand{\arraystretch}{2}
\begin{tabular}{|l|l|}\hline
$m$ &\ $p_{m,[2,1]}(\beta_1,\beta_2,W,Y)$ \\    \hline
1 &\ $ \beta_1$ \\  \hline
2 &\ $ \frac{1}{\sqrt{3}}\beta_2 $ \\ \hline 
3 &\ $ \frac{1}{3} \sqrt{2} \beta_1 \left(-3+W^2\right)$  \\ \hline 
4(i) &\ $ \frac1{\sqrt{5}}\left(\beta_2+6 \beta_1 Y\right) $ \\ %\hline 
4(ii) &\ $ \frac1{27 \sqrt{5}}\left(-54 \beta_2+12 \beta_2 W^2+W^4 \left(\cos [4 \theta]+\sqrt{3} \sin [4 \theta]\right)\right)$ \\ \hline 
5(i) &\ $ \frac1{\sqrt{33}}\left(\beta_1 \left(-3+2 W^2\right)+6 \beta_2 Y\right) $ \\ %\hline 
5(ii) &\ $ \frac{1}{81} \sqrt{\frac{2}{55}} (-\sqrt{3} W^5 \cos [5 \theta ]+3 (5 \sqrt{3} \beta_1 (27-18 W^2$ \\ 
&\ $+2 W^4)+W^5 \sin [5 \theta ])) $ \\ \hline 
\end{tabular}
%\captionsetup{justification=centerlast}
\caption{The polynomials $p_m$ for the first few $N=3$-body PPs for bosons in the total spin $|S|=\frac{1}{2}$, i.e. $\lambda=[2,1]$ channel. The primitive polynomials $\beta_1$ and $\beta_2$ are given by Eqs. \ref{beta1} and \ref{beta2}.  $\cos 4\theta,\sin 4\theta, \cos 5\theta, \sin 5\theta$ can all be decomposed into the elementary symmetric polynomials and primitive polynomials $W,Y,\beta_1$ and $\beta_2$.  }
\label{N21}
\end{table}

To illustrate the full procedure for PP construction with internal DOFs, we detail the case of $S=-1/2$ below. For $m=1$, 
\begin{eqnarray}
b^{1\dagger}_n\ket{0} &=& \mathcal{S} \left[\sum_{\sum_j n_j=n}(-\bar n_3) e^{-\frac1{3}W^2}\ket{\uparrow \downarrow \downarrow }\right]\notag\\
\nonumber &=& -e^{-\frac{1}{3}W^2}[(\bar n_2 + \bar n_3) \ket{\uparrow\downarrow\downarrow} \\
&+& (\bar n_1 + \bar n_3) \ket{\downarrow\uparrow\downarrow}+(\bar n_1 + \bar n_2) \ket{\downarrow\downarrow\uparrow}]\notag\\
&=& e^{-\frac{1}{3}W^2}[\bar n_1 \ket{\uparrow\downarrow\downarrow} +\bar n_2 \ket{\downarrow\uparrow\downarrow}+\bar n_3 \ket{\downarrow\downarrow\uparrow}], 
\label{b1S12}
\end{eqnarray}
while for $m=2$ we have 
\begin{eqnarray}
b^{2\dagger }_n|0\rangle &=& \mathcal{S} \left[\sum_{\sum_j n_j=n}((\bar n_3)^2+2\bar n_1\bar n_2) e^{-\frac1{3}W^2}\ket{\uparrow \downarrow \downarrow }\right]\notag\\
&= &e^{-\frac1{3}W^2} [((\bar n_2)^2 + (\bar n_3)^2+2\bar n_1 (\bar n_2+\bar n_3)) \ket{\uparrow\downarrow\downarrow}\notag\\ && +((\bar n_1)^2 + (\bar n_2)^2+2\bar n_3 (\bar n_1+\bar n_2)) \ket{\downarrow\downarrow\uparrow}\notag\\
&&+ ((\bar n_1)^2 + (\bar n_3)^2+2\bar n_2 (\bar n_1+\bar n_3)) \ket{\downarrow\uparrow\downarrow}], 
\label{b2S12}
\end{eqnarray}
Expressions for $S=1/2$ are obtained via $\ket{\uparrow} \leftrightarrow \ket{\downarrow}$.

For the present case of SU$(2)$ spins, i.e. $b=2$, there exists a nice closed-form generating function for the dimension of the spatial basis for each $m$. 
Define the generating function $Z_{[\lambda_1,\lambda_2]}(q)= \sum_m d({[\lambda_1,\lambda_2]};m)q^m $ where $ d({[\lambda_1,\lambda_2]};m)$ is the number of different degree $m$ polynomials with symmetry type $[\lambda_1,\lambda_2]$. It can be shown that for bosons, %(with $\lambda_1\geq\lambda_2$),
\begin{eqnarray}
 Z_{[\lambda_1,\lambda_2]}(q) &=& \frac{1-q}{ \prod_{m=1}^{\lambda_1}(1-q^m)\prod_{n=1}^{\lambda_2}(1-q^n)} \notag\\
 &-&\frac{1-q}{ \prod_{m=1}^{\lambda_1+1}(1-q^m)\prod_{n=1}^{\lambda_2-1}(1-q^n)}.
\end{eqnarray}
This is obtained\cite{davenport2012} by considering the dimensionality from two symmetry subsets, and then subtracting overlaps from the higher symmetry case $[\lambda_1+1,\lambda_2-1]$. The dimension is related to the q-binomial coefficient.
For $b>2$, however, the situation is much more complicated, involving Kostka coefficients which count the number of semisimple labelings of $\lambda$ with a given alphabet.

\section{A deeper look into the Wannier polarization}
\label{sec:wannpol}

\subsection{Relation to band topology}
As introduced in Sect. \ref{sec:wannierbasis}, maximally localized Wannier Functions (WFs) are the eigenfunctions of the operator $\hat X=P\text{exp}[ix\frac{2\pi}{L_x}]P$, where $\text{exp}[ix\frac{2\pi}{L_x}]$ is a unitary operator whose argument represents the (periodic) position in the $\hat x $ direction and $P$ is a projection operator onto the occupied bands. In 2-dimensional systems, application of this operator will yield WFs of the form $\psi(x,k_y)$ since $\hat X$ only acts in the $x$-direction. 

As $k_y$ is varied over a period of $2\pi$, the center of mass in the x-direction, i.e. polarization of the WFs shift by a total of $C_1$ number of sites, where $C_1$ is the Chern number of the system. When there are $n$ occupied bands $\ket{\phi_j(k_x,k_y)}$, $j=1,2,...,n$, there are $n$ WFs and hence $n$ polarizations $P^j(k_y)$. With the (non-abelian) gauge field given by $a^{jk}_x=-i\bra{\phi_j(k_x,k_y)}\partial_{k_x}\ket{\phi_k(k_x,k_y)}$, the $n$ polarizations $P^j(k_y)$ are given by the eigenvalues of (the logarithm of) the gauge-invariant Wilson loop operator  

\begin{equation}
\frac1{2\pi i}\log Pe^{i\int^{2\pi}_0 a^{jk}_x(k_x,k_y)dk_x}
\end{equation}
modulo an integer, where $P$ is the path-ordering operator. When there is only $n=1$ occupied band, we simply have 
\begin{equation}
P(k_y)=\frac1{2\pi }\int^{2\pi}_0 a_x(k_x,k_y)dk_x
\end{equation}
The relation between the winding (twisted boundary condition) of the polarization and $C_1$ can be understood as follows. Considering one occupied band for simplicity, 
\begin{eqnarray}
C_1&=& \frac{1}{2\pi}\int_{BZ} d^2k F_{xy}\notag\\
&=&\frac{1}{2\pi}\int_{BZ} Tr (\nabla \times a)\cdot dS\notag\\
&=&\frac{1}{2\pi}\oint_{C} Tr(a) \cdot dl
\end{eqnarray}
If we choose a gauge where $a_y = 0$, the contour $C$ which loops around the boundary of the BZ $(0,0)\rightarrow (2\pi,0)\rightarrow (2\pi,2\pi)\rightarrow (0,2\pi)\rightarrow (0,0)$ will have contributions only from the segments where $k_y=0,2\pi$. But these contributions are precisely $P(0)$ and $-P(2\pi)$. Hence $P(2\pi)-P(0)=C_1$. 

For $0<k_y<2\pi$, $\int^{2\pi}_0 a_x(k_x,k_y)dk_x$ also defines a loop in the torus BZ. However, this loop is now the boundary of only a portion of the BZ and thus does not encircle a quantized amount of $F_{xy}$. As we move this loop continuously around the BZ, the value of its line integral should vary continuously except when it crosses points where $a_x$ is undefined. Such is the case with the Dirac points as we will see later, where the polarization exhibits a jump. In general, the polarization is just the integral of the Berry flux $F_{xy}$ over the rectangle $[0,2\pi]\times [0, k_y]$. 

%The eigenvalue of the trace of an operator is the sum of the eigenvalues of the operator. Hence for $C_1=n \neq 0$, the total polarization at $k_y=0$ and $k_y=2\pi$ must differ by the integer $n$. Note that this analysis holds for any loop $C$ that contains the whole BZ, a fact that is consistent with the periodicity of the total polarization as $k_y$ varies.

\subsection{2-band case}

It is instructive to first study how the polarization changes in a 2-band system with one occupied band, i.e for the Dirac Hamiltonian
\begin{equation} 
H_{Dirac}=\sin  k_x \sigma_x+\sin  k_y \sigma_y +(m+\cos k_x+\cos k_y)\sigma_z 
\end{equation}
In this simplest case, it is possible to obtain an exact closed-form expression for the polarization near the jump, and thus study how the nature of singularities affect the change in polarization. 

The eigenvalues of $H_{Dirac}$ are $\pm \lambda$, where $\lambda= \sqrt{\sin k_x^2+\sin k_y^2+ (m+\cos k_x+\cos k_y)^2}$. These eigenvalues are not degenerate (gap does not close) except when $m=0,\pm 2$. These values of $m$ demarcate regimes with different Chern numbers: $C_1=\text{sgn}m$ for $0<|m|<2$, and $C_1=0$ otherwise. The eigenvector corresponding to the negative (occupied band) eigenvalue is \begin{equation} 
\phi=\frac{1}{N}(-\sin k_x + i\sin k_y, (m+\cos k_x+\cos k_y)+\lambda)^T
\end{equation}
with the normalization factor $N=\sqrt{2\lambda((m+\cos k_x+\cos k_y)+\lambda)}$. Singularities occur when the Berry gauge field $a_x=-iv^\dagger\partial_{k_x}v$ is undefined, either when $\lambda=0$ or $(m+\cos k_x+\cos k_y)+\lambda=0$, i.e  $m+\cos k_x+\cos k_y<0$ with $\sin k_x=\sin k_y=0$. In summary, the singularities, if any, occur at:
\begin{itemize}

\item  $m=0$: $a_x$ undefined at $(0,\pm\pi)$,$(\pm\pi,0)$ and $(\pm\pi,\pm\pi)$. The eigenvalues of $H_{Dirac}$ are degenerate at $(0,\pm\pi)$ and  $(\pm\pi,0)$, where also $H_{Dirac}=0$ causing the filled/occupied eigenspaces are ill-defined. As we will see later, this becomes more complicated when there is more than one occupied bands, since it is possible for the filled/occupied bands to be individually degenerate.   

\item $0<m<2$: $a_x$ undefined at $(\pm\pi,\pm\pi)$. The filled eigenspace is always well-defined, but the singularities of $a_x$ can lead to jumps in the polarization. The Hamiltonian is gapped everywhere in the (current) translationally invariant case.

\item $m=2$: $a_x$ undefined at $(\pm\pi,\pm\pi)$. The eigenvalues are degenerate and $H_{Dirac}=0$ at all of these points too.

\item $m>2$: $a_x$ is defined everywhere and $H_{Dirac}$ is everywhere gapped. There is no singularity to produce a nontrivial Chern number.

\item $-2<m<0$: $a_x$ undefined at $(0,\pm\pi)$,$(\pm\pi,0)$ and $(\pm\pi,\pm\pi)$. The filled eigenspace is always well-defined as $H$ is gapped everywhere.

\item $m=-2$: $a_x$ undefined at all nine points $(0,0)$, $(0,\pm\pi)$,$(\pm\pi,0)$ and $(\pm\pi,\pm\pi)$. $H_{Dirac}$ is not gapped only at $(0,0)$.

\item $m<-2$: $a_x$ is still undefined at all nine points $(0,0)$, $(0,\pm\pi)$,$(\pm\pi,0)$ and $(\pm\pi,\pm\pi)$. However, $H_{Dirac}$ is gapped everywhere.
\end{itemize}

It is important to note that having one or more singularities of $a_x$ does not gurantee a nontrivial Chern number, since the resultant polarization jumps may cancel.

\subsubsection{Calculation of Dirac polarization}

For brevity of notation, we shall for now denote $k_x, k_y$ by $x,y$. As there is only one occupied band, the polarization is just $\frac{1}{2\pi i}\int^{2\pi}_0 \phi^\dagger\partial_{x}\phi dx$. A few simplifications can be made. We write $\phi=(a+bi)/N$, where $a$ and $b$ are real vectors and $N$ is the abovementioned normalization factor. As $|\phi|^2=a^2+b^2=1$ is a constant, the real parts of $\phi^\dagger\partial_{x}\phi$ must disappear. Hence 
\begin{eqnarray}
\phi^\dagger\partial_{x}\phi&=&i ((b/N)\partial_{x}(a/N)-(a/N)\partial_{x}(b/N)) \notag\\
&=&\frac{i}{N^2}(b\partial_{x}a-a\partial_{x}b)\notag\\
&=&\frac{i}{N^2}(-\sin y\partial_{x}\sin x)\notag\\
&=&\frac{-i \sin y\cos x}{2\lambda((m+\cos k_x+\cos k_y)+\lambda)}
\label{derivation1}
\end{eqnarray}
Thus the exact integral expression for the polarization is 
\begin{eqnarray}
&& P(k_y)=-\frac1{2\pi}\int^{2\pi}_0\frac{\sin y\cos x}{\lambda(\lambda+m+\cos x+\cos y)}dx
\label{exactqah}
\end{eqnarray}

\begin{figure}[H]
\includegraphics[scale=0.4]{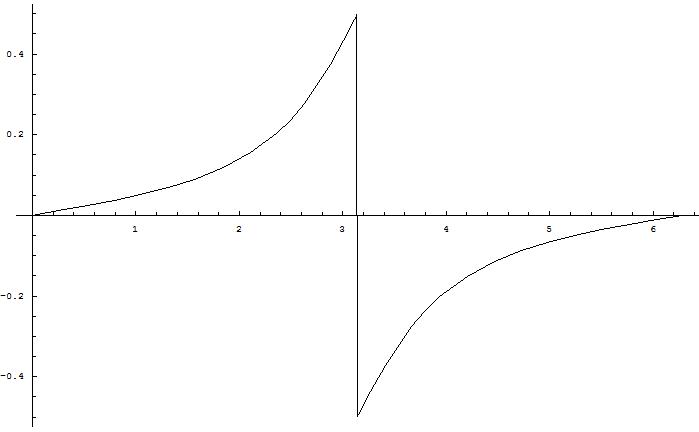}
\includegraphics[scale=0.4]{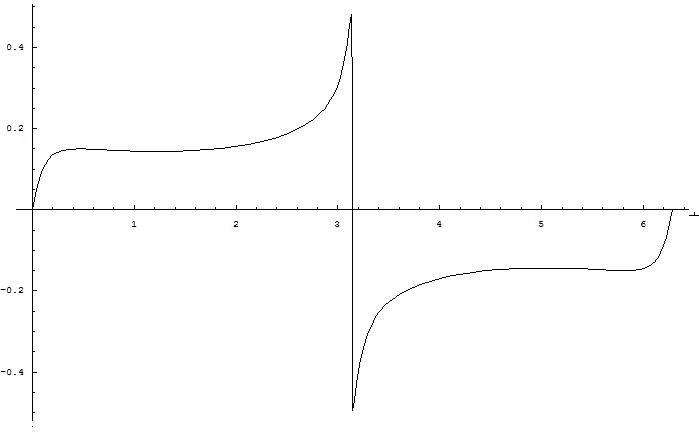}
\caption{Exact numerical plots of the polarization against $k_y=[0,2\pi]$ for $m=0.1,1$ (Left,Right) according to Eq. \ref{exactqah}.}
\label{QAHpol}
\end{figure}

The most interesting and universal aspect of this polarization curve is its jump at $k_y=\pi$. To study the jump in the regime $0<m<2$ shown above, we Taylor expand the integrand about $k_y = \pi$ so that $Y= k_y-\pi$ and $\cos k_y \approx -1+ \frac{Y^2}{2}$. This integrates to a closed-form albeit complicated formula for the polarization which is extremely accurate for $|Y|=|k_y-\pi|<1$. To make things more tractable, we shall also do an approximation on the integrated variable $X=x+\pi$. But here we have to be very careful as the integral \eqref{exactqah} has different limits for different values of $m$. For the current simplest case of $0<m<2$, where the only singularity in the BZ $[0,2\pi]^2$ occurs at $(\pi,\pi)$. We can thus expand about $X=x-\pi=0$, so that $\cos k_x \approx -1+ \frac{X^2}{2}$ and
\begin{eqnarray}\lambda&=& \sqrt{(m-2)^2+(m-1)(X^2+Y^2)+\left(\frac{X^2+Y^2}{2}\right)^2}\notag\\
&\approx&(2-m)\sqrt{1+\frac{m-1}{(m-2)^2}(X^2+Y^2)}\notag\\
&\approx&(2-m)(1+\frac{m-1}{2(m-2)^2}(X^2+Y^2)),
\end{eqnarray}

\begin{eqnarray}\lambda+(m+\cos x+\cos y)&=& \lambda+m-2+\frac{1}{2}(X^2+Y^2)\notag\\
&\approx&(2-m)(1+\frac{m-1}{2(m-2)^2}(X^2+Y^2))+m-2+\frac{1}{2}(X^2+Y^2)\notag\\
&\approx& \frac{1}{2(2-m)}(X^2+Y^2)
\end{eqnarray}
All in all, the polarization integral \eqref{exactqah} can be approximated to lowest nontrivial order in $X^2+Y^2$ by 
\begin{equation}
P(Y)=-\int^\pi_{-\pi}\frac{YdX}{\pi(X^2+Y^2)}=-\frac{1}{\pi}tan^{-1}(\pi/Y)
\end{equation}
or, in an analytically continued form, $Y=-\pi cot (\pi P(Y))$ where $Y=k_y-\pi$. This is an exceedingly simple expression obtained by expanding $\lambda$ up to first order in $X^2+X^2$; an analogous expansion up to the second order yields a much more accurate curve also expressible in closed-form, both of which are shown in the figure below.

\begin{figure}[H]
\includegraphics[scale=0.7]{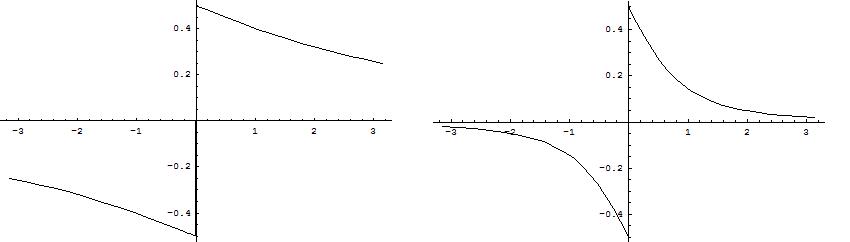}
\caption{Plots of the polarization against $Y=k_y-\pi$ for $m=1$, obtained by approximating $\lambda$ up to first/second (Left,Right) order in $X^2+Y^2$. The first order expansion reproduces only the correct size of the jump, but the second order approximation closely reproduces the exact curve in Fig. \ref{QAHpol}.}\end{figure}

%We see that the approximation is in very good agreement with the exact result in figure 1. To better understand why, we look at the behavior of $a_x$ in figure 4. The divergence near the origin produces the polarization peak of $\pm 0.5$. This behavior is generic; very similar plots show up in the different stages of approximation. Also note that this divergence is not physical as it involves an integer jump of $P(k_y)$. However, the nontrivial winding of $P(k_y)$ can be seen from the fact that the plot has an overall negative gradient in the $k_y$ direction. While the jump itself does not indicate a nontrivial winding, it has to be present to make an overall negative gradient possible. 

%As $m$ approaches other critical points like $m \rightarrow 0_+$, additional jumps of $P(k_y)$ can occur. This should not come as a surprise since more singularities of $a_x$ start to appear at $k_y=0$ and $k_y=2\pi$ as $m \rightarrow 0_+$.

\subsection{4-band case and Wannier function construction}

To illustrated the case of multiply occupied bands, we provide a 4-band model of 2-coupled Dirac fermions. Due to the half Hall conductivity from each Dirac fermion ~\cite{ryu2010, qi2008topological}, the Chern number of the system is $C_1=1$, giving rise to a total polarization winding of $1$ for both WFs. The model Hamiltonian is given by 
\begin{equation}
H_{4 band}=m I\otimes \sigma_z- \tau_z\otimes\sigma_x sink_x -\tau_z\otimes\sigma_y sink_y +\tau_x\otimes \mathbb{I} (2+cosk_x+cosk_y)  
\end{equation}
The $\tau$ and $\sigma$ Pauli matrices act on the layer and spin basis respectively. As before, $x$ and $y$ will be used in place of $k_x$ and $k_y$ for notational simplicity. 

\begin{figure}[H]
\includegraphics[scale=0.7]{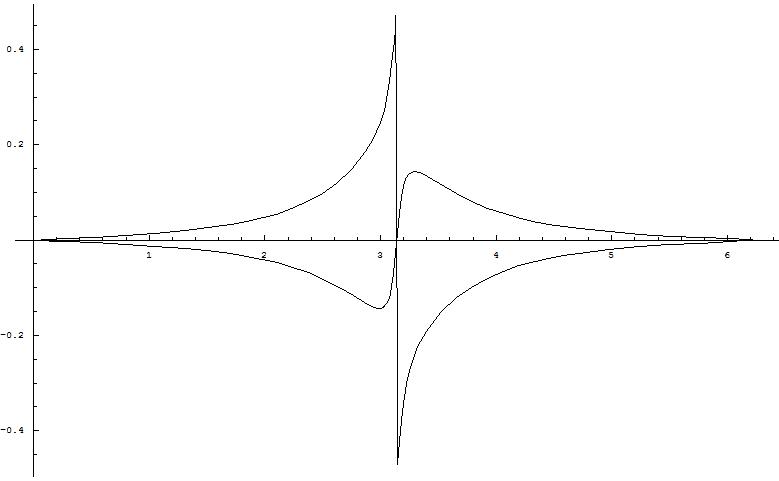}
\caption{Numerical plots of $P^1(k_y)$ and $P^2(k_y)$  against $k_y$ for $m=0.1$. $P^1$ exhibits a jump of $1$ site at $k_y=\pi$, but not $P^2$. Indeed the sum of their windings is $C_1=1$.}\label{pol4exact}
\end{figure}

The two occupied bands correspond to the two negative eigenenergies 
\begin{equation}
\lambda_1=-\sqrt{6 + \left( -4 + m \right) \,m - 2\,\left( -2 + m \right) \,\cos (y) + 2\,\cos (x)\,\left( 2 - m + \cos (y) \right) }\end{equation}
and 
\begin{equation}
\lambda_2=-\sqrt{6 + m\,\left( 4 + m \right)  + 2\,\left( 2 + m \right) \,\cos (y) + 2\,\cos (x)\,\left( 2 + m + \cos (y) \right) }.
\end{equation}
Their eigenvectors $\phi_1$ and $\phi_2$ are very complicated, but fortunately $a^{ij}_x\propto \phi_i \partial_x \phi_j=0$ for $i\neq j$. Hence we still have an abelian gauge field, albeit with two diagonal components which can be evaluated as in eq. \eqref{derivation1}. %After a great deal of simplification, the  $a^{11}_x$ and $a^{22}_x$ are found to be 
These Berry gauge fields can be integrated to produce the polarization curves $P^1(k_y)$ and $P^2(k_y)$ in Fig. \ref{pol4exact}. It turns out that for $m>0$, $P^1(k_y)$ has a discontinuity and thus nontrivial winding at $k_y=\pi$, while $P^2(k_y)$ is remains topologically trivial. These polarizations are related to those for $m<0$ via $P^1(k_y,m)=P^2(\pi-k_y,-m)$.

\subsubsection{Construction of the Wannier function}
 
In this rest of this appendix, we shall focus on the analytic construction of the Wannier function corresponding to $P^1(k_y)$ with nontrivial winding. According to Sect. \ref{sec:wannierbasis} and~\cite{qi2011},
\begin{eqnarray}
W^j(x,k_y)&=&\langle x\ket{W_{jk_y}}\notag\\
&=&\sum_{k_x,m,n}e^{-ik_x(P^j(k_y)-x)}[Pe^{i\int^{k_x}_0a^{mn}_x(p_x,k_y)dp_x}]_{mn}u^j_n\phi_m(k_x,k_y)
\label{WF}
\end{eqnarray}
where $P$ is again the path-ordering operator, and $P^j(k_y)$ the polarization of the $j^{th}$ WF. The $m,n$ indices label the occupied bands, and $u^j_n$ is the $n^{th}$ band matrix element of the $j^{th}$ eigenvector of the Wilson loop operator in the square parentheses. $\phi_m(k_x,k_y)$ refers to the $m^{th}$ occupied eigenstate of the Hamiltonian in momentum space. The path ordering operator $P$ in front of the Wilson loop operator is necessary because the $a^{mn}_x$ gauge field is a possibly nonabelian matrix, although it is trivially abelian in our case of interest. In the continuum limit, the sum can be replaced by an integral over $[0,2\pi)$. 

The complicated expression above is essentially a Fourier Transform of the occupied Bloch eigenstates $\phi_m(k_x,k_y)$ multiplied by a phase factor involving the Wilson loop operator. If the phase factor is absent, it will just reduce to an ordinary Fourier Transform centered at $x=P^j(k_y)$, as seen from the $e^{-ik_x(P^j(k_y)-x)}$ term. The phase factors conspire to produce a maximally localized WF in the x-direction. Hence $W^j(x,k_y)$ can be thought of as a judiciously chosen superposition of the occupied Bloch states that is \emph{maximally} localized around its Wannier Center $x_{CM}=P^j(k_y)$. As proven in~\cite{qi2011}, this form of the WF is indeed an eigenstate of $\hat X=Pe^{iX}P$. Note that since the above integrand is only periodic if $x$ is an integer, the value of the WF is only physical at integer lattice sites $x$. 
For our system with two coupled Dirac fermions, the vanishing of the off-diagonal $a^{ij}_x$ terms greatly simplifies the calculation. The eigenvectors $u^j$ are simply $(1,0)^T$ and $(0,1)^T$ and the Wilson Loop operator is abelian. Since the jump occurs for only $P^1$, as previously shown, only $a^{11}_x$ and $\phi_1(k_x,k_y)$ need to be considered. While both of these have rather complicated closed-form expressions, the analysis can be simplified near the jump, which is the most interesting region. It occurs at $k_y=\pi$, for which the main contributions to the polarization occur near $k_x=0$ and $k_x=\pi$. 

%However, we do not have the liberty of performing a Taylor approximation,In fact, the 4-component\footnote{$spin\otimes layer$} eigenstate $\phi_1(k_x,k_y)$ is singular in the $m\rightarrow 0$ and $ky\rightarrow \pi$ limit, precisely where it becomes interesting. Hence  which will in this case erroneously replace the jump by a steep and unphysical analytic function.  

%Fortunately, it turns out that important features of the WF at the jump can be understood without knowing the details of the Bloch eigenstate. We observe that the graph of the exponential part of the Wilson loop operator $\int^{k_x}_0a^{11}_x(p_x,k_y)dp_x$ is almost almost flat in the limit of $m\rightarrow 0$ and $ky\rightarrow \pi$, although the exact behavior depends on the direction the limits are taken. This can be derived analytically.    

\begin{figure}[H]
\centering
\includegraphics[scale=0.8]{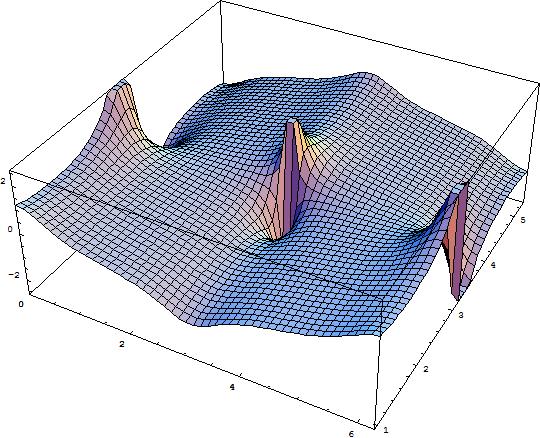}
\caption{Plot of $a^{11}_x$ over the entire BZ. The two singularities occur at $(\pi,\pi)$ and $(0,\pi)$, the latter which straddles the two edges of the BZ which are in reality glued together.}\label{fig:a11}\end{figure}

It can be shown that the gauge field is approximately 
\begin{equation} \tilde a^{11}(X,Y)\approx\frac{Y(-K+\sqrt{K^2+X^2+Y^2})}{2(X^2+Y^2)\sqrt{K^2+X^2+Y^2}}\end{equation}
near singularities $(0,\pi)$ or $(\pi,\pi)$, where $(X,Y)=(k_x,k_y-\pi)$ or $(k_x-\pi,k_y-\pi)$ are the coordinates centered around either singularity. $K$ is the effective average value of $2-m+\cos k_x + \cos k_y$ across the integral. For the Wilson loop operator we also compute
\begin{eqnarray} p(X,Y,K)&=&\frac1{2\pi}\int_{-\pi}^X\tilde  a^{11}_x(X',Y)dX'\notag\\
&=&\frac{1}{8}sgn(Y)+\frac{Y}{4\pi}\left(tan^{-1}\frac{X}{Y}+tan^{-1}\frac{K}{Y}+tan^{-1}\left(\frac{XK}{Y\sqrt{X^2+Y^2+K^2}}\right) \right) \notag\\ \end{eqnarray}
for which $p(X,\pi,K)$ is the polarization. It suffices to set $K=M-m=2$ near $(0,\pi)$ in the limit of small $m$. In the same limit, $K=-m$ near $(\pi,\pi)$. The exact value of the gauge field is very close to the sum of the above approximate expressions $\tilde a^{11}_x$ for $a^{11}_x$ over both singularities (Fig. \ref{fig:a11}). Summing over them, the phase in the Wilson loop operator becomes
\begin{eqnarray}
&&\int^{k_x}a_x^{11}(k'_x,Y)dk'_x \notag\\
&\approx &2\pi[p(k_x-\pi,Y,-m)-p(X,Y,2)]\notag\\
&\rightarrow& -\frac{3\pi}{4}sgn(Y)-\frac{1}{2}\left(tan^{-1}\frac{m}{Y}+tan^{-1}\frac{\pi-k_x}{Y}+tan^{-1}\left(\frac{m(k_x-\pi)}{Y\sqrt{m^2+Y^2+(k_x-\pi)^2}}\right)\right)\notag\\
\label{pax}\end{eqnarray}
for small $m$, $Y=k_y-\pi$.

It is clear that Eq. \ref{pax} evaluates to either $\pm \frac{\pi}{2} $ or $\pm \pi$ for $m\rightarrow 0$, depending on the value of $Y$ and whether the not necessarily small $k_x$ is smaller or larger than $\pi$. Hence, as $k_y-\pi=Y$ approaches the discontinuity\footnote{Note that the polarization is $\pm 0.5$ near $Y=0$.}, the Wannier function takes the form
\begin{eqnarray}
W^j(x,Y)%&=&\int_0^{2\pi}e^{-ik_x(-sgn(Y)/2-x)}e^{2\pi i P(k_x,Y)}\phi_1(k_x,Y)dk_x\notag\\
&=& A\int_0^{\pi}e^{-ik_x(-sgn(Y)/2-x)}\phi_1(k_x,Y)dk_x+B\int^0_{-\pi}e^{-ik_x(-sgn(Y)/2-x)}\phi_1(k_x,Y)dk_x\notag\\
\end{eqnarray}
where $A,B=\pm i$ or $\pm 1$. In other words, the WF very near the jump is expressible in terms of the Fourier Transform of the Bloch eigenstate over half the BZ. Different limits of $m,Y$ will correspond to different $A$ and $B$.%,and their physical relevance will be discussed in the following.

\begin{itemize}
\item If the $m\rightarrow 0$ limit is reached before $Y\rightarrow 0$: In this case, $\int^{k_x}a_x^{11}(k'_x,Y)dk'_x\rightarrow -\pi\text{sgn}(Y)$ for $0<k_x<\pi$ and $\int^{k_x}a_x^{11}(k'_x,Y)dk'_x\rightarrow -\frac{\pi}{2}\text{sgn}(Y)$ for $\pi<k_x<2\pi$. This gives $A=\pm 1 $ and $B=\pm i$. Since $k_y$ can be arbitrarily small near the jump, this is the limit of zero $m$, where the Chern number is undefined. Indeed, the polarization jumps by $\pm \frac{1}{2}$, with the sign undefined. The resultant WF looks discontinuous:
\begin{figure}[H]\centering
\includegraphics[scale=0.6]{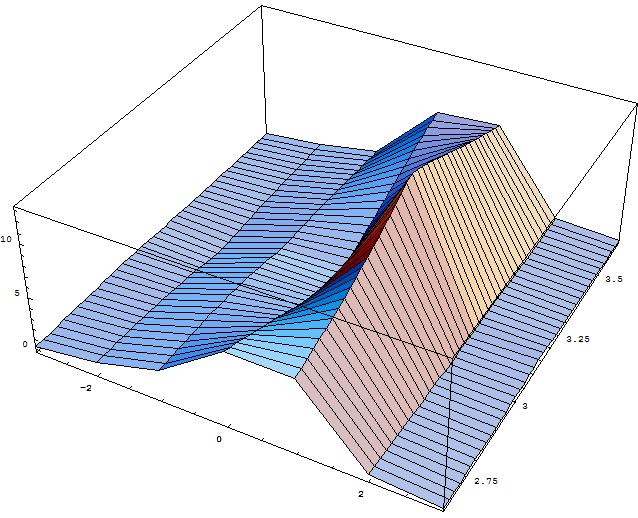}
\caption{WF $|W^1(x,Y)|^2$ when $m=0$, where $Y$ goes into the page.}\end{figure}

\item Now consider the limit where $Y\rightarrow 0$ first: The mass $m$ need not approach zero, but should be sufficiently small $(<0.2)$ so that our approximation of the effect of the singularities holds. First see what happens when $m>0$: 
\begin{eqnarray}
&&\frac1{2\pi}\int^{k_x}a_x^{11}(k'_x,Y)dk'_x\notag\\ 
&\rightarrow& -\frac{3}{8}\text{sgn}(Y)-\frac{1}{4\pi}\left(tan^{-1}\frac{m}{Y}+tan^{-1}\frac{\pi-k_x}{Y}+tan^{-1}\left(\frac{m(k_x-\pi)}{Y\sqrt{m^2+Y^2+(k_x-\pi)^2}}\right)\right)\notag\\
&\rightarrow &-\frac{3}{8}\text{sgn}(Y)-\frac{1}{8}(\text{sgn}(mY)+\text{sgn}((\pi-k_x)Y)-\text{sgn}((\pi-k_x)mY))\notag\\
&= &-\frac{3}{8}\text{sgn}(Y)-\frac{1}{8}(\text{sgn}(Y)+\text{sgn}((\pi-k_x)Y)-\text{sgn}((\pi-k_x)Y))\notag\\
&=&-\frac{1}{2}\text{sgn}(Y)
\end{eqnarray}
Hence $A=B=1$ in this case, and the WF near the jump is just given by the Fourier transform of the Bloch eigenstate $\phi_1$, which has to be continuous. %Although $W(x,k_y)$ is supposed to be maximally localized, here it does not get any more localized... It happens that here the WF is as unlocalized as it will be without the phase being fixed by being an eigenstate of the position operator. Intuitively, this does make sense as the jump is also where the x-position of the WF looks the most spread-out, straddling two sites. 

\begin{figure}[H]\centering
\includegraphics[scale=0.6]{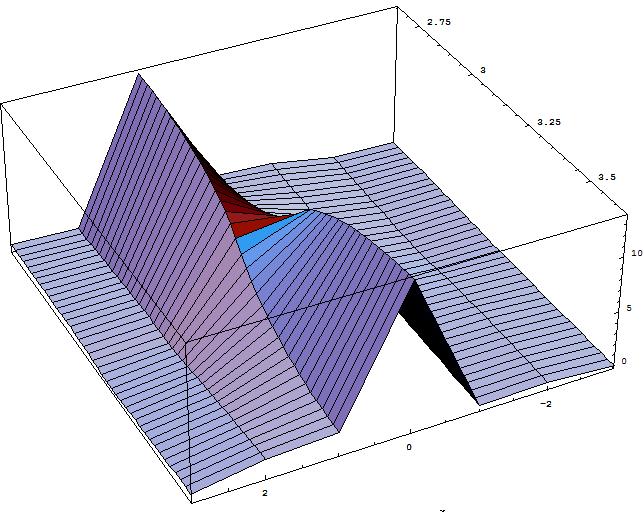}
\caption{WF $|W^1(x,Y)|^2$ near the jump for $m=0.01>0$. $Y$ goes into the page. The WF is most delocalized at the jump, straddling two sites.}\end{figure}

\item Next consider the limit where we also have $Y\rightarrow 0$ first, but with $m<0$: 
\begin{eqnarray}
&&\frac1{2\pi}\int^{k_x}a_x^{11}(k'_x,Y)dk'_x
\notag\\ &\rightarrow &-\frac{3}{8}\text{sgn}(Y)-\frac{1}{8}(\text{sgn}(mY)+\text{sgn}((\pi-k_x)Y)-\text{sgn}((\pi-k_x)mY))\notag\\
&= &-\frac{3}{8}\text{sgn}(Y)-\frac{1}{8}(-\text{sgn}(Y)+\text{sgn}((\pi-k_x)Y)+\text{sgn}((\pi-k_x)Y))\notag\\
&=&-\frac{1}{4}(\text{sgn}(Y))(1+\text{sgn}(\pi-k_x))
\end{eqnarray}
Hence $A=-1$ and $B=1$. It turns out that this has the effect of reversing the direction of jump, as it should. In this case a relative negative sign between the two halves of the Fourier integral of the Bloch Eigenstate has the effect of inverting it in the x-direction. 
\begin{figure}[H]
\centering
\includegraphics[scale=0.6]{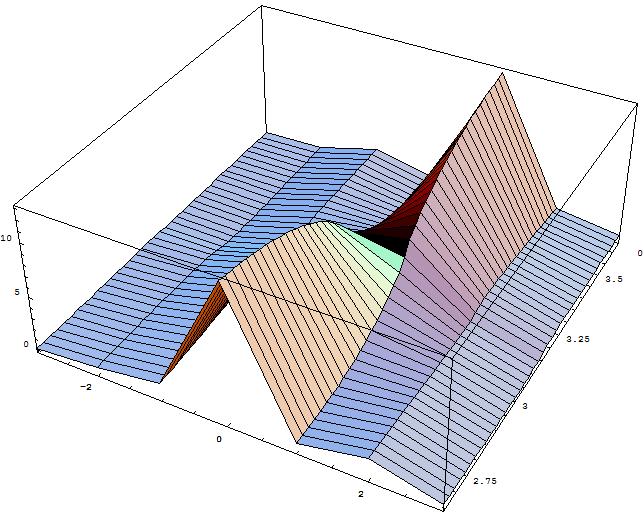}
\caption{WF $|W^1(x,Y)|^2$ near the jump for $-m=0.01<0$. $Y$ goes into the page. Note that its direction is opposite from that with $m>0$.}\end{figure}
\end{itemize}

\section{Pseudopotentials in terms of coherent states}
\label{ppcoherent}

In this appendix, we shall show how Eq. \ref{main}, where the PP $U^m$ is expressed in terms of the coherent state operators, simplifies to its usual definition in Eq. \ref{lag}. To start, it is first helpful to replace the gradient operators $\nabla_z^r$ by their conjugate variables $q$:
\begin{eqnarray}
U^m &\propto& \int dz_1dz_2 \sum_r v^m_r l_B^{2r}\nabla_z^r (c^\dagger_z c_z ) \nabla_z^r (c^\dagger_z c_z )\notag\\
&\propto& \sum_r \int d^2q v^m_r q^{2r} \rho(-q)\rho(q)\notag\\
&=& \sum_r v^m_r \int d^2q l_B^{2r}q^{2r} \int d^2z e^{iq\cdot z} c^\dagger_z c_z \int d^2z' e^{-iq\cdot z'}c^\dagger_{z'}c_{z'}\notag\\
&\propto & \sum_r v^m_r  \int d^2q l_B^{2r}q^{2r} \int d^2z \int d^2z' e^{iq\cdot (z-z')} c^\dagger_z c^\dagger_{z'}c_{z'}c_z+\text{quadratic}\notag\\
\end{eqnarray}
where $v^m_r$ are the coefficients of the $m^{th}$ Laguerre polynomial. Now we substitute the explicit expression of the coherent state operators from Eq. \ref{coherentdef}:
\begin{equation} c^\dagger_z =\sum_K e^{iz_2K}e^{-\pi A(z_1/C_1-K/(2\pi))^2}a^\dagger_K \end{equation}
where $a^\dagger_K$ creates the LLL Landau gauge wavefunction $\ket{\psi_K}$ given by Eq. \ref{LLLbasis}. With a slight abuse of notation of $z$ as being both a vector and complex number,
\begin{eqnarray}
&&U^m \\
&\propto& \sum_r \sum_{K_1...K_4}  v^m_r \int d^2q l_B^{2r}q^{2r} \int d^2z \int d^2z' e^{iq\cdot (z-z')} e^{iz_2K_1}e^{-\pi A(z_1/C_1-K_1/(2\pi))^2}\notag\\
&&e^{iz'_2K_2}e^{-\pi A(z'_1/C_1-K_2/(2\pi))^2}e^{-iz'_2K_3}e^{-\pi A(z'_1/C_1-K_3/(2\pi))^2}e^{-iz_2K_4}e^{-\pi A(z_1/C_1-K_4/(2\pi))^2}a^\dagger_{K_1} a^\dagger_{K_2} a_{K_3}a_{K_4} \notag\\
&\propto& \sum_r \sum_{K_1...K_4} v^m_r \int d^2q l_B^{2r}q^{2r} \int d^2z \int d^2z' e^{i(K_1-K_4)z_2}e^{i(K_2-K_3)z'_2}e^{iq\cdot (z-z')} e^{\frac{A}{C_1}(z_1(K_1+K_4)+z'_1(K_2+K_3))}\notag\\
&& e^{-A(x^2+x'^2)/(C_1^2l_B^2)}e^{-\frac{A}{4\pi}\sum_i K_i^2}a^\dagger_{K_1}a^\dagger_{K_2} a_{K_3}a_{K_4}\notag\\
&\propto& \sum_r \sum_{K_1...K_4} v^m_r \int d^2q l_B^{2r}q^{2r} \int d^2z \int d^2z' e^{iq\cdot (z-z')} e^{iK_1z_2-\frac{A}{2C_1}(z_1/l_B-l_B K_1)^2}e^{iK_2z'_2-\frac{A}{2C_1}(z'_1/l_B-l_B K_2)^2}\notag\\
&& e^{-iK_3z'_2-\frac{A}{2C_1}(z'_1/l_B-l_B K_3)^2}e^{-iK_4z_2-\frac{A}{2C_1}(z_1/l_B-l_B K_4)^2}a^\dagger_{K_1}a^\dagger_{K_2} a_{K_3}a_{K_4}\notag\\
&\propto &\frac{V_0}{4}\sum_{K_1...K_4}\int \frac{d^2q}{(2\pi)^2}\int d^2z d^2z' L_m(q^2l_B^2)e^{i q\cdot (z-z')}\psi^*_{K_1}(z)\psi^*_{K_2}(z')\psi_{K_3}(z')\psi_{K_4}(z)a^\dagger_{K_1}a^\dagger_{K_2} a_{K_3}a_{K_4} \notag\\
&=&\frac{V_0}{4\sqrt{2\pi l_B^2}}\int d^2r d^2r'\psi^\dagger(r)\psi^\dagger(r') L_m(l_B^2 \nabla^2)(\delta(r-r'))\psi(r')\psi(r)
\end{eqnarray}
where
\[\psi_{K=\frac{2\pi n }{L}}(r)=\frac{1}{\sqrt{\sqrt{\pi}Ll_B}}e^{-iKy}e^{-\frac{A}{2C_1}(\frac{x}{l_B}-l_BK)^2}\]
is the LLL Landau gauge wavefunction.

For $m=1$, the $r=0$ term vanishes due to fermionic antisymmetry. The $(\nabla_z c_z^\dagger\cdot \nabla_z c_z^\dagger) c_zc_z$ and $(\nabla_z c_z\cdot \nabla_z c_z) c_z^\dagger c_z^\dagger$ parts of the $r=1$ term also vanishes due to the same reason. Hence we are left with 
\begin{equation}
V^1 \propto \int dz_1dz_2 c_z(\nabla_z c_z^\dagger\cdot \nabla_z c_z)c_z^\dagger \propto \int dz_1dz_2 c_z^\dagger \nabla_z c_z^\dagger c_z \nabla_z c_z
\end{equation} 
which is the same manifestly local expression in \cite{qi2011}. The unimportant proportionality constant has the same units for all $m$, and can be deduced from dimensional analysis and normalization.

\section{The Coulomb Gauge condition for minimal coherent state spread}
\label{gaugederivation}

We perform the Euler-Lagrange minimization $\frac{\delta I}{\delta \theta}=0$ on $I[\theta]$ given by Eq. \ref{coherentI}. First, note that $I$ only depends explicitly on $\nabla_k \theta$, not $\theta$, because $|\nabla_k(e^{i\theta }f)|^2=|if\nabla_k\theta+\nabla f |^2$ where $f$ represents the remaining part of the integrand in $I[\theta]$ not containing $\theta$. Hence
\begin{eqnarray}
0&=&\frac{\delta I}{\delta \theta}=\nabla_k \cdot \frac{\partial I}{\partial (\nabla_k \theta)}\notag\\
&=&  \sum_m \int_{[-1/2,1/2]^2}d^2z \nabla_k\cdot (-i (e^{-E_m+ik_yz_2}\phi)^\dagger \nabla_k (e^{-E_m+ik_yz_2}\phi)+e^{-2E_m}\nabla_k \theta)\nonumber \\
&=&\sum_m\int_{-1/2}^{1/2} dz_1\int_{-1/2}^{1/2} dz_2 e^{-2E_m}(-2\partial_y E_m(z_2+i\partial_y E_m+\partial_y\theta)+\nabla_k\cdot \vec{a}\notag\\&&+a_y(iz_2-\partial_y E_m) +\frac{iA}{2\pi}+\nabla_k^2 \theta)  \nonumber \\
&=& \int_{-1/2}^{1/2} dz_2 \int_{-\infty}^{\infty}d\eta e^{-2\pi A\eta^2}\left(-2A\eta(z_2+iA\eta+\partial_y\theta)+\nabla_k\cdot\vec{a}+a_y(iz_2-\eta A) +\frac{iA}{2\pi}+\nabla_k^2 \theta\right)  \nonumber \\
&=& \int_{-1/2}^{1/2} dz_2 \int_{-\infty}^{\infty}d\eta e^{-2\pi A\eta^2}\left(-2iA^2\eta^2+\nabla_k\cdot\vec{a} +\frac{iA}{2\pi}+\nabla_k^2 \theta\right)  \nonumber \\
&=& \int_{-1/2}^{1/2} dz_2 \frac{1}{\sqrt{2}}\left(\nabla_k\cdot\vec{a} +\nabla_k^2 \theta+0\right)  \nonumber \\
&\propto & \nabla_k\cdot\vec{a} +\nabla_k^2 \theta \notag \\
&=&\nabla_k\cdot(\vec{a}+\nabla_k \theta)
\label{eq:el}
\end{eqnarray}
Some steps deserve explanation. I have defined $E_m=-\pi A \left(\frac{k_y}{2\pi}-\frac{z_1}{C_1}+m\right )^2=-\pi A \eta^2$ in line $3$, so that $\partial_y E_m = A\eta$. This combines $z_1$ and $m$ into one continuous variable $\eta$, thereby reducing the infinite sum into one simple gaussian integral. We also note that the factor multiplying $a_y$ in the third line disappears because it is linear in $\eta$ and $z_2$. This allows the final expression to be symmetric in $a_x$ and $a_y$.

The terms involving $A$ cancels nicely, implying that the optimal phase is independent of the aspect ratio $A$ used for the gaussian envelope. However there still exists a certain $A$ that will gives the minimal spread in the coherent state, and that will be derived in the subsection after the next.

Actually, the MLWF phase $\theta_W$ can also be derived via a conceptually identical E-L minimization approach~\cite{kivelson1982}. If we do not integrate over $k_y$ or sum over $m$, do not involve $z$ (i.e. just let $A=0$) and treat $k_y$ as a parameter, the minimal $\langle r^2\rangle $ will still be given by
\begin{equation}
\nabla_k\cdot(\vec{a}+\nabla_k \theta)
\end{equation}
This is obvious from the derivation of Eq.~\ref{eq:el}. WLOG, let us choose to work in gauges where $a_y=0$. Noting that $\nabla_k=\partial_{k_x}$ when $k_y$ is a parameter,

\begin{equation}\nabla_k^2 \theta = \partial^2_{k_x} \theta = -\partial_{k_x} a_x \end{equation}

The first term of $\theta_W$ (shown in Eq.~\ref{eq:XL}) solves this equation up to a term linear in $k_x$. The latter enforces the periodicity of $2\pi$ in $k_x$, and can thus be uniquely determined up to an irrelevant overall phase.

\section{The spread of the coherent state}
\label{spread}

Here is how the average spread of the coherent state can be computed:
\begin{eqnarray}
I[A]&=&\int_{[-1/2,1/2]^2} d^2z\langle r^2\rangle_{\psi_z}\notag\\
&=&\int_{[-1/2,1/2]^2} d^2z\sum_m \int d^2k \left|\nabla_k \left (e^{-E_m-ik_yz_2}e^{i\theta (k_x,k_y)}\phi(k_x,k_y)\right )\right |^2 \nonumber \\
&=&\int_{[-1/2,1/2]}dz_2 \int d^2k\int_{-\infty}^{\infty}d\eta  \left|\nabla_k \left (e^{-E_m-ik_yz_2}e^{i\theta (k_x,k_y)}\phi(k_x,k_y)\right )\right |^2 \nonumber \\
&=&\int_{-\frac{1}{2}}^{\frac{1}{2}}dz_2 \int d^2k\int_{-\infty}^{\infty}d\eta e^{-2\pi A \eta^2}(|\nabla_k \phi|^2 + (A\eta)^2 +z_2^2 + |\nabla_k \theta|^2 +2z_2 \partial_y \theta\notag\\
&& +\left[-i\vec{a}\cdot ((-A\eta +iz_2)\hat{y}+i\nabla_k \theta)+c.c.\right]  ) \nonumber \\
&=&\int_{-\frac{1}{2}}^{\frac{1}{2}}dz_2 \int d^2k\int_{-\infty}^{\infty}d\eta e^{-2\pi A \eta^2}\left(|\nabla_k \phi|^2 + (A\eta)^2 +z_2^2 + |\nabla_k \theta|^2 +2\vec{a}\cdot \nabla_k \theta  \right) \nonumber \\
&=&\int_{-\frac{1}{2}}^{\frac{1}{2}}dz_2 \int d^2k\int_{-\infty}^{\infty}d\eta e^{-2\pi A \eta^2}\left(|\nabla_k \phi|^2 + (A\eta)^2 +z_2^2 + \nabla_k\theta\cdot(\nabla_k\theta +2\vec{a})  \right) \nonumber \\
&=&\int_{-\frac{1}{2}}^{\frac{1}{2}}dz_2 \int d^2k\int_{-\infty}^{\infty}d\eta e^{-2\pi A \eta^2}\left(|\nabla_k \phi|^2 + (A\eta)^2 +z_2^2 + |\vec{a}_{new}|^2-|\vec{a}|^2 \right) \nonumber \\
&=&\int_{-\frac{1}{2}}^{\frac{1}{2}}dz_2 \int d^2k \frac{1}{\sqrt{2A}}\left(|\nabla_k \phi|^2  +z_2^2 + |\vec{a}_{new}|^2-|\vec{a}|^2 +\frac{A}{4\pi}\right) \nonumber \\
&=&\frac{1}{\sqrt{2A}}\int d^2k \left(\left[|\nabla_k \phi|^2-|\vec{a}|^2\right]   + |\vec{a}_{new}|^2+\frac{1}{12} +\frac{A}{4\pi}\right) 
\label{eq:r2}
\end{eqnarray}
where, as before, $E_m=-\pi A \left(\frac{k_y}{2\pi}-\frac{z_1}{C_1}+m\right )^2=-\pi A \eta^2$.

\section{$\Omega_1$ in terms of projected position operators}
\label{omega1}

Following Ref. \onlinecite{marzari1997}, we can also express $\Omega_1 $ as
\begin{eqnarray}
\Omega_1 &=& \int d^2k \left (|\nabla_k \phi|^2-|\vec{a}|^2\right ) \notag \\
&=& \int d^2k\left( \langle \nabla_k\phi|\nabla_k\phi \rangle - \langle \phi|\nabla_k \phi \rangle \langle \nabla_k \phi|\phi\rangle\right) \notag \\
&=& \int d^2k \langle \nabla_k \phi |Q|\nabla_k \phi \rangle\notag \\
&=& Tr(PxQx)+Tr(PyQy)\notag\\
&=& Tr(PxQ^2xP)+Tr(PyQ^2yP)\notag\\
&=&Tr[(PxQ)^2+(PyQ)^2]
\end{eqnarray}
where $P=|\phi\rangle\langle \phi|$ and $Q=I-P$. The sum of occupied bands is implied. The final expression is obviously a gauge-invariant quantity. The physical interpretation of $\Omega_1$ becomes clearer if we write
\begin{eqnarray}
\Omega_1 &=& Tr(PxQx)+Tr(PyQy)\notag\\
&=& Tr(Px^2) -Tr(PxPx)+ (x\leftrightarrow y) \notag\\
&=& Tr(P^2x^2)-Tr(P^2xP^2x) +(x\leftrightarrow y) \notag\\
&=& Tr(Px^2P)-Tr[(PxP)^2] +(x\leftrightarrow y) \notag\\
&=& Tr(Pr^2P)-Tr[\tilde{r}^2]
\end{eqnarray}
where $\tilde{x}=PxP$ is the operator whose eigenvectors are the MLWFs, and $\tilde{r}^2=\tilde{x}^2+\tilde{y}^2$. Hence $\Omega_1$ provides a certain measure of the spread of the state that exists within the subspace of occupied bands. This can be seen more clearly in the Wannier basis $|\vec R m\rangle=\frac{1}{4\pi^2}\int d^2k e^{-i\vec k \cdot \vec R} |\phi_m(\vec k )\rangle$, where~\cite{marzari1997}
\begin{equation} \Omega_1 = \sum_n \left[ \langle \vec 0n|r^2|\vec 0n\rangle -\sum_{\vec R m}|\langle \vec R m|\vec r|\vec 0 n\rangle|^2\right]
\end{equation}
We see that $\Omega_1$ is the mean-square spread of the WF (not necessarily maximally localized or even localized) minus a certain positive-definite quantity.

\section{Properties of the Mutual Information}

\subsection{Relation to the single-particle correlators}
\label{app:mutualcorr}

The mutual information as given by Eq. \ref{mutual} contains the two-site entanglement entropy, which depends on the two-site correlator $C_{xy}$ via Eq. \ref{Ixy00}. %It can be reduced to a much friendlier form by first expressing the two-site correlator in terms of single-site correlators. For that, we invoke Wick's Theorem:

%\begin{eqnarray}
%C_{xy}&=&\langle \b_x \b^\dagger_x \b_y \b^\dagger_y\rangle \notag\\
%&=&  \langle \b_x \b^\dagger_x \rangle\langle \b_y \b^\dagger_y \rangle-\langle \b_x \b^\dagger_y \rangle\langle \b_y \b^\dagger_x \rangle\notag\\
%&=& C_x C_y - C_{x-y}C^\dagger_{x-y}
%\end{eqnarray}
It is a matrix of single-particle propagators $[C_{xy}]_{ij}=\langle\beta_i \beta_j^\dagger\rangle$, where $i,j\in \{ x,y\}$ and $\b_x,\b_y$ are the bulk annihilation operators. Below, we shall write
\begin{equation}C_{xy}=\left(\begin{matrix}
 & C_x & C_{x-y} \\
 & C_{x-y}^\dagger  & C_y \\
\end{matrix}\right)=C_0 + V
\label{Cxy}\end{equation}
where $C_x$ and $C_y$ the on-site single-particle correlators, and $C_{x-y}$ is the (non-hermitian) single-particle propagator from site $y$ to $x$. The off-diagonal contribution $C_{x-y}$ decays rapidly for large $|x-y|$, and $I_{xy}$ can be accurately approximated in that limit. Below, we write $C_{xy}=C_0 + V$, where $C_0=C_x\otimes C_y$ and $V$ is a perturbation containing the off-diagonal parts $C_{x-y}$ and $C_{x-y}^\dagger$:
 
\begin{eqnarray}
I_{xy}&=&S_x+S_y-S_{xy}\notag\\
&=& S_x+S_y+\text{Tr}~((C_0+V)\log(C_0+V)+(1-C_0-V)\log(1-C_0-V))\notag\\
&\approx & S_x+S_y+\text{Tr}~((C_0+V)(\log C_0 + C_0^{-1}V-\frac{1}{2}(C_0^{-1}V)^2)\notag\\&&+(1-C_0-V)(\log (1-C_0) + (1-C_0)^{-1}V-\frac{1}{2}((1-C_0)^{-1}V)^2))\notag\\
&\approx& (S_x+S_y-\text{Tr}~(S_x\otimes S_y))+2 \text{Tr}~ V + \text{Tr}~ [V (\log C_0 + \log(1-C_0))]\notag\\&&+\frac{1}{2}\text{Tr}~ VC_0^{-1}V  +\frac{1}{2}\text{Tr}~ V(1-C_0)^{-1}V  \notag\\
&=& (S_x+S_y-S_x-S_y)+\frac{1}{2}\text{Tr}~ VC_0^{-1}V  +\frac{1}{2}\text{Tr}~ V(1-C_0)^{-1}V  \notag\\
&\approx& \frac{1}{2}\text{Tr}~[ V(C_0(1-C_0))^{-1}V  ]\notag\\
&\approx& \frac{1}{2}\text{Tr}~\left[C_{x-y}\frac{1}{C_y(\mathbb{I}-C_y)}C_{y-x}+(x\leftrightarrow y)\right]\notag\\
&\sim & \text{Tr}~ [C^\dagger_{y-x}C_{y-x}]
%&\sim& \frac{1}{2}\text{Tr}~ V^2
%&=& \frac{[a(1-a)-A^2][2b^2 +(|u|^2+|v|^2)]+2bA(1-2a)Re[u-v]}{(A^2-a^2)(A^2-(1-a)^2)}
\label{Ixy0}
\end{eqnarray}
where $C_x,C_y$ are the single-particle onsite correlators, and $C_{x-y}$ is the single-particle propagator between the two different sites $x$ and $y$. Note that we have implicitly assumed that $C_0$ and $V$ commute while going from lines 2 to 3, which holds in the IR limit.

\subsection{The mutual information in terms of the reduced density matrix}
\label{app:mutual}

The mutual information can also be understood as the Kullback-Liebler divergence\cite{cover1991} of the distributions described by $\rho_{\bold x \bold y}$ and $\rho_{\bold x}\rho_{\bold y}$, where $\rho_{\bold x}$ is the single-site reduced-density matrix (RDM) and $\rho_{\bold x \bold y}$ is the two-site RDM. Writing the trace over one site as $\text{Tr}~_{\bold x}$ and two sites as $\text{Tr}~_{\bold x \bold y}$, we have
\begin{eqnarray}
I_{\bold x\bold y} &=& S_{\bold x} + S_{\bold y} - S_{\bold x \bold y}\notag\\
&=& - \text{Tr}~_\bold y \rho_\bold x\log \rho_\bold x- \text{Tr}~_\bold x \rho_\bold y\log \rho_\bold y + \text{Tr}~_{\bold x \bold y} \rho_{\bold x \bold y}\log \rho_{\bold x \bold y}\notag\\
&=& - \text{Tr}~_{\bold x \bold y} \rho_{\bold x \bold y}\log \rho_{\bold x }-\text{Tr}~_{\bold x\bold y } \rho_{\bold y}\log \rho_{\bold y} + \text{Tr}~_{\bold x \bold y} \rho_{\bold x \bold y}\log \rho_{\bold x \bold y}\notag\\
&=& \text{Tr}~_{\bold x \bold y}\rho_{\bold x \bold y}\log \frac{\rho_{\bold x \bold y}}{\rho_\bold x \rho_\bold y}\notag\\
&=& \mathbb{E}\left[\log \frac{\rho_{\bold x \bold y}}{\rho_\bold x \rho_\bold y} \right]
\label{mutual2}
\end{eqnarray}
To maximize the mutual information $I_{\bold x \bold y}$, we need $\rho_{\bold x \bold y}$ and $\rho_\bold x \rho_\bold y $ to be as different as possible. Since the $\log$ function drops steeply below unity, we especially want to avoid situations where the (square root of the) single-site RDM eigenvalue is large compared to that of the two-site RDM. Since $\rho_\bold x = \text{Tr}~_\bold y \rho_{\bold x \bold y}$, the above-mentioned situation is more likely when the RHS contains a large number of contributions from states belonging to different $y$. Hence we conclude that a maximal $I_{\bold x \bold y}$ must have minimal spread of $\rho_{\bold x \bold y}$, i.e. have 2-particle states that are maximally entangled in the precise sense of Eq. \ref{mutual2}. Note that this scenario represents a hypothetical optimum, and may not be realized the physical systems that we have discussed.

\section{Correlators for two-band fermionic systems with particle-hole symmetry}
\label{app:twoband}

Here we derive the detailed results for two-band, particle-hole symmetric models studied in this work. They hold for generic two-band models at arbitrary chemical potential $\mu$, although we will ultimately use them only for the Dirac model at $\mu=0$.

Due to particle-hole symmetry, the onsite correlator (projector) $C_x$ takes the following form in spin space:
\begin{equation}C_x=\langle \b_x \b^\dagger_x\rangle=\left(\begin{matrix}
 & a & i A \\
 & -i A  & a \\
\end{matrix}\right)\label{Cx}\end{equation}
where $a$ and $A$ are real, up to inconsequential corrections at nonzero temperatures. At zero chemical potential $\mu$, we always have $a=\frac{1}{2}$. The single-particle propagator is slightly more complicated:
\begin{equation}C_{x-y}=\langle \b_x \b^\dagger_y\rangle=\left(\begin{matrix}
 & b & i u \\
 & i v  & b \\
\end{matrix}\right)\label{Cxy2}\end{equation}
When $\mu=0$, the equal-spin propagator $b=0$ and the unequal-spin propagators $u,v$ are real. At nonzero $\mu$, $u,v$ are not necessarily real but $b$ is still real. $|u|\neq|v|$ in general, since $C_{x-y}$ does not have to be hermitian in spin space alone. When the onsite propagators $C_x=C_y$, as in the case when they are related by translation symmetry in the same bulk layer, we obtain upon substituting Eqs. \ref{Cx} and \ref{Cxy2} into Eq. \ref{Ixy0}
\begin{eqnarray}
&&I_{xy}=\notag\\
&& \frac{[a(1-a)-A^2][2b^2 +(|u|^2+|v|^2)]+2bA(1-2a)Re[u-v]}{(A^2-a^2)(A^2-(1-a)^2)}\notag\\
\label{Ixy}
\end{eqnarray}
where $a,A,b,u,v$ are defined in Eqs. \ref{Cx},\ref{Cxy2} and \ref{Cxy}. At zero chemical potential, $b=0$ and $a=\frac{1}{2}$, and Eq. \ref{Ixy} simplifies further to
\begin{equation}
I_{xy}|_{\mu=0}=\frac{4(|u|^2+|v|^2)}{1-4A^2}
\label{Ixy2}
\end{equation}
Since the onsite contribution $A$ does not vary with the spatial displacement $\bold x -\bold y$, the decay properties of $I_{xy}$ at $\mu=0$ depends almost entirely on $u$ and $v$.

From Eq. \ref{Cx}, it is almost trivial to write down the expression of the single-site Entanglement entropy (EE):
\begin{eqnarray}
S_x&=&-\text{Tr}~[C_x \log C_x+ (\mathbb{I}-C_x)\log(\mathbb{I}-C_x)]\notag\\
&=&-2\sum_{\lambda_\pm}\left[\lambda_\pm \log \lambda_\pm+(1-\lambda_\pm)\log(1-\lambda_\pm)\right]
\label{entropy3}
\end{eqnarray}
where $\lambda_\pm=a\pm |A|$ are the eigenvalues of $C_x$. When $\mu=0$, $a=\frac{1}{2}$ and the EE is maximal at $S_x^{max}=\log 4$ at $A=0$. Indeed, without unequal-spin correlation ($A=0$), we have no information about what is happening to the other spin state. When $A$ is small, the maximal entropy is corrected according to $S_x\approx S ^{max}_x-2A^2=\log 4 -2A^2$. $S_x$ vanishes when $a=A=\frac{1}{2}$, which produces a pure eigenstate $\propto |-\rangle+|+\rangle$.

We now present the explicit forms of the correlators, denoted collectively as $G_{\bf{q}}$. Given a Hamiltonian $h_\bold q$, $G_\bold q(\tau)=e^{\tau(h_{\bold q}-\mu )}(\mathbb{I}+e^{\beta(h_{\bold q}-\mu)})^{-1}
$ (Eq. \ref{gq0}) is explicitly
\begin{eqnarray}
G_\bold q(\tau)&=&\frac{e^{-\tau \mu}}{2}\left(\cosh(\tau E_\bold q)\mathbb{I}+\frac{h_\bold q}{E_\bold q}\sinh(\tau E_\bold q)\right)\notag\\
&&\times\left(\left(1+\frac{\sinh \beta \mu}{\cosh \beta E_\bold q+\cosh \beta \mu}\right)\mathbb{I}-\frac{h_\bold q}{E_\bold q}\frac{\sinh\beta E_\bold q}{\cosh \beta E_\bold q +\cosh \beta \mu}\right)\notag\\
\label{gq}
\end{eqnarray}
In the limit of zero temperature $\beta\rightarrow \infty$, the equal-time correlator $G_\bold q=G_\bold q (\tau=0)$ tends to
\begin{eqnarray} G_\bold q&\rightarrow& G_\bold q|_{\mu=0}+ \frac{1}{2}\theta(|\mu|-E_\bold q)\left(\text{sgn}(\mu)\mathbb{I}+\frac{h_\bold q}{E_\bold q} \right)\notag\\
&=&\theta(\mu-E_\bold q)\mathbb{I}+\theta(E_\bold q-|\mu|)\frac{1}{2}\left(\text{sgn}(\mu)\mathbb{I}-\frac{h_\bold q}{E_\bold q}\right)
\label{mueq}
\end{eqnarray}
The physical interpretation of the above  is clear: When $E_\bold q>|\mu|$, $G_\bold q$ is exactly the same as in the $\mu=0$ case. For $E_\bold q<|\mu|$, $G_\bold q$ projects identically to either both bands or none depending on the sign of $\mu$.

For zero $\mu$ but nonzero temperature, we have
\begin{equation}
G_\bold q|_{\mu=0}=\frac{1}{2}\mathbb{I}-\frac{h_\bold q}{2E_\bold q}\tanh\frac{\beta E_\bold q}{2}
\label{gqtemp}
\end{equation}
Of course, This further reduces to the usual projection operator given by $\frac{1}{2}\left( \mathbb{I}-\frac{h_\bold q}{E_\bold q}\right)$ at zero-temperature.

\section{Derivation of the bulk mutual information for $(1+1)$-dim  critical boundary systems at $T=0$}
\label{app:critical}
We start from the following expression for the mutual information $I_{xy}$ between sites $x$ and $y$ (Eq. \ref{Ixy2}):
\begin{equation}
I_{xy}|_{\mu=0}=\frac{4(|u|^2+|v|^2)}{1-4A^2}\sim 8|u|^2\sim 8|v|^2
\end{equation}
where, as introduced in the main text and the previous appendix, $u$ and $v$ are the unequal-spin single particle propagators, and $A$ the unequal-spin onsite propagator which tends to zero beyond moderately large $n$. Below, we shall derive their asymptotic behavior in detail.

\subsection{Angular direction}
\label{app:critical1}

We evaluate Eq. \ref{uv} for a large angular interval of $\Delta x$ sites by deforming the contour around the branch cut from $z=0$ to $z=\infty$. (looking like a tight-lipped Pac-man):
\begin{eqnarray}
u\sim v &\sim&\frac{1}{2}\int_0^1 W_n^*(z^{-1})W_n(z)z^{2^n\Delta x}(\sqrt{z}-\sqrt{e^{2\pi i}z})\frac{dz}{z}\notag\\
&=&\int_0^1 W_n^*(z^{-1})W_n(z)z^{2^n\Delta x}\frac{1}{\sqrt{z}}dz\notag\\
&=&\int_0^1 (W_n^*(z^{-1})W_n(z)z^{2^n})z^{2^n(\Delta x-1)-1/2}dz\notag\\
&=&\int_0^1 Q(z)z^X dz
\label{u2}
\end{eqnarray}
where $W_{n}(z)=\frac{1}{\sqrt{2\pi}}D(z^{2^{n-1}})\prod_{j=1}^{n-1}C(z^{2^{j-1}})$ and $C(z),D(z)=\frac{1\pm z}{\sqrt{2}}$. We have decomposed the integrand into a term $Q(z)=W_n^*(z^{-1})W_n(z)z^{2^n}$ that does not have negative powers of $z$, and $z^X$ with $X=2^n(\Delta x-1)-1/2$ still very large. We next integrate by parts to get the asymptotic behavior of $u\sim v$:
\begin{eqnarray}
u\sim v&\sim &\int_0^1 Q(z)z^X dz\notag\\
&=&\frac{Q(1)}{X+1}-\frac{1}{X+1}\int_0^1Q'(z)z^{X+1}dz\notag\\
&=& \frac{Q(1)}{X+1}-\frac{Q'(1)}{(X+1)(X+2)}+...\notag\\
&=& 0-0 + \frac{Q''(1)}{(X+1)(X+2)(X+3)}+...\notag\\
&\sim & \frac{Q''(1)}{X^3}
\label{u3}
\end{eqnarray}
Here, we have stopped at the 2nd derivative of $Q$, because it is the lowest nonzero derivative at $z=1$. $Q(1)=Q'(1)=0$ because they each must contain at least one factor of $W_n(1)$ or $W_n^*(1)$, both of which are zero due to the presence of UV projectors $D(z^{2^{n-1}})$. We also truncate off higher derivative terms as they contain higher powers of $\frac{1}{X}$. As one may expect, $u$ or $v$ depends \emph{exclusively} on the behavior of the EHM basis at $z=1$ or $q=-i\log z = 0$, the IR point where criticality occurs.

Substituting the explicit form of $Q(z)$ and differentiating, we obtain
\begin{eqnarray}
u\sim v&\sim & \frac{Q''(1)}{X^3}\notag\\
&=& \frac{2^{2(n-1)}}{\pi}|C(1)^{n-1}D'(1)|^2\frac{1}{X^3}\notag\\
&=& \frac{2^{2(n-1)}}{\pi}\left|\sqrt{2}^{n-1}\left(-\frac{1}{\sqrt{2}}\right)\right|^2\frac{1}{X^3}\notag\\
&=& \frac{2^{3n-4}}{\pi}\frac{1}{(2^n\Delta x)^3}\notag\\
&=& \frac{1}{16\pi}\frac{1}{(\Delta x)^3}
\label{u4}
\end{eqnarray}
That $n$ drops out is not a coincidence, but a manifestation of scale invariance. Note that the presence of the critical point, which is a property of the Hamiltonian, only ensures that the first line of Eq. \ref{u2} will not evaluate to zero; the power-law decay rate is \emph{entirely} determined by the analytic properties of the chosen EHM basis at that IR point. In this case, the mutual information behaves asymptotically as
\begin{equation}
I_{xy}=I_n(\Delta x)\sim\frac{1}{32\pi^2}\frac{1}{(\Delta x)^6}
\end{equation}

\subsection{Radial direction}
\label{app:critical2}

Here, we evaluate Eq. \ref{uv2} for the mutual information between sites that are separated radially. When one of the layer lies at the UV boundary of the bulk (layer $1$), $u$ and $v$ between layers $1$ and $n$ can be easily evaluated viz.
\begin{eqnarray}
u,v&=&-i\oint_{|z|=1}\frac{dz}{z}W^*_{1}(z^{-1})W_{n}(z)z^{\mp \frac{1}{2}}\notag\\
&=&-i\oint_{|z|=1}\frac{dz}{z}\frac{1}{2\pi\sqrt{2^{n+1}}}(1-z^{-1})\frac{(1-z^{2^{n-1}})^2}{1-z}z^{\mp \frac{1}{2}}\notag\\
&=&i\oint_{|z|=1}\frac{dz}{z^{2\pm \frac{1}{2}}}(1-2z^{2^{n-1}}+z^{2^n})\notag\\
&=&i\oint_{|z|=1}\frac{dz}{z^{2\pm \frac{1}{2}}}-i\oint_{|z|=1}\frac{(2z^{2^{n-1}}-z^{2^n})dz}{z^{2\pm \frac{1}{2}}}\notag\\
\end{eqnarray}
The first integrand diverges when $z\rightarrow 0$, so the contour should be inverted about $|z|=1$ and closed at infinity. The second one diverges when $z\rightarrow \infty$ for $n\geq 2$, so the contour should be closed like a Pac-man.

Performing the resultant real integrals analogously to Eq. \ref{u2}, we obtain
\begin{equation}
|u|=\frac{1}{2\pi\sqrt{2^{n-1}}}
\end{equation}
\begin{equation}
|v|=\frac{1}{6\pi\sqrt{2^{n-1}}}
\end{equation}
The single-particle bulk propagator is not hermitian in spin-space, with $u\neq v$. The above are exact results, not asymptotic ones. However, exact results like these usually do not exist for more generic wavelet mappings. The resultant mutual information is
\begin{equation}
%- \log I_{(1,n)}= \log\frac{4(u^2+v^2)}{1-4A^2} \sim (n-1)\log\frac{1}{C(0)^2}+\log\frac{1-4A^2}{8 D_n(0)^2}+\text{small const.}
%\label{radial2}
I_{xy}=I(1,n)\sim 4(|u|^2+|v|^2)=\frac{10}{9\pi^2}\frac{1}{2^{n-1}}
\label{radialmutual}
\end{equation}

\section{The bulk mutual information for critical systems at $T\neq 0$ in any number of dimensions}
\label{app:temp}

In this appendix, we shall fill in the mathematical gaps in the derivations of various nonzero $T$ results for gapless systems. Since the mutual information $I_{xy}$ between sites $x$ and $y$ is (Eq. \ref{Ixy2}),
\begin{equation}
I_{xy}=\frac{4(|u|^2+|v|^2)}{1-4A^2}\sim 8|u|^2\sim 8|v|^2
\end{equation}
where $A$ trivially approaches zero beyond the first few layers $n$, we will just need to find the asymptotic behavior of the unequal-spin single particle propagators (two-point functions) $u\sim v$.

\subsection{Decay of nonzero $T$ correlators for generic critical dispersions}
\label{app:temp2}

We consider the energy dispersion Eq. \ref{criticaldispersion}

\begin{equation}  E_{\bold q} = \sqrt{\sum_j^D v_j^2 q_j^{2\gamma_j}} \end{equation}
In the $j^{th}$ direction, the decay rate of the correlators $u,v$  is given by the the imaginary part of the root $q=q_j$ of
\begin{equation}
v_j^2 q^{2\gamma_j} = -(m^2+\pi^2 T^2)
\end{equation}
nearest to the real axis, where $m^2=\sum_{i\neq j}^D v_i^2 q_i^{2\gamma_i}$ represents an effective mass from the momentum contributions from all the other directions. $m^2\ll T^2$ is always satisfied for a bulk layer sufficiently deep in the IR, i.e. of sufficiently large $n$. In terms of the variable $z=e^{iq}=e^{iq_j}$ or $q=q_j=\frac{z-z^{-1}}{2i}$ used previously, the decay rate is given by $-\log|z_0|$, where $z_0$ is the root of 
\begin{equation}
z^4 + 2(\alpha_j-1)z^2+1 =0
\end{equation} 
within the unit circle and closest to its boundary, with $\alpha_j=\frac{2e^{i\pi/\gamma}}{v_j^{2}}\left((\pi T)^2 + m^2\right)^{1/\gamma}$.  Solving the above equation and taking the imaginary part,
\begin{eqnarray}
|Im(q_j(m))|&=& \frac{\sin \frac{\pi}{2\gamma_j}}{v_j^{1/\gamma_j}}\left((\pi T)^2 + m^2\right)^{1/(2\gamma_j)}\notag\\
&\approx & U_j\left(1+ V_j\sum_{i\neq j}^D v_i^2 q_i^{2\gamma_i}\right)
\end{eqnarray}
where $U_j=\sin \frac{\pi}{2\gamma_j}\left(\frac{\pi T}{v_j}\right)^{\frac{1}{\gamma_j}}$ and $V_j=(2\gamma_j (\pi T)^2)^{-1}$. However, this is not the physical decay rate as it still depends on the other momenta. To obtain the physical decay rate, we integrate over the latter (taking $j=1$ without loss of generality):
\begin{eqnarray}
u\sim v &\sim& \int e^{i2^n \bold q \cdot \Delta \vec x}\tanh\frac{\beta E_{\bold q}}{2}d^D \bold q \notag\\
&\sim & \prod_{j=2}^D \int dq_j e^{-2^n |Im (q_1(m))|\Delta x_1 }e^{i2^n\sum_{j\geq 2}^D \Delta x_jq_j}\notag\\
&\approx & e^{-2^nU_1 \Delta x_1}\prod_{j=2}^D\left[\int dq e^{-(U_1V_1v_j^22^n \Delta x_1) q^2}e^{i2^n\Delta x_j q}\right]\notag\\
&\sim & e^{-2^nU_1 \Delta x_1}\prod_{j=2}^D e^{-\frac{1}{4U_1V_1 v_j^2}\frac{(2^n\Delta x_j)^2}{2^n\Delta x_1}}\notag\\
&= &  e^{-2^nU_1 \Delta x_1-\frac{2^n}{4U_1V_1 \Delta x_1}\sum_{j\geq 2}^D \frac{(\Delta x_j)^2}{v_j^2}}
\label{genericdecay1}
\end{eqnarray}
This is just Eq.\ref{multianisotropic2} for the general anisotropic case. For the isotropic linear Dirac case where $U_j=\frac{\pi T}{v}$ and $V_j=\frac{1}{2(\pi T)^2}$ for all $j$, Eq. \ref{genericdecay1} nicely simplifies to
\begin{eqnarray}
u\sim v &\sim &  exp\left(-2^nU_1 \Delta x_1-\frac{2^n}{4U_1V_1 \Delta x_1}\sum_{j\geq 2}^D \frac{(\Delta x_j)^2}{v_j^2}\right )\notag\\
&=&exp\left(-2^n\frac{\pi T}{v} \Delta x_1-\frac{2^n}{4\frac{\pi T}{v}\frac{1}{2(\pi  T)^2} \Delta x_1}\sum_{j\geq 2}^D \frac{(\Delta x_j)^2}{v^2}\right)\notag\\
&=&e^{-2^n\frac{\pi T}{v} \Delta x_1\left(1+\frac{\sum_{j\geq 2}^D (\Delta x_j)^2}{2\Delta x_1}\right)}\notag\\
&\approx & e^{-2^n\frac{\pi T}{v}\sqrt{\sum_j (\Delta x_j)^2}}\notag\\
&= & e^{-2^n\frac{\pi T}{v}|\Delta \vec x|}
\label{genericdecay2}
\end{eqnarray}
This shows that the correlators and hence mutual information decay isotropically in $\Delta \vec x$ for the isotropic linear Dirac model, at least in the neighborhood of $\hat e_j$, $j=1,2,...,D$. The nonlinearity of Eq. \ref{genericdecay1} is further explored in the main text around Fig. \ref{horizon_angle}.

\section{Calculational details for the imaginary time correlator}
\label{app:time}
Here we shall present the full derivations of the more involved results on the imaginary time correlator $C_n(\tau)$. We shall derive the results for an arbitrary chemical potential $\mu$, so as to illustrate the interesting continuous crossover of $C_n(\tau)$ as a nonzero chemical potential is introduced to a critical system. 

For large $\tau$ and $\mu\geq 0$, Eq. \ref{gq} and \ref{mueq} simplify to
\begin{equation}
G_{E_q< \mu}(\tau)= \frac{e^{-\mu \tau}e^{\tau E_q}}{2}\left(\mathbb{I}+\frac{h_q}{E_q}\right)
\label{gqtime1a}
\end{equation}
and
\begin{equation}
G_{E_q\geq  \mu}(\tau)= \frac{e^{-\mu \tau}e^{-\tau E_q}}{2}\left(\mathbb{I}-\frac{h_q}{E_q}\right)
\label{gqtime1b}
\end{equation}
Hence the full correlator is given by 
\begin{equation}
C_{n}(\tau)=\frac{e^{-\mu\tau}}{2}\int_{-\pi}^\pi dq |W_n(e^{iq})|^2 \left[\theta(E_q-\mu)e^{-\tau E_q}\left(\mathbb{I}-\frac{h_q}{E_q}\right)+\theta(\mu-E_q)e^{\tau E_q}\left(\mathbb{I}+\frac{h_q}{E_q}\right)\right]
\label{time}
\end{equation}
where $\theta$ is the Heaviside function. We only have to care about the extreme IR region of this integral, since $e^{-E_q\tau}=e^{-v_F|q|\tau}$ decays rapidly for large $\tau$ in a critical system.

To proceed further, we only need to understand the IR behavior of $W_n(z)=D(z^{2^{n-1}})\prod_{j=1}^{n-1}C(z^{2^{j-1}})$.
The chemical potential sets an energy scale that divides two qualitatively different regimes. When the layer $n$ is below above the energy scale of $\mu$, i.e. $n<n^*$ where $2^{n^*}=\frac{2\pi v_F}{\mu}$, $W_n(e^{iq})$ is effectively dominated by a sharp peak at $q_0=\frac{2\pi}{2^n}<\mu$. However, for $n>n^*$, $W_n(e^{iq})$ is governed by its analytic behavior near the IR point $z=1$. Perturbing away from the IR point with $z=e^{i(0+\Delta q)}\approx 1-i \Delta q$,
\begin{eqnarray}
|W_n(e^{i \Delta q})|^2&\approx & |W_n(1-i \Delta)|^2\notag\\
&=&W_n^*(1+i\Delta q)W_n(1-i \Delta q)\notag\\
&\approx &\frac{1}{2\pi} \left |\left(\prod_{j=1}^{n-1}C(1)\right)\right|^2\left |D'(1)\right|^22^{2(n-1)}(\Delta q)^{2}\notag\\
&=&\frac{2^{3n}}{32\pi} (\Delta q)^{2}
\label{u6}
\end{eqnarray}
Noting that the even part of $\frac{h_q}{E_q}$ is $\frac{|q|}{2}$, the correlator simplifies to (with $E_\nu=\mu$)
\begin{equation}
C_{n}(\tau)\approx\frac{e^{-\mu\tau}2^{3n}}{32 \pi}
\int_{0}^\pi dq  q^{2}\left[\theta( q-\nu)e^{-\tau v_F q}\left(\mathbb{I}-\frac{q}{2}\sigma_2\right)+\theta(\nu- q)e^{\tau v_F q}\left(\mathbb{I}+\frac{ q}{2}\sigma_2\right)\right]\\
\label{time2}
\end{equation}
This integral can be exactly solved in terms of incomplete Gamma functions. However, we just want to extract the relevant asymptotic behavior set by the scale $\mu \tau$. Since the correlator captures the IR behavior, it should remain invariant even if the upper limit of $\pi$ is replaced by an arbitrary $q_{cutoff}$ in the first term on the RHS. The following formulae come in handy: $\int_0^{\mu}q^\gamma e^{q\tau}dq \sim \frac{\mu^{\gamma+1}}{\gamma+1}$ for  $\mu\tau\ll 1$ and $\frac{e^{\mu \tau}\mu^\gamma}{\tau}$ for $\mu\tau \gg 1$. Also, $\int_\mu^{q_{cutoff}}q^\gamma e^{-q\tau}dq\sim\frac{\gamma!}{\tau^{\gamma+1}}$ for $\mu\tau\ll 1$ and $\frac{e^{-\mu \tau}\mu^\gamma}{\tau}$ for $\mu\tau\gg 1$.

For the case of chemical potential discussed in the main text, $\mu\tau=0$ and we always have
\begin{eqnarray}
C_n(\tau)|_{\mu=0}&\approx&\frac{e^{-\mu\tau}2^{3n}}{16 \pi} \frac{1}{v_F^3\tau^3}\left(\mathbb{I}+\frac{3}{2v_F\tau}\sigma_2\right)|_{\mu=0}\notag\\
&\sim &\frac{1}{16 \pi}\left(\frac{2^n}{v_F\tau}\right)^3\mathbb{I}
\label{time3}
\end{eqnarray}
The extension of this result to nonlinear dispersions will be discussed in Appendix \ref{genericdirac}. Eq. \ref{time3} may also be used for the short-time behavior when $\mu\neq 0$, as long as $\mu\tau\ll 1$.

\section{Extension to the generic critical $(1+1)$-dim Dirac Model and discussion of criticality }
\label{genericdirac}

In the main text, we have focused on the $M=1$ case of the critical (gapless) Dirac model 
\begin{equation} H_{Dirac}(k)=v_F[\sin k \sigma_1 + M(1-\cos k )\sigma_2]\end{equation}
where $M$ controls the relative weight between the $\sin k$ and $1-\cos k $ terms. These two terms have respectively linear and quadratic dispersions for small $|k|$, and here we study the effects of their interplay.

\begin{figure}[H]
\centering
\includegraphics[scale=1.25]{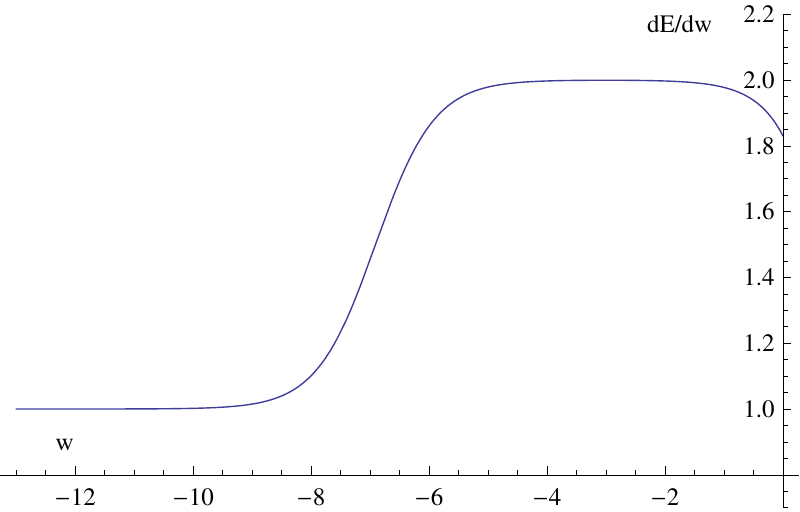}
\caption{$\frac{d\log E_w}{dw}$ for $M=2000$. We see that the dispersion of $E_k$ is linear ($\frac{dE_w}{dw}=1$) for $w<-\log M\approx -7.6$ or layer $n\approx 14$, but quadratic at momentum above that, till $k\sim O(1)$ where the dispersion levels off.} \label{fig:diracM}\end{figure}

A simple plot reveals that the dispersion $E_k=\sqrt{\sin^2 k + M^2(1-\cos k)^2}$ looks almost perfectly quadratic for $M>5$. However, the short linear region near $k=0$ is still expected to dominate the physics at the IR layers of the bulk. To see that this is indeed true, we explicitly calculate the order of the dispersion which is given by the derivative $\frac{d\log E_w}{dw}$, where $w=\log k$:
\begin{eqnarray}
\frac{d\log E_w}{dw}&=& \frac{e^w \left(-M^2+\left(-1+M^2\right) \cos\left[e^w\right]\right) \cot\left[\frac{e^w}{2}\right]}{-1-M^2+\left(-1+M^2\right) \cos\left[e^w\right]}
\end{eqnarray}
For small negative values of $w=\log k$, $\frac{d\log E_w}{dw}=2(1-M^2e^{-2w})=2\left(1-\frac{M^2}{k^2}\right)\approx 2$. But $\frac{d\log E_w}{dw}\rightarrow 1$ for large negative $w$. The transition region occurs at $k_c\approx \frac{1}{M}$, as shown in Fig. \ref{fig:diracM}. Note that a simple Taylor expansion will \emph{not} reveal a quadratic dispersion, because it is unable to concentrate on an exponentially small IR region.

The shape of the dispersion discussed above affects the bulk geometry profoundly at nonzero temperature. The emergent black hole radius behaves like $b\propto T^{\frac{1}{\gamma}}$, when $E_k\propto k^\gamma$. Hence $M$ sets the critical temperature which separates the $b\propto T$ and $b\propto \sqrt{T}$ regimes.

\subsubsection{Imaginary time correlator}

The zero-temperature asymptotics of the imaginary time correlation function are only dependent on the extreme IR behavior of the Hamiltonian. As evident in Eq. \ref{time}, a very large $\tau$ in the imaginary time correlator suppresses contributions from all but the lowest energy regime. As such, a finite $M$ should not change the long time behavior of $C_n(\tau)$.

This is however not true for an infinite $M$, which will produce a purely quadratic dispersion since $k_c\approx \frac{1}{M}=0$. %This does affect $C_n(\tau)$ nontrivially in the critical case.
Let us write $E_k=v_0k^\gamma$ for small $k$, where $\gamma=2$ here. Then Eq. \ref{time2} becomes
\begin{eqnarray}
\text{Tr}~ C_{n}(\tau)&\approx&\frac{2^{3n+1}}{32 \pi}
\int_{0}^\pi dq  q^{2}e^{-\tau v_0 q^\gamma}\notag\\
&=&\frac{2^{3n}}{16 \pi} \frac{\Gamma\left[\frac{3}{n}\right]}{\gamma (v_0\tau)^{\frac{3}{\gamma}}}
\end{eqnarray}
Hence a purely nonlinear dispersion of order $\gamma$ affects the long time correlator by changing the exponent in the power-law decay from $3$ to $3/\gamma$.
In our current context with the Dirac model, there exists a moderate imaginary time regime where the correlator decays like $\sim\tau^{-\frac{3}{2}}$. The duration of this regime becomes longer and longer as $M$ increases, till it finally becomes infinitely long at $M=\infty$.

\subsubsection{Spatial correlators and criticality}

In general, the power law decay of the spatial correlators depends on the EHM basis, and cannot be changed unless the critical point becomes degenerate. To be precise, the power law decay depends on the existence of a branch cut in the complexified correlator.

In our two-band case, the complexified correlator $\frac{h_z}{E_z}$, which is introduced in Section IV just before Table I, has nonzero elements given by
\begin{equation}
\sqrt{\frac{\sin k + iM(1-\cos k)}{\sin k - iM(1-\cos k)}}=\sqrt{\frac{i(z-z^{-1})-iM(2-(z+z^{-1}))}{i(z-z^{-1})+iM(2-(z+z^{-1}))}}
\end{equation}
and its reciprocal. They have square-root branch points at $z=1$, $z=\frac{M^2-1}{M^2+1}$ and $\frac{M^2+1}{M^2-1}$. To compute the correlator, we perform the contour integral around $|z|=1$ around the branch cut from $z=1$ to $z=\frac{M^2-1}{M^2+1}$, like what was done in Eq. \ref{u2}. The correlator matrix elements are thus proportional to
\begin{equation}
2\int_{\frac{1-M^2}{1+M^2}}^1 W^*_n(z^{-1})W_n(z)z^{2^n\Delta x}\sqrt{F(z)}\frac{dz}{z}
\end{equation}
where $F(z)=\sqrt{\frac{\left(z-\frac{1-M^2}{1+M^2}\right)}{\left(z-\frac{1+M^2}{1-M^2}\right)}}$, or its reciprocal. Via the same steps leading to Eq. \ref{u3}, we will eventually  find a $\sim\frac{1}{x^3}$ decay in the spatial correlator, \emph{as long as the $z=1$ branch point is present}.

The branch point at $z=1$ may disappear when it combines with another branch point. In our case, it happens when $M=\infty$. Then $F(z)$ becomes trivially equal to unity, and the correlator is identically zero.

There are other more interesting degenerate cases where we end up with a spatial correlator that decays \emph{exponentially}, even though the system is gapless. This happens when $F(z)$ has singularities within the unit circle, while the gapless point on the unit circle is not a branch point. An example is given by
\begin{equation}
H(k)=\sin^2 k \sigma_1 + (1-\cos k )\sigma_2
\end{equation}
whose singularities in $z=e^{ik}$ occur at $|i-1+i\sqrt{1+2i}|=0.346$, $1/0.346$ and $1$, with $1$ being a double root that cancels off in $\frac{h_z}{E_z}$. Hence its correlator decays like $\sim 0.346^{2^nx}$. Physically, the gapless point at $z=1$ is not critical because the two bands touch but do not intersect. The exponential decay arises from the effective mass scale due to the curvature of the dispersion.

For two-band models, gapless points of even order are always noncritical (degenerate). However, such points may be critical if there are more than two bands. In general, an $N$-band gapless point will be noncritical if the order of its dispersion is a multiple of $N$.

\section{Derivation of results for higher-dimensional critical systems at $T=0$}
\subsection{Decay of the imaginary time correlator for general $D$}
\label{app:multidimtime}

It is instructive to first perform the derivation for a $(2+1)$-dim boundary system. From Eq. \ref{time}, we have
\begin{eqnarray}
\text{Tr}~ C_n(\tau)%&=& \sum_\alpha C_n^{\alpha\alpha}(\tau)\notag\\
&=&\int_{-\pi}^\pi\int_{-\pi}^\pi dq_x dq_y |\tilde W_n(e^{iq_x})|^2 |\tilde W_n(e^{iq_y})|^2e^{- E_{\bold q} \tau}
\label{time2d}
\end{eqnarray}
where $\tilde W_n$ is the $(1+1)$-dim  bulk projector and $\text{Tr}~$ is a trace over the band indices, \emph{not} the $\upsilon$ indices (suppressed for now) labeling the $2^D-1$ bulk sectors containing various combinations of one-dimensional holographic basis vectors. For large $\tau$, it suffices to consider the contributions close to the IR point $\bold q=0$ where $W_n(e^{iq_j})$ is maximal and $E_{\bold q}\approx \sqrt{q_x^2+q_y^2}$. As explained in the main text, $\tilde W_n(e^{iq_j})$ either behaves like a constant or is linear in $q_j$ near $q_j=0$, depending on whether the IR or UV projector is chosen. In this appendix, we shall derive the forms of \emph{all} the terms in the correlator $\text{Tr}~ C^{\upsilon_1\upsilon_2}_n$, and not just the dominant terms.

Let us write the $\upsilon$ index in binary form $(\kappa_1,\kappa_2,...,\kappa_D)$, where $\kappa_j=0,1$ depending on whether the leading factor of $W_n^\upsilon(z)$ corresponds to an IR or UV projector. From Eq. \ref{u6} and the definition of the projectors in the main text, we know that $|W_n(e^{iq_j})|^2\approx \zeta_j^2 q_j^{2\kappa_j}$ for $q_j<\frac{2\pi}{2^n}$, where (letting $v_F=1$ for simplicity)
\begin{equation}
\zeta_j^2=\frac{1}{2\pi} 2^{(-1)^{\kappa_j}+(2\kappa_j+1)(n-1)}
\end{equation}
We next perform the integral in Eq. \ref{time2d} iteratively, starting from the integral over $q_x$:
\begin{equation}
\text{Tr}~ C(\tau)\approx \int_{-\pi}^\pi A_x^2A_y^2 k_y^{2\kappa_y} J_{k_y,\kappa_x}(\tau) dk_y
\label{time2d2}
\end{equation}
where $J_{k_y,\kappa_x}$ is an effective massive $(1+1)$-dim  correlator. For large $\tau> \frac{1}{m}$ and $\kappa_x=1$, it can be approximated by
\begin{eqnarray}
J_{k,\kappa_x=1}(\tau)&=& \int_{-\pi}^\pi  e^{-\sqrt{k^2+q^2}\tau} q^{2\kappa_x}dq\notag\\
&\approx & 2\int_{0}^\infty  e^{-\sqrt{k^2+q^2}\tau} q^{2\kappa_x}dq \notag\\
&= & 2\int_{k}^\infty  e^{-\epsilon\tau} (\epsilon^2-k^2)^{\kappa_x}\frac{\epsilon}{\sqrt{\epsilon^2-k^2}}d\epsilon \notag\\
&= & 2\int_{k}^\infty  e^{-\epsilon\tau} (\epsilon^2-k^2)^{\kappa_x-1}\frac{\epsilon(\epsilon^2-k^2)}{\sqrt{\epsilon^2-k^2}}d\epsilon \notag\\
&\approx  & 2\int_{k}^\infty  e^{-\epsilon\tau} (\epsilon^2-k^2)^{\kappa_x-1}\left(\epsilon^2-\frac{k^2}{2}\right)d\epsilon \notag\\
&=  & 2\int_{k}^\infty  e^{-\epsilon\tau} \left [(\epsilon^2-k^2)^{\kappa_x}+\frac{k^2}{2}(\epsilon^2-k^2)^{\kappa_x-1}\right]d\epsilon \notag\\
&=& e^{-k\tau}\frac{(2+k\tau)^2}{\tau^3}\notag\\
&=& \frac{e^{-k\tau}}{\tau^3}Q_{\kappa_x=1}(\tau k)
\label{time2d3}
\end{eqnarray}
where $Q_{\kappa}$ is a $2\kappa$-th degree polynomial with constant term $(2\kappa)!$. The approximation from line 1 to 2 is extremely accurate for large $\tau$, while that from line 4 to 5 is valid for for extremely small $k$. This is the regime that contributes most to $\text{Tr}~ C_n(\tau)$, because $J_{k,\kappa_x}$ is suppressed by at least like $e^{-k\tau}$ where $\tau$ is large. For small $k$, the integrand does not decay fast, and indeed it is the regime where $u\gg k$ that contributes most to the integral. %We have left the dependence on $\kappa_x$ explicit throughout most of the derivation, since higher $\kappa_x$..
The other case, $J_{k,\kappa_x=0}(\tau)$, resist all known approaches of analytical approximation. However, it is obvious that it behaves asymptotically like
\begin{equation}  J_{k,\kappa_x=0}(\tau)\sim \frac{e^{-k\tau}}{\tau} \end{equation}
from the relation $(\partial_\tau^2-k^2)J_{k,\kappa_x=0}(\tau)=J_{k,\kappa_x=1}(\tau)$, i.e. with $Q_{\kappa_x=0}(\tau k)\sim \text{const}$.

For a critical system in $D+1$ dimensions, we just have to replace
$E_q$ by $v_F\sqrt{\sum_{j=1}^{D} q_j^2}$, and substitute that into Eq. \ref{time2d}. Now let's define $T_j(\tau)$ to be the integrand of $\text{Tr}~ C_n(\tau)$ with the first $j$ dimensions integrated over. Our goal is to find the asymptotic behavior of $T_D(\tau)$. We have
\begin{equation}
T_i(\tau)\approx e^{-P_i \tau }\left[\prod^i_{j=1}\int_{-\pi}^{\pi}dq_j \zeta^2_j q_j^{2\kappa_j}\right]
\end{equation}
where $P_i=\sqrt{\sum_{j=1}^i q_j^2}$. We perform the integral over last variable using the same approximations (valid for large $\tau$) as in Eq. \ref{time2d3}:
\begin{eqnarray}
T_i(\tau)&\approx &\left[\prod^{i-1}_{j=1}\int_{-\pi}^{\pi}dq_j \zeta^2_j q_j^{2\kappa_j}\right]\int_{-\pi}^{\pi}dq \zeta^2_i q^{2\kappa_i}e^{-P_i \tau  }\notag\\
&= &2\zeta^2_i\left[\prod^{i-1}_{j=1}\int_{-\pi}^{\pi}dq_j \zeta^2_j q_j^{2\kappa_j}\right]\int_{0}^{\infty}dq  q^{2\kappa_i}e^{-\sqrt{P_{i-1}^2+q^2} \tau  }\notag\\
&\approx  &2\zeta^2_i\left[\prod^{i-1}_{j=1}\int_{-\pi}^{\pi}dq_j \zeta^2_j q_j^{2\kappa_j}\right]Q_{\kappa_i}(\tau P_{i-1} )\frac{e^{-P_{i-1} \tau  }}{\tau^{2\kappa_i+1}}\notag\\
&= &2\zeta^2_i\tilde Q_{\kappa_i}\left(-\tau \frac{d}{d\tau} \right)\left(\frac{1}{\tau^{2\kappa_i+1}}T_{i-1}(\tau)\right)\notag\\
&= &2^{i-1}\left[\prod_{j=2}^{i}\zeta^2_j\tilde Q_{\kappa_j}\left(-\tau \frac{d}{d\tau} \right)\right]\left(\frac{1}{\tau^{2(\sum_{j=2}^i\kappa_j)+i-1}}T_{1}(\tau)\right)\notag\\
&= &2^i (2\kappa_1)!\prod_{j=1}^{i}\zeta^2_j\left[\prod_{j=2}^{i}\tilde Q_{\kappa_j}\left(-\tau \frac{d}{d\tau} \right)\right]\frac{1}{\tau^{2(\sum_{j=1}^i\kappa_j)+i}}\notag\\
&\propto &\prod_{j=1}^{i}\zeta^2_j\frac{1}{\tau^{2(\sum_{j=1}^i\kappa_j)+i}}\notag\\
&\sim& \prod_{j=1}^{i}\left(\frac{2^{n-1}}{\tau}\right)^{2\kappa_i+1}
\label{timeD}
\end{eqnarray}
On line 4, the tilde in $\tilde Q_{\kappa}\left(-\tau \frac{d}{d\tau} \right)$ denote normal ordering of the $\tau$ and $\frac{d}{d\tau}$ operators, i.e. $Q_{\kappa_j}$ and its products will have all $\tau$ moved to the left of all $\frac{d}{d\tau}$. When going from the sixth to seventh line, we note that each operator $(-\tau^n)\frac{d^n}{d\tau^n}\rightarrow \frac{(n+\kappa)!}{\kappa !}$ when acting on expressions of the form $1/\tau^{\kappa+1}$, without incurring additional factors of $1/\tau$.

Hence
\begin{equation}
%\text{Tr}~ C(\tau)\sim \frac{1}{\tau^2}\prod_{j=1}^D |D^{\kappa_j}_n(1)|^2\frac{2^{(2\kappa_j+1)(n-1)}}{\tau^{2\kappa_j}}
\text{Tr}~ C_n(\tau)\sim \frac{1}{v_F^D\tau^D}\prod_{j=1}^D \frac{2^{(2\kappa_j+1)(n-1)}}{(v_F\tau)^{2\kappa_j}}
\label{generaltau}
\end{equation}
, after restoring $v_F$. Evidently, the leading terms occur when $\kappa_j=1$ for just one $j$, and are zero for the others. Hence, we have
\begin{equation}
%\text{Tr}~ C(\tau)\sim \frac{1}{\tau^2}\prod_{j=1}^D |D^{\kappa_j}_n(1)|^2\frac{2^{(2\kappa_j+1)(n-1)}}{\tau^{2\kappa_j}}
\text{Tr}~ C_n(\tau)\sim \left(\frac{2^n}{v_F \tau}\right)^{D+2}
\label{generaltau2}
\end{equation}
This is the main result for the imaginary time correlator in the multidimensional critical case.%, and we see that \textbf{$\text{Tr}~ C (\tau)$ only depends the Hamiltonian through $v_F$, and on the wavelet basis through the $\kappa_j$ of each direction $j$.}

For the multidimensional massive case with a fixed mass $m$, $P_i$ in Eq. \ref{timeD} is replaced by $P_i=\sqrt{m^2+\sum_{j=1}^i q_j^2}$. Hence $T_1(\tau)$ also contains a mass and the third last line of Eq. \ref{timeD} becomes \[2^i \left[\prod_{j=1}^{i}\zeta^2_j\tilde Q_{\kappa_j}\left(-\tau \frac{d}{d\tau} \right)\right]\frac{e^{-m\tau}}{\tau^{2(\sum_{j=1}^i\kappa_j)+i}},\]
valid for $m\tau<1$ due to the approximations in Eq. \ref{time2d3}. Hence

Eq. \ref{generaltau} becomes
\begin{eqnarray}
&&\text{Tr}~ C_n(\tau)|_{0<m\tau<1}\notag\\
&\sim& \frac{e^{-m\tau}}{v_F^2\tau^2}\prod_{j=1}^D \frac{2^{(2\kappa_j+1)(n-1)}}{(v_F\tau)^{2\kappa_j}} + \text{higher orders of $1/\tau$}
\label{generaltaum}
\end{eqnarray}
and
\begin{equation}
\text{Tr}~ C_n(\tau)|_{m\tau\gg 1}\sim e^{-m\tau} \times \text{weak dependence on powers of  $1/\tau$}
\label{generaltaum2}
\end{equation}
The nonzero  mass correlator is exponentially suppressed by $e^{-m\tau}$, though for small $m\tau$ we still see a subleading power law in $\tau$, albeit with a different power from that of the massless case.

\subsection{Nonuniversal properties of $(2+1)$-dim critical model}
\label{app:nonuniversal2d}

Here we illustrate how a different choice of model in $(2+1)$-dim can affect certain quantities but not others. We consider the model
\begin{eqnarray}
H(q_x,q_y)&= &d(q)\cdot \sigma\notag\\
&=&\sin q_x \sigma_1+\sin q_y \sigma_2 +(\cos q_x-\cos q_y)\sigma_3
\end{eqnarray}
which is also gapless at $q=0$. Its correlator in momentum space is
\begin{eqnarray}
G_q&=&\frac{1}{2}\left( \mathbb{I}-\hat d(q)\cdot \sigma \right)\notag\\
&=&\frac{1}{2}\left(\begin{matrix}
 & 1 -\frac{\cos q_x -\cos q_y}{E_q}& -\frac{\sin q_x -i\sin q_y}{E_q} \\
 & -\frac{\sin q_x +i\sin q_y}{E_q} & 1+\frac{\cos q_x -\cos q_y}{E_q} \\
\end{matrix}\right)
\label{gq2d}
\end{eqnarray}
where $E_q= \sqrt{2(1-\cos q_x \cos q_y)}\rightarrow \sqrt{q_x^2+q_y^2}=|q|$ near criticality. When considered as a $(1+1)$-dim  correlator depending on $q_x$($q_y$) alone, it behaves like a massive correlator with mass $q_y$($q_x$), as can be seen from its poles at $\pm i\cosh^{-1}\sec q_y$ (and vice versa for $q_x\leftrightarrow q_y$).

Since the single-site correlator $C_x$ is given by
\begin{equation}C_x=\int_{-\pi}^\pi\int_{-\pi}^\pi dq_x dq_y |W_n(e^{iq_x})|^2 | W_n(e^{iq_y})|^2 G_{\bold q}\end{equation}
where $ |W_n(e^{iq_j})|^2$ is even about $q=0$, we see that the off-diagonal components, being odd in $q_x$ or $q_y$, must disappear. This is different from the $1+1$-dim Dirac model, where $d_2(q)$, which is not odd in $q$,  plays an important role in the decay of the correlator. In the current $(2+1)$-dim case, it is $d_3$ that controls the decay of the correlator.

Juch as importantly, note that the nonconstant part of the diagonal terms are odd under the interchange $q_x\leftrightarrow q_y$. Hence $G_q\propto \frac{1}{2}\mathbb{I}$ due to the symmetry betwee the wavelet bases $W_n(e^{iq_x})$ and $W_n(e^{iq_y})$. With the off-diagonal part $A$ (defined previously) vanishing rigorously, the single-site entropy is always maintained at exactly $S_x=\log 4$, the same universal limiting value in the $(1+1)$-dim  case. Evidently, the small $n$ (UV) behavior of the entropy depends nonuniversally on the details of the model.  %If not, $S_x$ may saturate at an arbitrary value between $0$ and $2\log 2$ when $n\rightarrow 0$. This dissimilarity from the $1+1$ D case is mainly due to the different definition of the Dirac Hamiltonian.

%Now let's consider the 2-site correlator $C_{xy}$ at equal $n$, i.e.

%\begin{equation}
%C_{\Delta x, \Delta y}&=&\int_{-\pi}^\pi\int_{-\pi}^\pi dk_x dk_y |W_n(e^{ik_x})|^2 |\tilde W_n(e^{ik_y})|^2 e^{i2^n (k_x\Delta x+k_y \Delta y)} G_k
%\label{Cx2d}
%\end{equation}
%The mutual information of a critical $2+1$ D system is still given by Eq. \ref{Ixy}, reproduced here for clarity:

%\begin{eqnarray}
%I_{xy}&=& \frac{[a(1-a)-A^2][2b^2 +(|u|^2+|v|^2)]+2bA(1-2a)Re(u-v)}{(A^2-a^2)(A^2-(1-a)^2)}
%\end{eqnarray}

%But some of $a,A, b,u ,v$ will be more complicated than in the $1+1$ D case.

\section{Relationship between real and imaginary time correlators}
\label{sec:time}

Our discussion of the EHM will not be complete without a proper discussion of the real time correlator, which is arguably of more direct physical significance. However, its oscillatory nature makes it unsuitable as a definition of bulk distance. Here, we shall discuss its mathematical and physical significance with the imaginary time correlator.
\subsection{Critical case with linear dispersion}

When there is a linearly dispersive critical point $E_q = v_F q$, Galilean invariance is restored and there is symmetry between space and time. Restricting ourselves again to $(1+1)$-dimensions, the bulk correlator within a layer is explicitly given by
\begin{eqnarray}
C(n,n,\Delta x, \Delta t)&=&\sum_q  |W_{n}(q)|^2 e^{i2^nq\Delta x}G_q(-i\Delta t)\notag\\
&=&\sum_q   |W_{n}(q)|^2 e^{i2^nq\Delta x}e^{i E_q \Delta t}G_q\notag\\
&=& \sum_q  |W_{n}(q)|^2 e^{iq[2^n\Delta x+ v_F \Delta t ]}G_q
\label{realtime1}
\end{eqnarray}
which depends symmetrically on $2^n \Delta x$ and $v_F \Delta t$. From Eq. \ref{u4}, we deduce that the magnitude of the real time correlator behaves like
\begin{equation}
|C(n,-i\Delta t)|=u|_{\Delta x =2^{-n}v_F \Delta t}\sim \frac{1}{16\pi}\left(\frac{2^n}{v_F \Delta t}\right)^3
\end{equation}
which is identical to that of the imaginary time correlator\footnote{Its \emph{phase}, however, exhibits a nonuniversal and generally non-oscillatory behavior that depends on the details of the model}. This conclusion is consistent with the result obtained by a naive Wick rotation, since extra factors of $i$ in a power-law do not affect the decay behavior.

When the dispersion is nonlinear, Galilean invariance is lost and the correct result cannot be simply obtained via Wick rotation, even if the system is still critical.

\subsection{Non-critical cases}

When a mass scale is present, the energy $E_q$ is bounded below by a value $m$, i.e. $E_q=m+ \epsilon_q$ where $\epsilon_q\geq 0$. Hence, the real time bulk correlator
\begin{eqnarray}
C(n_1,n_2,0, \Delta t)
&=&\sum_q  W_{n_1}^*(q)W_{n_2}(q) e^{i E_q \Delta t}G_q\notag\\
&=&e^{im\Delta t}\sum_q  W_{n_1}^*(q)W_{n_2}(q) e^{i \epsilon_q \Delta t}G_q
\label{realtime2}
\end{eqnarray}
acquires an oscillatory phase with frequency $m$, a result again consistent with Wick-rotating the exponential decay $e^{-m\tau}$ behavior of the imaginary time correlator.

\section{The behavior of geodesic distance}
\label{app:geodesics}
Some results of this subsection and the next can also be found in Ref. \onlinecite{qi2013}. Here, we reproduce them for completeness.

\subsection{AdS space}

Due to its remarkable symmetry, Anti-de-Sitter space can be embedded in a flat Minkowski spacetime one dimension higher. As such, it inherits the simple metric structure of the latter, which yields simple expressions for geodesic distances.

For brevity, lets only consider the Euclidean $(2+1)$-dim AdS space, since its higher dimensional analogues will give rise to very similar expressions. We parametrize the space by coordinates $w=(\rho,\theta,\tau)$, and embed it in a 4-d Minkowski spacetime as the locus of $X^ax^b\eta_{ab}=X\cdot X = R^2$, where $R$ is the AdS radius, $\eta=diag(1,-1,-1,-1)$ and
\begin{eqnarray}
X^\mu(w)&=& \left(\sqrt{\rho^2+R^2}\cosh \frac{ \tau}{\sqrt{R^2+\frac{L^2}{4\pi^2}}},\sqrt{\rho^2+R^2}\sinh\frac{ \tau}{\sqrt{R^2+\frac{L^2}{4\pi^2}}},\rho \cos \theta, \rho \sin \theta\right)\notag\\
\label{geod1}
\end{eqnarray}
Here $\tau$ appears with a rescaling factor of $\frac{1}{\sqrt{R^2+\frac{L^2}{4\pi^2}}}$, instead of the more conventional $\frac{1}{R}$. This is to ensure that the rescaled AdS metric
\begin{equation}
\frac{1+\frac{\rho^2}{R^2}}{1+\frac{L^2}{4\pi^2R^2}}d\tau^2 + \frac{d\rho^2}{1+\frac{\rho^2}{R^2}}+\rho^2 d\theta^2\rightarrow d\tau^2 + \rho^2 d\theta^2
\label{adsmetric}
\end{equation}
is $O(2)$-invariant at $\rho=\frac{L}{2\pi}\gg R$.
From well-known properties of the Minkowski metric, the geodesic distance between points $w_1$ and $w_2$ is given by
\begin{equation}
d_{12}=R\cosh^{-1}\left(\frac{X(w_1)\cdot X(w_2)}{R^2}\right)
\label{geod2}
\end{equation}
From Eq. \ref{geod2}, we find the geodesic distance corresponding to an angular displacement %$\Delta x = \rho\Delta \theta$ to be
to be
\begin{eqnarray}
d^{min}_{\Delta \theta}&=&R\cosh^{-1}\left(\frac{\rho^2 + R^2 - \rho^2 \cos \Delta \theta}{R^2}\right)\notag\\
&=& R\cosh^{-1}\left(1+\frac{2\rho^2}{R^2}\sin^2\frac{\Delta \theta}{2}\right)\notag\\
&\sim & 2R\log \frac{\rho \sin\Delta \theta}{R}\notag\\
&\approx & 2R\log \frac{\rho \Delta \theta}{R}
%&=& 2R \log \frac{|\Delta x|}{R}
\label{ads1}
\end{eqnarray}
for $\rho \gg R$. In the scale-invariant case where $\Delta x =\rho \Delta \theta$, and we also have $d^{min}_{\Delta x}\sim 2R \log \frac{|\Delta x|}{R}$.

For a radial displacement with $\Delta \rho = |\rho_2-\rho_1|$, we have
\begin{eqnarray}
d^{min}_{\Delta \rho}&=&R\cosh^{-1}\left(\frac{\sqrt{(R^2+\rho_1^2)(R^2+\rho_2^2)}-\rho_1\rho_2}{R^2}\right)\notag\\
&\approx &R\cosh^{-1}\left(\frac{\rho_1}{R}\frac{\rho_2}{R}\left(\left(1+\frac{R^2}{2 \rho_1^2}\right)\left(1+\frac{R^2}{2 \rho_2^2}\right)-1\right)\right)\notag\\
&\approx& R\cosh^{-1}\left(\frac{1}{2}\left(\frac{\rho_1}{\rho_2}+\frac{\rho_2}{\rho_1}\right)\right)\notag\\
&=& R \left|\log\frac{\rho_1}{\rho_2}\right|\notag\\
&=& R|\Delta(\log \rho)|,
\label{ads2}
\end{eqnarray}
also for $\rho_1,\rho_2 \gg R$.

The geodesic distance in the imaginary time direction is given by
\begin{eqnarray}
d_\tau &=&R\cosh^{-1}\left(\left(\frac{\rho^2}{R^2}+1\right)\cosh\frac{2\pi \tau}{L^2 + 4\pi^2 R^2}-\frac{\rho^2}{R^2}\right)\notag\\
&\approx &R\cosh^{-1}\left(\frac{\rho^2}{R^2}\left(\cosh\frac{2\pi \tau}{L^2 + 4\pi^2 R^2}-1\right)\right)\notag\\
&\approx& R\cosh^{-1}\left(\frac{\rho^2}{R^2}\frac{(2\pi)^2 \tau^2}{2(L^2 + 4\pi^2R^2)}\right)\notag\\
&\sim& 2R\log \frac{2\pi \rho \tau}{RL}
\label{ads3}
\end{eqnarray}
where we have used $\rho\gg R$ while going from line $1$ to $2$, and $t\ll L$ from line $2$ to $3$ and $\rho \tau \gg RL$ from line $3$ to $4$.

\subsection{Geodesics near a black hole horizon}
\label{app:rindler}

We start from a metric
\begin{equation}
ds^2= V(r)d\tau^2 + \frac{dr^2}{V(r)}+ r^2 d\Omega^2 
\end{equation}

with $V(r_0)=0$, which admits a horizon at $r=r_0$. This horizon is a null surface within which $ds^2=0$. Intuitively, the geodesics between two points infinitesimally close to the horizon must necessarily wrap around the horizon (i.e. with with $r$ being constant), since the cost of radial displacements $\Delta r $ diverges to infinity at the horizon. Hence we always have $\Delta s \approx r_0 \Delta \theta$ near a horizon.

To explore the near-horizon geometry more rigorously, we switch to Rindler coordinates valid near the horizon $r=r_0$. We define
\begin{equation}
\rho = 2 \text{sgn}(r-r_0)\sqrt{\frac{|r-r_0|}{|V'(r_0)|}}
\end{equation}
so that, to order $O(\rho^2)$, the metric becomes

\begin{eqnarray}
ds^2|_R&=& \rho^2\left(1+\frac{V''(r_0)}{8}\rho^2\right)dT^2 + \left(1-\frac{V''(r_0)}{8}\rho^2\right)d\rho^2 + \left(r_0\pm \frac{|V'(r_0)|\rho^2}{4}\right)^2 d\Omega^2\notag\\
\end{eqnarray}

where $T=\frac{V'(r_0)}{2}\tau$ and the $\pm$ sign refers to the region immediately outside and inside the horizon respectively. 

Upon dropping the 2nd order terms, we recover the usual Rindler metric which just describes a plane with $T$ taking the role of the polar angle. To avoid a conical singularity, we require that $T$ has period $2\pi$, i.e. that the period of $\tau$ and thus $\beta= T^{-1}$ is of the value $\frac{4\pi}{V'(r_0)}$.

We now specialize to an example most relevant to the main text, which is the near-horizon geometry of a $(2+1)$-dim BTZ black hole. The horizon occurs at $r=b$, with another parameter $R$ setting the overall length scale. The horizon must also occur at a $D+1$-th order zero of $V(r)$. Hence we have
\begin{equation}
V(r)=\frac{r^3-b^3}{R^2r}
\end{equation}
with $V'(b)=\frac{3b}{R^2}$ and $V''(b)=0$. That the second derivative is identically zero is unique to $D=2$ dimensions. After some tedious derivation, the geodesic distance at Rindler radius $\rho_0$ between two points separated by $\Delta \theta$ is
\begin{eqnarray}
&&\Delta \lambda\notag\\ &=&\sqrt{\frac{2}{3}}\frac{R}{\sqrt{1+\eta^2}}\cosh^{-1}\left[ \sqrt{1+\eta^2}\cosh \left(\sqrt{\frac{3}{2}} \frac{\sqrt{1+\eta^2}}{R}b\Delta \theta\right)\right]
\end{eqnarray}
where $b$ is the horizon radius and
\begin{equation} \eta=\sqrt{\frac{3}{2}}\frac{\rho_0}{R}\text{ sech}\sqrt{\frac{3}{2}}\frac{b\Delta\theta}{R}
\end{equation}
A moment of calculation reveals that the above reduces to $\Delta\lambda = b\Delta\theta$ as we approach the horizon where $\eta\propto \rho_0\rightarrow 0$.

\section{Simple formula for entanglement entropy of free fermions}
\label{entformula}

The goal of this appendix is to derive the formula \begin{equation}S=-\text{Tr}~[C \log C + (\mathbb{I}-C)\log (\mathbb{I}-C)]\end{equation} for the entanglement entropy of free fermions, starting from the definition $S=-\text{Tr}~\rho\log\rho$ where $\rho$ is the reduced density matrix. 
\subsection{First derivation: Via entanglement Hamiltonian}

By definition, 
\begin{equation}
\rho=\frac{e^{-H_E}}{Z}
\end{equation}
where $H_E$ is the entanglement Hamiltonian and $Z$ enforces the normalization $\text{Tr}~ \rho=1$. The latter can also be interpreted as a partition function with respect to the entanglement ``temperature'' $\beta_E=1$. By \cite{peschel2002} or the appendix of \cite{leeye2015}, $H_E$ is related to the single-particle correlator $C$ via
\begin{equation} 
C=\frac1{1+e^{H_E}}
\end{equation}
for free fermion systems. Writing the \emph{single-particle} eigenvalues of $H_E$ as $\epsilon_l$, we have (performing a trace over many-body states)
\begin{eqnarray}
S&=& -\text{Tr}~\rho\log\rho \notag\\
&=& \text{Tr}~\rho (H_E + \log Z)\notag\\
&=& Z^{-1}\text{Tr}~H_E e^{-H_E} + (\text{Tr}~\rho)\log Z\notag\\
&=& \prod_{l} (1+e^{-\epsilon_l})^{-1}\left[\sum_l \epsilon_l e^{-\epsilon_l}\prod_{j\neq l}(1+e^{-\epsilon_j})\right]+ \log\prod_{l} (1+e^{-\epsilon_l}) \notag\\
&=& \sum_l\left[\frac{\epsilon_l }{1+e^{\epsilon_l}}+\log(1+e^{-\epsilon_l})\right]
\end{eqnarray}
This useful intermediate formula expresses the entanglement entropy $S$ in terms of the single-particle entanglement eigenenergies $\epsilon_l$. Continuing the derivation, 
\begin{eqnarray}
S&=& \sum_l\left[\frac{\epsilon_l }{1+e^{\epsilon_l}}+\log(1+e^{-\epsilon_l})\right]\notag\\
&=& \sum_l\left[\frac{\epsilon_l }{1+e^{\epsilon_l}}+\log(1+e^{\epsilon_l})-\epsilon_l\right]\notag\\
&=& \sum_l\left[\left(\frac1{1+e^{\epsilon_l}}+\frac1{1+e^{-\epsilon_l}}\right)\log(1+e^{\epsilon_l})+\epsilon_l\left(\frac1{1+e^{\epsilon_l}}-1\right)\right]\notag\\
&=& \sum_l\left[\frac1{1+e^{\epsilon_l}}\log(1+e^{\epsilon_l})+\frac1{1+e^{-\epsilon_l}}\log(1+e^{-\epsilon_l})\right]\notag\\
&=& -Tr \left[C\log C + (\mathbb{I}-C)\log (\mathbb{I}-C)\right]
\end{eqnarray}
Note that the trace on the last line is performed only over single-particle states.

\subsection{Second derivation: via direct counting}

Here, we give a more general derivation \emph{not} limited to fermions, but to any type of particle whose states can be empty, singly occupied, doubly occupied etc. up to $(r-1)$-occupied~\cite{lee2014exact}. Call the single-particular correlator (occupation) eigenvalues $c_{a_j}$, where $j=1,...,N$ label the particles and $a_j=1,...,r$ labels the state occupancy. Note that $\sum_j^rc_j=1$. In the case of fermions, for instance, $r=2$ and we have $c_1=\frac1{1+e^{\epsilon}}$, $c_2=1-\frac1{1+e^{\epsilon}}$.

The eigenvalues of the reduced density matrix $\rho$ are given by $\prod_{\{a_j\}}c_{a_j}$ over all rearrangements of the $N$ labels $a_j$. Hence 
\begin{eqnarray}
S&=& -Tr \rho\log\rho\notag\\
&=& -\left(\prod_j \sum_{a_j}\right) c_{a_1}...c_{a_N}\log (c_{a_1}...c_{a_N})\notag\\
&=& -\sum_{k=1}^N\left(\prod_j \sum_{a_j}\right) c_{a_1}...c_{a_N}\log c_{a_k}\notag\\
&=& -\sum_{k=1}^N\left(\prod_j \sum_{a_j}\right) c_{a_1}...c_{a_{k-1}}c_{a_{k+1}}...c_{a_N}(c_{a_k}\log c_{a_k})\notag\\
&=& -\sum_{k=1}^N\sum_{a_k} 1^{N-1}(c_{a_k}\log c_{a_k})\notag\\
&=& -N \sum_{j=1}^r c_j \log c_j
\end{eqnarray}
For $r=2$, this of course reproduces the desired result $S=-Tr \left[C\log C + (\mathbb{I}-C)\log (\mathbb{I}-C)\right]$.

\section{Relationship between the Wannier polarization and entanglement spectrum of free fermions}
\label{wannES}

Here we shall elaborate on an interesting relationship between the Wannier polarization and entanglement spectrum (ES) of free fermions, quantities used respectively in Parts I and II for the study of topological properties.  Encoded in both of these quantities are, in essence, the extent of real space non-locality of the eigenstates due to quantum constraints. The Wannier functions are not perfectly localized because their constituent occupied eigenstates do not form a complete basis; the behavior of their polarization further reveal the constraints due to nontrivial (quantum) band topology. In a related vein, quantum entanglement occurs due to the innate non-commutativity of position and momentum, which is a purely quantum property.

The relationship between the Wannier polarization spectrum and the entanglement spectrum will be clarified through an explicit interpolation between them\cite{marzari1997, qi2011,alexandradinata2011,huang2012,lai2014}. This interpolation gives a physically intuitive picture relating wavefunction localization and their entanglement properties. Central to understanding this interpolation is a detailed analysis of the decay properties of the Wannier and entanglement eigenfunctions. Specifically, the imaginary gap that controls the wavefunction locality is shown to also provide a rigorous lower bound for the decay rate of the ES. It also gives a natural explanation for the edgestate-entanglement spectrum correspondence that has already garnered significant interest in the study of topological condensed matter systems\cite{qi2012,hughes2011,li2008,thomale2010,yao2010,turner2010,fidkowski2010,PhysRevB.86.045117,hermanns2014}, most recently involving the holographic topological insulator~\cite{gulee2015} introduced in Sect. \ref{sec:holotopo}. While the following results do not require the system to be translationally invariant, if the latter condition holds the essential behavior of the ES can be directly expressed in terms of the complex-analytic properties of the lattice Hamiltonian, the same properties that govern the spatial rate of decay of the Wannier functions.

As an offshoot from this interpolation, we will also obtain estimates for the asymptotic spacing between entanglement energies for generic free-fermion lattice systems. Despite being asymptotic results, these estimates are in practice quite accurate beyond the first one or two eigenenergies. The entanglement spacing is of potential importance in the estimating the truncation error incurred in computations based on the density-matrix renormalization group (DMRG) approach \cite{white1992}, whose complexity are drastically reduced by discarding irrelevant degrees of freedom in a Schmidt decomposition. Such methods have since evolved into sophisticated algorithms that exploit the short-range entanglement between Matrix Product States (MPS) for one dimensional systems, and Projected Entangled-Pair states (PEPS) for higher dimensional systems\cite{kraus2010,dubail2013,wahl2013,wahl2014}. With them, the physical properties of a large class of systems are computed with hirtherto unattained accuracy and efficiency\cite{peschel1999,schollwock2005}.

%We have derived asymptotic bounds on the behavior of the Entanglement Spectrum of free-fermion lattice systems, and showed how it is related to the Wannier function decay rate which is in princple exactly computable. The asymptotic bound can be made precise when exact results are known at certain points in parameter (transverse momentum) space, as demonstated in our example. Although we have only explicitly worked out the case with two bands, the general case follows directly, with the eigenenergies from each occupied band having its own spectral flow. Similarly, our results can be extended to higher dimensions by treating the momentum in each additional transverse direction as a parameter. 

%This paper is organized as follows. In section \ref{sec:foundations}, we shall introduce the entanglement spectrum and Wannier polarization spectrum of free fermion systems, and suggest how they may be related. Following that will be Section \ref{sec:main}, where we present Eq. \ref{key}, our key result for the asymptotic ES spacing. We shall illustrate it through a toy example involving the Dirac Model, and prove it in detail via an interpolation between the Wannier operator and the entanglement projector. Finally, we shall discuss further applications of our results to the study of Block Toeplitz matrices in Section  \ref{sec:toeplitz}.

\subsection{Theoretical review}
\label{sec:foundations}

\subsubsection{Entanglement Spectrum}

Consider a free-fermion system described by a Hamiltonian $H=\sum_{i,j}f_i^\dagger h_{ij}f_j$, with $f_i$ annihilating a fermion at site $i$. To study its entanglement properties, we introduce a real-space partition by defining a subregion $A$ and its complement $B=\bar{A}$ in the system. With these regions, we can define a reduced density matrix (RDM) $\rho_A$ by partially tracing out the degrees of freedom (DOFs) of region $B$:
\begin{eqnarray}
\rho_A={\rm tr_B}\left[\ket{G}\bra{G}\right]\,,
\end{eqnarray}
where $\ket{G}$ is the ground state of the system, and $\rho=\ket{G}\bra{G}$ its full density matrix. For free fermion systems, a crucial simplification follows from the fact that all multi-point correlation functions obey Wick's theorem. This allows the  RDM $\rho_A$ to be always expresssed in the following Gaussian form\cite{peschel2002}:
\begin{eqnarray}
\rho_A=e^{-H_E},~H_E=\sum_{i,j\in A}f_i^\dagger {h_E}_{ij}f_j
\end{eqnarray}
where $h_E$, known as the single-particle ``entanglement Hamiltonian'', has a role superficially resembling that of a physical Hamiltonian at finite temperature. Furthermore, $h_E$ can be determined from the two-point correlation function ${C}_{ij}=\bra{G}f_i^\dagger f_j\ket{G},~i,j\in A$ via
\begin{equation}
h_E=\log\left(C^{-1}-\mathbb{I}\right)
\label{peschel}
\end{equation}
with $\mathbb{I}$ the identity matrix. $C$, being the correlator\footnote{In this appendix, $C$ refers to the correlator and not the Chern number.} \emph{within} subsystem $A$, is obtained by projecting $P$, the correlation matrix of the whole system, onto the subsystem $A$.  Writing $R=\sum_{i\in A}\ket{i}\bra{i}$ as the projection operator\cite{huang2012edge,huang2012,alexandradinata2011,lee2014prp} that implements the entanglement cut onto $A$, we obtain%\footnote{To be more precise, $RPR$ is a matrix with the dimension of the total number of sites, while $C$ is its truncation to the $A$ subsystem. Since they only differ by trivial zero eigenvalues in the $B$ subsystem, we will simply identify $C$ with $RPR$ in the following.}
\begin{equation}
\hat C= RPR\,.\notag
\label{rpr}
\end{equation}
Now, $P$ is also a projection operator, since it projects onto the occupied states via $P=\sum_n\theta(-\lambda_n)\ket{n}\bra{n}$. Here $\ket{n}$ and $\lambda_n$ are the eigenstates and eigenvalues of the single particle Hamiltonian $h$, and $\theta(x)$ is the step function. 
For instance, $P=\sum_k \theta(-\epsilon_k)$ for a Fermi sea,  
while $P=\frac{1}{2}\left(\mathbb{I}-\hat d(k)\cdot \sigma \right)$ for a two-band free-fermion lattice Hamiltonian $H(k)=d(k)\cdot \sigma$, where $\sigma_i$, $i=1,2,3$ are the Pauli matrices. 

Although $P$ and $R$ do not generically commute, the eigenvalues of the operators $RPR$ and $PRP$ are in fact equal because both $P$ and $R$ are projectors. This useful little fact was shown in Refs. \onlinecite{huang2012edge,huang2012,lee2014prp}, and in fact holds for generic basis-independent combinations of $P$ and $R$. To facilitate the Entanglement-Wannier interpolation that we shall introduce shortly, we shall henceforth identify the correlator $C$ with
\begin{equation}
\hat C'=PRP
\label{prp}
\end{equation}
with the entanglement spectrum, i.e. the eigenspectrum of $h_E$, completely determined by the eigenspectrum of $\hat C$ or $\hat C'$ via Eq. \eqref{peschel} or Eq. (\ref{prp}). 

\subsubsection{Wannier Polarization Spectrum}

We next define the Wannier polarization spectrum in a way slightly different from Sect. \ref{sec:wannierbasis}. The Wannier functions $\ket{\psi}$ are defined as the eigenfunctions of the \emph{Wannier operator}\cite{kivelson1982} 
\begin{equation}
\hat W= PXP
\label{wannier}
\end{equation}
where $P$ is the projectors onto the occupied bands as before, and $X=\frac{x}{L}$ is the position operator that takes values between $0$ and $1$, where $L$ is the length of the system in the direction of $x$. The eigenvalues of $\hat W$ form the \emph{Wannier polarization spectrum}, which physically correspond to the centers of mass of the corresponding Wannier functions (WFs) $\psi(x)=\langle x|\psi\rangle$, as plotted in Fig. \ref{comparespectra}. Essentially, the latter are the `best possible' localized orbitals formed from the occupied DOFs, and will reduce to delta function peaks when there are no unoccupied bands, i.e. when $P$ is trivial. That the WFs are indeed maximally localized has been shown in various sources like Refs. \onlinecite{kivelson1982,marzari1997}. For our purposes, their optimal localization allows us to uniquely determine their real-space decay rate which we shall utilize extensively later on. Note that a periodic version of $\hat W$, i.e. with $X=e^{\frac{2\pi i x}{L}}$, was used in Section \ref{sec:wannierbasis} as well as frequently in the literature~\cite{qi2011,lee2013,lee2014,yu2011}, in order to be consistent with the periodicity of the system. In our case, however, it is more convenient to use the aperiodic version from Eq. \ref{wannier} since we will be studying the physics near the entanglement cut.

\subsubsection{Comparison of $\hat C'$ and $\hat W$}

\begin{figure*}
\includegraphics[scale=0.39]{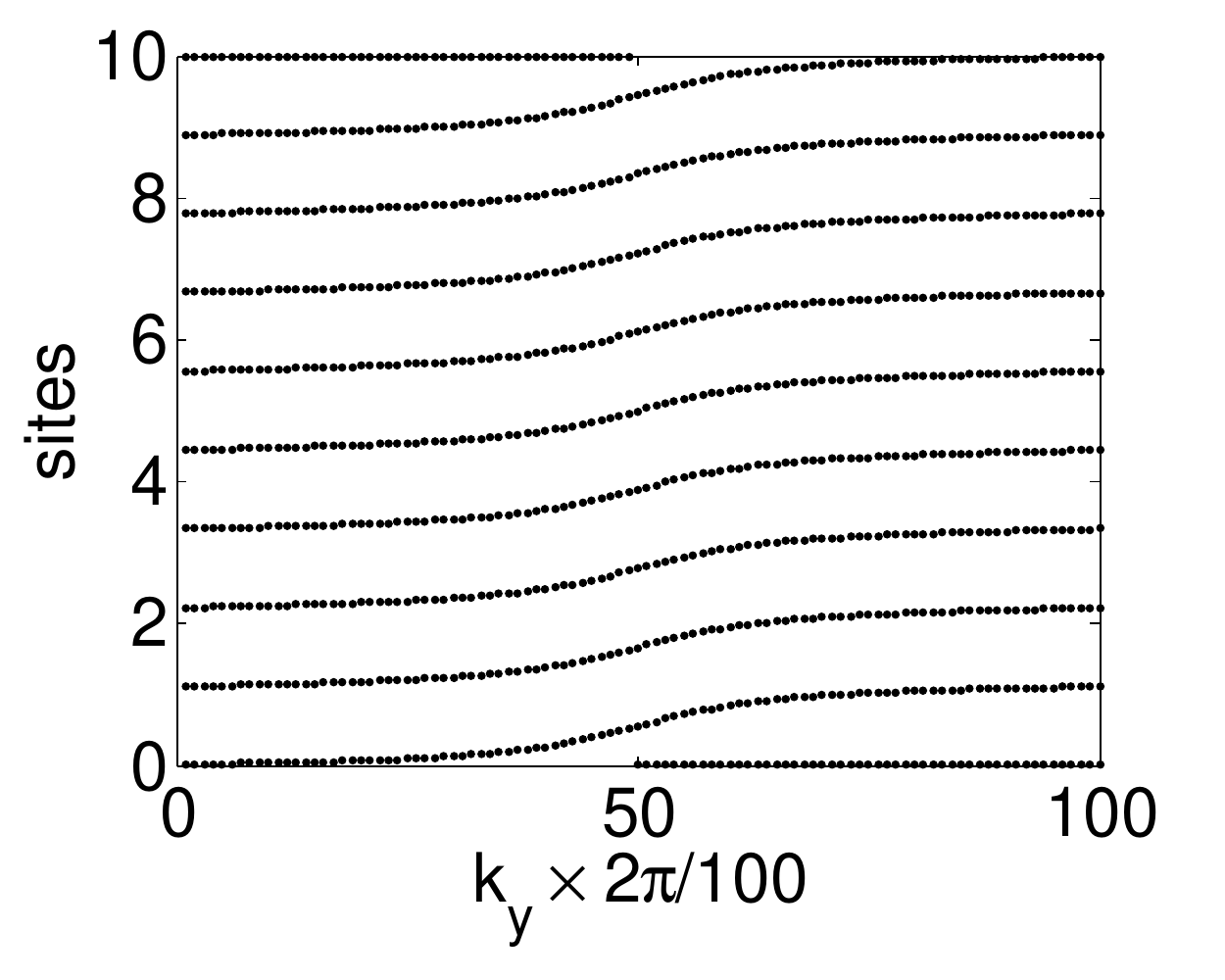}
\includegraphics[scale=0.39]{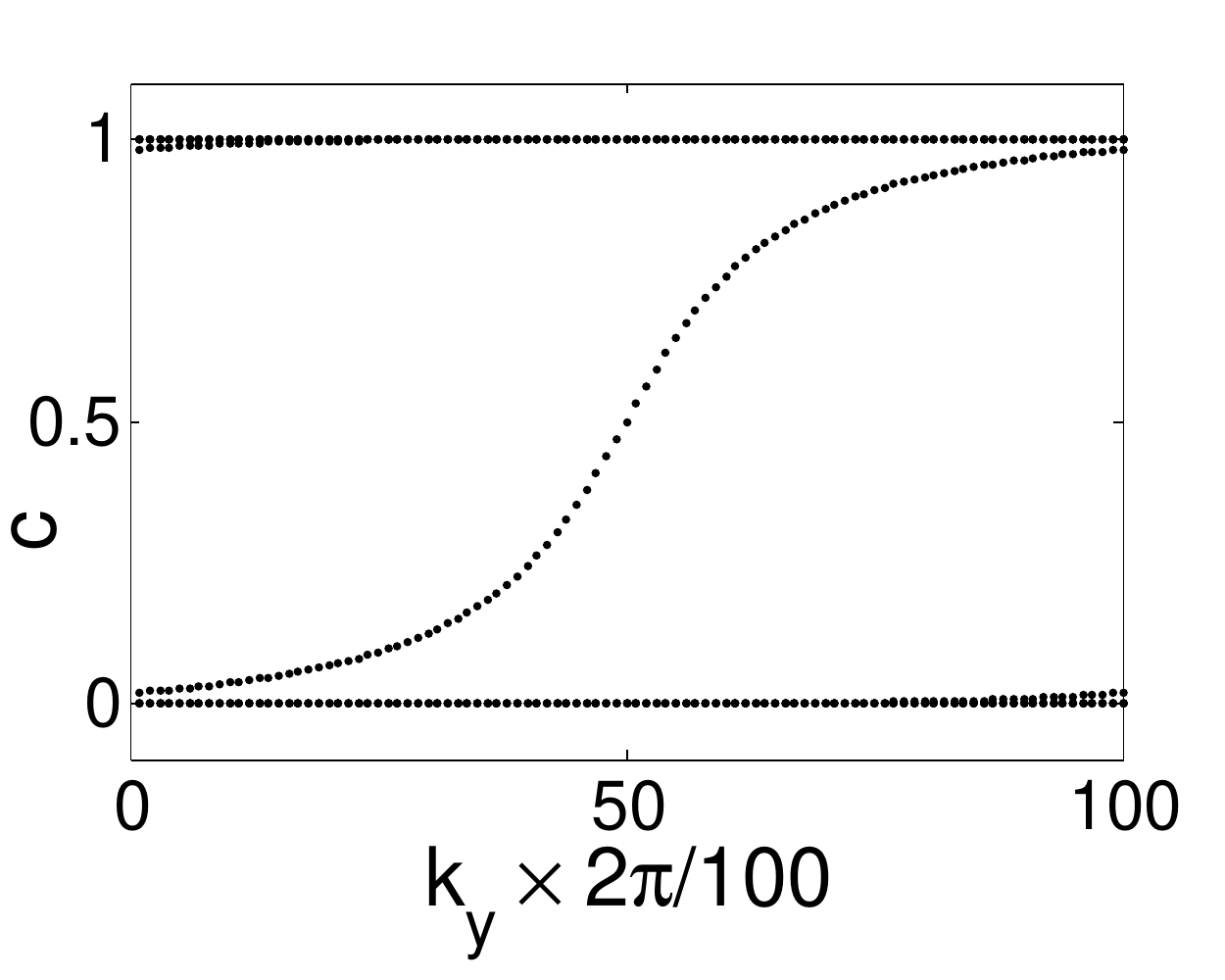}
\includegraphics[scale=0.39]{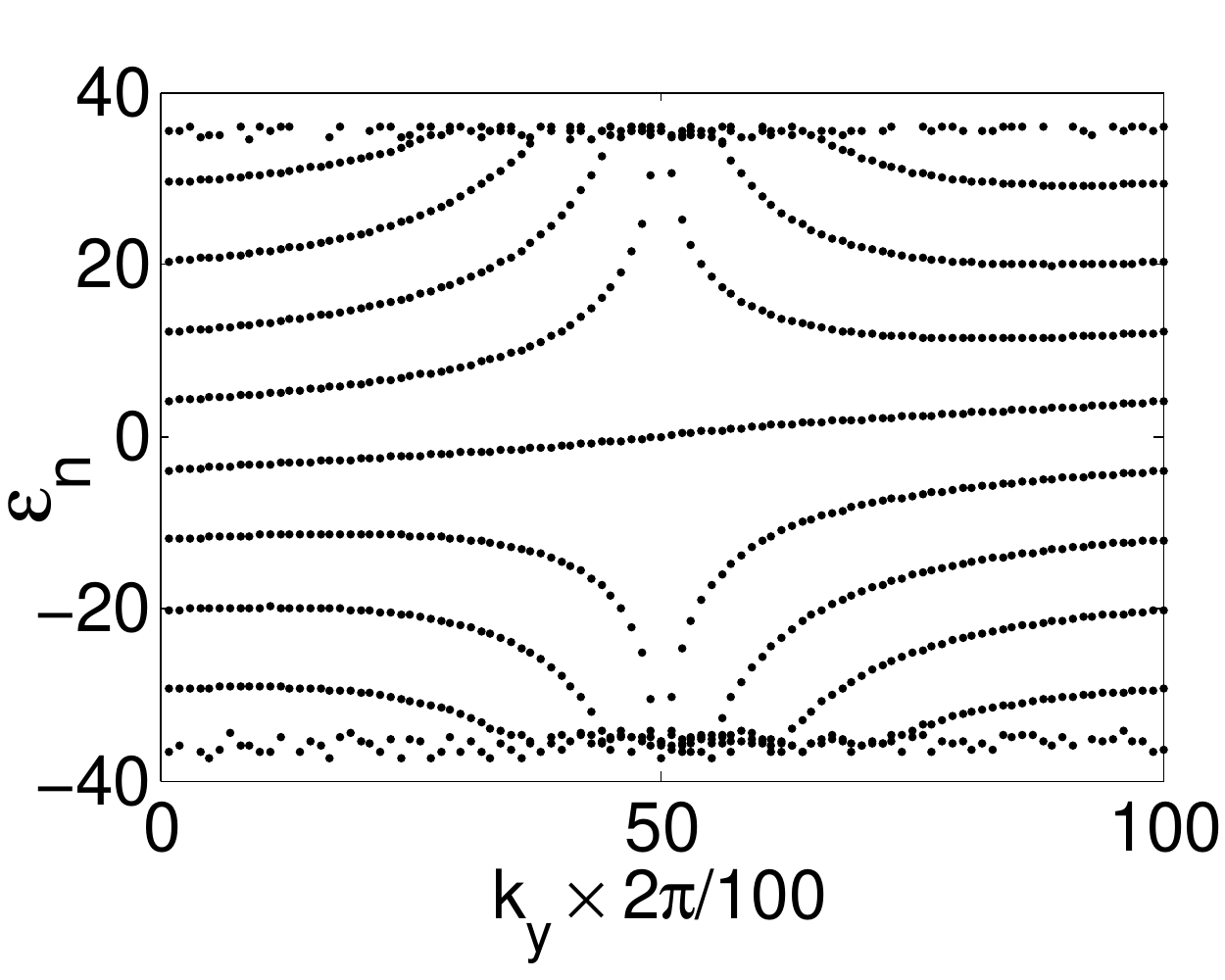}
\caption{a)  The spectrum (Wannier polarization) of $\hat W$ for the Dirac model given by Eq. \ref{dirac} with $m=1$. We see a spectral flow of $C_1=1$ site per $2\pi$ period of $k_y$. b) The spectrum of $\hat C'$ for the same model. Except one eigenvalue belonging to the edge state which also exhibits an analogous spectral flow from $c=0$ to $1$, the rest stay exponentially close to $0$ and $1$, i.e. are exponentially contained in one entanglement partition. c) Plot of the ES $\epsilon=\log(c^{-1}-1)$, which shows the eigenvalues $c$ very near $0$ or $1$ more clearly. The ES looks suggestively similar to the Wannier polarization, with the same spectral flow, except that the eigenvalue spacings depend on $k_y$. For clarity, we have used open boundary conditions (BCs), so that only one edge state appears. Periodic BCs will be used in subsequent plots.}
\label{comparespectra}
\end{figure*}

Evidently, the entanglement correlator $\hat C'=PRP$ and the Wannier Polarization operator $\hat W=PXP$ assume similar mathematical forms although their physical interpretations are quite different. Their only difference is that $R$ is a step function in real space, while $X$ is a linear function. Their spectra are compared in Fig. \ref{comparespectra}. In the rest of this paper, we shall explore in depth the implications of interpolating between these two operators. 

\subsection{Main results of the Entanglement-Wannier interpolation}
\label{sec:main}

Here, we consider a generic D-dimensional free-fermion system, and define the entanglement cut and the Wannier operator to be along the same direction. For now we shall assume that the system is translationally invariant before the cut, so  the crystal momenta is well-defined in the perpendicular directions, and will  collectively denoted as the $k_\perp$ parameter. The result for broken translational symmetry will be discussed at the end of this section.

\subsubsection{The key result}

Our \emph{key} result is that the entanglement spectrum inherits the spectral flow of the Wannier polarization spectrum, but with the gap between eigenvalues related to the imaginary gap of the system. This will be shown via the interpolation between $\hat C'$ and $\hat W$ in Section \ref{interpolation}. Quantitatively, we write
\begin{equation}
\epsilon_{n,a}(k_\perp)\approx [n+X_a(k_\perp)]f(g(k_\perp)), ~f(g)>2g
\label{key}
\end{equation}
where $\epsilon_{n,a}$ is the $n^{th}$ entanglement eigenenergy corresponding to the band/edge $a$, and $X_a(k_\perp)$ is its Wannier polarization (center-of-mass). $f(g)$ is a monotonically increasing function bounded below by $2g$, where $g(k_\perp)$ is the decay rate of the Wannier functions (WFs) that can be rigorously computed. 

Let us first briefly comment on the salient features of result Eq. \ref{key}. It states that the ES is approximately equally spaced, as shown in Fig. \ref{fig:dirac}, with the spacing depending monotonically only on the Wannier decay rate $g$. Physically, $g$ characterizes the maximal possible localization of the wavefunction using the available (occupied) states. Since entanglement measures the corresponding quantum uncertainty behind a real-space cut, it should depend monotonically with the amount of the wavefunction 'leaking' through the cut, which is quantified by $g$.   

The WS inherits a spectral flow from the Wannier polarization $X_a$. This flow arises inevitably due to a topological charge pumping mechanism, and has already been thoroughly studied in other works\cite{yu2011,huang2012edge,huang2012,alexandradinata2011}. In the following Section \ref{interpolation}, we shall justify this inheritance of spectral flow through an interpolation argument between $\hat C'$ and $\hat W$. Our simple interpolation argument provides yet another `proof' of the edge state - entanglement spectrum correspondence explored in some other works mentioned in the introduction, together with important quantitative estimates of the decay properties of the ES.

\subsubsection{Example: 2-D Dirac Model}%\label{sec:model}

To make the above statements more concrete, we shall study the example of a 2D Dirac model with band Hamiltonian
%Plotted in Fig. \ref{comparison} are both types of spectra of a 2-D Dirac model
\begin{equation}
H_{Dirac}(k)=d(k)\cdot\sigma
%\label{dirac}
\end{equation}
where $\sigma_i$, $i=1,2,3$ are the Pauli Matrices, and $d(k_x,k_y)=(m+\cos k_x+\cos k_y,\sin k_x,\sin k_y)$. This is among the simplest model that exhibits a nontrivial 1-parameter spectral flow\cite{hermanns2014} due to nontrivial topology when $|m|<2$. WLOG, we shall assume that the cut be normal to the x-direction, so that $k_\perp=k_y$ is a good quantum number.

\begin{figure}[H]
\includegraphics[scale=.43]{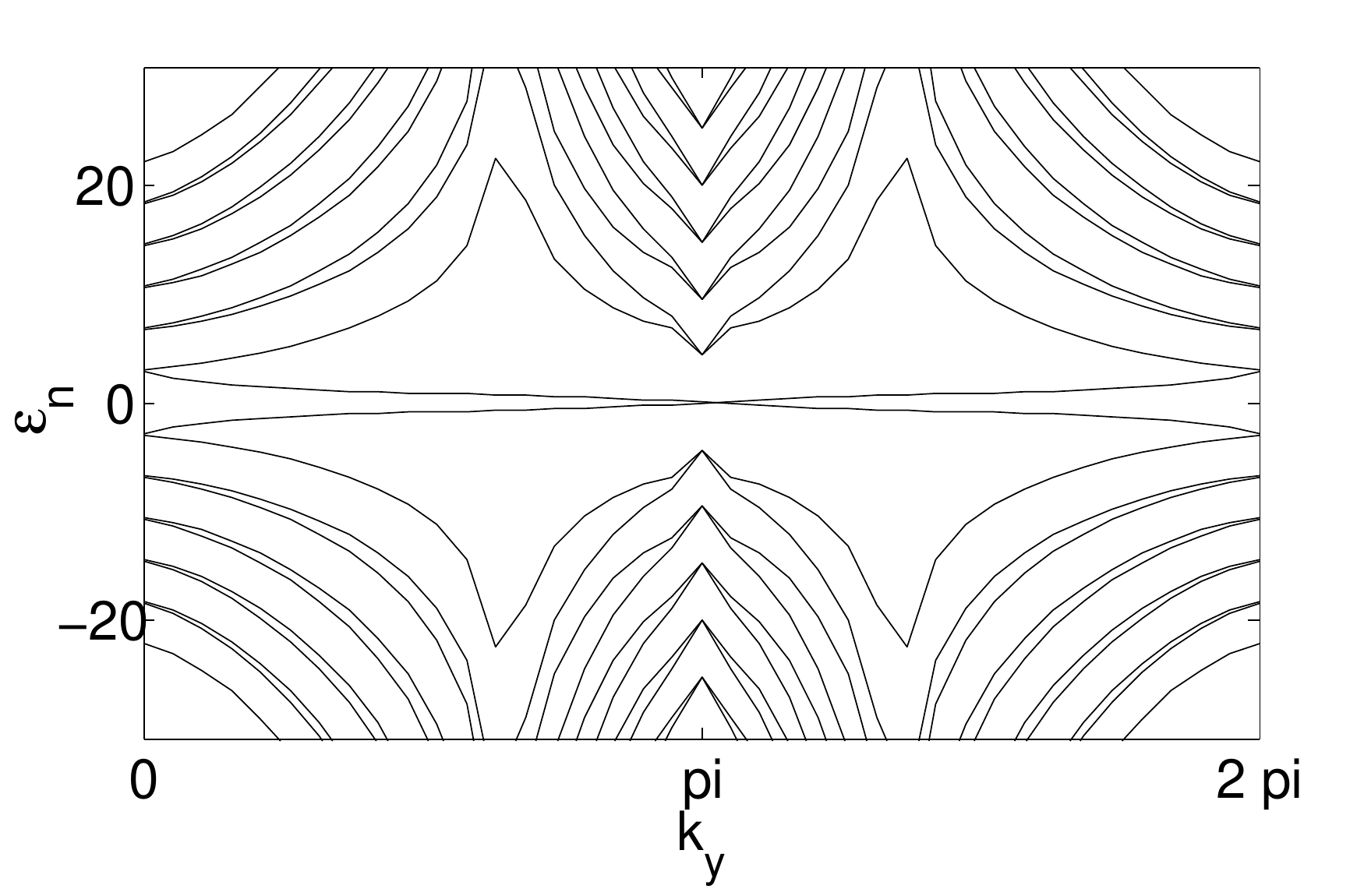}
\includegraphics[scale=.43]{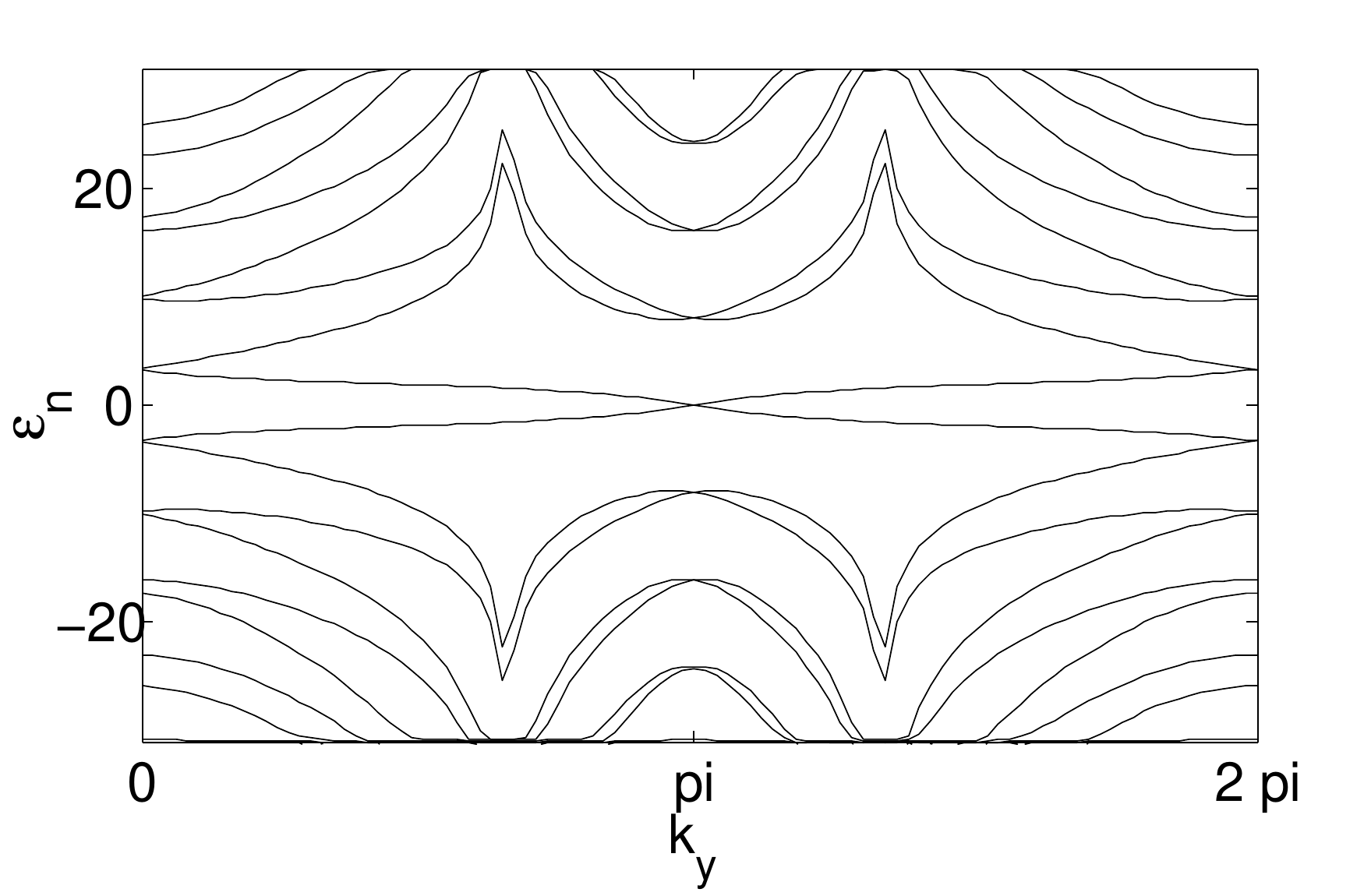}
\caption{Analytical (Eq. \ref{key}) (Left) and numerical (Right) results for the entanglement spectrum $\epsilon_n$ for the Dirac model with $m=0.5$. The x-axis represents $ky\in[0,2\pi]$ while the y-axis represents the entanglement energies.
Only the first few eigenvalues are plotted. The spectra agree qualitatively, and in fact exactly at $k_y=0$ and $\pi$. }
\label{fig:dirac}
\end{figure}

In Fig. \ref{fig:dirac}, the analytic approximation to the entanglement eigenvalues $\epsilon_n(k_y)$ from Eq. \ref{key} is compared against exact numerical results. We see that they agree rather well, especially for those further from zero. This is encouraging, because the decay rate $g$ in Eq. \ref{key} is exact in the asymptotic limit of large $n$, which is numerically inaccessible.

To first order, $f(g)$ may be (rather accurately) represented by a simple linear ansatz
\begin{equation}
f(g(k_y))= (2+A)g(k_y)+J
\label{linearansatz}
\end{equation}
where $A$ and $J$ are parameters that can be exactly determined by exactly evaluating ES at $k_y=0$ or $\pi$. These are two points where there exists exact analytic results for the Block Toeplitz Matrices corresponding to the ES\cite{its2005,its2009}. This will be derived in detail in Appendix \ref{2bandtoeplitz}. The physical interpretations of $A$ and $J$ will be discussed in the context of the interpolation argument in Section \ref{interpolation}.

We also observe spectra flow of the entanglement eigenvalues in Fig. \ref{fig:dirac}, which is present due to the nontrivial topology of $H_{Dirac}$. The two sets of ES eigenvalues, one advancing by one site and one receding by one site, represent the ES spectral flow of the two entanglement cuts. Similar observations are also discussed at length in Refs. \onlinecite{qi2008topological,hughes2011}. Like the Wannier polarization, the ES shifts by $C_1$ sites upon one periodic evolution of $k_y$, where $C_1$ is the Chern number ($C_1=1$ here) of the Hamiltonian\cite{yu2011,huang2012edge,huang2012,alexandradinata2011}. As suggested by the $X_a(k_\perp)$ term in Eq. \ref{key}, this spectra flow is inherited from the Wannier polarization, which we shall quantitatively derive in Section \ref{app:dirac}.

\subsection{The Entanglement-Wannier Interpolation}
\label{interpolation}

In this subsection, we shall present the interpolation between $\hat C'$ and $\hat W$ in detail, and justify our main result Eq. \ref{key}.

\subsubsection{Definition of the interpolation}

The interpolation operator (which is slightly different from that in Ref. \onlinecite{huang2012}) is given by 
\begin{equation}
\hat W(s)=PX(s)P=P[\tilde AX_A\tilde A+\tilde B X_B\tilde B]P\,,
\end{equation}
where $\tilde A$ and $\tilde B$ are respectively the projectors onto regions $A$ and $B$. $X_A$ and $X_B$ are their position operators given by 
\begin{equation} X_A(s)= \frac{s}{L}x \label{XA}\end{equation}
\begin{equation} X_B(s)= 1-\frac{s}{L}(L-x) \label{XB}\end{equation}

Here $\tilde A X_A\tilde A$ acts on region $A$ which includes sites $x=1,\dots,L_A$ while $\tilde BX_B\tilde B$ acts on region $B$ which includes sites $x=L_A+1,\dots,L_x$. $ X(s=1)=X$ is just the usual equally-spaced position operator linearly assigning values between $x=0$ and $1$, while $X(s=0)=R$ is the coarse-grained position operator taking only values of $0$ and $1$ in regions $A$ and $B$ respectively. Hence $\hat W(1)$ is the Wannier operator while $\hat W(0)$ is the correlator corresponding to the Entanglement Hamiltonian.

%Before continuing on the rest of the argument, the reader is invited to refer to Appendix \ref{sect:graphical} to develop some intuition via some numerical results of the simple case where $N=2$.

\subsubsection{Evolution under the interpolation}

Now, we study how exactly the Wannier functions morph into the eigenstates of the Entanglement Hamiltonian, so as to understand the relation between Wannier polarization and the entanglement spectrum.

First, we note an important property of the maximally localized WFs, which is that they decay exponentially, i.e. $\psi(x)\sim e^{-g|x|}$ asymptotically. Their decay rate $g$ is related to the imaginary bandgap between the occupied and empty bands, which will be elaborated later.
For most realistic Hamiltonians, $g\sim O(1)$, so their WFs have very small exponential tails within a few sites of their peaks. This implies that most of the WFs, except for those straddling the cut, will be mostly contained in one region, with an exponentially small tail in the other.

%The idea is to look at the parts of the WF $\psi(x)$ above regions A and B separately, 
As such, we can gain some insight by analyzing the contributions of the $\psi(s)$ from each region separately, where $\psi(s)$ is the eigenstate of the operator $\hat W(s)$:
\begin{equation}
\psi(s)=\psi_A(s)\oplus \psi_B(s)
\end{equation}
where $\psi_A(s)$ and $\psi_B(s)$ are nonzero only in region A and B respectively. When
$s=1$, $\psi(s=1)$ is just the Wannier function. 

Let us explore what happens when $s$ is interpolated from $1$ to $0$. For definiteness, suppose that $\psi(1)$ is mostly contained in region A, i.e $\psi_A(1)$ differs by an exponentially small extent from an eigenstate of $PX_AP$. Then 
%Write the WF , where  Then $B_2\psi_1=B_1\psi_2=0$. Thus for a WF with CM\footnote{Since the interpolation operator $PR(s)P$ is not the usual position operator when $s\neq 1$, its eigenstates (Wannier functions (WFs)) will have CMs that differ from their polarizations. The polarization is defined as the eigenvalue of $PR(s)P$, while the CM is just the average position of the WF $\psi$ defined by $\sum_x x |\psi(x)|^2$.} and polarization in region A (and not on its boundary),
\begin{eqnarray} 
\hat W(s) \psi &= &PX_A(s)P\psi_A\oplus PX_B(s)P\psi_B\notag\\
&\approx&  x_A(s)\psi_A\oplus PX_B(s)P\psi_B\notag\\
&=& x_A(s)PX_B(s)P\psi
\end{eqnarray}
As we tune $s\rightarrow 0$, $\psi_A(s)$ will be modified to an exponentially small extent. %still remain an approximate eigenstate of $PX_A(1)P$. 
This is because the operator $X_A(s)$ remains linear in $x$, and variations of $s$ merely correspond to a rescaling of coordinates\footnote{Note that this will not be true for the edge states which straddle both regions and are not approximate eigenstates of $X_A$ or $X_B$ alone}. A rescaling just introduces a scalar multiplier, and does not change the eigenstates. %Since $X_1(s)$ and $X_1(1)$ looks the same to $\psi_1$ after a rescaling of coordinates, the shape of $\psi_1$ should not change. %Of course, its polarization (eigenvalue of $PX_1(s)P$) will still approach $0$ as $s\rightarrow 0^+$ to reflect the rescaling of coordinates. 

%...For $s=1$, $R(1)$ is just the usual position operator and we already know how its WF $\psi_1$ will exponentially decay into region B. Since only an exponentially small portion of it lies outside region A, it will be an approximate eigenvector of $PB_1X_1(1)B_1P$, the operator $X_1(s=1)$ projected onto region A and the occupied band. This means that with exponentially small error, $\psi_1$ is also the WF of the operator $PX_1(1)P$ in region A alone. 

At the end of the interpolation $s=0$, $\hat W(0)$ is just the projector onto the occupied states in region $B$:
\begin{eqnarray}
\langle \hat C'\rangle &=&\langle \psi(0) |\hat W(0)|\psi(0)\rangle \notag\\
&=& \langle \psi_A(0) |\bar A X_A(0) \bar A|\psi_A(0) \rangle +  \langle \psi_B(0) |\bar B X_B(0) \bar B|\psi_B(0) \rangle \notag\\
&=&0+\langle \psi_B(0) |\bar B X_B(0) \bar B|\psi_B(0) \rangle \notag\\
&=& \langle \psi_B (0)| \psi_B(0)\rangle%\notag\\
%&\sim & \langle \psi_B (1)| \psi_B(1)\rangle
\end{eqnarray} 

%As discussed in the beginning, the polarization $r_n$ of the $n^{th}$ WF $\psi=\psi_1+\psi_2$ takes the form
%\[r_n=|\psi_1|^2r_a + |\psi_2|^2r_b \]

%with $a=0$ and $b=1$. When $s\rightarrow 0^+$, $r\rightarrow 0^+$ due to the rescaling. Hence the contribution to $r_n$ will be solely from $|\psi_2|^2$ (Remember that we are always talking about the case where most of the WF is concentrated in region A). 

While we do not yet understand how $\psi_B(s)$ evolves with the interpolation, we know that it should be approximately proportional to its value $\psi_B(1)$ at the start of the interpolation, which can be rigorously computed. Since $\psi(s=1)\sim e^{-g|x|}$ where $x$ is the displacement from its center of mass (COM), $\langle \psi_B (1)| \psi_B(1)\rangle=\int_B dx |\psi(s=1)|^2\sim e^{-2gn}$, where $n$ is number of sites the COM of $\psi$ is from the entanglement cut. The error from approximating $\psi_B(0)$ by $\psi_B(1)$ also scales like (a small power of) $e^{-gn}$. 
Hence
\begin{eqnarray}
\langle \hat C'\rangle &\approx & \langle \psi_B (0)| \psi_B(0)\rangle\notag\\
&\sim & e^{-f(g)n}
\end{eqnarray} 
where $f(g)>2g$ takes into account \emph{both} the decay rate of $2g$ from $\psi_B(1)$ before the interpolation, and an additional error introduced by the interpolation. 

\begin{figure}[H]
\includegraphics[scale=0.53]{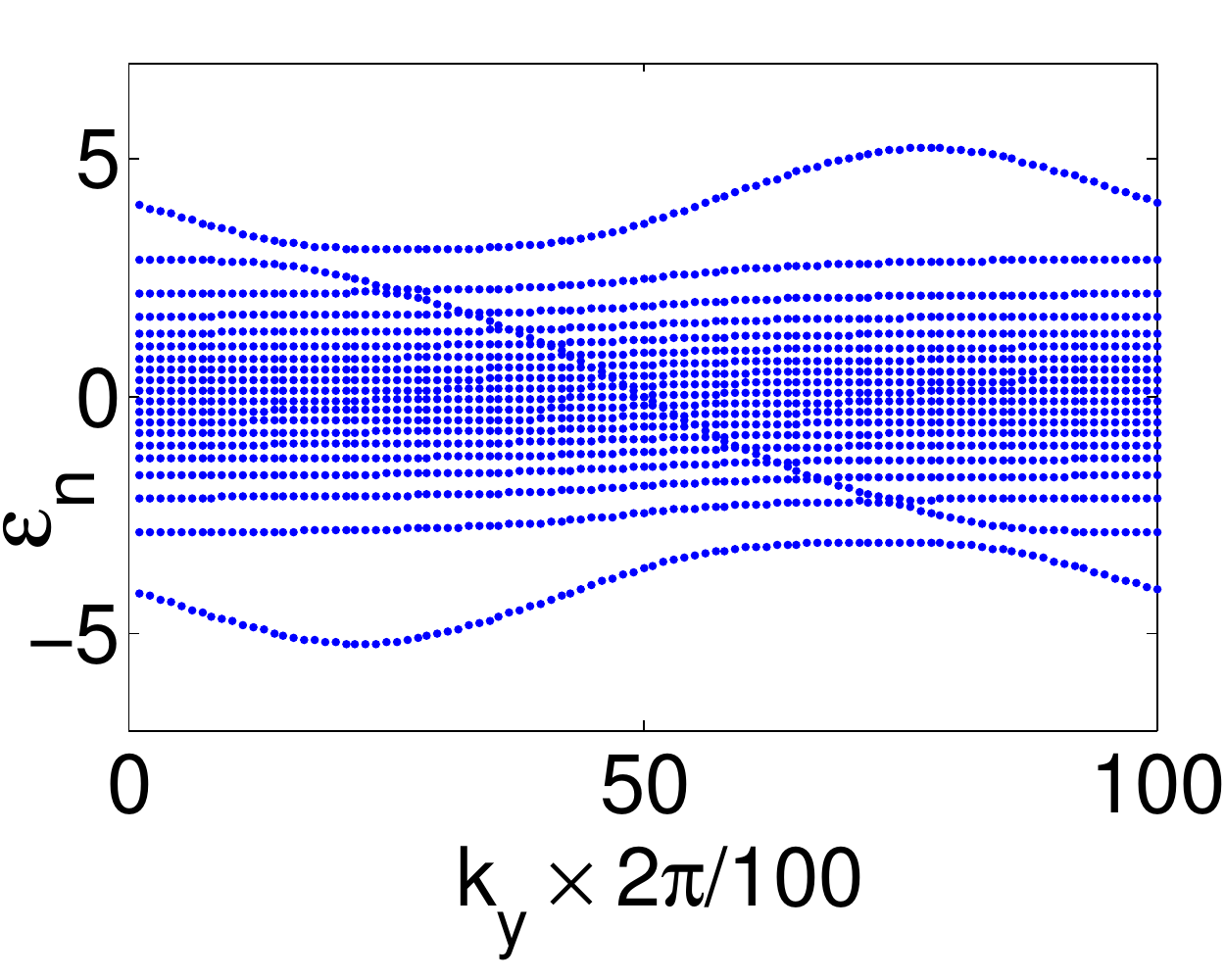}
\includegraphics[scale=0.52]{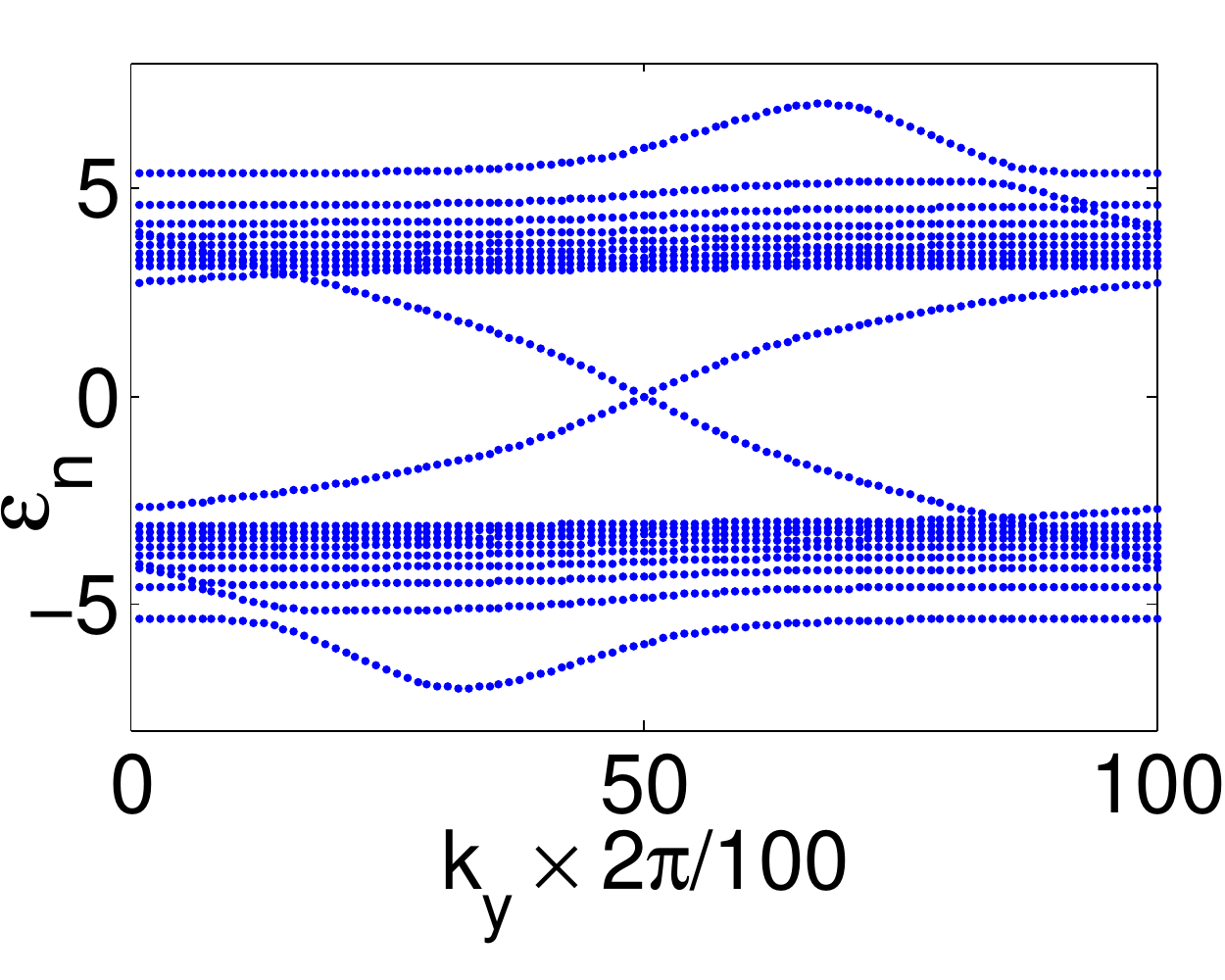}
\includegraphics[scale=0.6]{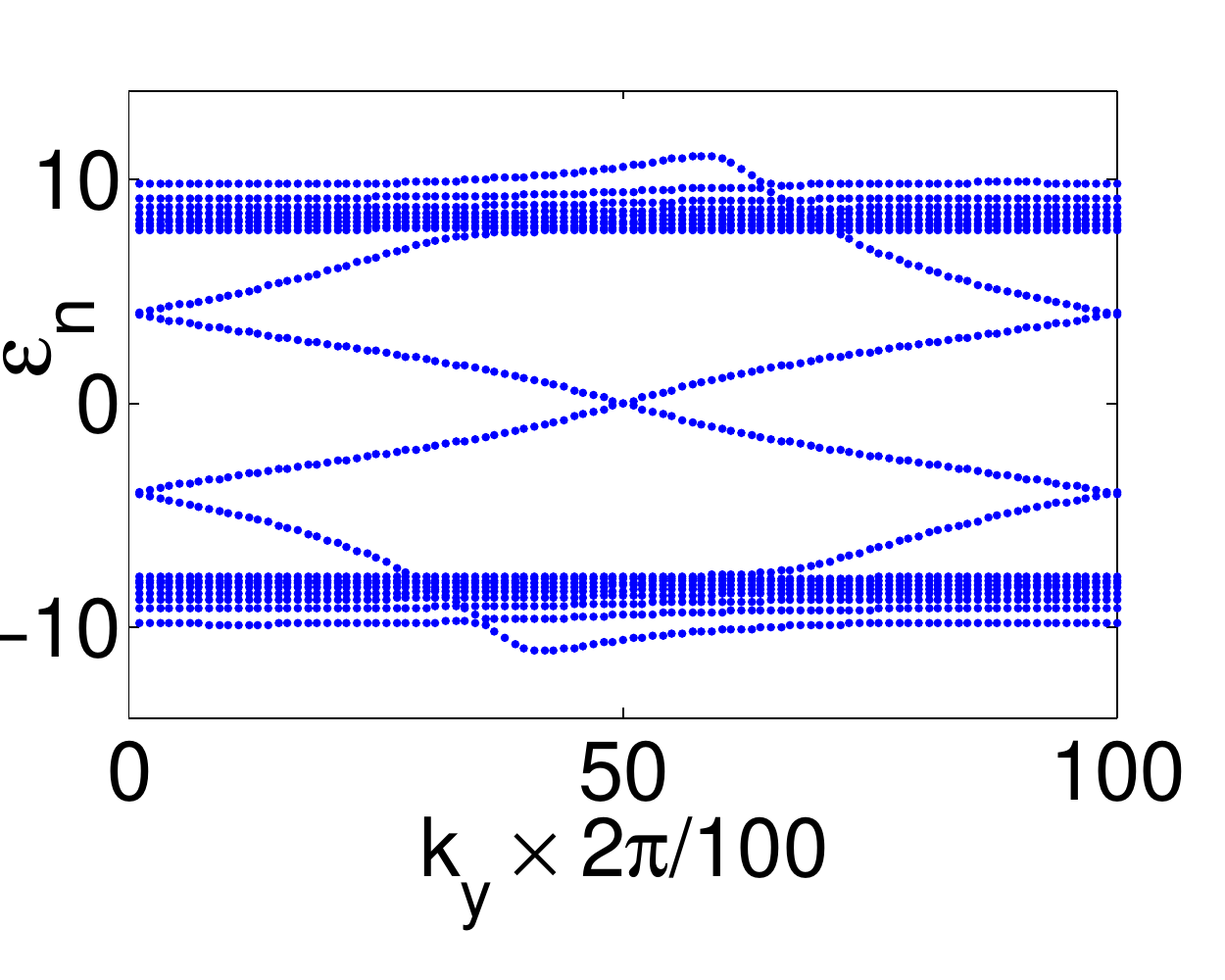}
\includegraphics[scale=0.57]{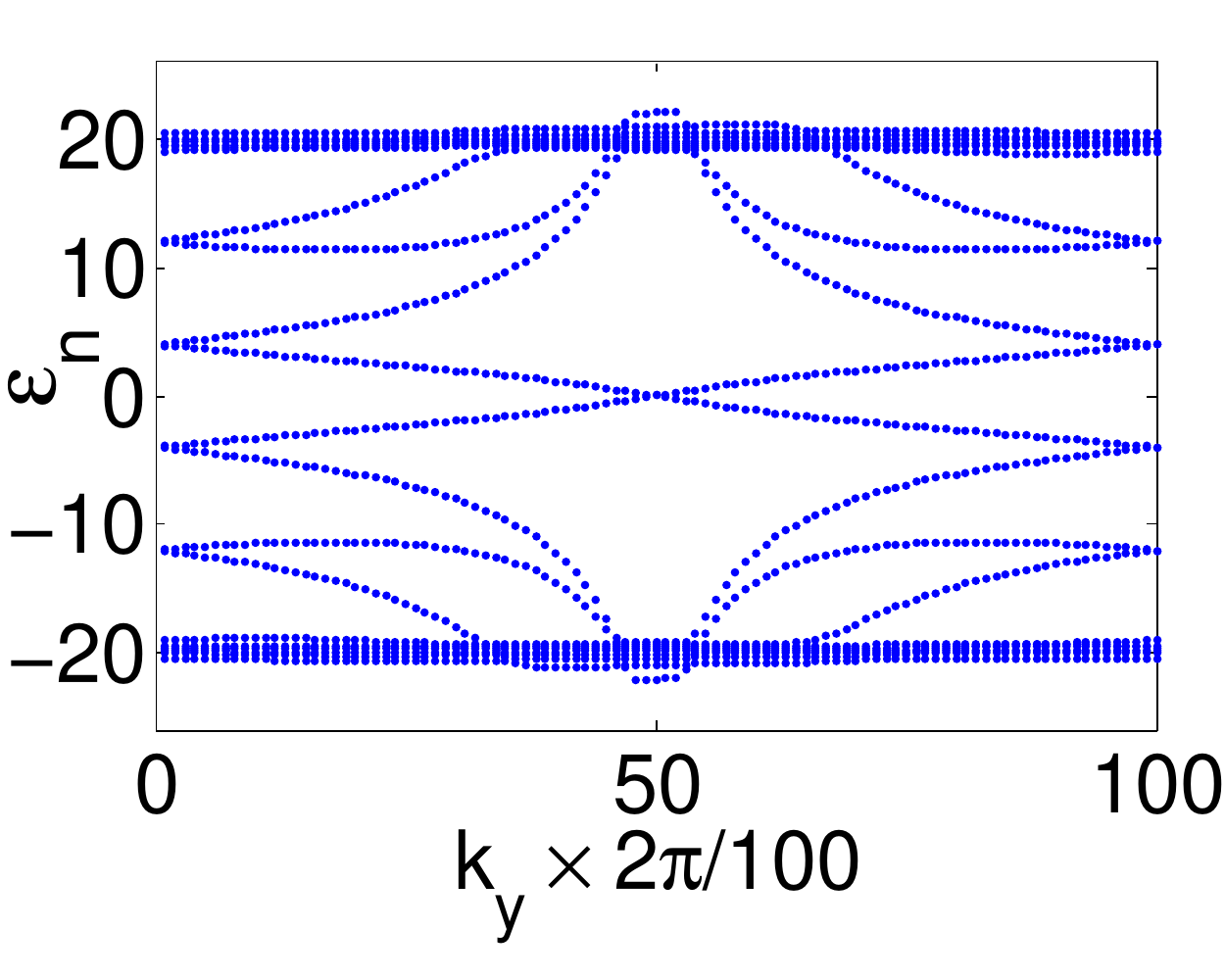}
\caption{(Top Left to bottom Right) Plots of $\log(w_s^{-1}-1)$, where $w_s$ are the eigenvalues of $\hat W(s)$ for $s=1,10^{-1},10^{-3}$ and $10^{-8}$. We see how the Wannier spectrum ($s=1$) evolves into the ES ($s=0$). As $s$ is decreased, $\hat W(s)$ tends towards a step function, and $w_s$ tends towards $0$ or $1$ at a rate dependent on $g(k_y)$, the rate of decay of the WFs. For $s>0$, the $w_s$'s are only exponentially spaced  exponentially spaced beneath a certain length scale set by the finite gradient of $\hat W(s)$. }
%Wannier polarization of $PR(s)P$ for the Dirac model as a function of $k_y\in [0,2\pi]$ for $s=1,10^{-1},10^{-3},10^{-6},0$. The bands split into two regions as $s$ decreases and $R(s)$ become more "coarse-grained". The edge state from region A to B remains in its position while that from B to A beomes increasingly symmetrical with the former.  These exponential spacings unravel as $s$ is decreased to zero. Graphs with any $s <10^{-15}$ will look the same due to numerical limitations.}
\label{fig:interpolation}
\end{figure}

Since the above interpolation is never singular, we expect a one-to-one correspondence between the Wannier spectrum and the Entanglement spectrum. Since an WF exists above each site, away from the cut the entanglement energies are, from Eq. \ref{peschel},
\begin{eqnarray}
\epsilon_n &\sim & \log(e^{f(g)n}-1)\notag\\
&\sim & f(g)n
\label{corrr}
\end{eqnarray} 
Although this linear dependence on $n$ strictly holds only for asymptotically large $n$, it holds true to better than $99\%$ for $n>2$, as evident in numerical computations (Figs. \ref{comparespectra} and \ref{fig:interpolation}). Analogous results hold when $\psi$ were mostly localized in region $B$ instead. 

In the above, it was assumed that each WF $\psi(1)$ were exactly localized $n$ sites away from the cut. In general, this may be not true, especially for topologically nontrivial systems\cite{soluyanov2011,yu2011}. We then have to replace $n$ by $n+X_a$, where $X_a$ is the Wannier polarization (shift of COM) of band $a$, yielding Eq. \ref{key}:
\begin{equation}
\epsilon_{n,a}(k_\perp)\approx [n+X_a(k_\perp)]f(g(k_\perp))
\end{equation}
where $k_\perp$ contains the momentum components transverse to the normal of the cut. %The Wannier polarization, which is intimately related to the band topology of the system, has already been introduced earlier (indicate where). %and we refer the reader to Refs. (cite) for excellent treatments. 

\subsubsection{The Wannier decay rate $g$ elaborated}
\label{fourieranalytic}

The decay rate $g$ and hence the lower bound for the spacing of the ES can be determined precisely. A result in Fourier analysis, which will be proved shortly, states that if
\begin{equation}
\psi(x)=\int dk e^{ikx}e^{i\theta(k)}\psi(k) 
\label{fourierresult}
\end{equation}
then $\psi(x)$ decays like $\psi(x)\sim e^{-gx}$, where $e^{i\theta(k)}\psi(k)$ has a singularity at $Im(k)=g$, but is analytic for $Im(k)<g$. When $\theta(k)$ is chosen such that $\psi(x)$ is maximally localized, both $\theta(k)$ and $\psi(k)$ depend explicitly\cite{qi2011,lee2013} only on the projector to occupied bands $P(k)$. Thus $g$ is just the distance from the real k-axis where $P(k)$ ceases to be analytic, i.e. when the gap between the occupied and unoccupied bands closes. Intuitively, a complex momentum entails a real-space decaying wavefunction because $|\psi(k)|\sim |e^{i(Re(k)+i(Im(k)))x}|\sim e^{-Im(k)x}$.

When there are only two bands, $H(k)=\sigma\cdot d(k)$ and $P(k)=\frac{1}{2}(\mathbb{I}-\hat d\cdot \sigma)$. So $g$ is simply $g=min(|Im(k_0)|)$ where $|d(k_0)|=0$. This is explicitly worked out for the Dirac Model later in this appendix. When there are more than two bands, the projector may not always be expressible in closed-form\footnote{If there are five or more bands, a closed-form expression almost certainly does not exist due to the Abel-Ruffini theorem, save for some special cases.}. However, $g$ will always be a well-defined quantity that can be obtained numerically.

In more than one dimension, a different decay rate $g(k_\perp)$ can exist for each dimension, with $k_\perp$ denoting the momenta from the other directions. 

Now let's proceed to prove that the Fourier coefficients $\psi(x)$ of Eq. \ref{fourierresult} decay like $\psi(x)\sim e^{-gx}$. This is an important theorem that our key result Eq. \ref{key} prominently relies on. A similar proof can already be found in for instance Refs. \onlinecite{leeandy2014} or \onlinecite{kohn1959}, though in different contexts. Here, we shall reproduce it in a way tailored to our context.

Since $\psi(k)$ is an eigenfunction of the Hamiltonian $h(k)$ (up to a phase factor), it belongs to a degenerate eigenspace when there the gap closes. Consider the analytic continuation (with abuse of notation) of $\psi(k)$ into $\psi(z) =\psi(e^{ik})$:
\begin{equation}  \psi(z)= \sum_{x\ge 0} \frac{\psi(x)}{2} \left(z^x + \frac{1}{z^x}\right) \label{fourierseries}\end{equation}

Due to the Theorem of Monera and the fact that $h(z)$ is real on the unit circle $|z|=1$, $\psi(z)$ necessarily has a singularity (pole or branch point) inside the unit circle. Let $z_0$ be the singularity of largest magnitude inside the unit circle. We want to show that 

\begin{equation} |\psi(x)|\sim |z_0|^x =e^{-gx}\label{decaykproof}\end{equation}
up to a proportionality factor, where $|z_0| <1$.  In particular, there is a constant $C$ such that $|\psi_x|<C|z_0|^x$. This is a known result\cite{kohn1959,he2001}, and in the next paragraph we sketch a simple derivation suitable for our context. 

Since $\psi(z)$ is analytic for $|z|>|z_0|$ within the unit circle, the series Eq. \ref{fourierseries} must converge in that region.   As Eq. (\ref{decaykproof}) must hold for some value of $|z_0|$ for this series to converge at all inside the unit circle, %let us assume this holds and determine the required value of $|z_0|$.  
for $z$ such that $|z_0| <|z|<1$, \begin{equation}
|\psi(z)| < \sum_{x\ge 0} \frac{|\psi(x)|}{|z|^x} < C \sum_{x\ge 0} \left|\frac{z_0}{z}\right|^x < \infty
\end{equation}In addition, $\psi(z)$ fails to be analytic at $z_0$, so the above series must diverge when $|z|=|z_0|$.  % Evidently, $|z_0|=\lambda$. 
This implies that $|\psi(x)|$ must asymptotically decay like $|z_0|^x$, thus proving Eq. (\ref{decaykproof}).

\subsubsection{Generalization to systems without translational symmetry}

We have previously focused only on translationally invariant systems with a well-defined Wannier decay rate. However, the gist of the Wannier Interpolation argument still holds true without requiring translation invariance at all. Wannier functions have well-defined decay rates even in the absence of translation symmetry, i.e. in a magnetic field where it is broken down to the magnetic translation subgroup. 

In this general setting, Eq. \ref{key} is modified to
\begin{equation}
\epsilon_{n,a_\perp}\approx \tilde f(n+X_{a_\perp}) 
\end{equation}
where $\tilde f$ is a generically nonlinear function. Here, $a_\perp$ refers to the residual collection of good quantum numbers, which can still include a transverse momentum $k_\perp$ if translation symmetry is not broken in that direction. From Eq. \ref{corrr}, the form of $\tilde f$ depends precisely on the decay behavior of the Wannier function at each position $n$ away from the cut. For instance, the orbitals of a system in a magnetic field possess a Gaussian profile, so $\tilde f$ should be a quadratic function.

Another possibility is that the entanglement cut itself does not respect translational symmetry. This can happen, for instance, when the cut is composed of an inhomogeneous distribution of real space intervals. Such is exactly the case in the context of the Exact Holographic mapping, where the various layers are composed of linear combinations of the original real space sites. Layer cuts, which implement local partitionings in scale/energy space, respect only the residual translational symmetry within the lower layer involved. The entanglement spectra corresponding to such cuts in one-dimensional systems will be shown in Appendix. \ref{ES1dim}.

\subsection{Details on the Entanglement Spectrum of the Dirac model} 
\label{app:dirac}
Here we derive and explore the ES of the Dirac model in more detail. There are two main mathematical quantities to determine from Eq. \ref{key}: They are the monotonically increasing function $f(g(k_y))$, and the Wannier polarization $X_\pm (k_y)$, where $\pm$ label the ES corresponding to the two entanglement cuts.

\subsubsection{Determination of linear ansatz parameters}

To a first approximation, the function $f(g)$ is given by
\begin{equation}
f(g)= (2+A)g+J
\label{linearansatzapp}
\end{equation}
where $A$ and $J$ are parameters. In the case of the Dirac models, there indeed exists two points $k_y=0$ and $\pi$ where the entanglement spectra can be rigorously solved. $A$ and $J$ can thus be obtained by fitting Eqs. \ref{key} and \ref{linearansatzapp} with the exact results.

At $k_y=0$ or $\pi$, the Hamiltonian is given by $H=\sigma\cdot d$, where $d(k)=(m+\cos k_x+\cos k_y,\sin k_x,\sin k_y)=(m\pm 1 + \cos k_x,\sin k_x,0)$, i.e. only $d_1$ and $d_2$ are nonzero. In such cases, there exists exact analytic results for the asymptotic spacing between entanglement eigenvalues  %and we can use eq.\eqref{noncritical_spacing}:
\begin{eqnarray}
&&\lim_{n\rightarrow \infty}(\epsilon_{n+1}-\epsilon_n)\notag\\
&=&(3-sgn(-m\mp 1-2))\frac{\pi}{2}\frac{I(\sqrt{1-|(m\pm 1 )/2|^2})}{I(|(m\pm 1 )/2|)}\notag\\%=(2+A)[h(k_y)]+J
\label{toeplitzspacing0}
\end{eqnarray}
where the $\mp$ refers to $k_y=0$ or $k_y=\pi$ and $I(\kappa)=\int_0^1 \frac{dx}{\sqrt{(1-x^2)(1-\kappa^2x^2)}}$ is the complete elliptic integral of the first kind\cite{jeffreys1999book,stone2009book}. This impressive result from Eq. \ref{noncritical_spacing} will be derived shortly after; for now we just mention that $A$ and $J$ can be by comparing it with $\epsilon_{n+1}-\epsilon_n\approx (2+A)g(k_y)+J$, where $k_y=0$ or $\pi$. 

%Here $\gamma=1$ and $p=-m\mp 1 $, so $k=|p/2|=|(m\pm 1 )/2|$. 
To find $g(0)$ and $g(\pi)$, we solve for $h(k_0,k_y)=d_1^2+d_2^2=0$, and identify $g$ with $Im(k_0)$. It is easily shown that the gap closes at complex $k_0=(1+sgn(P))\pi/2 + i\cosh^{-1}|P|$, where 
\begin{equation} P=\frac{2+m^2+2m\cos k_y}{2(\cos k_y+m)}=\frac{2+m^2\pm 2m}{2(m\pm 1)}\end{equation} 
Hence $g(0)=\cosh^{-1}\frac1{2}\left(\frac{1}{m+1}+m+1\right)$ and $g(\pi)=\cosh^{-1}\frac1{2}\left(\frac{1}{m-1}+m-1\right)$, and $A,J$ can be easily obtained.

\subsubsection{The exact ES for certain 2-band Hamiltonians through Toeplitz Matrices}
\label{2bandtoeplitz}

We now discuss some known exact results for the eigenspectrum of 2-band models, with the goal of obtaining Eq. \ref{toeplitzspacing0}. Consider $d_1(k_x),d_2(k_x)$ (with $k_y$ as a parameter) of the form

\begin{equation} d_1(k_x) = \cos k_x - \alpha/2 \end{equation} 
\begin{equation} d_2(k_x) = \gamma \sin k_x \end{equation}

with $\gamma\neq 0$ and $\alpha>0$ so the system is gapped. For our Dirac model at $k_y=0$ or $\pi$, $\alpha=-2m\mp 2$ and $\gamma=1$. Although the exact eigenspectrum of \emph{block} Toeplitz matrices are notoriously hard to compute, in the current case a brilliant solution was found by \onlinecite{its2008}. The asymptotic (large $L_A$) spacing between the eigenvalues were found to be
\begin{equation}
\lim_{n\rightarrow \infty}(\epsilon_{n+1}-\epsilon_n)=(3-sgn(\alpha-2))\frac{\pi}{2}\frac{I(\sqrt{1-\kappa^2})}{I(\kappa)} 
\label{noncritical_spacing}
\end{equation}

which tends towards a constant, unlike those of critical 1-D systems which goes like $\propto \frac{1}{\log L_A}$. Here 
\begin{itemize}
\item $\kappa=\sqrt{\alpha^2/4+\gamma^2-1}/\gamma$ if $4(1-\gamma^2)<\alpha^2<4$
\item $\kappa=\sqrt{(1-\alpha^2/4-\gamma^2)/(1-\alpha^2/4)}$ if $\alpha^2<4(1-\gamma^2)$
\item $\kappa=\gamma/\sqrt{\alpha^2/4+\gamma^2-1}$ if $\alpha>2$
\end{itemize}
with $\kappa'=\sqrt{1-\kappa^2}$. For the Dirac model, the first (third) case applies when $(m\pm 1)^2<1$ ($(m\pm 1)^2>1$).  
As a bonus, we also have exact expression for the entanglement entropy
\begin{equation} S_A=\frac{1}{6}\left(log\frac{\kappa^2}{16\kappa'}+\left(1-\frac{\kappa^2}{2}\right)\frac{4I(\kappa)I(\kappa')}{\pi}\right)+log2 \end{equation}
for $\alpha<2$, and 
\begin{equation} S_A=\frac{1}{12}\left(log\frac{16}{\kappa^2\kappa'^2}+\left(\kappa^2-\kappa'^2\right)\frac{4I(\kappa)I(\kappa')}{\pi}\right) \end{equation}
for $\alpha>2$. All these results can be obtained via a detailed analysis of the pole positions of $\phi(z)$ in Eq. \ref{toeplitzpoly}. 
Note that $S_A$ tends to a constant asymptotically, unlike in the critical case. The Entanglement Entropy of the whole system will then by proportional to the length of the cut $L_y$, in agreement with well-known area laws\cite{wolf2006,verstraete2006,plenio2005,li2008,swingle2010}. 

\section{Relation between entanglement spectra and  Toeplitz matrices}
\label{sec:toeplitz}

In the previous appendix, the entanglement spectrum of a free fermion system was shown to be continuously related to its Wannier polarization. While that was based on an ingenious physical argument, it relied on certain exact mathematical results for quantitative predictions. These mathematical results concern (block) Toeplitz matrices, objects which we will describe below.

Block Toeplitz matrices are finite matrices $T$ with translational invariance along each diagonal, i.e. $T_{ij}=T_{i-j}$, where each $T_{ij}$ is also a matrix which represents the internal DOFs belonging to each site. A Toeplitz matrix can be characterized its symbol, which is defined as the fourier transform along one of its rows (or columns): 
\begin{equation} g(k)=\frac{1}{2\pi}\int_{-\pi}^{\pi}T_x e^{ikx}dx \end{equation}
Loosely speaking, $g(k)$ is the 'momentum-space' representation of the matrix $T$, and a singular $g$ contains a momentum-space branch point which can be interpreted as a momentum-space projector. 

The entanglement spectra of free fermions correspond to the eigenspectra of those Block Toeplitz matrices with symbols that are singular, i.e. with discontinuities or singularities. Such matrices are ubiquitous in diverse areas of physics, whenever there are translationally-invariant systems with internal DOFs and abrupt truncations. They appear, for instance, in various spin chain models\cite{its2005, its2008, keating2005}, dimer models\cite{basordimer2007}, impenetrable bose gas systems\cite{ffapps} and full counting statistics pertaining to certain non-equilibrium phenomena involving quantum noise \cite{ivanov2010phase,noneq}. But due to considerable mathematical difficulties, there has been no known explicit result for the asymptotic eigenspectra of such Toeplitz matrices, except for the simplest few cases\cite{its2008}, one of which will be described below. 

To illustrate how Toeplitz matrices appear in the calculation of entanglement spectra, we review a simple class of 2-band Hamiltonians whose ES have been analytically studied\cite{its2009}. Consider a Hamiltonian given by $H=\sigma\cdot d$, so that the projector to the occupied band $P$ is given by $P=\frac{1}{2}(\mathbb{I}-\hat d\cdot\sigma)$. If only $d_1$ and $d_2$ are nonzero, the eigenvalues $\hat \epsilon$ of $P$ can be expressed as the roots of $Det(i\hat\epsilon I + \frac{I-\hat \Gamma'}{2})$, where
\begin{eqnarray} 
\hat \Gamma'=\left( \begin{array}{cc} 
0& \frac{d_1-id_2}{\sqrt{d_1^2+d_2^2}}  \\ 
\frac{-d_1-id_2 }{\sqrt{d_1^2+d_2^2}}& 0 \end{array}\right)=\left( \begin{array}{cc} 
0& \sqrt{\frac{d_1-id_2}{d_1+id_2}}  \\ 
-\sqrt{\frac{d_1+id_2}{d_1-id_2}}& 0 \end{array} \right)
\end{eqnarray}
For the purpose of calculating the entanglement spectrum with a cut parallel to the $y$-direction, we consign $k_\perp=k_y$ to an external parameter and consider the analytic properties of $k_x$. As $k_x$ is periodic, $d_1\pm i d_2$ will be a function of $e^{\mp ik_x}$. We can analytically continue $\hat \Gamma'$ to complex values of $k_x$ by letting it be a meromorphic function of $z=e^{ik_x}$. Since the analytic continuation is unique, $e^{-ik}\rightarrow 1/z$ not just on the unit circle where $k_x$ is real, but over the whole complex plane. For local Hamiltonians, we can write $\hat \Gamma'$ in terms of a polynomial $p(z)$ and \emph{winding function} $\phi(z)$ via 
\begin{eqnarray} 
\hat \Gamma'=\left( \begin{array}{cc} 
0& \sqrt{\frac{p(z)}{z^{2n}p(1/z)}} \\ 
-\sqrt{\frac{p(1/z)}{z^{-2n}p(z)}}& 0 \end{array}\right)=\left( \begin{array}{cc} 
0& \phi(z)  \\ 
-\frac{1}{\phi(z)}& 0 \end{array} \right)%\notag\\
\label{toeplitzpoly}
\end{eqnarray}  
This is exactly the matrix in eq. 43 of Ref.~\cite{its2009}. If we want to find the entanglement entropy, we will need to project the $\hat \Gamma'$ onto region A, i.e. find the eigenvalues of $R\Gamma'R$. This can be done by fourier transforming $\hat \Gamma'$ onto real space and taking open boundary conditions. Mathematically, the real space $\hat \Gamma'(x,x')$ is a finite $2L_A \times 2L_A$ block Toeplitz matrix generated by the symbol $\hat \Gamma'(z)$ i.e.
\begin{equation}\hat \Gamma'_{ij}(x,x')=\oint \frac{\hat \Gamma'_{ij}(z)dz}{z^{x-x'+1}} \end{equation}
where $2$ is the dimension of the internal degrees of freedom and $L_A$ is the number of sites in region $A$. 

Clearly, the ES of generic Hamiltonians with $N$ bands must be given by the eigenvalues of analogous Block Toeplitz Matrices with $N\times N$ blocks. The finite size of the Toeplitz matrix mathematically implements the entanglement cuts\footnote{Note that a large but finite Toeplitz Matrix will still have a qualitatively different spectrum as a truly infinite matrix.}.

\subsection{Primer on Toeplitz Matrices}
\label{app:history}

%This is not your typical appendix that contains the nitty-gritty details of long convoluted derivations. 

While the eigenspectrum of Toeplitz matrices without internal DOFs can be obtained rather easily through methods like Wiener-Hopf factorization%(cite)
, those of \emph{Block} Toeplitz matrices (i.e. with internal DOFs) are much more elusive. Below is an overview of the history and development of Toeplitz Matrices, so as to put their applications and known results in better perspective. 

Toeplitz matrices are finite-sized matrices that have translational symmetry along each diagonal.  They appear in a wide variety of applications, from the thermodynamic limit of the 2D classical Ising model and its generalizations\cite{ising1,ising2,ising3}, various spin chain models\cite{its2008, its2009,keating2005}, dimer models\cite{basordimer2007}, impenetrable bose gas systems\cite{ffapps} to full counting statistics  and certain non-equilibrium phenomena\cite{ivanov2010phase,noneq}. In a celebrated result by Potts and Ward\cite{ising4}, the spin-spin correlator of the 2D Ising model is expressed as a Toeplitz determinant. In other settings, the asymptotic limits of the eigenvalues of Toeplitz matrices are essential in the calculation of the entanglement spectrum and entropy, such as the XX and XY quantum spin chains and their equivalent free-fermion problems\cite{its2009}. Of more exigent physical importance is the use of Toeplitz determinants in computing the correlation functions of dimer models that arise in high-temperature superconductors\cite{basordimer2007}. Such models, which are equivalent to certain 2D Ising models\cite{kasteleyn1961statistics,kasteleyn1963dimer,stephenson1964ising, fisher1966dimer}, have been used to study the possibility of realizing Anderson's RVB liquid in valence-bond dominated phases\cite{fendley2002classical,moessner2003,rokhsar1988}.  More recently, Toeplitz matrices have also been studied in the context of quantum noise, for instance through the calculation of the full counting statistics of 1-D fermions\cite{abanov2011quantum} or their non-equilibrium interactions via bosonization in the framework of the Keldysh action formalism\cite{noneq}. 
 
%Roughly speaking, they represent systems with open boundary conditions, whose edges can give rise to nontrivial phenomena that persist even in the thermodynamic limit. 
In these abovementioned applications, quantities of physical interest are usually computed in the thermodynamic limit, where the size of the finite Toeplitz matrices tend to infinity. In this limit, however, the finite Toeplitz matrices do \emph{not} converge to truly infinite Toeplitz matrices whose spectra can be trivially obtained. Intuitively, this is because finite Toeplitz matrices, no matter how large, will always contain "edges" that nontrivially modify the original spectrum and eigenvectors. This fact is prominently illustrated in the exemplary case of topological insulators, where the Toeplitz matrix is taken to be the real-space Hamiltonian. When the Toeplitz matrix is made finite by imposing open boundary conditions, the nontrivial edge eigenstates that appear have distinct energy dispersions from those bulk eigenstates belonging to the original infinite Toeplitz matrix.

As such, a lot has been studied about the asymptotic properties of Toeplitz matrices. The Szeg\"{o} limit theorem\cite{szego} which dates back to 1915 first related the the asymptotics of the determinant of a Toeplitz matrix $T_{ij}=T_{i-j}$ to its symbol $g(k)=\frac{1}{2\pi}\int^{\pi}_{-\pi} T_x e^{ikx}dx$, a quantity that has been introduced in more detail in Section \ref{sec:toeplitz}. Physically, the symbol represents the fourier-space operator corresponding to the Toeplitz matrix representing the truncated real-space version of the same operator. Subsequently, this fundamental 1915 result was extended to the so-called Strong Szeg\"{o} limit theorem requiring much less restrictive assumptions by Kac, Baxter, Hirschman and others\cite{kac,baxter,hirschman}. This result, however, still required the symbol to be continuous with zero winding number. These constraints were relaxed by a series of breakthroughs that follow, thereby opening up the important class of Toeplitz matrices with singular symbols to physical applications\cite{fh2,widom1973,widom1974,widom1975,basor1978,szego2}. Such Toeplitz matrices can physically represent, for instance, flattened Hamiltonians acting as projectors to eigensubspaces. In fact, most of the previously mentioned physical applications rely heavily on a class of singular Toeplitz matrices of the Fisher-Hartwig type.

However, relatively little is known about the asymptotic eigenvalue distribution of general Block Toeplitz matrices, i.e. those with matrix-valued symbols $g^{ab}(k)$. They are generalizations of the abovementioned Toeplitz matrices to admit "internal degrees of freedom" which, not surprisingly, will contain vastly richer structure. For instance, Block Toeplitz matrices can represent lattice systems with more than one band, thereby allowing for the possibility of nontrivial topological phenomena\cite{thouless1982}. Exact results for the asymptotic eigenvalue distribution only exists for a special class of $2\times 2$ Block Toeplitz matrices\cite{its2008,basordimer2007}, as already reviewed in Sect. \ref{sec:toeplitz}. No result on the full asymptotic eigenvalue distribution of general $N \times N$ Block Toeplitz matrices exists to our knowledge, although there has been asymptotic results on the arithmetic mean of their eigenvalues\cite{gutierrez2012}.

Despite their ubiquity, finding the asymptotics of generic Toeplitz matrices remain a notoriously difficult task. The the authors' knowledge, rigorous asymptotic results are not known for the spectra of generic Block Toeplitz matrices with distcontinuous fourier transforms along the diagonals, i.e, those with singular symbols. These are exactly the types of Toeplitz matrices appearing in the entanglement Hamiltonians of free-fermion systems. 

As such, it is our hope that our asymptotic bounds on the ES derived via Wannier interpolation will provide some helpful hints on the spectral properties of generic Block Toeplitz Matrices, even those not originally appearing in an entanglement calculation.

\section{Entanglement spectrum under 1-dimensional EHM}
\label{ES1dim}

We have seen in Appendix \ref{wannES} how the entanglement spectrum (ES) of a free fermion system can be computed, and how it relates to the Wannier polarization of the same system. Previously, all the cuts are real-space cuts. But with the Exact Holographic Mapping, one can also define layer cuts separating regions at different scales. Here, we shall study how these layer cuts can affect the ES.

To recap, the free-fermion entanglement Hamiltonian $H_E$ is given by $(\mathbb{I}+e^{H_E})^{-1}=C=PRP$, where $P$ and $R$ are projectors that act in momentum and real space respectively. $P$ projects to the occupied bands and implements the Fermi surface, if any. $R$ traditionally implements the real-space cut. Now, however, we will generalize $R$ to include layer cuts\footnote{Actually, layer cuts are just complicated combination of real space cuts, as the layers are defined by wavelet bases that are local in real space.}. 

\subsection{When only a layer cut is present}
We first focus on the case where there is only a layer cut, and no real-space cut. Nontrivial entanglement energies appear because the layer cut breaks the translational symmetry of the system by enlarging the unit cells. Mathematically, this promotes the $C=PRP$ correlator to a Block Toeplitz matrix (see Appendix \ref{sec:toeplitz}) with a highly nontrivial spectrum, with the blocks containing information on translational symmetry breaking.

\begin{figure}[H]
\includegraphics[scale=.34]{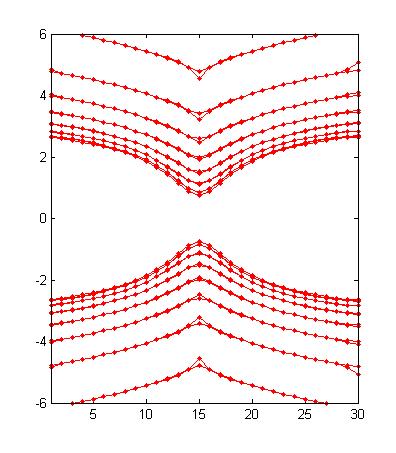}
\includegraphics[scale=.34]{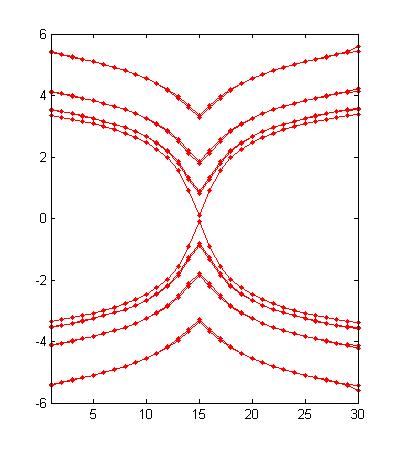}
\includegraphics[scale=.34]{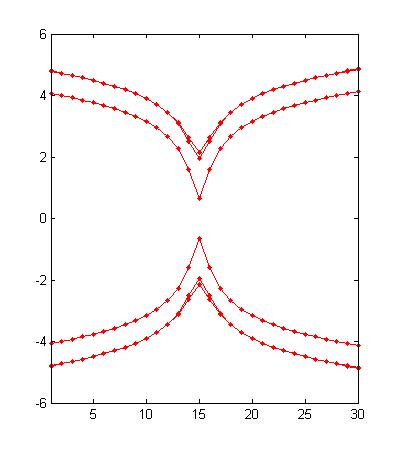}
\caption{a) Plots of the entanglement spectrum against $k_y$ for the Dirac model Eq. \ref{dirac} in $D=2$ dimensions, with $M=-0.1$. The EHM is done in the $x$-direction. Layers traced out are layer $1$ to layer $3,4,5$, from left to right respectively. Like in the $2$-dimensional EHM results of Sect. \ref{sec:holotopo}, we also observe a gap closure when the layer cut is performed at a layer between the two Chern number density peaks in the UV and IR (small and large layers $n$). }
\label{fig:layerentanglement}
\end{figure}

\begin{figure}[H]
\includegraphics[scale=.34]{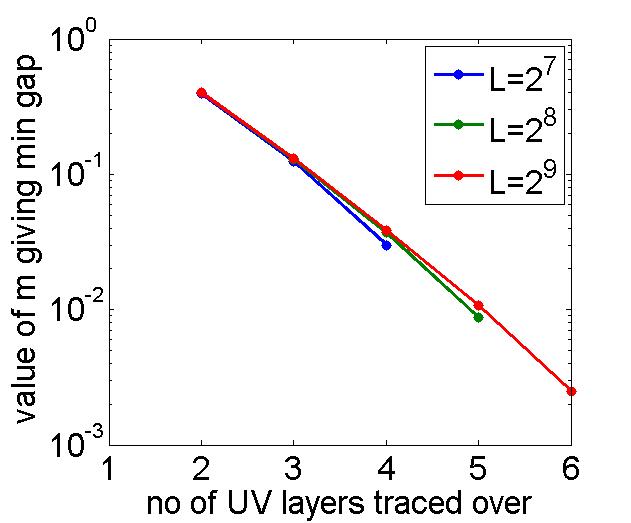}
\includegraphics[scale=.38]{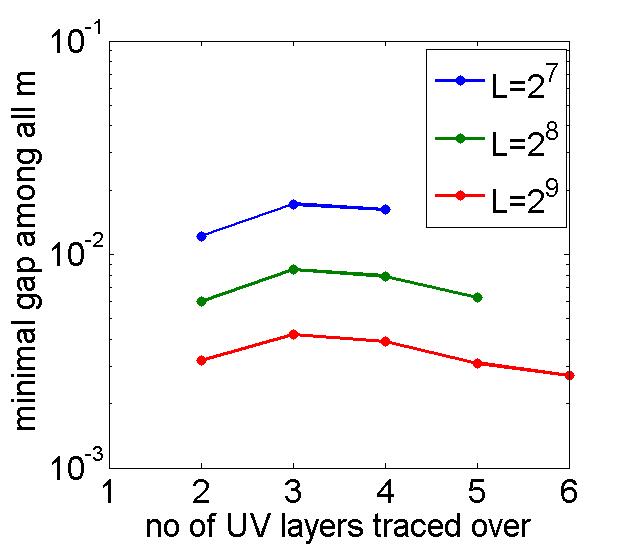}
\caption{For each position of layer cut, the gap $|M|$ that gives the minimal entanglement gap is explored. Left) The value of the gap that gives the minimal gap for various cuts. We see very little error from finite system size. Note the exponential dependence on $n$, corresponding to a behavior between $.25^n$ to $.32^n$, consistent with results on the bulk holographic topological insulator in Sect. \ref{sec:holotopo} and Appendix. \ref{Canalytic}. Right) The minimal gap at the optimal $M$ used in the previous plot. Evidently, the minimal gap is almost exactly inversely proportional to system size. It varies only very weakly with position of cut. Note that the use of the minimal gap avoids issues with incommensurability effects. }
\label{fig:layerentanglementgap}
\end{figure}

\subsection{When both real space and layer cuts are present}

In this case, we will see entanglement eigenenergies in addition to the usual ones described as length in Appendix \ref{wannES}. These additional states occur at the additional boundary introduced by the layer cut. 

\begin{figure}[H]
\includegraphics[scale=.32]{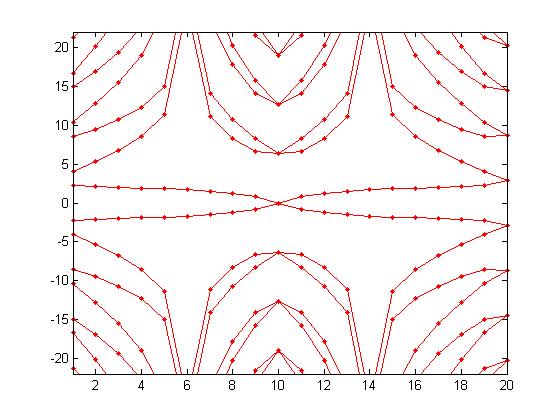}
\includegraphics[scale=.32]{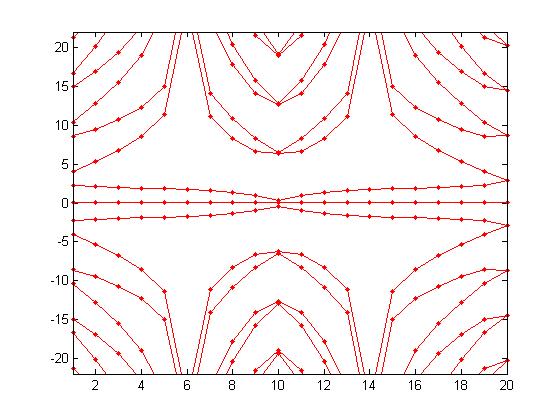}
\includegraphics[scale=.4]{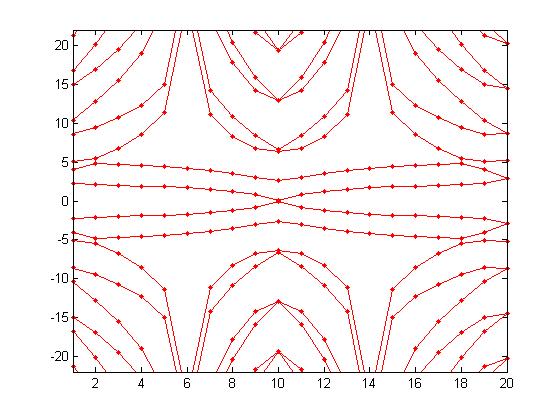}
\includegraphics[scale=.4]{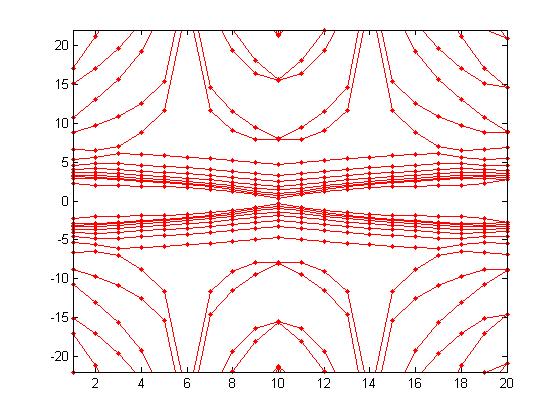}
\caption{Entanglement spectrum of Dirac model with $M=-0.3$ and $L=2^6$ sites with the following layers traced out: none, 7 only, 5 to 7, 3 to 7 from left to right. As compared to the ES with only a real space cut in Appendix \ref{wannES}, we see the appearance of additional low energy states due to the layer cut. Note that the higher entanglement eigenenergies are not affected by the cut, and must thus also correspond to states in the UV. }
\label{fig:layerent}
\end{figure}

\section{Entanglement entropy of critical (1+1) dimensional systems}
\label{criticalEE}
This appendix contains a derivation of the entanglement entropy (EE) of critical systems, which scales like the bulk geodesic distance in the bulk geometry of a critical system from part II. This is consistent with expectations from the Ryu-Takanayagi formula. The derivation will be largely based on the conformal field theory (CFT) approach from Ref.~\cite{calabrese2009}. 

We can calculate the EE via the ``replica'' trick. Given a (1+1)-dim system spatially partitioned into regions A and B, we have 
\begin{equation}
S_A=-Tr(\rho_A \log  \rho_A)=-\lim_{n\rightarrow 1}\partial_n Tr \rho_A^n 
\end{equation}
with \begin{equation}Tr \rho_A^n=e^{(1-n)S_R(n)}\end{equation} where $S_R(n)$ is the Renyi entropy. $Tr\rho_A^n$ can be expanded as a path integral over real space $x$ and imaginary time $\tau$. The $n$ copies of $\rho_A$ corresponds to $n$ complex sheets. When we perform the partial trace over $A$, we are equivalently joining up the sheets such that the fields on the $j^{th}$ and and $(j+1)^{th}$ sheets obey $\phi_j(x,\tau=\beta^-)=\phi_{j+1}(x,\tau=0^+)$ for $x\in A$ only. This looks topologically like $n$ sheets successively connected to each other via branch cuts $(x,\tau=0)$ with $x\in A$. Call this n-sheeted Riemann surface $R_n$. Further details of this construction can be found in \cite{calabrese2009} and many other papers. The above construction can be summarized by writing 
\begin{equation} 
Tr\rho^n_A|_{R_n}=\langle \Phi(x_1)\bar \Phi(x_2)\rangle_C
\label{rhoA}
\end{equation}
where $C$ is the complex plane and $\langle \Phi(x_1)\bar \Phi(x_2) \rangle_C$ is taken over one of the $n$ sheets with $\Phi$, $\bar \Phi$ being branch-point twist fields that implement cyclic permutations $\Phi:\phi_i\rightarrow \phi_{i+1}$ and $\bar \Phi:\phi_i\rightarrow \phi_{i-1}$. All the $i$'s are taken mod $n$.

Eq. \ref{rhoA} can be evaluated by computing the stress-energy tensor $T$ on $R_n$ in two different ways. One way is a path integral approach with twist fields, like in the above, while another way is by direct conformal transformation from the complex plane. With the path integral approach,
\begin{equation} T|_{R_n}=\frac{\langle \Phi(x_1)\bar \Phi(x_2) T_j\rangle_C }{\langle \Phi(x_1)\bar \Phi(x_2)\rangle_C}=\frac{\langle \Phi(x_1)\bar \Phi(x_2) T_j\rangle_C }{Tr\rho^n_A|_{R_n}}
\label{Trn}
\end{equation}
where $j$ represents one of the $n$ copies of the complex sheets.

To evaluate the stress-energy tensor $T$ by conformal transformation, one considers $T(z)$, the holomorphic part of the $T$ with $z\in C$. We know that $T(z)=0$ due to global conformal invariance. Now, let's map $C$ to the n-sheeted surface parametrized by $w=x+i\tau \in R_n$ via $z=(\frac{w-x_1}{w-x_2})^{1/n}$. In other words, we implement branch cuts from $0$ to $\infty$ on the Riemann surface $z^n$, and these branch cuts correspond to our original region A from $w=x_1$ to $w=x_2$. Through standard manipulations with the Schwarzian derivative,%A direct calculation of $T_j$, the stress-energy tensor on the $j^{th}$ sheet gives
\begin{eqnarray}
T(w)|_{R_n}&=&T(z)(\frac{dz}{dw})^2 + \frac{c}{12}\{z,w\}\notag\\
&=& \frac{c}{12}\{z,w\}\notag\\
&=& \frac{c}{12}\frac{z'''z'-\frac{3}{2}z''^2}{z'^2}\notag\\
&=&\frac{c(n-1/n)}{24n}\frac{(x_1-x_2)^2}{(w-x_1)^2(w-x_2)^2}
%&=&  \frac{\langle \Phi(x_1)\bar \Phi(x_2) T_j(w)\rangle_C }{\langle \Phi(x_1)\bar \Phi(x_2)\rangle_C}
\end{eqnarray}
where $c$ is the central charge. This can be compared with Eq. \ref{Trn} via the conformal Ward identity involving the stress-energy tensor:
\begin{eqnarray}
\langle \Phi(x_1)\bar \Phi(x_2) T_j(w)\rangle_C &=& n\left(\frac{\partial_{x_1}}{w-x_1}+\frac{h_\Phi}{(w-x_1)^2}+\frac{\partial_{x_2}}{w-x_2}+\frac{h_{\bar \Phi}}{(w-x_2)^2}\right)\langle \Phi(x_1)\bar \Phi(x_2)\rangle_C \notag\\
&=& n\left(\frac{\partial_{x_1}}{w-x_1}+\frac{h_\Phi}{(w-x_1)^2}+\frac{\partial_{x_2}}{w-x_2}+\frac{h_{\bar \Phi}}{(w-x_2)^2}\right)|x_1-x_2|^ {-2n\Delta}\notag\\
\label{Trn2}
\end{eqnarray}
where $\Delta=h_\Phi=h_{\bar\Phi}$ is the scaling dimension. The factor of $n$ arises from summing over $n$ identical copies of $T$. Comparing the two equations \ref{Trn} and \ref{Trn2}, we obtain
\begin{equation}\Delta=\frac{c}{12n}\left(n-\frac{1}{n}\right)
\end{equation}
which implies that
\begin{eqnarray} Tr\rho^n_A|_{R_n}&=&\langle \Phi(x_1)\bar \Phi(x_2)\rangle_C\notag\\
&\propto&|x_2-x_1|^{-2n\Delta}\notag\\
&\sim &l^{-\frac{c}{6}\left(n-\frac{1}{n}\right)}
\end{eqnarray}
From this, 
\begin{equation} S_A=-\lim_{n\rightarrow 1}\partial_n Tr \rho_A^n =\frac{c}{3}\log l +Y \end{equation}

where $Y$ is a non-universal constant. This result holds for a subregion A with length $l$ embedded in the (infinite) real line.

\subsection{Results for nontrivial boundary conditions}

When boundary conditions are imposed on the critical system, the entanglement entropy can still be computed straightforwardly by mapping onto the new geometry through a conformal map. Under the map $w\rightarrow v$, $Tr\rho^n_A=\langle \Phi(x_1)\bar \Phi(x_2)\rangle_C$ is transformed into
\begin{equation} \langle \Phi(v_1,\bar v_1)\bar \Phi(v_2,\bar v_2)\rangle =|w'_1(v_1)w'_2(v_2)|^{n\Delta} \langle \Phi(w_1,\bar w_1)\bar \Phi(w_2,\bar w_2)\rangle\end{equation}
since the twist fields $\Phi$ and $\bar \Phi$ have scaling dimension $n\Delta$. As a simple illustration, consider the case where the system has a finite periodicity of $L$. We can perform the mapping $v=\frac{L}{2\pi i}\log  w$ so that each sheet in the $w$ plane becomes an infinite cylinder with circumference $L$. This leads to the transformation
\begin{eqnarray}
Tr \rho^n_A&\rightarrow& |w'_1(v_1)w'_2(v_2)|^{n\Delta} \langle \Phi(w_1,\bar w_1)\bar \Phi(w_2,\bar w_2)\rangle\notag\\
&=& \left(\frac{2\pi}{L}\right)^{2n\Delta}\langle \Phi(w_1,\bar w_1)\bar \Phi(w_2,\bar w_2)\rangle\notag\\
&=& \left(\frac{2\pi}{L}\right)^{2n\Delta}|w_1-w_2|^{-2n\Delta}\notag\\
&=& \left(\frac{2\pi}{L}\right)^{2n\Delta}|e^{2\pi i v_1/L}-e^{2\pi i v_2/L}|^{-2n\Delta}\notag\\
&=& \left(\frac{2\pi}{L}\right)^{2n\Delta}|2 sin \left(\frac{\pi (v_2-v_1)}{L}\right)|^{-2n\Delta}\notag\\
&=& \left(\frac{L}{\pi} sin \left(\frac{\pi l}{L}\right)\right)^{-2n\Delta}
\end{eqnarray}
so that 
\begin{equation}
S_A=\frac{c}{3} \log  \left(\frac{L}{\pi}sin\frac{\pi l}{L}\right)+ Y' 
\label{EEperiodic}
\end{equation}
for a subregion A of length $l$ in a finite system with periodicity $L$.% If open boundary conditions are used instead, we can simply reason tha
%\[S_A=\frac{c}{6} \log  \left(\frac{L}{\pi}sin\frac{\pi l}{L}\right)+ Y'/2 \]

%because there is now one instead of two interfaces between the two subregions. This can be rigorously shown by doing another conformal transformation, as mentioned in \cite{calabrese}.

\section{Entanglement spectrum of a critical 1D system}
\label{sec:criticalspacing}

In Appendix. \ref{wannES}, the entanglement energies of a gapped free fermion system was shown to be equally spaced. Here, we shall complement that result by showing that the entanglement energies of a gapless (critical) free fermion system is also equally spaced, albeit with a spacing that decreases with system size. The derivation will be largely based on ~\cite{calabrese2008}.

First, let's label the eigenvalues of the reduced density matrix $\rho$ as $\lambda_1=e^{-t_1},\lambda_2=e^{-t_2}, \dots$. For now we make no assumption about its distribution, other than the usual constraint $0<\lambda_i <1$ so that $0<t_i<\infty$ for all $i$. Then define the distribution function
\begin{equation}
P(t)=\sum_i \delta(t-t_i) =\sum_i \lambda\delta(\lambda-\lambda_i)
\end{equation}
The entanglement spectrum will be uniformly spaced if we can show that $P(t)$ is equal to the multiplicity of the eigenvalues times a constant, something that was not explicitly shown in \cite{calabrese2008}. To proceed, we first bring $P(t)$ to a more palatable form involving $f(\lambda-i\epsilon)$, a quantity analogous to a Green's function:
\begin{eqnarray}
P(t)&=&\lambda \sum_i \delta(\lambda-\lambda_i)\notag\\
&=&\int d\lambda'\delta(\lambda'-\lambda) \sum_i \delta(\lambda-\lambda_i)\notag\\
&=&\lim_{\epsilon\rightarrow 0+}Im\int \frac{d\lambda'\lambda'\sum_i \delta(\lambda'-\lambda_i)}{\lambda-i\epsilon-\lambda'}\notag\\
&=&\lim_{\epsilon\rightarrow 0+}Im\frac{1}{\pi}\sum_i \frac{\lambda_i}{(\lambda-i\epsilon)-\lambda_i}\notag\\
&=&\lim_{\epsilon\rightarrow 0+}Im[f(\lambda-i\epsilon)]
\end{eqnarray}
where
\begin{eqnarray}
f(z)&=& \frac{1}{\pi}\sum_i \frac{\lambda_i}{z-\lambda_i}\notag\\
%&=& \frac{1}{\pi}\sum_i \frac{\lambda_i}{z(1-\frac{\lambda_i}{z})}\\
&=& \frac{1}{\pi}\sum_{n=1}^{\infty} \sum_i\lambda_i^n z^{-n}\notag\\
&=& \frac{1}{\pi}\sum_{n=1}^{\infty} (Tr \rho_A^n) z^{-n}\notag\\
&\approx & \frac{1}{\pi}\sum_{n=1}^{\infty} (L^{-\frac{c}{6}(n-1/n)}) z^{-n}\notag\\
&=& \frac{1}{\pi}\sum_{n=1}^{\infty} (e^{-\frac{c}{6}(\log  L)(n-1/n)}) z^{-n}\notag\\
&=& \frac{1}{\pi}\sum_{n=1}^{\infty} \sum_{k=0}^{\infty} \frac{(c/6 (\log  L))^k}{k!}\frac{(L^{-c/6}/z)^n}{n^k}\notag\\
&=& \frac{1}{\pi}\sum_{k=0}^{\infty} \frac{(c/6( \log  L))^k}{k!}Li_k(L^{-c/6}/z)
\end{eqnarray}
The fourth line was obtained from a CFT result from the previous appendix. Since $\lim_{\epsilon\rightarrow 0+}Im[Li_k(p+i\epsilon)]=\pi \frac{(\log  p)^{k-1}}{\Gamma(k)}$ where $\Gamma(k)=(k-1)!$,
\begin{eqnarray}
P(t)=P(-\log  \lambda)&=&\lim_{\epsilon\rightarrow 0+}Im[f(\lambda-i\epsilon)]\notag\\
&=&\lim_{\epsilon\rightarrow 0+}\frac{1}{\pi}\sum_{k=0}^{\infty} \frac{(c/6( \log  L))^k}{k!}Im Li_k(L^{-c/6}/\lambda + i\epsilon)\notag\\
&=&\lim_{\epsilon\rightarrow 0+}\sum_{k=1}^{\infty} \frac{(c/6( \log  L))^k}{k!}\frac{(\log  (L^{-c/6}/\lambda))^{k-1}}{(k-1)!}+\frac{1}{\pi}\sum_{n=1}^{\infty}(L^{c/6}/(\lambda-i\epsilon))^{-n} \notag\\
&=&\frac{\theta(L^{-c/6}-e^{-t})}{\left(\frac{t}{c/6(\log  L)}-1\right)}I_1\left(2\sqrt{(t-\frac{c}{6}\log  L)\frac{c}{6}\log  L}\right)+ e^{-t}\delta(L^{-c/6}-e^{-t})\notag\\
\label{Pt}
\end{eqnarray}
where $\theta(L^{-c/6}-e^{-t})$ is the Heaviside function and $I_1$ is the first modified Bessel function of the first kind. The delta function arises from the $k=0$ term in the k-summation. We see that the largest eigenvalue is $\lambda_{max}=L^{-c/6}$. Now, let's show that $P(t)$ is asymptotically a constant times the multiplicity of the eigenvalues. 

In Section \ref{entformula}, we have seen how the eigenvalues $\lambda$ of the RDM $\rho$ are related to $c_j$, the eigenvalues of the single-particle correlator:
\begin{equation} \lambda=\prod_{a\in A}c_a \prod_{b\in B}c_b \end{equation}
where each $c_j=c_a$ or $c_b$ can either correspond to an occupied or unoccupied state. To proceed, we shall \emph{assume} for now that the entanglement spectrum $\epsilon_j$ is indeed asymptotically equally spaced with spacing $k$, and check if that is true. With this assumption the overwhelming majority of the $c_j$'s are approximately
\begin{equation}
c_j=\frac1{1+e^{\epsilon_j}}=\frac1{1+e^{kj}}\approx e^{-kj}
\end{equation}
Then a typical RDM eigenvalue will look like 
\begin{equation}
\lambda \sim e^{-(j_1+...j_p)k}\prod^{l_A-p}_q (1-e^{-j_q k})\approx e^{-(j_1+...j_p)k},\end{equation}
where the $j$'s are integers. Hence the multiplicity of a $\lambda = e^{-t}$ should also be asymptotically $Q(n=t/k)$, where $Q(n)$ is the number of restricted partitions of integer $n$, i.e. number of unordered ways to express $n$ as a sum of distinct integers. Asymptotically, $\log  Q(n)\sim \pi \sqrt{\frac{n}{3}}-\frac{3}{4}\log  n$. 

Hence we just need to take the ratio $\log (P(t)/Q(t/k))$ and check if $\log (P(t)/Q(t/k))\rightarrow \log (O(1))$ asymptotically for a certain $k$. It is well-known that $I_1(x)\sim \frac{e^x}{\sqrt{2\pi x}}$ asymptotically. Letting $x= 2\sqrt{(t-\frac{c}{6}\log  L)\frac{c}{6}\log  L}$ in Eq. \ref{Pt}, we find that 
\begin{eqnarray}
&&\log(P(t=kn)/Q(n))\notag\\
&\sim& \frac{1}{2}\log\left(\frac{\sqrt{\frac{c}{6}\log L}}{4\pi}\right)-\frac{3}{4}\log\left(k-\frac{\frac{c}{6} \log L}{n}\right)+2\sqrt{\left(kn-\frac{c}{6}\log L\right)\frac{c}{6}\log L}-\pi \sqrt{\frac{n}{3}}\notag\\
\label{expapprox}
\end{eqnarray}
If we set the RHS to $\log(O(1))\sim 0$ and take the large $n$ and $L$ limit, the logarithmic terms can be dropped and we can solve for $k$. Indeed, 
\begin{equation}
c_j \sim L^{-c/6}e^{-\frac{\pi^2}{2\log L}j} = \lambda_{max}e^{-\frac{\pi^2}{2\log L}j}
\end{equation}
i.e. the entanglement spectrum has an asymptotic spacing of $k=\frac{\pi^2}{2\log L}$. Note that this is only true in the large $n$ and $L$ limit, and $c_0 \neq \lambda_{max}$, no matter how suggestive that may look. %Due to the logarithmic dependence of $L$, we need $L >10^5$ to justify dropping the log terms. 

\begin{figure}[H]
\includegraphics[scale=0.8]{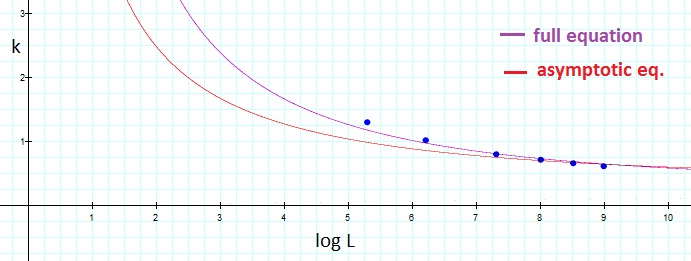}
\caption{The exponential separation $k$ (y-axis) against $\log  L$ (x-axis) for a free fermion system. The red curve represents the asymptotic result $k=\frac{\pi^2}{2\log  L}$ while the purple curve represents the equation \eqref{expapprox}. The numerical results (blue dots) agree well with Eq. \ref{expapprox} for $\log L>5$, and with the asymptotic result for $\log L>7$. }\end{figure}

\section{Analytic properties of the Chern number density}
\label{Canalytic}

The Chern number density $C(n)$ is given by the integral in  Eq. \ref{Cn}. As shown numerically in Fig. \ref{numerical}, $C(n)$ peaks at some $n$ in the IR, with rapidly decaying tails on either side. Here, we shall attempt to find accurate analytic approximations for its tails. Such approximations will be universal, unlike those of the UV peak which we will hence not analyze.

%\subsection{Approximating analytic expression for $C(n)$ in the IR}
Near the $\Gamma$ point, the integrand can be approximated by replacing the central rectangular 'peak' of $Y(k_x,k_y,n)$ with a circular peak in polar coordinates: $Y(k_x,k_y,n)\rightarrow Y(r,r,n)$ where $r=\sqrt{k_x^2+k_y^2}$. To further simplify our calculations, we shall further approximate its radial profile by truncating terms of $4^{th}$ order or higher in $r$. This is sufficient for capturing the essential \emph{quantitative} behavior of the tails of $C(n)$, as we will justify a posteriori. Some algebra leads us to 
\begin{equation}
C(n)=\frac{1}{4\pi}\int d\theta \int r dr \left(1-\frac{ 4^n}{6}r^2\right)4^{n-1}\alpha r^2 F_{xy}(r,\theta)
\label{Cnpolar}
\end{equation}
where $\alpha$ corrects for the error introduced by the different shapes of the original (rectangular) and approximated (circular) peak of $Y(k_x,k_y,n)$.  Due to scale invariance, $\alpha$ takes approximately the same value for $n$ on the same side of the IR peak of $C(n)$. However, it takes different values on different sides of the peak: $\alpha_{L}\approx 0.73$ and $\alpha_{R}\approx 2$ for $n$ on the UV and IR side of the $C(n)$ peak respectively. 

%For generic wavelet bases, the $4^n$ factor in Eq. \ref{Cnpolar} will be replaced by $4^{\kappa n}$, where $\kappa$ is the order of $D(z)$ at $z=1$, while the $\frac{1}{6}$ factor will be replaced by its subleading order cocefficient.

\subsection{Dirac model}
\begin{figure}[H]
\center
\includegraphics[scale=1.6]{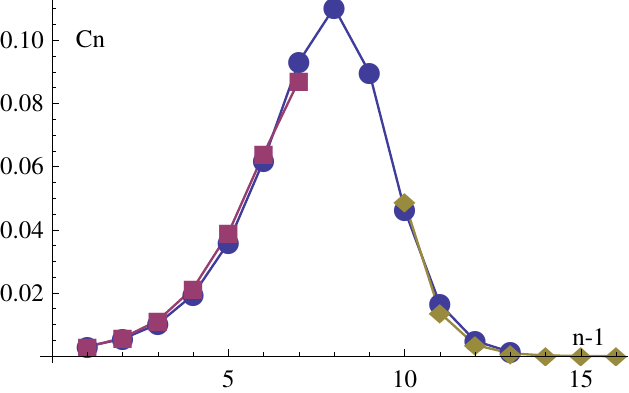}
\caption{Plot of the exact numerically computed $C(n)$ (blue) for the $m=0.01$ Dirac point, vs. its analytical approximation from Eqs. \ref{cnleft} and \ref{cnright} (red and yellow for the left and right tails respectively). We see excellent agreement in both regimes.% b) Analytical plots of $C_n$ from Eq. \ref{Cndirac} for $M=10^{-1}$ to $10^{-4}$ (left to Right). }
}
\label{fig:Cndirac}
\end{figure}

For the Dirac model with $d$-vector $\vec d = v_F(\sin k_x, \sin k_y, 2+m-\cos k_x - \cos k_y)$, the Berry curvature is given by
\begin{eqnarray}
F_{xy}&=& \frac{d\cdot (\partial_{k_x}d \times \partial_{k_y}d)}{|d|^3}\notag\\
&=& \frac{-\cos k_x-\cos k_y +(2+m) \cos k_x \cos k_y }{(\sin^2 k_x + \sin^2 k_y +(2-m -\cos k_x -\cos k_y )^2)^{3/2}}\notag\\
&\approx & \frac{-m }{(m^2+k_x^2+k_y^2)^{3/2}}\notag\\
&= & \frac{-m }{(m^2+r^2)^{3/2}}
\label{Fxydirac}
\end{eqnarray}
The 2nd line is exact, while the 3rd line is very accurate near the Dirac point at $k=(0,0)$. Substituting Eq. \ref{Fxydirac} into Eq. \ref{Cnpolar}, we obtain with more algebra
\begin{equation}
C(n)=\alpha(\sqrt{\mu(1+\mu)}-\mu(4\mu+3))
\label{Cndirac}
\end{equation}
where $\mu=\frac{m^24^n}{6v_F^2}$ and $\alpha=\alpha_{L}=0.73$ for $\mu<1$, and  $\alpha=\alpha_{R}=2$ for $\mu>1$. Note that $C(n)$ depends only on a \emph{single} parameter $\mu$, consistent with expectations from scale-invariance. From Fig. \ref{fig:Cndirac}, we see that Eq. \ref{Cndirac} agrees excellently with exact numerical results, ad hoc as its assumptions seem. From now, we shall set $v_F=1$ for simplicity.

To the UV (Left) side of the peak, we have from Eq. \ref{Cndirac} 
\begin{equation}
C(n)\sim \alpha(\sqrt{\mu}-3\mu)\approx 0.3\times 2^n m,
\label{cnleft} 
\end{equation}
while on the IR (Right) side, 
\begin{equation}
C(n)\sim \frac{\alpha}{8\mu}\approx \frac{1.5}{m^2 4^n} 
\label{cnright}
\end{equation}
with the $4^{-n}$ decay expected from the shrinking of the wavelet basis beyond the scale of the Dirac peak. The peak position is approximately located where these two curves intersect: 
\begin{eqnarray}
n_{peak}\approx \log_2 \frac{\sqrt[3]{5}}{m}
\label{peak}
\end{eqnarray}

In the case of a more general model, the location of the IR peak will be linearly related to $-\log m$, although the coefficients from $\alpha$ will have to be modified.
%As mentioned, for a different wavelet basis, we will have $\mu\propto M^2 4^{\kappa n}$, with a different proportionality constant. But behavior from Eq. \ref{Cndirac} will remain essentially the same, being characteristic of the Dirac point. 
The key conclusions from these calculations are that 
\begin{itemize}
\item The IR peak, which corresponds to the Dirac cone in the boundary Dirac Hamiltonian, is localized (exponentially decaying) in the emergent direction indexed by $n$.
\item The shape of the IR peak is remains invariant upon tuning of $m$, but its location is shifted as $\sim -\log_2m$.
\end{itemize}

\subsection{D-wave model}

\begin{figure}[H]
\includegraphics[scale=1.5]{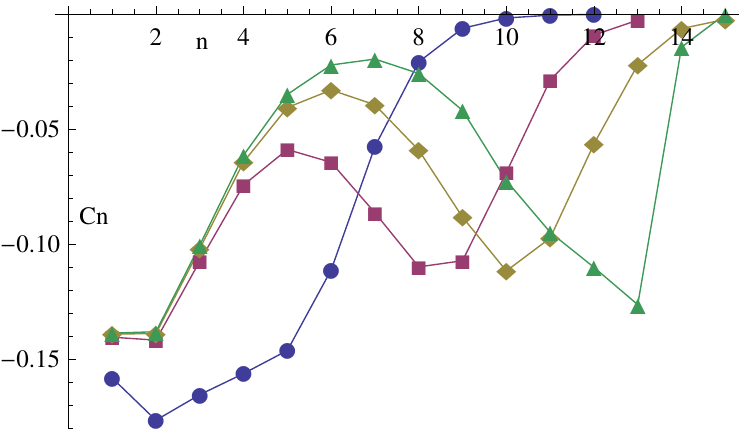}
\caption{a) Plot of the exact numerically computed $C_n$ for a d-wave point for $2-M=10^{-1},10^{-2},10^{-2.5},10^{-3}$ (Left to Right). We see that although the peak shifts right as $M\rightarrow 2^{-}$, the UV contributions never disappear, unlike the case with a Dirac point.  }
\label{fig:Cndwave}
\end{figure}

To illustrate a qualitatively different example from the Dirac model, let's consider the d-wave model $d=v_F(\sin k_x-\sin k_y, \sin k_x + \sin k_y , 2 \cos k_x \cos k_y -M)$ with a $d$-vector winding of $2$ around $\Gamma$. Its Berry curvature is 
\begin{eqnarray}
F_{xy}&=& \frac{4\cos^2 k_y\sin^2 k_x + 4\cos^2 k_x\sin^2 k_y+4\cos^2 k_x\cos^2 k_y}{(2(\sin^2 k_x + \sin^2 k_y) +(M -2\cos k_x \cos k_y )^2)^{3/2}}
\label{Fxydwave}
\end{eqnarray}
From this, one obtains the Chern number density $C(n)$ plotted in Fig. \ref{fig:Cndwave}. It exhibits similar qualitative behavior as the $C=1$ Dirac case, except that each of the two peaks carry a Chern number of $1$ instead.
%The major difference here is that the \textbf{UV contribution of a single D-wave point does not vanish} as the gap vanishes with $M\rightarrow 2$. This is unlike the Dirac-point, whose UV weight disappears as the gap closes (of course, it won't disappear if there is another dirac point elsewhere). A detailed inspection reveals that $F_{xy}>0$ for the D-wave point (unlike the Dirac point), prohibiting the cancellation of the UV contributions as the gap closes. This is possibly due to the different $r^C e^{iC\theta}$ behavior of the d-vector near the IR Dirac point. 

%\nocite{*}
    %\bibliographystyle{plain}
    \bibliographystyle{unsrt}

%\bibliography{suthesis}

%\bibliography{fci,localself,paper,TI,pseudopotentials,entanglement,ehm}

%\begin{thebibliography}{1}
%\begin{thebibliography}{10}

\bibliography{thesis,paper}
%\bibitem{vonklitzing} K. von Klitzing. Rev. Mod. Phys., 58:519, 1986. 1

%\end{thebibliography}

%\onlinesignature

\end{document}